\documentclass[a4paper,fleqn,usenatbib]{mnras}

\usepackage{newtxtext,newtxmath}

\usepackage[T1]{fontenc}
\usepackage{ae,aecompl}

\usepackage{graphicx}	
\usepackage{amsmath}	
\usepackage{amssymb}	
\usepackage{subfigure}
\usepackage{float}
\usepackage{natbib} 
\usepackage{hyperref}
\usepackage{tabularx}
\usepackage{supertabular}
\usepackage{multicol}        
\usepackage{multirow}
\usepackage{lscape}
\usepackage{longtable}
\usepackage{appendix}
\usepackage{epstopdf}
\usepackage{comment}
\usepackage[T1]{fontenc}
 \usepackage{footnote}

\def\blue#1   {{\textcolor{blue}{#1}}\,}
\def\red#1    {\textcolor{red}{#1}\,}
\def\new#1    {{\bf #1 }\,}
\def\cut#1    {\sout{#1}\,}

\def\cm3{cm$^{-3}$\,}

\title[Opacities of dense gas tracers]{Opacities of Dense Gas Tracers In Galactic Massive Star Forming Regions}

\author[S. Liu et al.]{
Shu Liu,$^{1}$\thanks{E-mail: liushu@nao.cas.cn}
Junzhi Wang,$^{2}$\thanks{E-mail: junzhiwang@gxu.edu.cn}
Fei Li,$^{3,4}$
Jingwen Wu,$^{5,1}$
Zhi-Yu Zhang,$^{3,4}$
Di Li,$^{1,6,7}$
\newauthor{
Ningyu Tang,$^{8}$
and Pei Zuo$^{1,9}$}
\\
$^{1}$National Astronomical Observatories, Chinese Academy of Sciences, Beijing 100101, People's Republic of China\\
$^{2}$Guangxi Key Laboratory for Relativistic Astrophysics, Department of Physics, Guangxi University, Nanning 530004, People's Republic of China\\
$^{3}$School of Astronomy and Space Science, Nanjing University, Nanjing 210023, People's Republic of China\\
$^{4}$Key Laboratory of Modern Astronomy and Astrophysics (Nanjing University), Ministry of Education, Nanjing 210023, People's Republic of China\\
$^{5}$University of Chinese Academy of Sciences, Beijing 100049, People's Republic of China\\
$^{6}$Key Laboratory of Radio Astronomy and Technology, Chinese Academy of Sciences, Beijing 100101, People's Republic of China\\\
$^{7}$Research Center for Intelligent Computing Platforms, Zhejiang Laboratory, Hangzhou 311100, People's Republic of China\\\
$^{8}$Department of Physics, Anhui Normal University, Wuhu, Anhui 241002, People's Republic of China\\
$^{9}$Kavli Institute for Astronomy and Astrophysics, Peking University, Beijing 100871, People's Republic of China\\
}

\date{Accepted 2023 August 08. Received 2023 July 13; in original form 2022 June 09}

\pubyear{2023}

\begin{document}
\label{firstpage}
\pagerange{\pageref{firstpage}--\pageref{lastpage}}
\maketitle

\begin{abstract}
Optical depths of dense molecular gas are commonly used in Galactic and extragalactic studies to constrain the dense gas mass of the clouds or galaxies. 
The optical depths are often obtained based on 
spatially unresolved data, especially in galaxies, which may affect the reliability of such measurements. We examine such effects in spatially resolved Galactic massive star forming regions. Using the 10-m SMT telescope, we mapped HCN and H$^{13}$CN 3-2, HCO$^+$ and H$^{13}$CO$^+$ 3-2 toward 51 Galactic massive star forming regions, 30 of which resulted in robust determination of  spatially-resolved optical depths. Conspicuous spatial variations of optical depths have been detected within each source.  
We first obtained opacities for each position and calculated an optical-thick line intensity-weighted average,
then averaged all the spectra and derived a single opacity for each region.
The two were found to agree extremely well, with a linear least square correlation coefficient of  0.997 for the whole sample.

\end{abstract}




\section{Introduction}


Dense molecular gas is a key to understand star formation in galaxies.  Observations have demonstrated that stars,
especially massive stars, are essentially and exclusively formed in the dense
cores of giant molecular cores \citep[GMCs,][]{2008ASPC..390...52E}. Low-$J$ CO lines trace
the total amount of molecular gas content, not sensitive to the dense cores with volume densities higher than 10$^4$ \cm3.
The transitions of molecules with large dipole moment, such as HCN, HCO$^+$, HNC, and CS, which have high  critical density $n_{\rm crit} > 10^4$ \cm3,  are
tracers  of dense gas. With
observations of HCN 1-0 toward 65 galaxies, \citet{2004ApJS..152...63G} 
found a strong linear correlation between the luminosities of HCN
1-0 and Infrared emission. This correlation was found to extend to
Galactic dense cores \citep[e.g.][]{2005ApJ...635L.173W}, and possibly to high-$z$ galaxies and
QSOs as well \citep[e.g.][]{2007ApJ...660L..93G}. Further observations of CS $J$=5-4
in 24 IR-bright galaxies show such linear correlation still valid for the gas
as dense as $n_{\rm H_2} \sim 10^6$ \cm3 \citep{2011MNRAS.416L..21W}, which was  supported by 
HCN 4-3 and CS 7-6 survey toward 20 nearby star-forming galaxies  \citep{2014ApJ...784L..31Z}.  
Multiple line single pointing observations of HCN 1-0, HCO$^+$ 1-0, HNC 1-0, and CS 3-2 toward a sample of 70 galaxies also showed  similar linear relationships \citep{2021MNRAS.503.4508L}. 
Spatially resolved observations for local galaxies were also performed to study dense gas fraction and related star formation, such as,  HCN 1-0, HCO$^+$ 1-0, CS 2-1, $^{13}$CO 1-0, and C$^{18}$O toward  inner region of four local galaxies with  ALMA and IRAM 30 m \citep{2018ApJ...858...90G}, HCN 1-0, HCO$^+$ 1-0, HNC 1-0, and CO isotopologues toward a number of nearby galaxies in the IRAM large program EMPIRE \citep{2018MNRAS.475.3909C, 2017ApJ...836L..29J, 2019ApJ...880..127J},  HCN 1-0 and CO 1-0 toward  M51 with the 50 meter Large millimeter telescope (LMT) \citep{2022ApJ...930..170H}, and CO isotopologues investigation within the CLAWS programme \citep{2022A&A...662A..89D}.

However, because of the large dipole moments and high column density, such
dense gas tracers are normally optically thick both in Galactic GMC cores and in
galaxies. It is of any case hard to convert from luminosity of these
tracers to dense gas mass, which  is similar to the issue of the standard
conversion factor of CO with several times or even more than 10
times uncertainty in different galaxies \citep[e.g.,][]{2012MNRAS.421.3127N, 2012ApJ...751...10P}. 
Multiple transitions of molecular lines are powerful to derive the physical
properties (volume density, temperature, etc.) of dense gas in galaxies, which 
had been  made for nearby starbursts and ULIRGs (e.g., M~82, NGC~253 in \citealt{1992ApJ...399..521N}, 
ARP~220, NGC~6240 in \citealt{2009ApJ...692.1432G}), with large uncertainties. 

Optically thin dense gas tracers, such as isotopic lines, are necessary for better
 understanding dense gas properties and chemical evolution in the Galaxy \citep{1990ApJ...357..477L, 1992A&ARv...4....1W, 1994ARA&A..32..191W, 1994LNP...439...72H, 2005ApJ...634.1126M}. 
In recent years, increasing studies of CO isotopologues have been carried out in other galaxies \citep{2010A&A...522A..62M, 2014A&A...565A...3H, 2015ApJ...801...63M, 2017MNRAS.466...49J, 2017ApJ...836L..29J, 2019ApJ...880..127J, 2018MNRAS.475.3909C, 2021MNRAS.504.3221D}. 
Assuming a reasonable isotopic abundance ratio, one  can 
 determine the optical depths of dense gas tracers, such as HCN and CS lines, with the intensity ratio of dense gas tracers and their isotopic lines \citep{2014ApJ...796...57W,2016MNRAS.455.3986W, 2020MNRAS.494.1095L}.   However, despite the uncertainty of isotopic abundance in different galaxies, there are still other problems for determining the optical depths of dense gas tracers. One important effect is that we can only assume one value of  optical depth  in one galaxy if we do not have spatial resolution, while optical depths  should vary at different regions, which had been seen in M 82 along major axis \citep{2022ApJ...933..139L}.  
 
 Thus, the derived optical depth of one dense gas tracer  in each galaxy with one pair of dense gas tracer and its isotopic line, is only an averaged value within the observed region.  Since it is a non-linear    relation between optical depth and line ratio,  the typical optical depth in one region with internal spatial distribution of different optical depth, may not be well constrained by line ratio of spatially integrated fluxes  for both dense gas tracer and  its isotopic line. The best way to study this effect is mapping a sample of massive star forming regions in the Milky Way with both dense gas tracers and their isotopic lines.   The detailed description for deriving optical depths with two ways will be presented in Section 3.
 
In this paper, the observations and data reduction are described in Section 2, while the methods of calculating optical depths in each sources are presented in Section 3. Then, the main results and discussions are given in Section4 and Section 5,  and a brief summary is presented in Section 6.

\section{Observations and data reduction}


The sample presented in this study is a subset of massive star forming regions with parallax distances from \citet{2014ApJ...783..130R} with strong ($>$0.5 K) H$^{13}$CN 2-1 emissions, which was detected by the Institut de Radioastronomie Millim\' etrique (IRAM) 30-m telescope in June and October 2016 \citep{2024inprep}, for the guarantee of strong H$^{13}$CN 3-2 emission.

The observations were carried out using the Arizona Radio Observatory (ARO) 10-m Submillimeter Telescope (SMT) on Mt. Graham, Arizona, during several observing runs in 2017 March to May, 2017 December, 2018 January, 2018 March, 2018 May and 2018 October to November. Four molecular lines were observed with 1.3 mm ALMA band 6 receiver. HCN 3-2 with rest frequency of 265.886431 GHz and HCO$^+$ 3-2 with rest frequency of 267.557633 GHz were tuned in the upper sideband (USB) simultaneously. The isotopologues, H$^{13}$CN 3-2 with rest frequency of 259.011787 GHz and H$^{13}$CO$^+$ 3-2 with rest  frequency of  260.255342 GHz were observed simultaneously also in the upper sideband (USB). The Forbes Filter Banks (FFB) backend was setup with 512 -MHz bandwidth for each line  and 1 MHz channel spacing, which corresponds to $\sim$ 1.15 km s$^{-1}$ at 260 GHz, with the spatial resolution of $\sim$27.8$''$. 

For each source, the on-the-fly (OTF) mode was used to cover 2$'\times2'$ regions for HCN and HCO$^+$ 3-2, and 1.5$'\times1.5'$ for H$^{13}$CN  and H$^{13}$CO$^+$ 3-2, respectively  (Table~\ref{tab:source}).  
The telescope time on each source for the tuning of  HCN and HCO$^+$ 3-2 is around 40 minutes for most of the samples except for 9 sources as 80 minutes. For the isotopologues, the time on the strong ones 
are about 1.5 hours or 3 hours while on the other moderately weak cores are about 4.5 hours, even up to 6 hours for two cores due to the weather conditions.
The off points were chosen as azimuth off  30$'$ away from the mapping centers. 
The observation information as on-source time, system temperatures ($T_{\rm sys}$) and rms of HCN and H$^{13}$CN 3-2 for each source are shown in Table~\ref{tab:obs}. The information of HCO$^+$ 3-2 and H$^{13}$CO$^+$ 3-2 are not exhibited, since what are comparable to those of HCN and H$^{13}$CN 3-2, respectively.

The OTF data of HCN and HCO$^+$ 3-2 for each source were re-gridded to the final maps with the step of 15$''$. The map of H$^{13}$CN and H$^{13}$CO$^+$ 3-2 are  gridded to match the center and each position of the spatially resolved HCN and HCO$^+$ 3-2 data, respectively.
The antenna temperature $T_A^*$ was converted to main beam brightness temperature $T_{mb}$ using $T_{mb} = T_A^* /{\eta}_b$ . The main beam efficiency ${\eta}_b$ is 0.77. 
Typical noise levels were 0.09 K for HCN and HCO$^+$  3-2, as well as 0.03 K for H$^{13}$CO$^+$ 3-2 at the frequency spacing of 1 MHz in the unit of $T_A^*$, respectively.

A total of 51 sources were mapped. However, only sources with enough spatially resolved data points to obtained reliable H$^{13}$CN and H$^{13}$CO$^+$ 3-2 signals,  are selected for final analysis.  The selected 30 sources with basic parameters are listed in Table~\ref{tab:source}. All of the data were reduced with the \texttt{CLASS} software package in GILDAS\footnote{http://www.iram.fr/IRAMFR/GILDAS}. For each line of each source, we first took a quick look at main regions with emission and  obtained an averaged spectrum within this region to determine the line velocity range. Such velocity ranges were used as ``mask'' with  ``set window'' in \texttt{CLASS} when  baseline subtractions were done with first order polynomial. Then we used ``print area'' in \texttt{CLASS} to obtain the velocity integrated fluxes for each pixel, with the same value used for baseline subtraction, which was fixed the spectra within the map for each line of each source.


\begin{table*}
    \begin{center}
      \caption{Sample of high mass star forming regions. }\label{tab:source}
      \begin{tabular}{lccccccccc}
      \\
    \hline
    \hline
source name    & Alias & RA(J2000)     & DEC(J2000)       & D    & D$_{\rm GC}$  & $v_{\rm LSR}$  \\
     &       &              &          &kpc  & kpc  & km s$^{-1}$  \\
     \hline
  G005.88-00.39  &                    & 18:00:30.31  & -24:04:04.50  &  3.0  &  5.3 &   9.3 \\
  G009.62+00.19  &                    & 18:06:14.66  & -20:31:31.70  &  5.2  &  3.3 &   4.8 \\
  G010.47+00.02  &                    & 18:08:38.23  & -19:51:50.30  &  8.5  &  1.6 &  65.7 \\
  G010.62-00.38  & W 31               & 18:10:28.55  & -19:55:48.60  &  5.0  &  3.6 &   4.3 \\
  G011.49-01.48  &                    & 18:16:22.13  & -19:41:27.20  &  1.2  &  7.1 &  10.5 \\
  G011.91-00.61  &                    & 18:13:58.12  & -18:54:20.30  &  3.4  &  5.1 &  36.1 \\
  G012.80-00.20  &                    & 18:14:14.23  & -17:55:40.50  &  2.9  &  5.5 &  36.3 \\
  G014.33-00.64  &                    & 18:18:54.67  & -16:47:50.30  &  1.1  &  7.2 &  22.6 \\
  G015.03-00.67  & M 17               & 18:20:24.81  & -16:11:35.30  &  2.0  &  6.4 &  19.7 \\
  G016.58-00.05  &                    & 18:21:09.08  & -14:31:48.80  &  3.6  &  5.0 &  60.0 \\
  G023.00-00.41  &                    & 18:34:40.20  & -09:00:37.00  &  4.6  &  4.5 &  78.4 \\
  G027.36-00.16  &                    & 18:41:51.06  & -05:01:43.40  &  8.0  &  3.9 &  93.1 \\
  G029.95-00.01  & W 43S              & 18:46:03.74  & -02:39:22.30  &  5.3  &  4.6 &  98.2 \\
  G035.02+00.34  &                    & 18:54:00.67  & +02:01:19.20  &  2.3  &  6.5 &  52.9 \\
  G035.19-00.74  &                    & 18:58:13.05  & +01:40:35.70  &  2.2  &  6.6 &  34.2 \\
  G037.43+01.51  &                    & 18:54:14.35  & +04:41:41.70  &  1.9  &  6.9 &  44.3 \\
  G043.16+00.01  & W 49N              & 19:10:13.41  & +09:06:12.80  & 11.1  &  7.6 &   5.4 \\
  G043.79-00.12  & OH 43.8$-$0.1      & 19:11:53.99  & +09:35:50.30  &  6.0  &  5.7 &  44.6 \\
  G049.48-00.36  & W 51 IRS2          & 19:23:39.82  & +14:31:05.00  &  5.1  &  6.3 &  60.7 \\
  G049.48-00.38  & W 51M              & 19:23:43.87  & +14:30:29.50  &  5.4  &  6.3 &  55.8 \\
  G069.54-00.97  & ON 1               & 20:10:09.07  & +31:31:36.00  &  2.5  &  7.8 &  12.2 \\
  G075.76+00.33  &                    & 20:21:41.09  & +37:25:29.30  &  3.5  &  8.2 &  -1.5 \\
  G078.12+03.63  & IRAS 20126+4104    & 20:14:26.07  & +41:13:32.70  &  1.6  &  8.1 &  -3.3 \\
  G081.87+00.78  & W 75N              & 20:38:36.43  & +42:37:34.80  &  1.3  &  8.2 &   9.8 \\
  G109.87+02.11  & Cep A              & 22:56:18.10  & +62:01:49.50  &  0.7  &  8.6 &  -9.8 \\
  G121.29+00.65  & L 1287             & 00:36:47.35  & +63:29:02.20  &  0.9  &  8.8 & -17.1 \\
  G123.06-06.30  & NGC 281            & 00:52:24.70  & +56:33:50.50  &  2.8  & 10.1 & -30.4 \\
  G133.94+01.06  & W 3OH              & 02:27:03.82  & +61:52:25.20  &  2.0  &  9.8 & -46.8 \\
  G188.94+00.88  & S 252              & 06:08:53.35  & +21:38:28.70  &  2.1  & 10.4 &   3.8 \\
  G232.62+00.99  &                    & 07:32:09.78  & -16:58:12.80  &  1.7  &  9.4 &  17.3 \\
           \hline
      \end{tabular}
  \end{center}
\end{table*}


\begin{table*}
    \begin{center}
         \footnotesize 
      \caption{Observation details.}\label{tab:obs}
      \begin{tabular}{lccccccc}
      \\
    \hline
    \hline
    &  \multicolumn{3}{c}{$\textrm{HCN}$} & $\ $ &\multicolumn{3}{c}{$\textrm{H}^{13}\textrm{CN}$}   \\
        \cline{2-4} \cline{6-8}
Source name    & On-source time & ${\langle T_{\rm sys}\rangle}^a$     & rms$^b$    &  & On-source time & ${\langle T_{\rm sys}\rangle}^a$  & rms$^b$   \\
     &  min     &     K         &    K      & & hr  & K & K \\
     \hline
G005.88-00.39  &   40     &      297   &    0.096   & &   1.5   &    368  &   0.066   \\  
G009.62+00.19  &   40     &      312   &    0.069   & &   1.5   &    339  &   0.061   \\  
G010.47+00.02  &   40     &      336   &    0.069   & &   3.0   &    370  &   0.047   \\  
G010.62-00.38  &   40     &      540   &    0.117   & &   4.5   &    308  &   0.025   \\  
G011.49-01.48  &   40     &      340   &    0.076   & &   3.0   &    239  &   0.023   \\  
G011.91-00.61  &   40     &      453   &    0.103   & &   3.0   &    313  &   0.033   \\  
G012.80-00.20  &   40     &      305   &    0.057   & &   1.5   &    277  &   0.049   \\  
G014.33-00.64  &   80     &      289   &    0.058   & &   1.5   &    305  &   0.047   \\  
G015.03-00.67  &   80     &      355   &    0.052   & &   1.5   &    221  &   0.026   \\  
G016.58-00.05  &   80     &      309   &    0.052   & &   6.0   &    272  &   0.023   \\  
G023.00-00.41  &   80     &      358   &    0.057   & &   3.0   &    258  &   0.031   \\  
G027.36-00.16  &   40     &      351   &    0.088   & &   4.5   &    266  &   0.020   \\  
G029.95-00.01  &   40     &      405   &    0.091   & &   1.5   &    223  &   0.026   \\  
G035.02+00.34  &   40     &      348   &    0.077   & &   3.0   &    238  &   0.027   \\  
G035.19-00.74  &   40     &      278   &    0.062   & &   3.0   &    306  &   0.036   \\  
G037.43+01.51  &   80     &      413   &    0.060   & &   1.5   &    236  &   0.030   \\  
G043.16+00.01  &   40     &      250   &    0.059   & &   1.5   &    238  &   0.031   \\  
G043.79-00.12  &   40     &      283   &    0.070   & &   4.5   &    243  &   0.030   \\  
G049.48-00.36  &   40     &      262   &    0.077   & &   3.0   &    240  &   0.035   \\  
G049.48-00.38  &   40     &      303   &    0.073   & &   3.0   &    270  &   0.040   \\  
G069.54-00.97  &   80     &      293   &    0.054   & &   3.0   &    201  &   0.031   \\  
G075.76+00.33  &   40     &      293   &    0.074   & &   3.0   &    234  &   0.048   \\  
G078.12+03.63  &   40     &      330   &    0.070   & &   4.5   &    200  &   0.022   \\  
G081.87+00.78  &   40     &      313   &    0.076   & &   3.0   &    206  &   0.033   \\  
G109.87+02.11  &   40     &      324   &    0.077   & &   3.0   &    213  &   0.025   \\  
G121.29+00.65  &   40     &      307   &    0.091   & &   4.5   &    209  &   0.023   \\  
G123.06-06.30  &   80     &      293   &    0.039   & &   4.5   &    238  &   0.021   \\  
G133.94+01.06  &   80     &      346   &    0.072   & &   6.0   &    204  &   0.024   \\  
G188.94+00.88  &   40     &      367   &    0.087   & &   3.0   &    203  &   0.027   \\  
G232.62+00.99  &   80     &      307   &    0.064   & &   4.5   &    238  &   0.031   \\
      \hline         
      \end{tabular}
\\ \begin{flushleft}
{\raggedright {\bf Notes:} $a.$ Averaged system temperature. $b.$ rms for all lines were obtained under the frequency resolution of 1 MHz.}
\end{flushleft}
  \end{center}     
\end{table*}


 \begin{figure*} 
    \centering
  \includegraphics[width=3.05in]{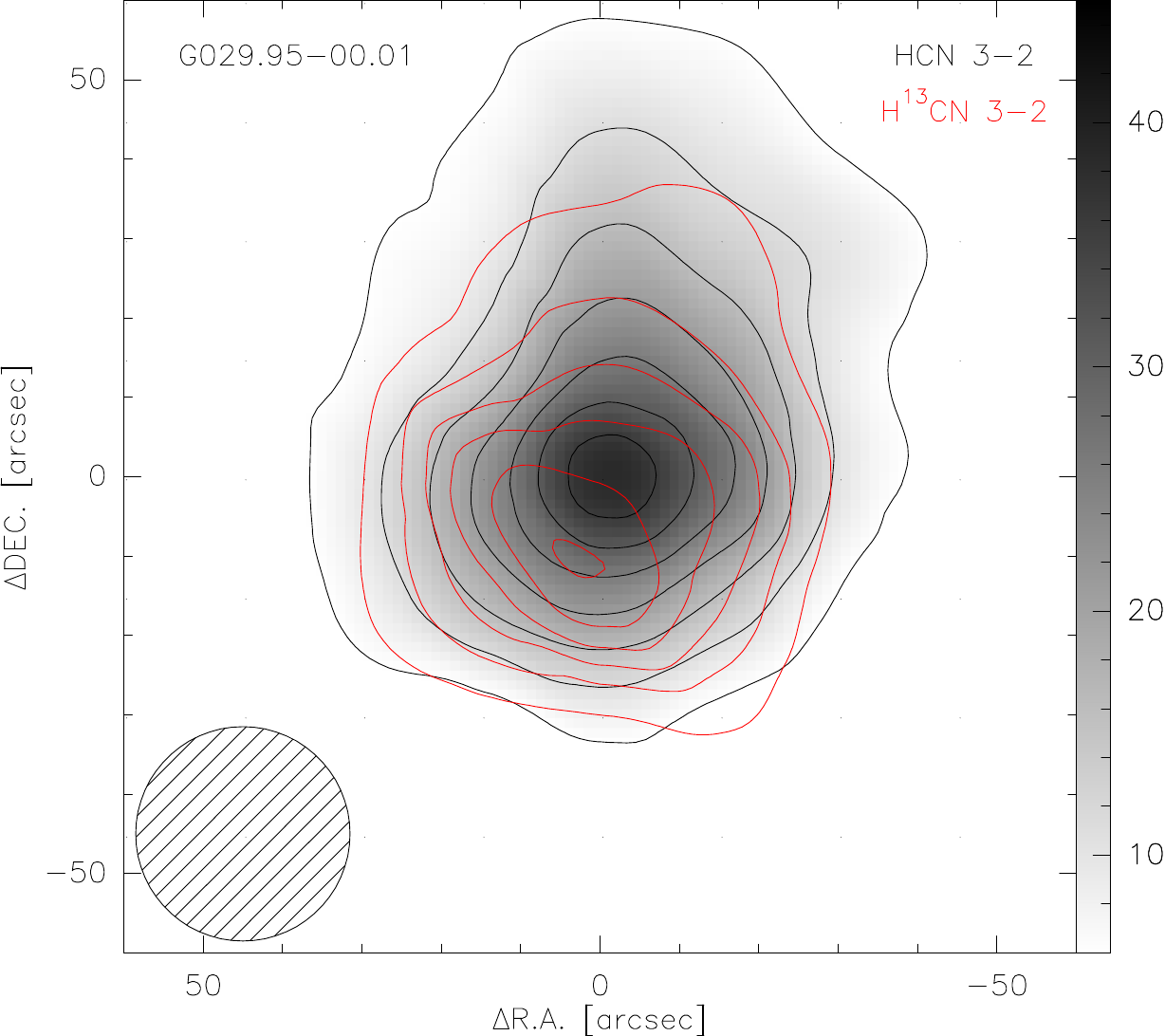}
   \includegraphics[width=2.8in]{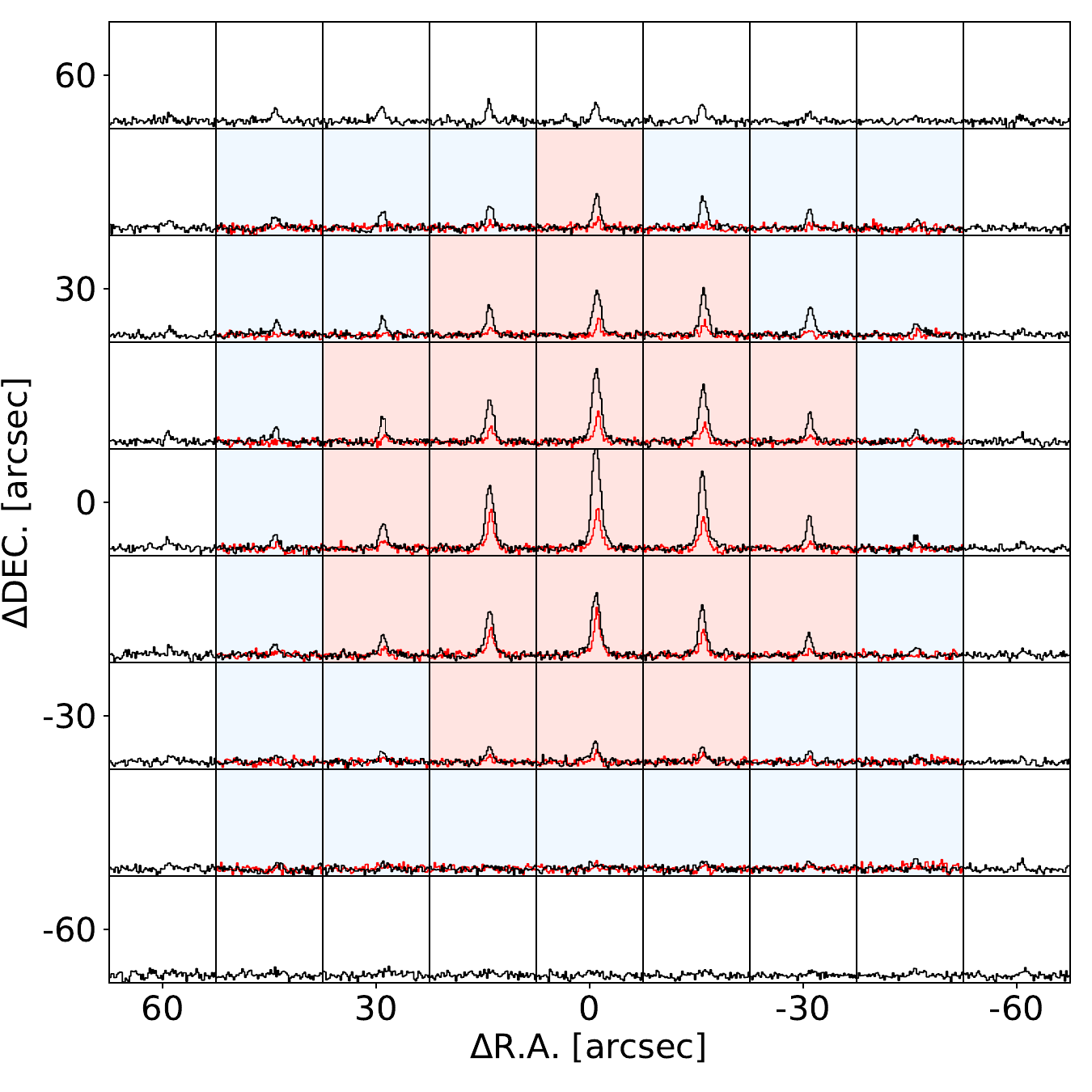}
       \includegraphics[width=3.1in]{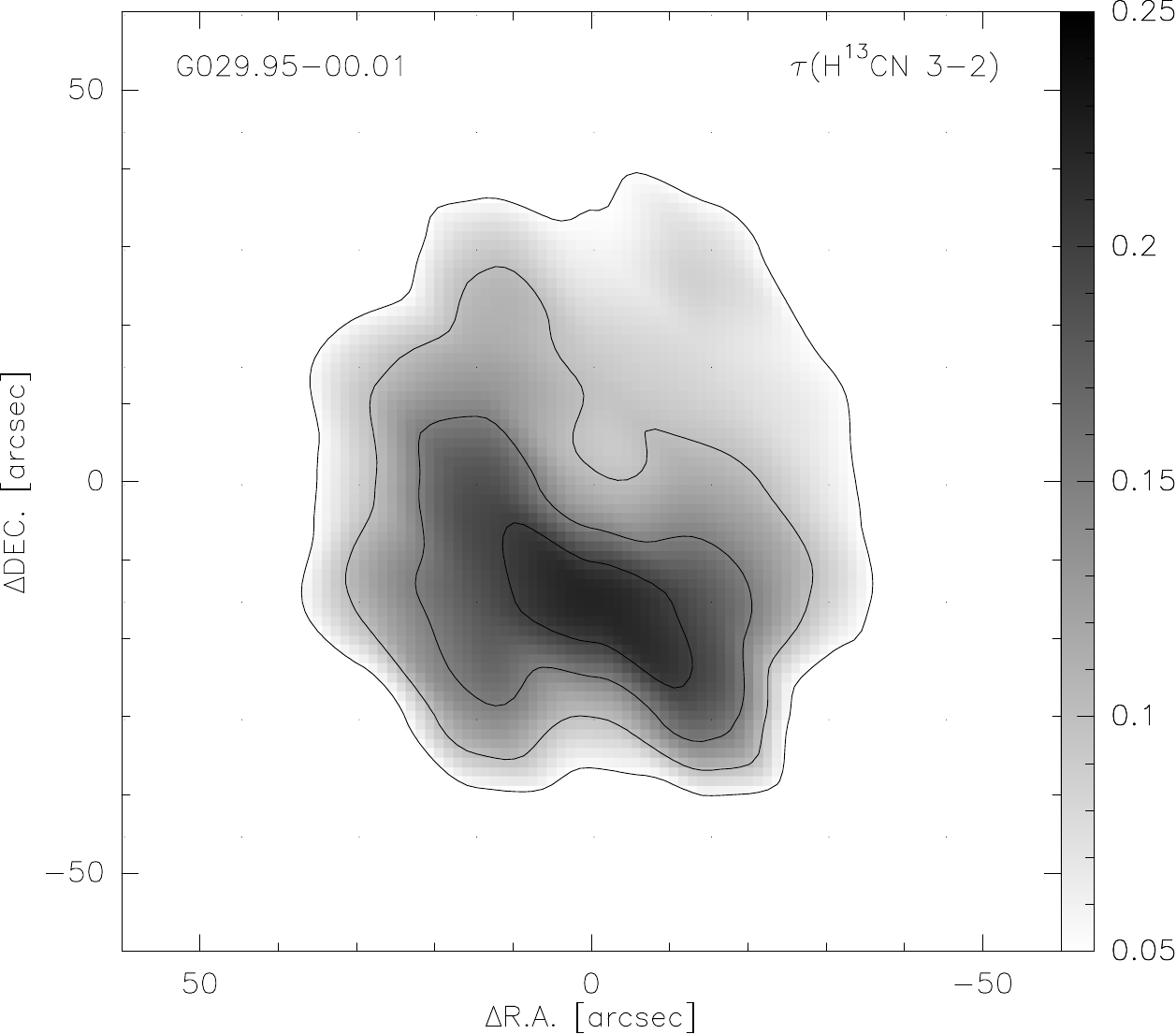}
       \includegraphics[width=2.75in]{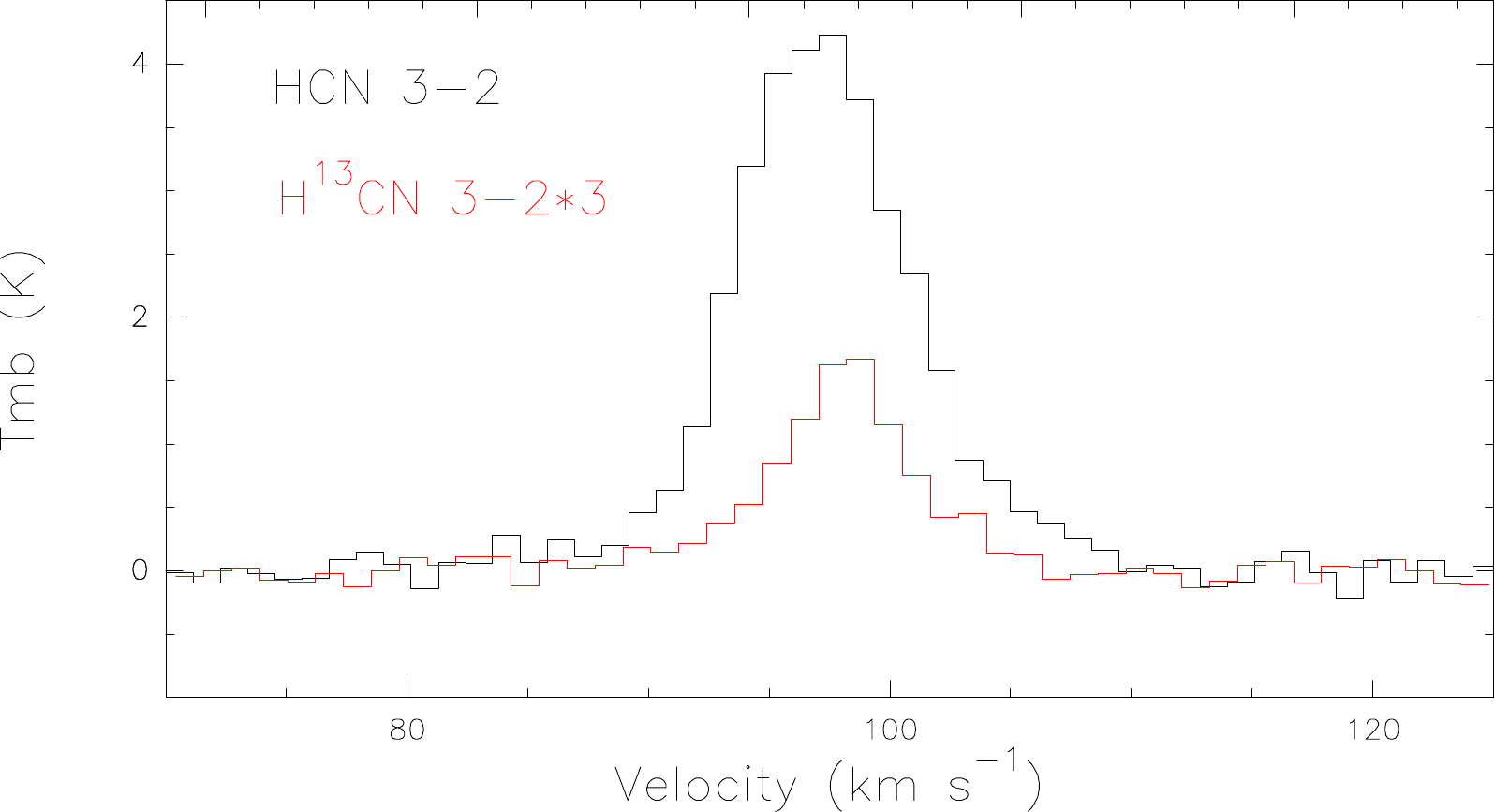}
 \caption{The data reduction results of G029.95-00.01 as an example. 
               {\it Top left:} The velocity integrated intensity maps of HCN and H$^{13}$CN 3-2, with the data  from OTF observation in May 2018 with 10-m SMT. 
               The mapping size of HCN 3-2 is 2$'\times2'$, while it is 1.5$'\times1.5'$ for H$^{13}$CN 3-2, with a beam size of $\sim$ 27.8$''$.
               The grey scale and the black contour with levels starting from 6 K km s$^{-1}$ in step of 5 K km s$^{-1}$ show the observed HCN 3-2. 
               The red contour with levels starting from 0.8 K km s$^{-1}$ in step of 0.8 K km s$^{-1}$ represents H$^{13}$CN 3-2.
               {\it Top right:} Grid-map of HCN 3-2 (black) with a size of 2$'\times2'$, overlapped by gridded spectra of H$^{13}$CN 3-2 (red) with a map size of 1.5$'\times1.5'$. 
               The flux intensities of H$^{13}$CN 3-2 are multiplied by 3. 
               22 positions marked in light pink located near the center part are selected with 3 $\sigma$ level of H$^{13}$CN 3-2,                
               and spectra within where are adopted for calculating the spatially resolved $\tau$. 27 positions marked in light blue are 
               selected on a criteria such as H$^{13}$CN 3-2 spectra with emission signals but not up to 3 $\sigma$, spectra with where 
               are used for calculating $\tau$ of the ``middle'' part of a source.
               The 32 most out part of spectra are data points of HCN 3-2 for ``Averaged (outside)'' $\tau$ estimation.
                {\it Bottom left:} The spatially resolved $\tau(\textrm{H}^{13}\textrm{CN})$ of G029.95-00.01 is demonstrated by black contour with levels 
               starting from 0.05 in step of 0.05. 
                {\it Bottom right:} The spectra of HCN (black) and H$^{13}$CN 3-2 (red) at center position of G029.95-00.01. 
                }       
 \label{fig:g02995}
\end{figure*}

\section{The Method}

\subsection{General description}
For the pair lines of HCN and H$^{13}$CN, as well as HCO$^+$ and H$^{13}$CO$^+$, with the assumption of  same filling factors and excitation temperatures, the optical depth of each  isotopologue can be estimated using 
\begin{equation}
    \frac{\int I^{13}}{\int I^{12}}=\frac{1-e^{-\tau^{13}}}{1-e^{-\tau^{12}}} .
 \label{eq:int-ratio}
\end{equation}
where $I^{13}$ and $I^{12}$ are measured velocity integrated fluxes, while $\tau^{13}$ and $\tau^{12}$ are the optical depths for each line.

With the spatially resolved maps of HCN and HCO$^+$ 3-2, and their isotopologues H$^{13}$CN and  H$^{13}$CO$^+$  3-2, we use two methods to derive optical depths of dense gas tracers and their isotopologues  to study whether they are consistent with each other.  For both methods, the same assumption of abundance ratio as 40  is adopted 
for HCN and H$^{13}$CN, as well as for HCO$^+$ and H$^{13}$CO$^+$ lines. Even $^{12}$C/$^{13}$C varies in different regions of the Milky Way \citep{1994ARA&A..32..191W}, taking the reasonable value of 40 will not affect our main results and conclusions, since we are comparing the relative optical depths with different methods instead of the absolute ones.
 The first method is deriving the spatially resolved optical depths for each position with H$^{13}$CN 3-2 (or H$^{13}$CO$^+$ 3-2) above 3$\sigma$ level, and averaging the optical depths weighted by velocity integrated HCN 3-2 (or HCO$^+$ 3-2) fluxes. The second way is to average the data of HCN 3-2 (or HCO$^+$ 3-2) and H$^{13}$CN 3-2 (or H$^{13}$CO$^+$  3-2)  in the positions used in the first method, to obtain a pair of  HCN/H$^{13}$CN or HCO$^+$/H$^{13}$CO$^+$ 3-2 line ratio, which will be used to derive optical depth for each source. 

The detailed description of both methods is presented as following.

{\bf Method 1: Average of spatially resolved ${\tau}$}

After data reduction, we get the grid-map and 
derive the velocity integrated fluxes of HCN and HCO$^+$ 3-2, as well as H$^{13}$CN and H$^{13}$CO$^+$ 3-2 of one source respectively, from which we can get the 
line ratio of H$^{13}$CN/HCN 3-2 and H$^{13}$CO$^+$/HCO$^+$ respectively for each position. 
The uncertainties of the velocity integrated intensities for one pair of lines 
are estimated with ${\sigma}_{rms}\times\sqrt{{\delta}v{\Delta}V}$, where the ${\sigma}_{rms}$ is from the baseline fitting for the center point of the spectra for each map. Since the system temperature and weather conditions nearly do not vary during each OTF mapping, and the effective on source time at each position within the final  grid-map  is almost the same,  the  expected noise level at each position should be almost the same. We also checked the distribution of noise level  from the baseline fitting  for several sources, which provided promising results as expected. 

We select the positions with velocity integrated intensity of H$^{13}$CN 3-2 and H$^{13}$CO$^+$ 3-2 greater than 3${\sigma}$, respectively, from the grid-map as reliable signal, count the positions and mark them as ``spatially resolved'', which contain 80$\%$ to 95$\%$ of the total isotopologue flux for the observed core regions.
Then we can obtain the spatially resolved $\tau(\textrm{H}^{13}\textrm{CN})$ and $\tau(\textrm{H}^{13}\textrm{CO}^+)$  for the certain source.

Finally we derive the averaged $\tau(\textrm{H}^{13}\textrm{CN})$ and $\tau(\textrm{H}^{13}\textrm{CO}^+)$ weighted by velocity integrated intensities of HCN and HCO$^+$ 3-2 respectively, and 
take them as the typical $\tau(\textrm{H}^{13}\textrm{CN})$ and $\tau(\textrm{H}^{13}\textrm{CO}^+)$ of this clump. 


{\bf Method 2: ${\tau}$ from the averaged line ratios}

By adopting the same selected positions for each source as described in Method 1, we obtain the spatially averaged spectrum for each pair of lines. 
Then, the obtained  H$^{13}$CN/HCN 3-2 and H$^{13}$CO$^+$/HCO$^+$ 3-2 line ratios for the region can be used to derive the optical depths of  
H$^{13}$CN 3-2 and H$^{13}$CO$^+$ 3-2 respectively, which is similar to that in galaxies without spatial information. 
The derived $\tau(\textrm{H}^{13}\textrm{CN})$ and $\tau(\textrm{H}^{13}\textrm{CO}^+)$, as well as $\tau(\textrm{HCN})$ and $\tau(\textrm{HCO}^+)$ by Method 2 are marked with ``Averaged (center)''. 

To each source, in order to guarantee that  the exact same data  were used in Method 1 and 2,  same velocity range and masking, which were derived by ``set mode x'' and ``set window'' in \texttt{CLASS} respectively, were adopted for all spectra of each source, including those in different positions for Method 1 and the averaged one for Method 2. 

Since the relation between optical depth ratio and line ratio is non-linear (see Equation~\ref{eq:int-ratio}), while the spatially integrated fluxes for the line emissions are simply linear collection, optical depths calculated with the two methods using the same data do not guarantee to provide the same results.

\subsection{Using one source as an example}

 The detailed calculation of the optical depth is described 
with one source --- G029.95-00.01 and one pair of lines --- H$^{13}$CN and HCN 3-2 as an example. 
The velocity integrated intensity map of H$^{13}$CN 3-2 as red contour overlaid on that of HCN 3-2 as grayscale and black contour, is presented in the {\it top left} panel of Figure~\ref{fig:g02995}. Note that there is an offset of peak position between H$^{13}$CN 3-2  and HCN 3-2. 
 22 positions with  H$^{13}$CN 3-2 emission above 3$\sigma$  level  were selected from this source, which were marked in light pink on the grid-maps for HCN and H$^{13}$CN 3-2 as demonstrated in the {\it top right} panel of Figure~\ref{fig:g02995}.  
  All the selected spectra of H$^{13}$CN 3-2 were also checked by eye to confirm data quality. The spatial distribution of H$^{13}$CN 3-2 optical depth is presented in  the {\it bottom left} panel of Figure~\ref{fig:g02995}, with contour levels from 0.05 to 0.25. 
 Based on  examination of the spectra, the ``Spatially resolved'' and ``Averaged'' $\tau$ of H$^{13}$CN 3-2 obtained respectively by two methods were listed in Table~\ref{tab:hcn}.  
 $\tau(\textrm{H}^{13}\textrm{CN})$ derived from the spatially resolved information is 0.1158$\pm$0.0002, while the ``Averaged'' $\tau(\textrm{H}^{13}\textrm{CN})$ is 0.1157$\pm$0.0002.  For the 22 positions located at ``center'' of the core, the intensity flux of H$^{13}$CN is 34.99$\pm$0.86 K km s$^{-1}$ and taking 86.4\% of the total H$^{13}$CN flux, while the ``center'' HCN only contains 64.8\% flux within the 2$'\times2'$ region.  
The uncertainty of ``Averaged'' $\tau(\textrm{H}^{13}\textrm{CN})$ is calculated from the intensity flux error propagation, while  the same value is adopted as uncertainty of the ``Spatially resolved'' $\tau(\textrm{H}^{13}\textrm{CN})$.  
For the region with detectable emission above 3$\sigma$  level of H$^{13}$CN 3-2, $\tau$ derived from the two methods for G029.95-00.01 are generally agreed well with each other.


 Meanwhile, we also obtained the averaged  $\tau(\textrm{H}^{13}\textrm{CN})$ for the positions 
with  H$^{13}$CN 3-2  data included in 1.5 $'\times1.5'$ mapping area, but the intensities of which are  less than 3${\sigma}$ level and not used for calculation of the spatially resolved $\tau(\textrm{H}^{13}\textrm{CN})$.  Since the re-grid step is 15$''$, there are $7\times7=49$ positions in total within1.5 $'\times1.5'$ 
isotopologue mapping area. 
Besides the selected 22 positions out of 49 worked for $\tau$ examination by two methods, 
there are 27 positions left, 
which are marked in light blue with the same area on grid-maps of HCN 3-2 and H$^{13}$CN 3-2 in Figure~\ref{fig:g02995} and labeled as ``middle'' part.

Even though no significant signal of  individual  H$^{13}$CN 3-2 emission in these 27 positions, the averaged or stacked spectrum of H$^{13}$CN 3-2  do have about 5$\sigma$ detection. 
We obtained $\tau$ by using Method 2 for the 27 positions without significant signal of H$^{13}$CN 3-2 emission.  
The result is shown in Table~\ref{tab:hcn} and marked as ``Averaged (middle). The $\tau(\textrm{H}^{13}\textrm{CN})$ of the 27 positions is 0.0512$\pm$0.0005, 
which is about half of $\tau(\textrm{H}^{13}\textrm{CN})$ derived from the area of ``center part''. 
 
 The intensity flux of H$^{13}$CN and HCN 3-2 for the 27 ``middle'' positions are 5.5$\pm$1.1 K km s$^{-1}$ and 95.6$\pm$3.8 K km s$^{-1}$ respectively, taking about 13.6\% and 19.5\% of the total flux in each case.

Since the HCN 3-2  mapping size is 2$'\times2'$,  there are $9\times9=81$ positions with re-grid step  of 15$''$. In fact, there are 32 positions have HCN 3-2 data and without H$^{13}$CN 3-2 data, which are marked as ``outside'' positions. It is impossible to calculate even for the averaged optical depths of H$^{13}$CN and HCN 3-2 there. However, since H$^{13}$CN 3-2 emission is quickly decreasing from center to the outside and even the ``middle'' 27 positions only contain $\sim$13.6\% flux within the  1.5 $'\times1.5'$ region, we can neglect the contribution of  H$^{13}$CN 3-2 emission in ``outside'' positions when counting total H$^{13}$CN 3-2  flux in this molecular core. But the HCN 3-2 emission can still be detected in such region. 
The stacked intensity flux of the ``outside'' for HCN emission is 67.2$\pm$3.9 K km s$^{-1}$ and contains 13.7\% of the total HCN 3-2 flux. The contribution of the outside HCN 3-2 emission is listed in Table~\ref{tab:hcn} and marked as ``Averaged (outside)''.

In addition, we also calculated the averaged optical depths of  H$^{13}$CN and HCN 3-2 of this molecular core, which derived from the flux ratio of total H$^{13}$CN emission for 1.5$'\times1.5'$ and HCN emission for 2$'\times2'$. 
The total flux of HCN, obtained by averaged spectrum in the 2$'\times2'$ regions,  is 490.1$\pm$6.1 K km s$^{-1}$, approximately equals to the sum of flux from ``center'', ``middle'' and ``outside'' parts. Also, the total flux of H$^{13}$CN as 40.5$\pm$1.6 K km s$^{-1}$ is similar  to the sum of  fluxes from ``center'' and ``middle''. By taking the H$^{13}$CN/HCN 3-2 ratio as 0.0826$\pm$0.0023 and using Method 2, we obtained the $\tau(\textrm{H}^{13}\textrm{CN})$ as 0.0832$\pm$0.0003, which was moderately lower than that from the ``center'' part. The results are listed in Table~\ref{tab:hcn} and marked as ``Averaged (whole)''.  For a short summary of the optical depths obtained in different regions, including ``center'', ``middle'' and ``outside''  parts,  the derived values are decreasing from center to the outside. The results of  ``middle'' and ``outside''  parts are just for showing the optical depth distribution in individual sources itself, which will not be used for the discussions in next section.

 For another pair of lines --- HCO$+$ and H$^{13}$CO$^+$, the same procedures are adopted for the optical depths calculation in different conditions, respectively. The results are listed in Table~\ref{tab:hcop}.





\begin{figure*} 
    \centering
  \includegraphics[height=2.5in,width=2.7in]{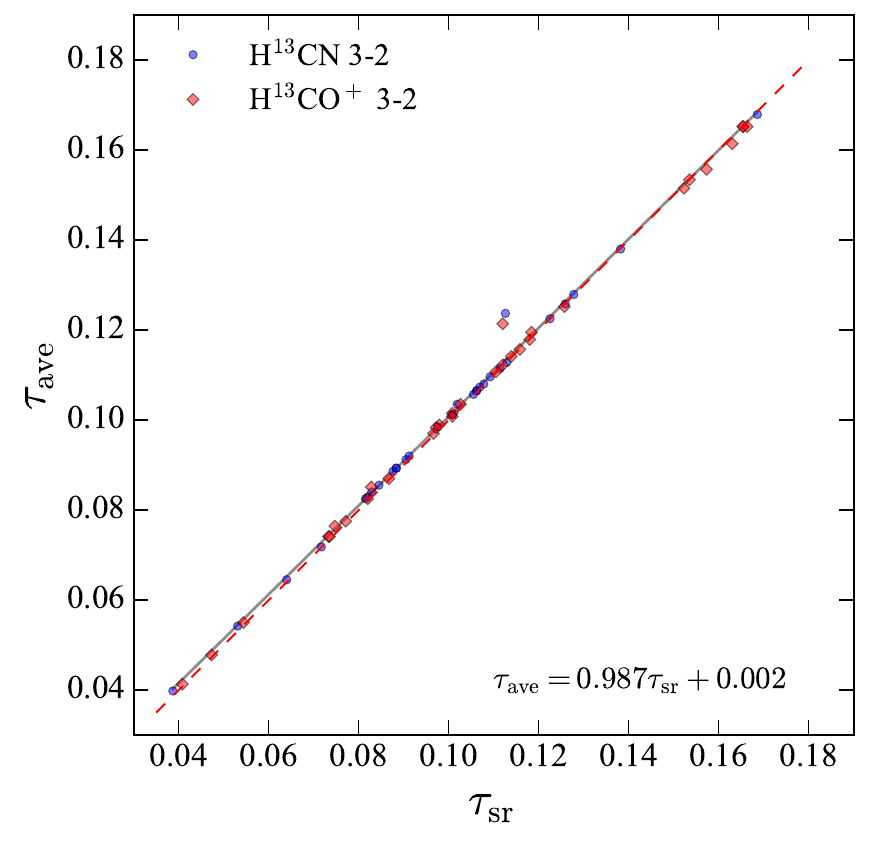}
 \caption{The relation of ``Spatially resolved'' and ``Averaged'' optical depths for 30 sources. The blue filled circle and red diamond represent 
 H$^{13}$CN 3-2 and H$^{13}$CO$^+$ 3-2, respectively. Black solid line shows the fitting result, while the dashed red line is the  function of ``y=x''.}       
 \label{fig:plot-iso1}
\end{figure*}


\begin{figure*} 
    \centering
   \includegraphics[height=2.5in,width=2.7in]{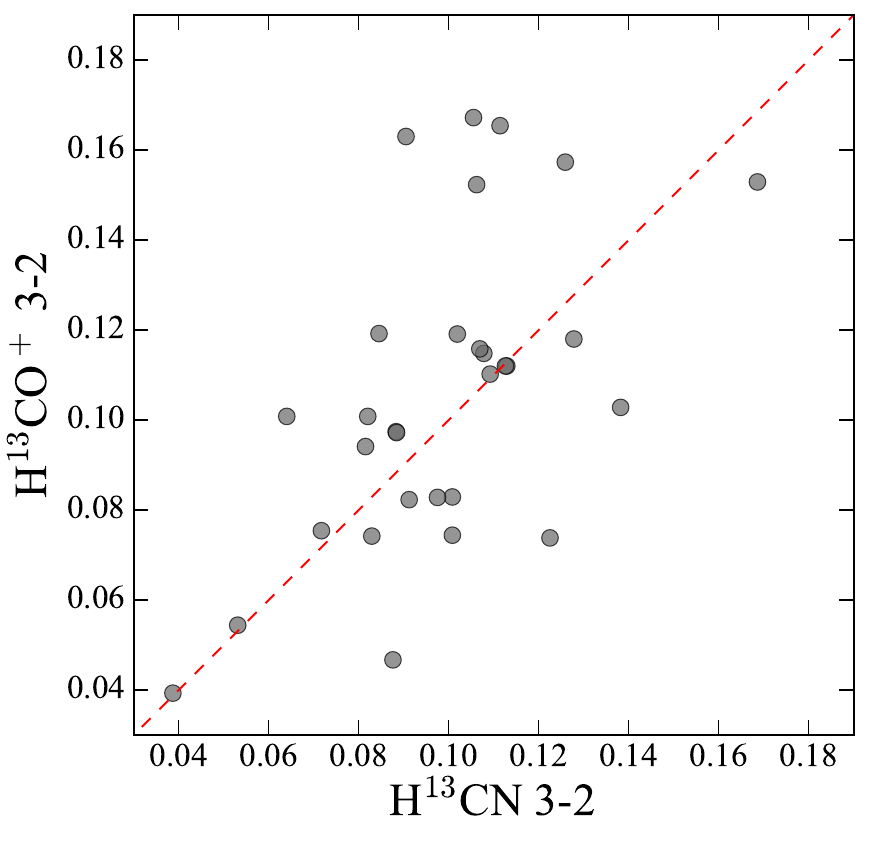}
 \caption{$\tau(\textrm{H}^{13}\textrm{CN})$ v. s. $\tau(\textrm{H}^{13}\textrm{CO}^+)$ for the same sources. The dashed red line is the  function of ``y=x''. }       
 \label{fig:plot-iso2}
\end{figure*}


\begin{table*}
    \begin{center}
      \caption{Optical depth derived from HCN 3-2 of G029.95-00.01.}\label{tab:hcn}
      \begin{tabular}{cccccc}
      \\
    \hline
    \hline
Range & $I(\textrm{HCN})$   & $I(\textrm{H}^{13}\textrm{CN})$     & $I(\textrm{H}^{13}\textrm{CN})/I(\textrm{HCN})$      & $\tau(\textrm{H}^{13}\textrm{CN})$    & $\tau(\textrm{HCN})$   \\
          &   K km s$^{-1}$  &   K km s$^{-1}$   &              &          & \\
     \hline
Spatially resolved (22, HCN weighted)  &    &    &		                            & 0.1158$\pm$0.0002  & 4.633$\pm$0.008 \\
     \hline
Averaged (center, 22)	             & 317.4$\pm$2.8  & 34.99$\pm$0.86 & 0.1102$\pm$0.0015 & 0.1157$\pm$0.0002 & 4.628$\pm$0.008 \\
Averaged (middle, 27)      & 95.6$\pm$3.8 & 5.5$\pm$1.1 & 0.0571$\pm$0.0094 & 0.0512$\pm$0.0005 & 2.048$\pm$0.020 \\
Averaged (outside, 32)        & 67.2$\pm$3.9 & --- & --- & --- & ---				\\
      \hline
Averaged (whole)                   & 490.1$\pm$6.1 & 40.5$\pm$1.6 & 0.0826$\pm$0.0023 & 0.0832$\pm$0.0003 & 3.328$\pm$0.012 \\
           \hline
      \end{tabular}
  \end{center}
\end{table*}


\begin{table*}
    \begin{center}
      \caption{Optical depth derived from HCO$^+$ 3-2 of G029.95-00.01. }\label{tab:hcop}
      \begin{tabular}{cccccc}
      \\
    \hline
    \hline
Range & $I(\textrm{HCO}^+)$   & $I(\textrm{H}^{13}\textrm{CO}^+)$     & $I(\textrm{H}^{13}\textrm{CO}^+)/I(\textrm{HCO}^+)$      & $\tau(\textrm{H}^{13}\textrm{CO}^+)$    & $\tau(\textrm{HCO}^+)$   \\
          &   K km s$^{-1}$  &   K km s$^{-1}$   &              &          & \\
     \hline
Spatially resolved (19, HCO$^+$ weighted)  &    &    &		                            & 0.1069$\pm$0.0001  & 4.276$\pm$0.004 \\
     \hline
Averaged (center, 19)	             & 240.3$\pm$2.0  & 24.74$\pm$0.43 & 0.1030$\pm$0.0009 & 0.1073$\pm$0.0001 & 4.292$\pm$0.004 \\
Averaged (middle, 30)                  & 137.4$\pm$2.8 & 6.36$\pm$0.61 & 0.0463$\pm$0.0035 & 0.0367$\pm$0.0001 & 1.468$\pm$0.004 \\
Averaged (outside, 32)                      & 98.6$\pm$2.8 & --- &--- & --- & --- \\
      \hline
Averaged (whole)                   & 476.3$\pm$5.2 & 31.12$\pm$0.68 & 0.0653$\pm$0.0007 & 0.0619$\pm$0.0001 & 2.476$\pm$0.004 \\
      \hline
      \end{tabular}
  \end{center}
\end{table*}


\onecolumn
\begin{landscape}
        \footnotesize 
\begin{longtable}{lcllllccllll}
\caption{Optical depths of H$^{13}$CN 3-2 and H$^{13}$CO$^+$ 3-2 for 30 sources.}\label{tab:tau-all}\\
\hline\hline
    &  \multicolumn{5}{c}{$\tau(\textrm{H}^{13}\textrm{CN})$} & $\ $ &\multicolumn{5}{c}{$\tau(\textrm{H}^{13}\textrm{CO}^+)$}   \\
    \cline{2-6} \cline{8-12}
  Source name & Pts$^{\dag}$ & Spatially resolved &  Ave (center) &  Ave (whole) &  Ave (middle) &  &Pts$^{\dag}$ & Spatially resolved & Ave (center) &   Ave (whole) & Ave (middle)  \\
\hline
G005.88-00.39   & 20 & 0.1523$\pm$0.0002 & 0.1515$\pm$0.0002 & 0.1187$\pm$0.0005 & 0.0398$\pm$0.0002 & & 28 & 0.1062$\pm$0.0001 & 0.1065$\pm$0.0001 & 0.0951$\pm$0.0002 & 0.0626$\pm$0.0005 \\                
G009.62+00.19   & 12 & 0.1654$\pm$0.0006 & 0.1652$\pm$0.0005 & 0.1143$\pm$0.0010 & 0.0748$\pm$0.0015 & & 15 & 0.1114$\pm$0.0003 & 0.1115$\pm$0.0003 & 0.0722$\pm$0.0006 & 0.0492$\pm$0.0002 \\    
G010.47+00.02   & 18 & 0.1663$\pm$0.0005 & 0.1652$\pm$0.0005 & 0.1242$\pm$0.0012 & 0.0809$\pm$0.0015 & & 14 & 0.1055$\pm$0.0005 & 0.1057$\pm$0.0005 & 0.0771$\pm$0.0006 & 0.0748$\pm$0.0009 \\    
G010.62-00.38   & 31 & 0.1535$\pm$0.0001 & 0.1534$\pm$0.0001 & 0.1282$\pm$0.0002 & 0.0714$\pm$0.0004 & & 30 & 0.1686$\pm$0.0001 & 0.1679$\pm$0.0001 & 0.1359$\pm$0.0001 & 0.0527$\pm$0.0006 \\    
G011.49-01.48   & 12 & 0.0771$\pm$0.0001 & 0.0775$\pm$0.0001 & 0.0336$\pm$0.0002 & ---               & & 22 & 0.0717$\pm$0.0001 & 0.0718$\pm$0.0001 & 0.0542$\pm$0.0003 & 0.0398$\pm$0.0004 \\               
G011.91-00.61   & 13 & 0.1139$\pm$0.0003 & 0.1141$\pm$0.0003 & 0.0912$\pm$0.0003 & 0.0352$\pm$0.0004 & & 15 & 0.1078$\pm$0.0005 & 0.1080$\pm$0.0005 & 0.0920$\pm$0.0004 & 0.0405$\pm$0.0003 \\                    
G012.80-00.20   & 46 & 0.1026$\pm$0.0001 & 0.1035$\pm$0.0001 & 0.0722$\pm$0.0001 & 0.0478$\pm$0.0006 & & 48 & 0.1382$\pm$0.0001 & 0.1380$\pm$0.0001 & 0.1027$\pm$0.0001 & $<$0.0764 \\                    
G014.33-00.64   & 12 & 0.1180$\pm$0.0004 & 0.1179$\pm$0.0004 & 0.0863$\pm$0.0006 & 0.0665$\pm$0.0009 & & 25 & 0.1278$\pm$0.0002 & 0.1279$\pm$0.0002 & 0.0806$\pm$0.0001 & 0.0455$\pm$0.0003 \\                    
G015.03-00.67   & 28 & 0.0733$\pm$0.0001 & 0.0741$\pm$0.0001 & 0.0756$\pm$0.0001 & 0.0962$\pm$0.0002 & & 34 & 0.1225$\pm$0.0001 & 0.1225$\pm$0.0001 & 0.1221$\pm$0.0001 & 0.1019$\pm$0.0004 \\                 
G016.58-00.05   & 11 & 0.1120$\pm$0.0003 & 0.1122$\pm$0.0003 & 0.0890$\pm$0.0011 & 0.0825$\pm$0.0016 & &  9 & 0.1129$\pm$0.0003 & 0.1128$\pm$0.0003 & 0.0790$\pm$0.0011 & 0.0672$\pm$0.0011 \\              
G023.00-00.41   & 11 & 0.1630$\pm$0.0006 & 0.1614$\pm$0.0006 & 0.0939$\pm$0.0009 & 0.0699$\pm$0.0011 & & 14 & 0.0905$\pm$0.0002 & 0.0912$\pm$0.0002 & 0.0409$\pm$0.0002 & 0.0176$\pm$0.0001 \\                            
G027.36-00.16   & 12 & 0.0867$\pm$0.0002 & 0.0870$\pm$0.0002 & 0.0855$\pm$0.0008 & 0.0661$\pm$0.0018 & & 21 & 0.1008$\pm$0.0002 & 0.1012$\pm$0.0002 & 0.0909$\pm$0.0006 & 0.0600$\pm$0.0013 \\                    
G029.95-00.01   & 22 & 0.1158$\pm$0.0002 & 0.1157$\pm$0.0002 & 0.0832$\pm$0.0003 & 0.0512$\pm$0.0005 & & 19 & 0.1069$\pm$0.0001 & 0.1073$\pm$0.0001 & 0.0619$\pm$0.0001 & 0.0367$\pm$0.0001 \\    
G035.02+00.34   & 10 & 0.1120$\pm$0.0003 & 0.1214$\pm$0.0003 & 0.0768$\pm$0.0009 & 0.0348$\pm$0.0006 & & 11 & 0.1126$\pm$0.0005 & 0.1237$\pm$0.0005 & 0.1008$\pm$0.0014 & 0.1088$\pm$0.0024 \\    
G035.19-00.74   & 22 & 0.0828$\pm$0.0002 & 0.0851$\pm$0.0002 & 0.0638$\pm$0.0003 & 0.0550$\pm$0.0004 & & 29 & 0.0975$\pm$0.0002 & 0.0985$\pm$0.0002 & 0.0645$\pm$0.0003 & 0.0260$\pm$0.0002 \\    
G037.43+01.51   & 10 & 0.0966$\pm$0.0004 & 0.0970$\pm$0.0004 & 0.0722$\pm$0.0006 & 0.0508$\pm$0.0007 & & 20 & 0.0815$\pm$0.0001 & 0.0825$\pm$0.0001 & 0.0626$\pm$0.0002 & 0.0298$\pm$0.0004 \\    
G043.16+00.01   & 30 & 0.0979$\pm$0.0001 & 0.0989$\pm$0.0001 & 0.0680$\pm$0.0001 & 0.0295$\pm$0.0002 & & 34 & 0.0883$\pm$0.0001 & 0.0893$\pm$0.0001 & 0.0645$\pm$0.0001 & ---               \\    
G043.79-00.12   &  8 & 0.1008$\pm$0.0004 & 0.1015$\pm$0.0004 & 0.0939$\pm$0.0011 & 0.0922$\pm$0.0024 & &  8 & 0.0820$\pm$0.0002 & 0.0829$\pm$0.0002 & 0.0573$\pm$0.0005 & 0.0188$\pm$0.0003 \\    
G049.48-00.36   & 36 & 0.1105$\pm$0.0001 & 0.1107$\pm$0.0001 & 0.0764$\pm$0.0001 & 0.0123$\pm$0.0001 & & 41 & 0.1092$\pm$0.0001 & 0.1096$\pm$0.0001 & 0.0745$\pm$0.0001 & 0.0321$\pm$0.0003 \\    
G049.48-00.38   & 32 & 0.1573$\pm$0.0002 & 0.1557$\pm$0.0002 & 0.1134$\pm$0.0002 & 0.0619$\pm$0.0003 & & 34 & 0.1259$\pm$0.0002 & 0.1258$\pm$0.0002 & 0.0897$\pm$0.0002 & 0.0180$\pm$0.0001 \\       
G069.54-00.97   & 19 & 0.1257$\pm$0.0007 & 0.1252$\pm$0.0007 & 0.0864$\pm$0.0012 & 0.0199$\pm$0.0005 & & 33 & 0.1019$\pm$0.0001 & 0.1035$\pm$0.0001 & 0.0748$\pm$0.0001 & 0.0348$\pm$0.0004 \\    
G075.76+00.33   & 21 & 0.0408$\pm$0.0001 & 0.0413$\pm$0.0001 & 0.0214$\pm$0.0001 & 0.0165$\pm$0.0001 & & 17 & 0.0387$\pm$0.0001 & 0.0398$\pm$0.0001 & 0.0192$\pm$0.0001 & 0.0260$\pm$0.0001 \\    
G078.12+03.63   &  9 & 0.0473$\pm$0.0001 & 0.0478$\pm$0.0001 & 0.0180$\pm$0.0001 & ---               & & 23 & 0.0876$\pm$0.0002 & 0.0886$\pm$0.0002 & 0.0748$\pm$0.0003 & 0.0493$\pm$0.0005 \\          
G081.87+00.78   & 34 & 0.0972$\pm$0.0001 & 0.0983$\pm$0.0001 & 0.0672$\pm$0.0001 & ---               & & 44 & 0.0884$\pm$0.0001 & 0.0893$\pm$0.0001 & 0.0706$\pm$0.0001 & 0.0188$\pm$0.0001 \\        
G109.87+02.11   & 32 & 0.0747$\pm$0.0001 & 0.0764$\pm$0.0001 & 0.0520$\pm$0.0001 & 0.0199$\pm$0.0001 & & 47 & 0.1008$\pm$0.0001 & 0.1012$\pm$0.0001 & 0.0775$\pm$0.0001 & 0.0535$\pm$0.0011 \\    
G121.29+00.65   & 16 & 0.0735$\pm$0.0002 & 0.0741$\pm$0.0002 & 0.0462$\pm$0.0003 & 0.0375$\pm$0.0003 & & 20 & 0.0829$\pm$0.0001 & 0.0840$\pm$0.0001 & 0.0504$\pm$0.0002 & 0.0283$\pm$0.0002 \\    
G123.06-06.30   &  9 & 0.1008$\pm$0.0004 & 0.1008$\pm$0.0004 & 0.0718$\pm$0.0008 & 0.0451$\pm$0.0007 & & 17 & 0.0640$\pm$0.0001 & 0.0645$\pm$0.0001 & 0.0470$\pm$0.0002 & 0.0584$\pm$0.0010 \\    
G133.94+01.06   & 26 & 0.1184$\pm$0.0002 & 0.1195$\pm$0.0002 & 0.0909$\pm$0.0002 & 0.0733$\pm$0.0003 & & 34 & 0.0845$\pm$0.0001 & 0.0855$\pm$0.0001 & 0.0672$\pm$0.0001 & 0.0352$\pm$0.0001 \\          
G188.94+00.88   & 10 & 0.0544$\pm$0.0002 & 0.0550$\pm$0.0002 & ---               & ---               & & 26 & 0.0531$\pm$0.0001 & 0.0542$\pm$0.0001 & 0.0367$\pm$0.0001 & ---               \\      
G232.62+00.99   &  6 & 0.0820$\pm$0.0008 & 0.0825$\pm$0.0008 & 0.0839$\pm$0.0024 & 0.0947$\pm$0.0023 & & 17 & 0.0912$\pm$0.0004 & 0.0920$\pm$0.0004 & 0.0825$\pm$0.0006 & 0.0733$\pm$0.0007 \\     
           \hline
\end{longtable}
\footnotesize{$^{\dag}$ Number of positions adopted for the calculation of spatially resolved optical depth.}\\
\end{landscape}
\twocolumn

\section{Results}

Following  methods described in \S3, the parameters of  two pairs of dense gas tracers of 30 clouds are derived and listed in Table~\ref{tab:tau-all}. Only optical depths of  H$^{13}$CN 3-2 and H$^{13}$CO$^+$ 3-2 are listed, while HCN 3-2 and HCO$^+$ 3-2 informations are not included, since they can be simply obtained by multiplying by 40  based on the assumption.  However,  the abundance ratio of $^{12}$C/$^{13}$C can vary from less than 40 to be even close to 100 \citep{1994ARA&A..32..191W, 2005ApJ...634.1126M}.  Since HCN 3-2 and HCO$^+$ 3-2  are mainly optically thick, the derived optical depths of H$^{13}$CN 3-2 and H$^{13}$CO$^+$ 3-2 are almost not related to the assumed abundance ratio, except for some sources with  line ratios of HCN/H$^{13}$CN 3-2 or HCO$^+$/H$^{13}$CO$^+$ 3-2  close to 40.  In the extreme cases with the line ratios greater than 40, there will be no solution for  optical depths under such assumption, which are listed as `$-$' for out regions of several sources  in Table~\ref{tab:tau-all}.  Note that  the ``Spatially resolved'' and ``Averaged (center)'' ones, which are used for the scientific discussions in next section, do have solution for all sources. Consequently, the derived  optical depths of H$^{13}$CN 3-2 and H$^{13}$CO$^+$ 3-2  are reliable, even if the assumption of abundance ratio is not  accurate. 
Since the assumed abundance ratio will only strongly affect the absolute optical depths of HCN 3-2 and HCO$^+$ 3-2 in each source, while we are focusing on the relative different between the optical depths of H$^{13}$CN 3-2 and H$^{13}$CO$^+$ 3-2 derived with  two methods, we do not use different  assumed abundance ratio in each source.

Column 2 -- 5 in Table~\ref{tab:tau-all}  contain information on $\tau(\textrm{H}^{13}\textrm{CN})$, which are the numbers of positions selected for the spatially resolved $\tau(\textrm{H}^{13}\textrm{CN})$, the  $\tau(\textrm{H}^{13}\textrm{CN})$ values of ``spatially resolved'', and the ``Averaged (center)'', ``Averaged (whole)'', ``Averaged (middle)'' , respectively. 
While the column 6 -- 10 represent the the numbers of positions selected for the spatially resolved $\tau(\textrm{H}^{13}\textrm{CO}^+)$ and the 
four kinds of $\tau(\textrm{H}^{13}\textrm{CO}^+)$. 
Note that ``Averaged (whole)''  $\tau(\textrm{H}^{13}\textrm{CN})$  in G188.94+00.88, ``Averaged (middle)'' $\tau(\textrm{H}^{13}\textrm{CN})$   in G011.49-01.48, G078.12+03.63, G081.87+00.78, and G188.94+00.88, and   ``Averaged (middle)'' $\tau(\textrm{H}^{13}\textrm{CO}^+)$ in G043.16+00.01 and G188.94+00.88, are marked as `$-$', with no solutions  caused by high line ratio of HCN/H$^{13}$CN  3-2 or  HCO$^+$/H$^{13}$CO$^+$ 3-2, which are described before.  Since there is only one point left for the calculation of ``Averaged (middle)'' $\tau(\textrm{H}^{13}\textrm{CO}^+)$ in G12.80-00.20, only upper limit is obtained. We checked other sources with data  points used for calculating  $\tau(\textrm{H}^{13}\textrm{CN})$ or $\tau(\textrm{H}^{13}\textrm{CO}^+)$  larger than 35, which means that only several data points are used to derive  ``Averaged (middle)''  optical depths. All sources do have  H$^{13}$CN or H$^{13}$CO$^+$  3-2 signals after averaging the spectra. Special notes need to be mentioned for G049.48-00.36 and  G049.48-00.38, which are very  close to each other and contaminate the outside regions each other. Therefore, the data points used for   ``Averaged (middle)''  are less than 49 subtracted by that used in  ``Averaged (middle)''. Also, the value of  ``Averaged (whole)''  in these two sources can also be over-estimated.


With the data points located in same region,  $\tau(\textrm{H}^{13}\textrm{CN})$ (or  $\tau(\textrm{H}^{13}\textrm{CO}^+)$) derived from the two different methods described in \S 3, i. e. the ``Spatially resolved'' and ``Averaged (center)''   $\tau(\textrm{H}^{13}\textrm{CN})$ (or  $\tau(\textrm{H}^{13}\textrm{CO}^+)$), are generally consistent with each other. 
Besides, both the ``Averaged (whole)'' $\tau(\textrm{H}^{13}\textrm{CN})$ and $\tau(\textrm{H}^{13}\textrm{CO}^+)$ are significantly smaller than those $\tau^{13}$ of the ``spatially resolved'' or ``Averaged (center)''. 
The reason is that the optical depths calculated by ``Averaged (whole)'' contains more outer points where have no detectable H$^{13}$CN or H$^{13}$CO$^+$ 3-2 emission. 

The velocity integrated intensity maps of two pairs of lines, as well as the spatially resolved $\tau$ of H$^{13}$CN 3-2 and H$^{13}$CO$^+$ 3-2 of the 30 sources are also presented in {Appendix~\ref{sec:figs}}.



\section{Discussion}
\subsection{The consistency of optical depths derived with two different methods}

The key question for this paper is to test whether  the optical depths calculated with two methods described in \S3 are consistent with each other or  not. With the information in Table~\ref{tab:tau-all},  optical depths of H$^{13}$CN 3-2 derived with the two different methods agree well with each other, within the range from $\sim$0.04 in G75.76+00.33  to $\sim$ 0.17 in G010.47+00.02 and G010.62-00.38, with dynamical range of more than 4.  The results of the relation between the optical depths derived with two methods for H$^{13}$CO$^+$ 3-2 are similar to that of H$^{13}$CN 3-2, while the optical depths of H$^{13}$CN 3-2 and H$^{13}$CO$^+$ 3-2 in each source do not show strong  correlation, with correlation coefficient of 0.549.
 The plot for optical depths of H$^{13}$CN and H$^{13}$CO$^+$ 3-2 derived with these methods are presented in Figure~\ref{fig:plot-iso1},  where H$^{13}$CN and H$^{13}$CO$^+$ 3-2 in one source are considered as two independent data points. The dashed red   line is a simple  function of ``y=x'', while the  black line is the linear least-square fitting result of ``y=0.987x+0.002'' with correlation coefficient of 0.997, which is close to that of  ``y=x''.  
 Thus, optical depths for such lines of dense cores, derived from integrated flux ratios of dense gas tracers and their isotopologues at different positions,
can generally represent the typical optical depths of these sources, without necessary systematic corrections.  For  Galactic dense cores without spatial resolved information, typical optical depths of those core derived from line ratios of dense gas tracers and their isotopic lines are acceptable, especially for global properties of  large sample surveys.  
Previous studies with dense gas tracers such as CO, HCN, HCO$^+$ and HNC and their isotopic lines in a small number of bright galaxies \citep{1992ApJ...399..521N, 2011A&A...528A..30C, 2011MNRAS.418.1753J, 2014ApJ...796...57W,2016MNRAS.455.3986W,2020MNRAS.494.1095L, 2017MNRAS.466...49J}, the optical depths derived without spatially resolved data, which is similar to method 2, can be considered as a good approximation. Based on our results (see Table~\ref{tab:hcn} and Table~\ref{tab:tau-all}), the differences of derived optical depths with the two methods were within 3\%, while the uncertainties of  isotopic abundance ratio and  measured line fluxes in galaxies are normally larger than 10\%. Thus,  
 the assumption of isotopic abundance ratio and the measurements of lines  can cause larger uncertainties than that caused by the  simple assumption of uniform optical depth in different regions. 

However, even the optical depths obtained with two methods generally  agree well with each other, there can be up to 8\% differences to each other in a few sources, such as H$^{13}$CN 3-2 in G035.02+00.34.  So for galaxies without spatial resolved line ratio of dense gas tracers and their isotopologues, the accuracy with such simple assumption should be tested with spatially resolved observations for several local galaxies, such as Arp 220, M82, and NGC 253.  
CN 1-0 and $^{13}$CN 1-0 observations toward nuclear regions of three nearby galaxies, NGC 253, NGC 1068, and NGC 4945, with ALMA were done recently \citep{2019A&A...629A...6T}, which mainly focused on the $^{12}$C/$^{13}$C ratio instead of optical depth distribution.

The re-grid steps of the observations are 15$''$, while the FWHM of beamsize is about 27.8$''$. On the other words,  the final maps are over-sampled, which means the emission obtained in a certain position is not independent with neighboring positions.  Especially, emission can be seen toward several positions even for a point source, with the same line profile.  However, normally only 5 positions with offsets (-15$''$,  0$''$), (0$''$,  -15$''$), (0$''$, 0$''$), (0$''$, 15$''$) and (15$''$, 0$''$), can have detectable lines.  Based on the information listed in Table~\ref{tab:tau-all}, no source can be treated as point source.  The spatially integrated fluxes of the lines listed in Table~\ref{tab:hcn}   need to be corrected by the fraction of sampling size and beamsize.  However, the absolute line fluxes are not used for any scientific discussion, while only line ratios are used.

\subsection{Assumptions of deriving optical depths with isotopic line ratios}

There are several assumptions during the calculation of line optical depths  with isotopic line ratios, some of which are implicit and may not be true, such as uniform optical depth within observing beam, uniform excitation conditions within observing beam, no influence of continuum emission to the line radiative transfer, and the same value of  molecular isotopic abundance ratio as  that of isotopic ratio.

In fact, due to limited spatial resolution and sensitivity,  the spatially resolved optical depths obtained  with mapping observations of dense gas tracers and their isotoplogues, are averaged for both of the lines within the telescope beam.  Thus, the optical depth distribution has already been smoothed.  Sensitive high resolution observations  of such line pairs are essential for deriving detailed optical depth distribution.  Such methods can also be applied to CO lines and their isotopologues, in Galactic  molecular clouds as well as molecular gas in galaxies.  

 We used the fluxes of HCN and HCO$^+$ 3-2 respectively as weights when calculating the averages from spatially resolved optical depths. Since the optically thin isotopic lines can better  trace gas mass than optically thick lines, isotopic lines are better choice than the optically thick lines if mass weighted optical depths of molecular gas were expected. However, as a dimensionless number, optical depth is  the property of one given line at particular frequency (or velocity), instead of the property for a cloud with given mass.  The main purpose of focusing on optical depths is to evaluate the self absorption of optically thick lines.  The exact number itself of spatially averaged optical depth for one optically thin  line is almost useless.  For optically thin case, the important parameters, such as beam averaged column density and gas mass, are only related to the velocity integrated fluxes, excitation conditions (i.e., partition function), etc,  while  the measured velocity integrated fluxes are affected by excitation temperature, and the combination of optical depth and filling factor.  Therefore, it is not necessary to discuss the spatially averaged optical depth for one particular  optically thin line.   Optical depths of H$^{13}$CN and H$^{13}$CO$^+$ 3-2 discussed in this paper are worked for the information of  HCN and HCO$^+$ 3-2, respectively. With the assumption of HCN/H$^{13}$CN and  HCO$^+$/H$^{13}$CO$^+$ abundance ratio, HCN 3-2 (or HCO$^+$ 3-2) optical depths can be simply derived from  optical depths  of H$^{13}$CN 3-2 (or H$^{13}$CO$^+$ 3-2).  Thus, the typical  optical depths of H$^{13}$CN 3-2 (or H$^{13}$CO$^+$ 3-2) from averaging different positions are reflecting those of HCN 3-2 (or HCO$^+$ 3-2). For spatially averaged optical depths of  HCN 3-2 (or HCO$^+$ 3-2), the derived values can be used for the self-absorption  property of HCN 3-2 (or HCO$^+$ 3-2).  For this reason, weighting with HCN 3-2 (or HCO$^+$ 3-2) fluxes is the only choice to derive typical optical depths with spatially resolved optical depths.  In fact, when the line ratios of a pair of lines, such as HCN/H$^{13}$CN 3-2, are used to derive the beam averaged optical depth,  weighting with HCN 3-2 fluxes has been used as an implicit assumption.


\subsection{The properties of dense gas tracers in these cores}

The dynamical range  of  H$^{13}$CN and H$^{13}$CO$^+$ 3-2  optical depths  derived in these cores is from about 0.04 in G075.76+00.33 to about 0.17 in G010.47+00.02 (see Table~\ref{tab:tau-all} and Figure~\ref{fig:plot-iso1}). The main reason of  lack of sources with H$^{13}$CN and H$^{13}$CO$^+$ 3-2 optical depths higher than 0.17 is due to the properties of  such Galactic massive cores.  Sources with high optical depth may exist as late stage compact cores with high column and volume density. On the other hand, the limitation of low optical depth part in our sample is the observational bias, since only sources with observable  H$^{13}$CN and H$^{13}$CO$^+$ 3-2 emissions are selected. Deep mapping observations toward sources with weak  H$^{13}$CN and H$^{13}$CO$^+$ 3-2 emissions may extend the low optical depth part. The lowest optical depth of  H$^{13}$CN and H$^{13}$CO$^+$ 3-2  will be limited by the real isotopic abundance ratio, because the line ratio will be the abundance ratio when the main isotopic line, i.e., HCN and HCO$^+$ 3-2, are almost optically thin. 

The assumption of same excitation condition of molecules within one beam is always used  when adopting line ratio of one pair of isotopic lines, such as HCN/H$^{13}$CN 3-2.  However, the gradients of excitation  temperatures  of molecular lines  often exist   in dense cores.   Line profiles with self absorption features of HCN and HCO$^+$ 3-2, are observed in the center of several sources, which can cause over-estimation of optical depth.  Nevertheless, since we are focusing on  the optical depth derived with two different methods instead of  discussing gas properties in this paper, such effect can be  ignored.

The spatially resolved optical depths  of  H$^{13}$CN 3-2 and  H$^{13}$CO$^+$ 3-2 in each source are presented in  Figure~\ref{fig:plot-iso2}, as $x$ axis and $y$ axis, respectively,  with  ``y = x'' as red dash line. With a non-linear coefficient of 0.549, no clear relation between  optical depths  of  H$^{13}$CN 3-2 and  H$^{13}$CO$^+$ 3-2 can be found. So, even though  optical depths  of both  H$^{13}$CN 3-2 and  H$^{13}$CO$^+$ 3-2 were used in discussing the difference between two methods, the parameters of  H$^{13}$CN 3-2 and  H$^{13}$CO$^+$ 3-2  in each source are independent, which means  that there are 60  data points instead of 30 in Figure~\ref{fig:plot-iso1}.  The large variation of optical depth ratio between H$^{13}$CN and H$^{13}$CO$^+$ 3-2  may be caused by different critical densities as well as possible astrochemical process for HCN and HCO$^+$, while $^{12}$C/$^{13}$C isotopic ratio should not be the reason.

\section{Summary and conclusion remarks}

With the mapping results of 30 Galactic massive  star forming regions for  two pairs of lines as  HCN and H$^{13}$CN 3-2, as well as HCO$^+$ and H$^{13}$CO$^+$ 3-2 
using the 10-m SMT telescope, we derived optical depths for these sources with two methods to test if the assumption for an uniform optical depth in different positions will strongly affect the typical optical depth in spatially unresolved sources.  

The main findings are summarized as follows:

1.  Spatially resolved optical depths of H$^{13}$CN and H$^{13}$CO$^+$ 3-2 are derived for all 30 sources with the line ratios of  HCN/H$^{13}$CN 3-2 and  HCO$^+$/H$^{13}$CO$^+$ 3-2. Significant spatial variation in all sources are found, which indicate that the assumption of uniform optical depth in different positions within one source is incorrect. 

2.  Based on the spatially resolved optical depths of H$^{13}$CN and H$^{13}$CO$^+$ 3-2,   we derived the optical depths of H$^{13}$CN 3-2  and H$^{13}$CO$^+$ 3-2 for these sources  by two ways, the average of spatially resolved 
$\tau_{sr}$ and the $\tau_{ave}$ from the averaged line ratios, while the latter is widely used in the extragalactic studies.  The optical depths obtained with two methods are agree well with each other, from about 0.04 to 0.17, with fitting result of $\tau_{ave}$=0.987$\tau_{sr}$+0.002 and coefficient of 0.997. 


3. No clear relation is found for the optical depths of H$^{13}$CN and H$^{13}$CO$^+$ 3-2 in the same sources, which means that  HCN/HCO$^+$ optical depth ratio can vary in different sources caused by chemical and/or physical conditions.

Even though the assumption  for an uniform optical depth in different positions of  spatially unresolved sources is incorrect, it is a good approximation to derive typical optical depth of molecular lines with isotopic line ratios. Further spatially resolved studies for pair of isotopic lines in  galaxies are necessary to do such test. 



\section*{Acknowledgements}
We would like to acknowledge the help of the staff of the SMT 10-m telescope for assistance with the observations. 
We also acknowledge Li Xiao, Hui Shi and Yan Duan for assisting with the remote observations.
We are grateful to the anonymous referee for the constructive report that improved this paper. 
This work is supported by the National Natural Science Foundation of China grant No. 11988101, 12173067, 12103024, 11725313, 12041302, 12041305, 12173016, 11903003 
and the fellowship of China Postdoctoral Science Foundation 2021M691531.  
Z.-Y Zhang acknowledge the Program for Innovative Talents, Entrepreneur in Jiangsu, 
acknowledge the science research grants from the China Manned Space Project with NOs.CMS-CSST-2021-A08 and CMS-CSST-2021-A07.
N.-Y. Tang is sponsored by Zhejiang Lab Open Research Project (NO. K2022PE0AB01), Cultivation Project for FAST Scientific Payoff and Research Achievement of CAMS-CAS, National key R\&D program of China under grant No. 2018YFE0202900  and the University Annual Scientific Research Plan of Anhui Province (NO.2022AH010013).

\section*{Data availability}
The derived optical depths of H$^{13}$CN and H$^{13}$CO$^+$ 3-2 for 30 sources are listed in Table~\ref{tab:tau-all}. 
The data underlying this paper will be shared on a reasonable request to the corresponding authors.









\appendix

\section{Figures of the 30 dense cores}
\label{sec:figs}

The figures presented below show the entirety data of the 30 dense cores, included the velocity integrated intensity maps of two pairs of lines, as well as the spatially resolved $\tau$ of H$^{13}$CN 3-2 and H$^{13}$CO$^+$ 3-2.


 \begin{figure*} 
    \centering
  \includegraphics[width=3.05in]{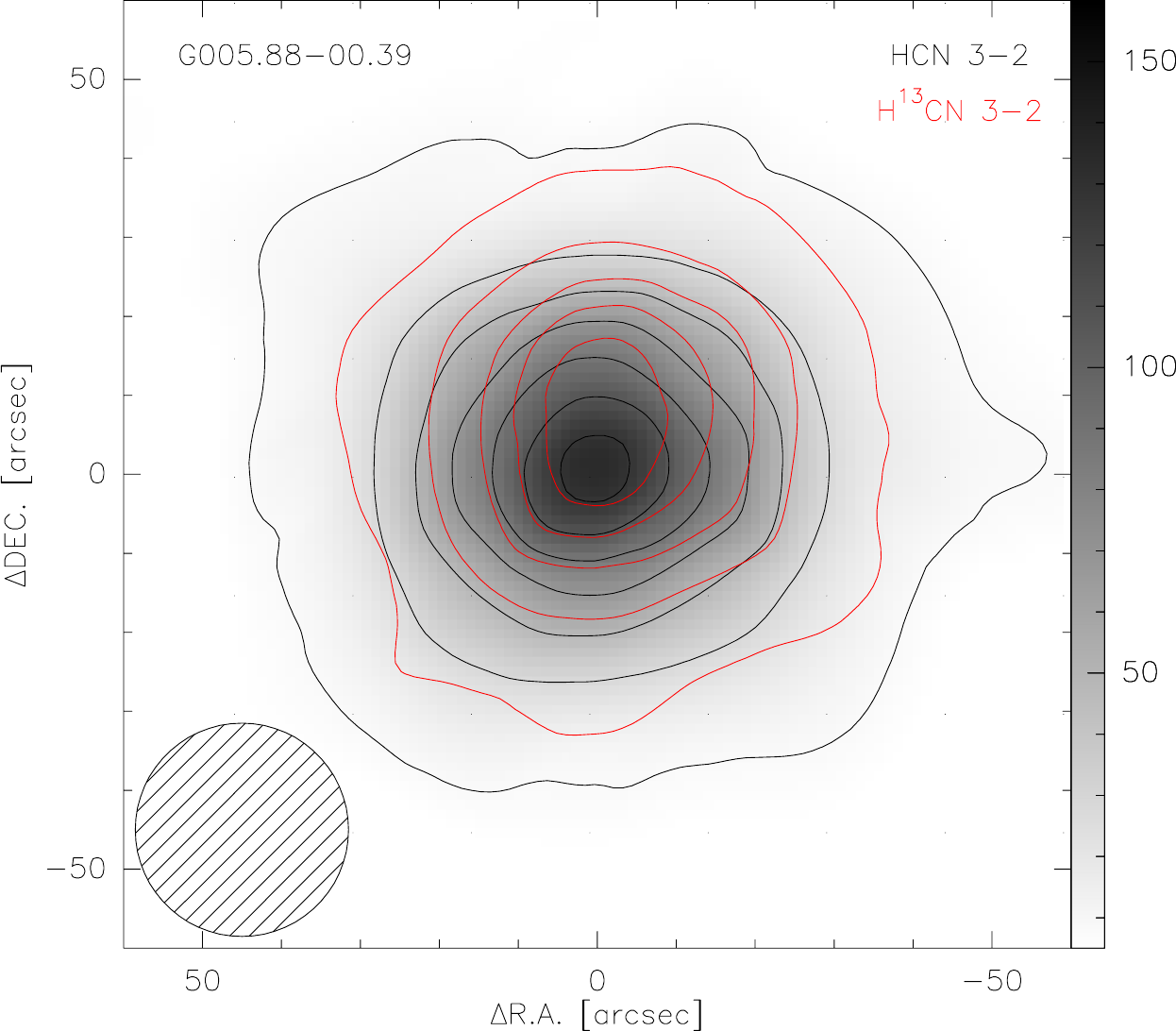}
   \includegraphics[width=3.03in]{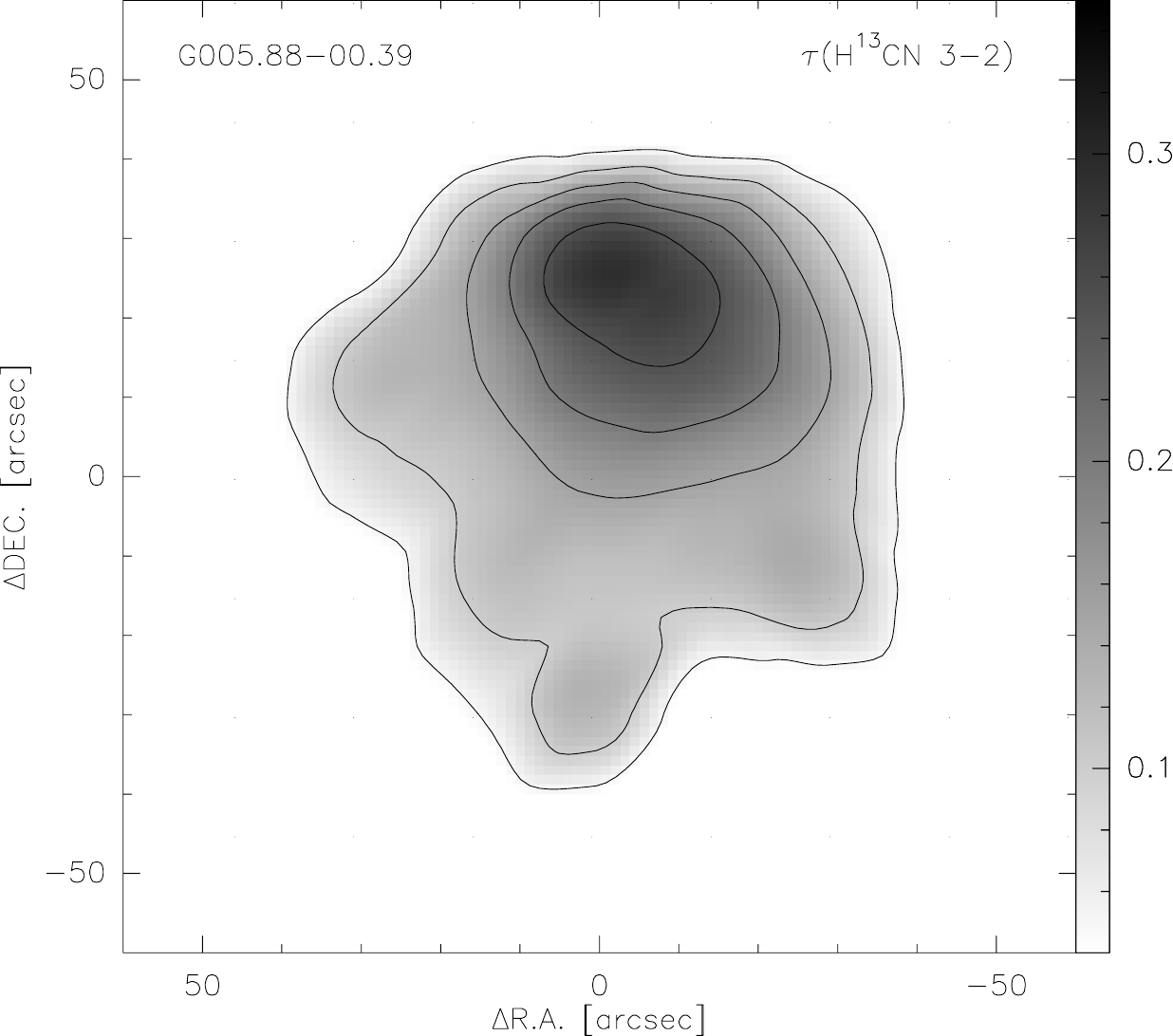}
       \includegraphics[width=3.05in]{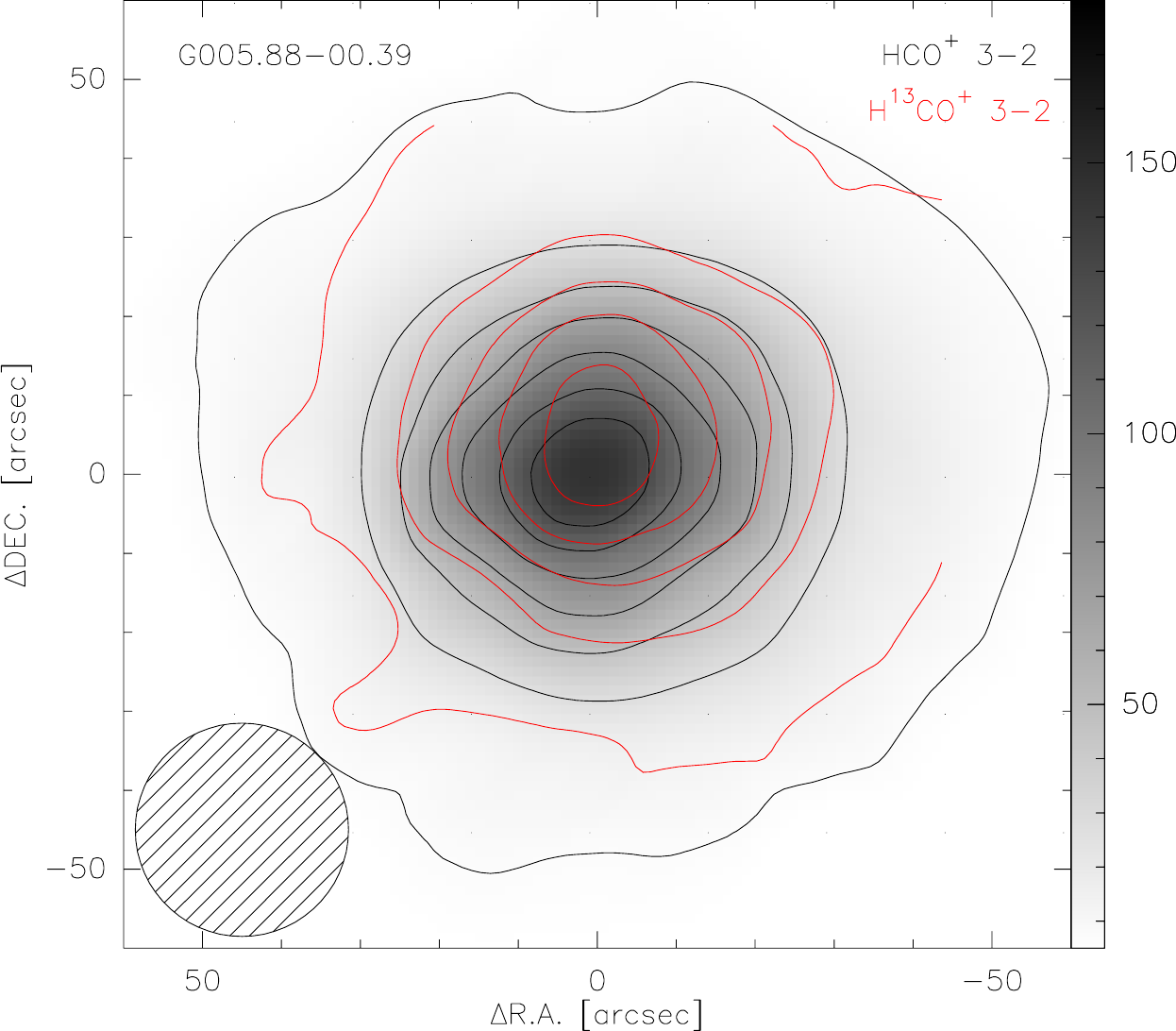}
       \includegraphics[width=3.08in]{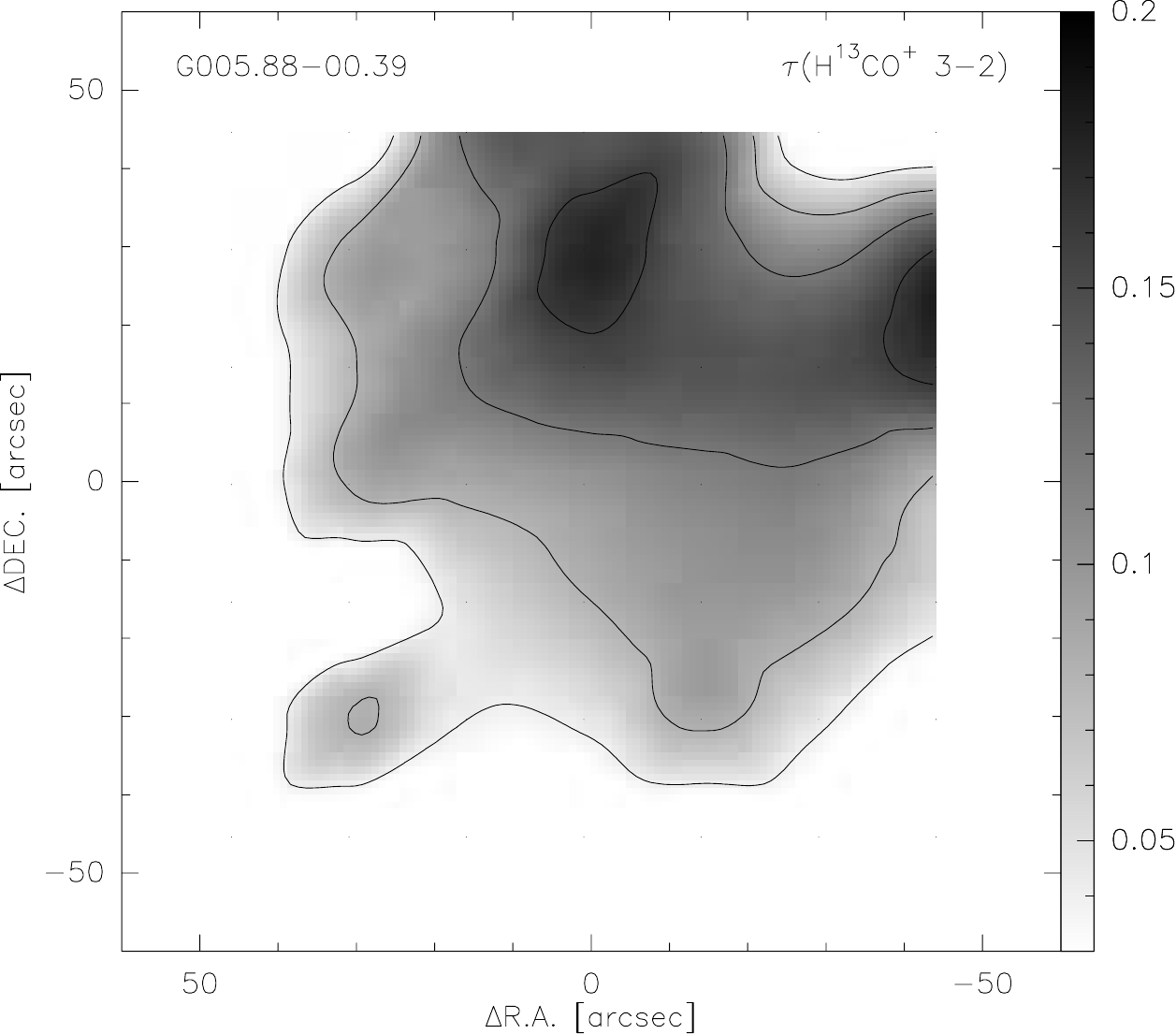}
 \caption{The data reduction results of G005.88-00.39. 
               {\it Top left:} The velocity integrated intensity maps of HCN and H$^{13}$CN 3-2. 
               The mapping size of HCN 3-2 is 2$'\times2'$, while it is 1.5$'\times1.5'$ for H$^{13}$CN 3-2, with a beam size of $\sim$ 27.8$''$.
               The grey scale and the black contour with levels starting from 8 K km s$^{-1}$ in step of 20 K km s$^{-1}$ show the observed HCN 3-2. 
               The red contour with levels starting from 1.4 K km s$^{-1}$ in step of 4 K km s$^{-1}$ represents H$^{13}$CN 3-2.
               {\it Top right:} The spatially resolved $\tau(\textrm{H}^{13}\textrm{CN})$ of G005.88-00.39 is demonstrated by black contour with levels 
               starting from 0.05 in step of 0.05. 
                {\it Bottom left:} The velocity integrated intensity maps of HCO$^+$ and H$^{13}$CO$^+$ 3-2. 
               The mapping size of HCO$^+$ 3-2 is 2$'\times2'$, while it is 1.5$'\times1.5'$ for H$^{13}$CO$^+$ 3-2, with a beam size of $\sim$ 27.8$''$.
               The grey scale and the black contour with levels starting from 7 K km s$^{-1}$ in step of 20 K km s$^{-1}$ show the observed HCO$^+$ 3-2. 
               The red contour with levels starting from 1.4 K km s$^{-1}$ in step of 4 K km s$^{-1}$ represents H$^{13}$CO$^+$ 3-2.
                {\it Bottom right:} The spatially resolved $\tau(\textrm{H}^{13}\textrm{CO$^+$})$ of G005.88-00.39 is demonstrated by black contour with levels 
               starting from 0.04 in step of 0.04. 
                }       
 \label{fig:g00588}
\end{figure*}


 \begin{figure*} 
    \centering
  \includegraphics[width=3.05in]{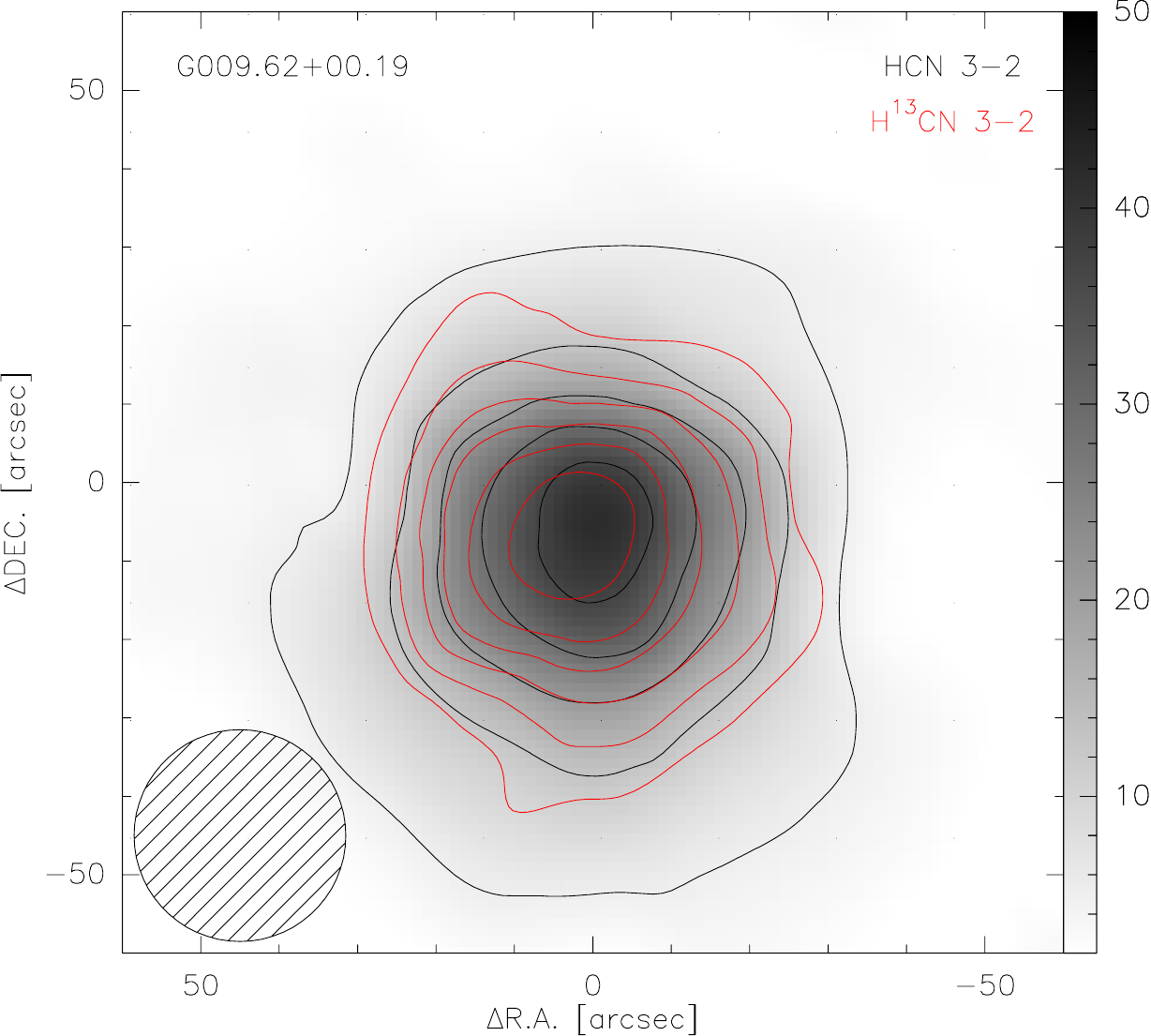}
   \includegraphics[width=3.03in]{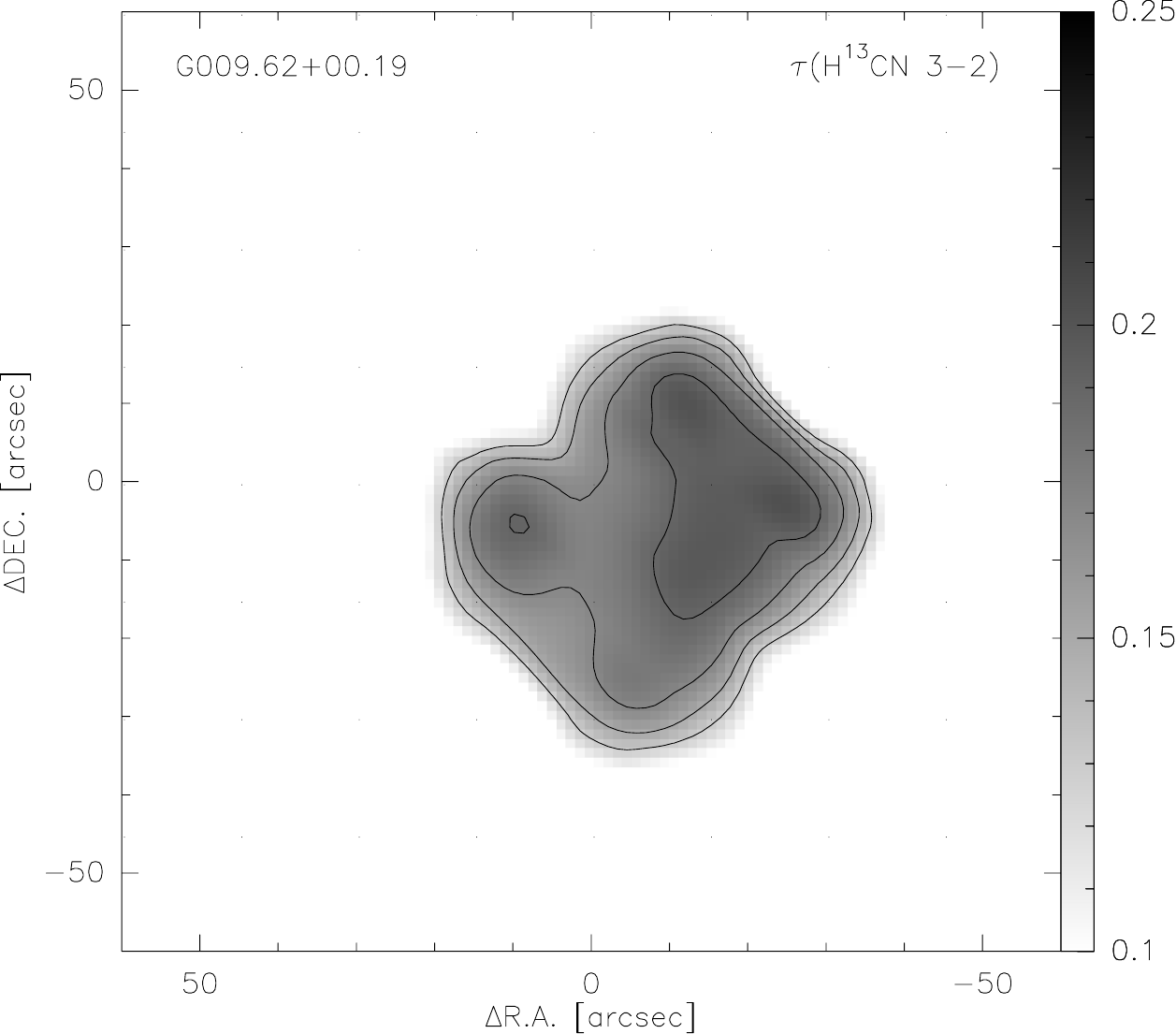}
       \includegraphics[width=3.05in]{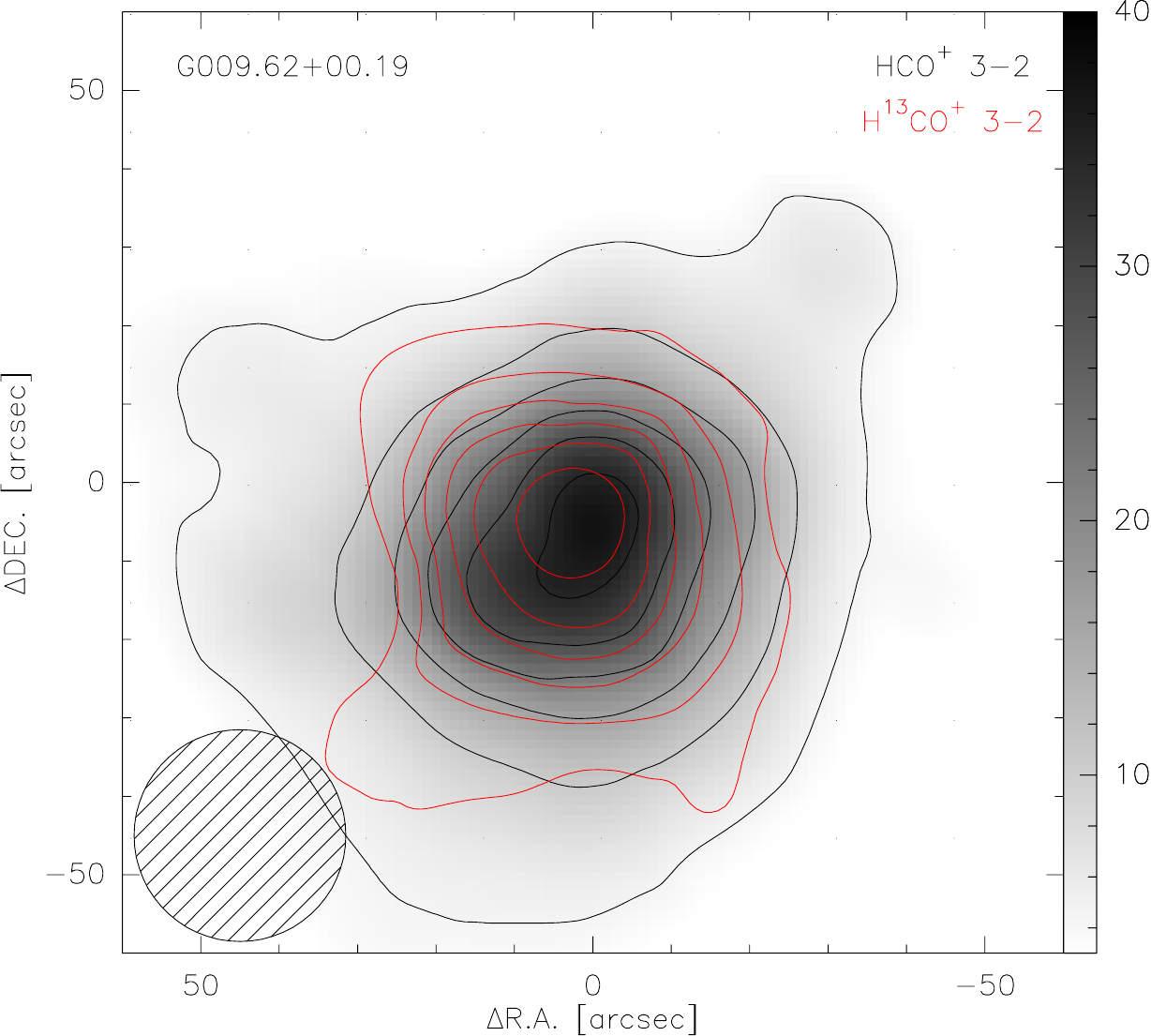}
       \includegraphics[width=3.08in]{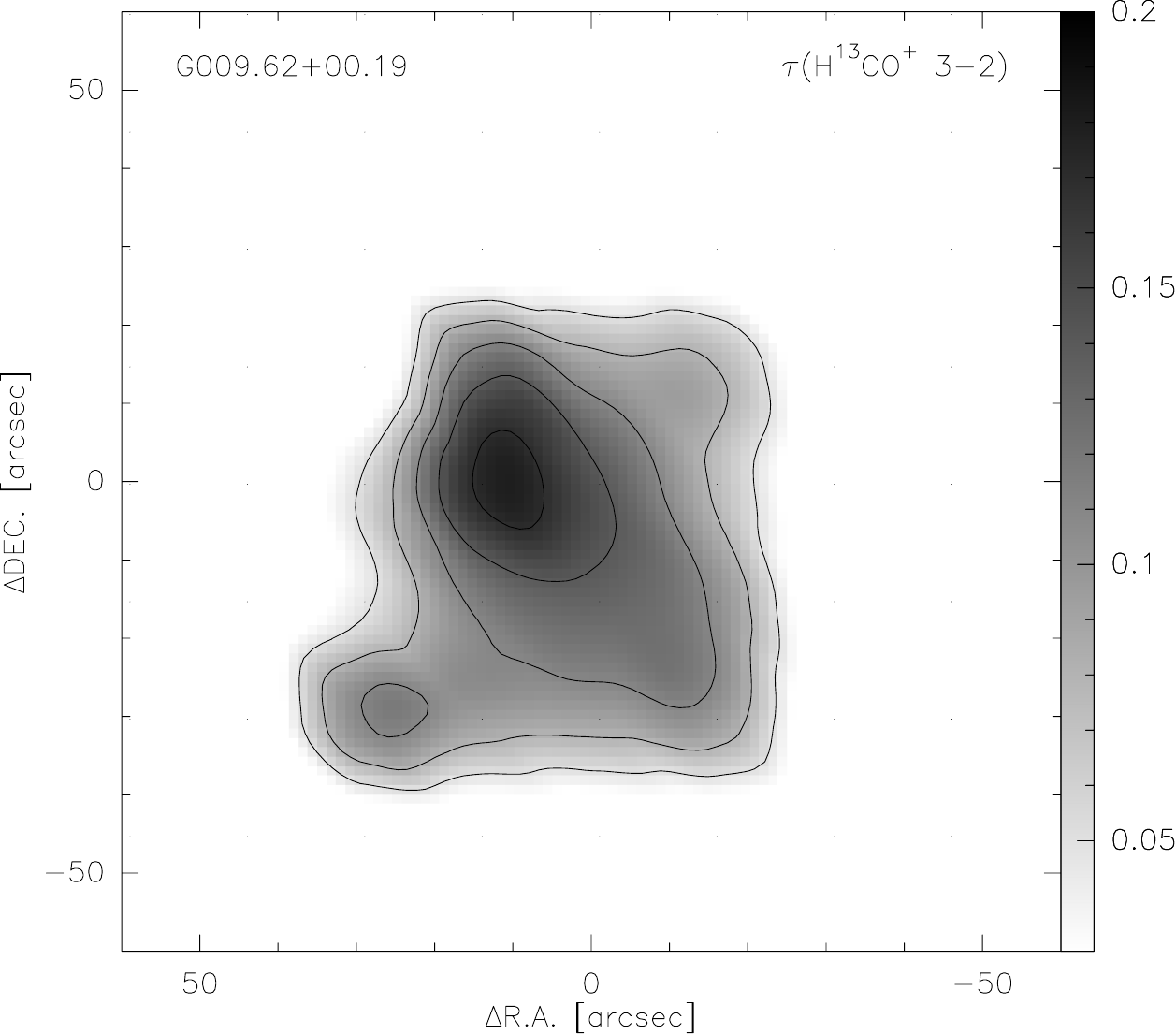}
 \caption{The data reduction results of G009.62+00.19. 
               {\it Top left:} The velocity integrated intensity maps of HCN and H$^{13}$CN 3-2. 
               The mapping size of HCN 3-2 is 2$'\times2'$, while it is 1.5$'\times1.5'$ for H$^{13}$CN 3-2, with a beam size of $\sim$ 27.8$''$.
               The grey scale and the black contour with levels starting from 4 K km s$^{-1}$ in step of 8 K km s$^{-1}$ show the observed HCN 3-2. 
               The red contour with levels starting from 1 K km s$^{-1}$ in step of 1 K km s$^{-1}$ represents H$^{13}$CN 3-2.
               {\it Top right:} The spatially resolved $\tau(\textrm{H}^{13}\textrm{CN})$ of G009.62+00.19 is demonstrated by black contour with levels 
               starting from 0.13 in step of 0.2. 
                {\it Bottom left:} The velocity integrated intensity maps of HCO$^+$ and H$^{13}$CO$^+$ 3-2. 
               The mapping size of HCO$^+$ 3-2 is 2$'\times2'$, while it is 1.5$'\times1.5'$ for H$^{13}$CO$^+$ 3-2, with a beam size of $\sim$ 27.8$''$.
               The grey scale and the black contour with levels starting from 4 K km s$^{-1}$ in step of 6 K km s$^{-1}$ show the observed HCO$^+$ 3-2. 
               The red contour with levels starting from 0.6 K km s$^{-1}$ in step of 0.8 K km s$^{-1}$ represents H$^{13}$CO$^+$ 3-2.
                {\it Bottom right:} The spatially resolved $\tau(\textrm{H}^{13}\textrm{CO$^+$})$ of G009.62+00.19 is demonstrated by black contour with levels 
               starting from 0.05 in step of 0.03. 
                }       
 \label{fig:g00962}
\end{figure*}


 \begin{figure*} 
    \centering
  \includegraphics[width=3.05in]{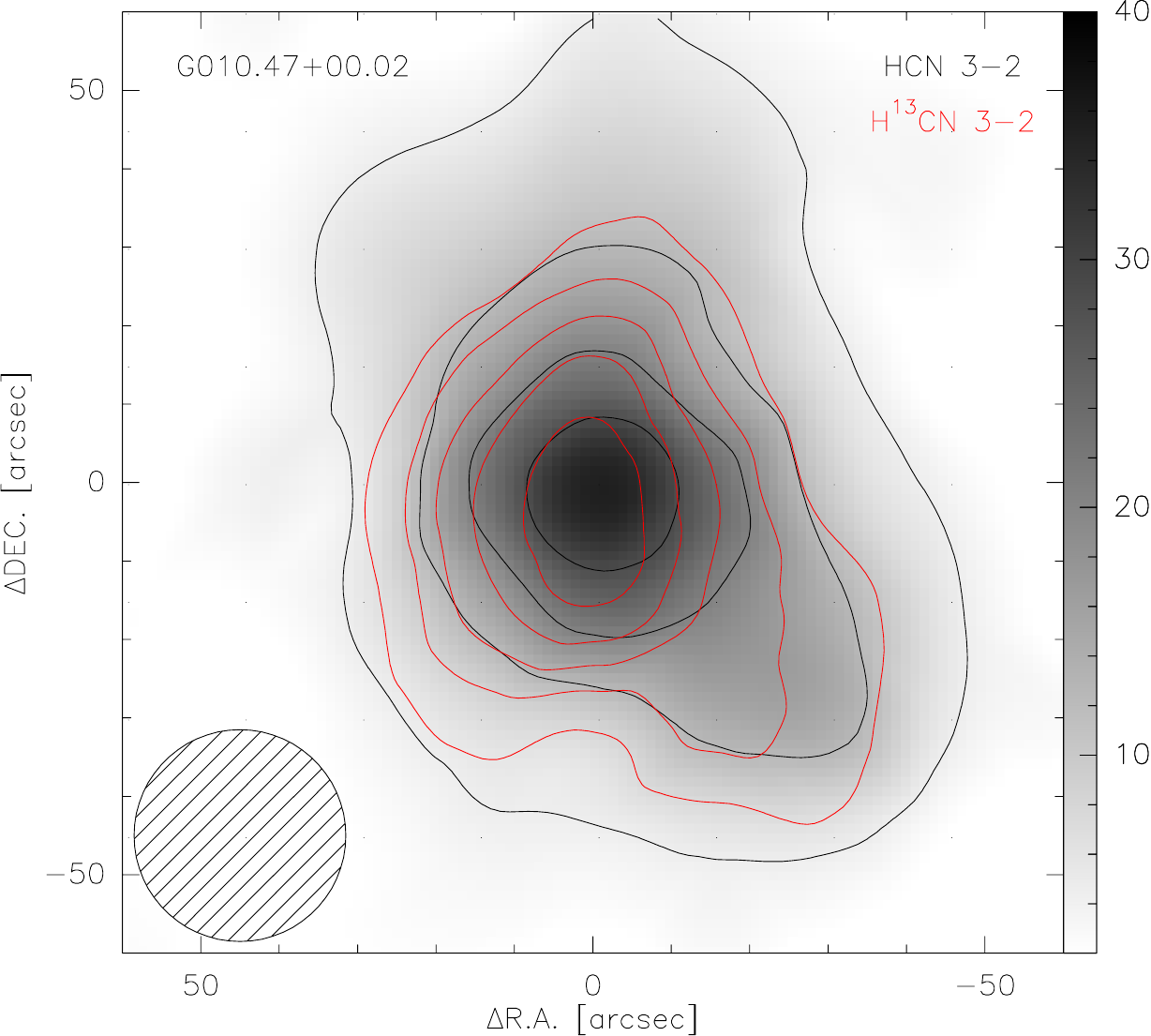}
   \includegraphics[width=3.03in]{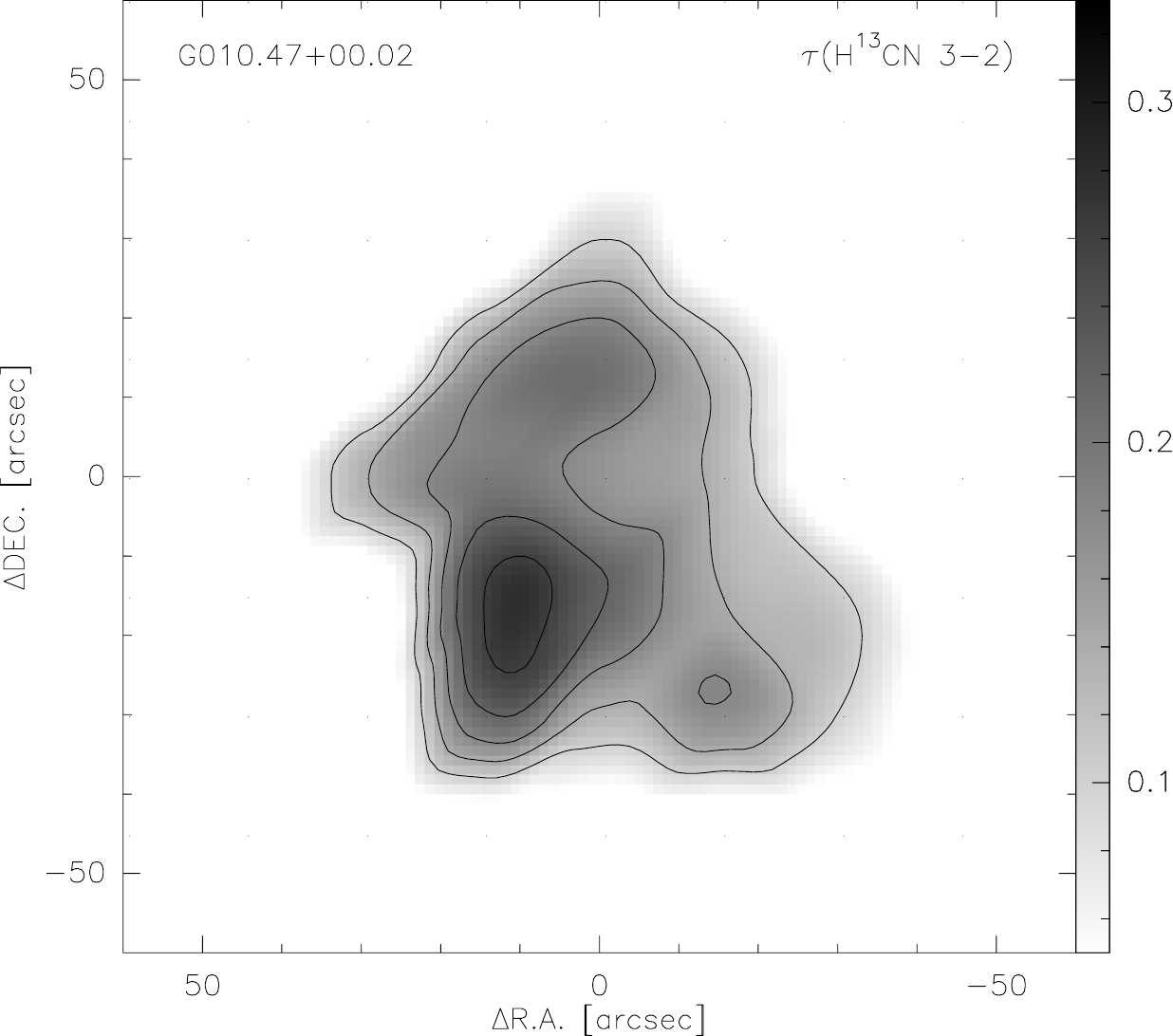}
       \includegraphics[width=3.05in]{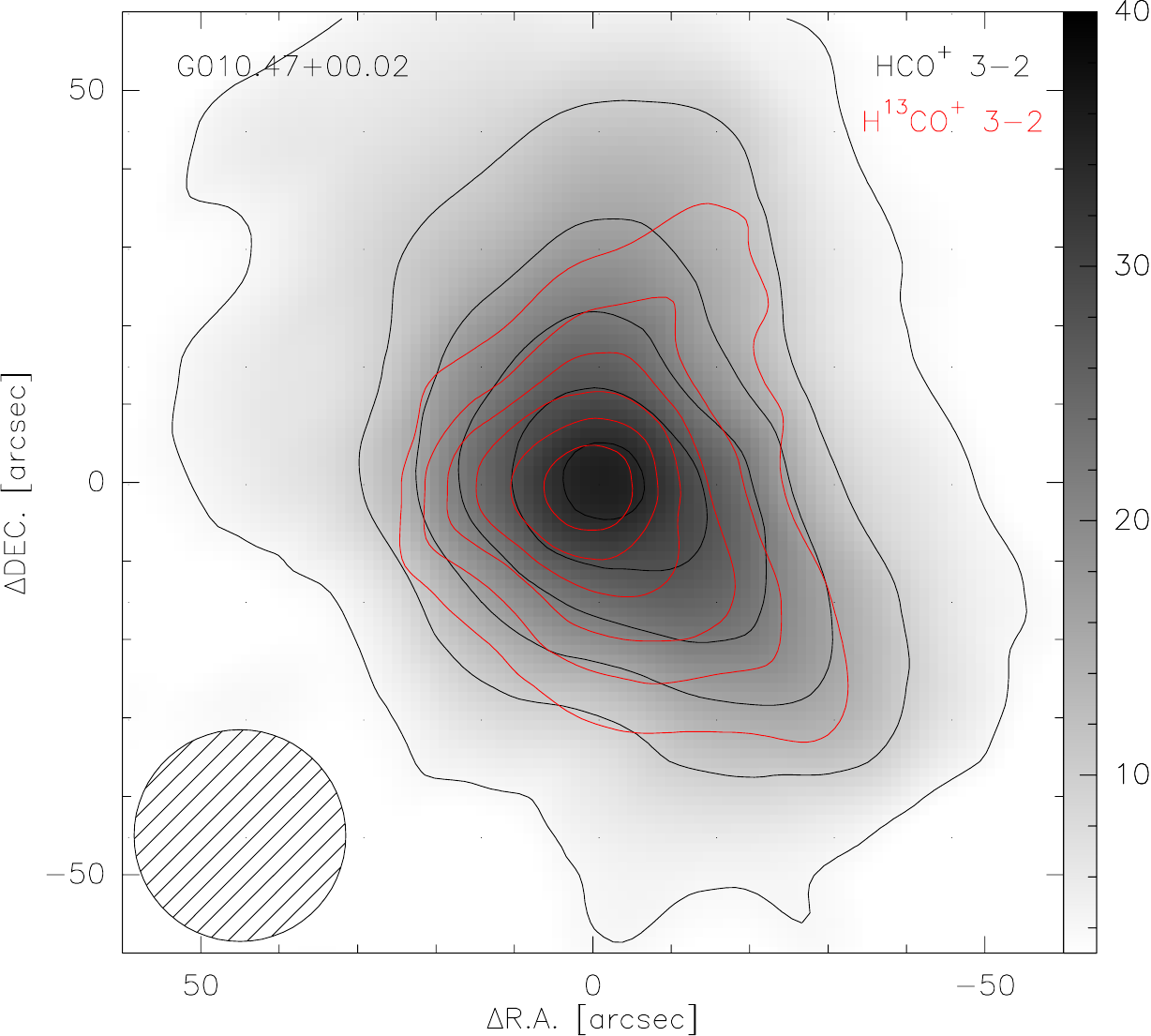}
       \includegraphics[width=3.08in]{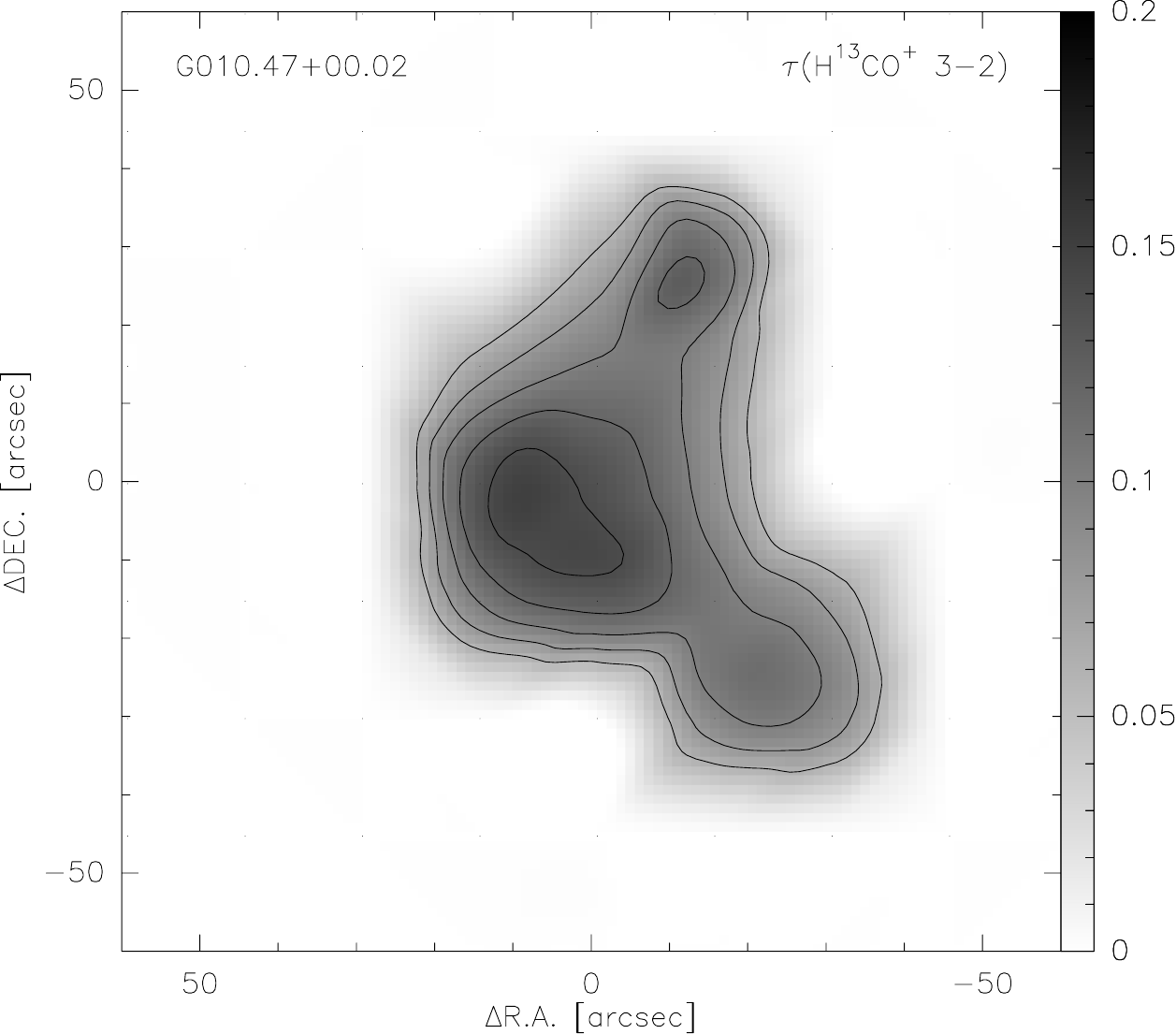}
 \caption{The data reduction results of G010.47+00.02. 
               {\it Top left:} The velocity integrated intensity maps of HCN and H$^{13}$CN 3-2. 
               The mapping size of HCN 3-2 is 2$'\times2'$, while it is 1.5$'\times1.5'$ for H$^{13}$CN 3-2, with a beam size of $\sim$ 27.8$''$.
               The grey scale and the black contour with levels starting from 3.5 K km s$^{-1}$ in step of 8 K km s$^{-1}$ show the observed HCN 3-2. 
               The red contour with levels starting from 0.85 K km s$^{-1}$ in step of 1 K km s$^{-1}$ represents H$^{13}$CN 3-2.
               {\it Top right:} The spatially resolved $\tau(\textrm{H}^{13}\textrm{CN})$ of G010.47+00.02 is demonstrated by black contour with levels 
               starting from 0.1 in step of 0.04. 
                {\it Bottom left:} The velocity integrated intensity maps of HCO$^+$ and H$^{13}$CO$^+$ 3-2. 
               The mapping size of HCO$^+$ 3-2 is 2$'\times2'$, while it is 1.5$'\times1.5'$ for H$^{13}$CO$^+$ 3-2, with a beam size of $\sim$ 27.8$''$.
               The grey scale and the black contour with levels starting from 4 K km s$^{-1}$ in step of 6 K km s$^{-1}$ show the observed HCO$^+$ 3-2. 
               The red contour with levels starting from 1.2 K km s$^{-1}$ in step of 0.6 K km s$^{-1}$ represents H$^{13}$CO$^+$ 3-2.
                {\it Bottom right:} The spatially resolved $\tau(\textrm{H}^{13}\textrm{CO$^+$})$ of G010.47+00.02 is demonstrated by black contour with levels 
               starting from 0.06 in step of 0.02. 
                }       
 \label{fig:g01047}
\end{figure*}


 \begin{figure*} 
    \centering
  \includegraphics[width=3.05in]{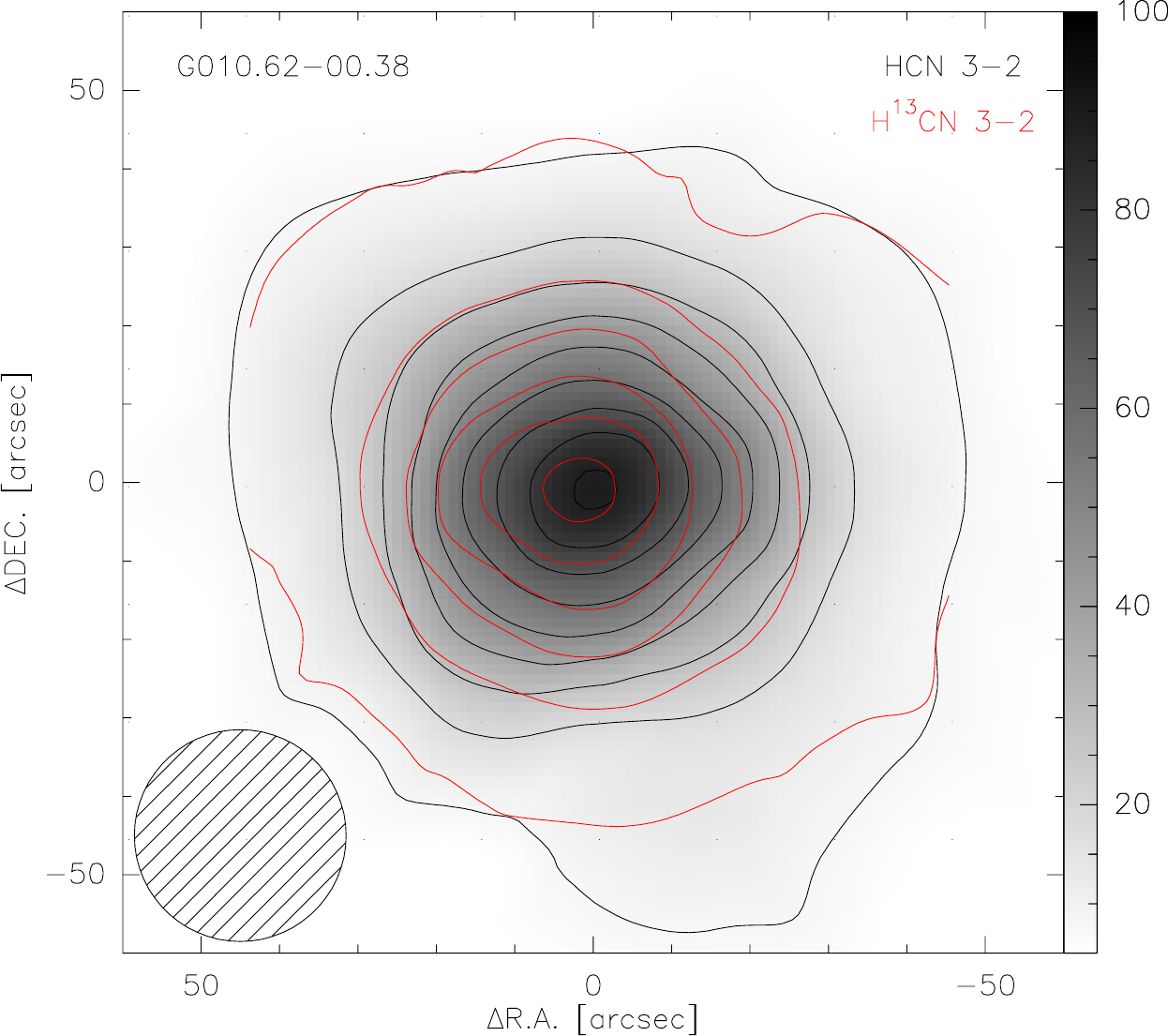}
   \includegraphics[width=3.03in]{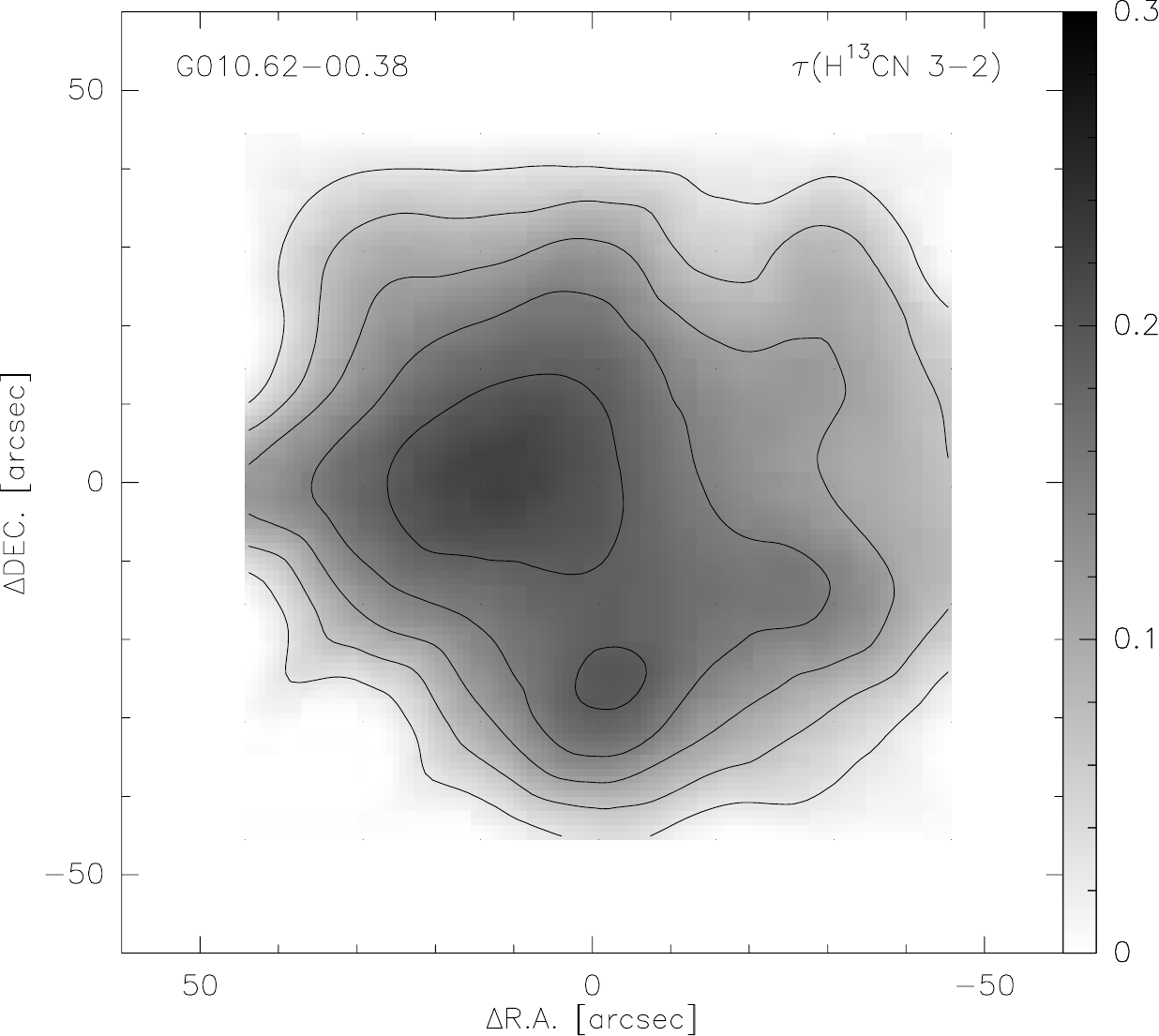}
       \includegraphics[width=3.05in]{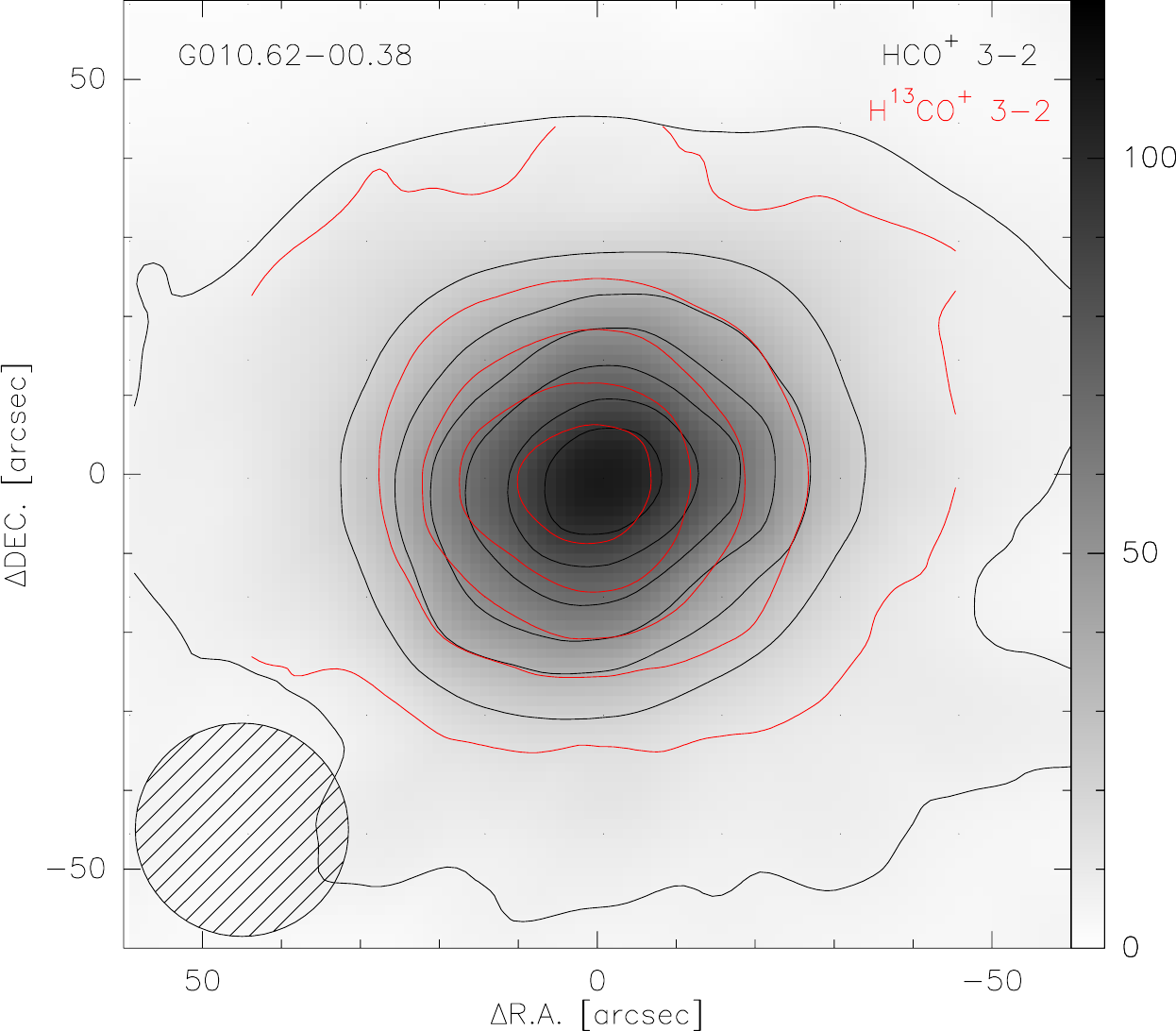}
       \includegraphics[width=3.08in]{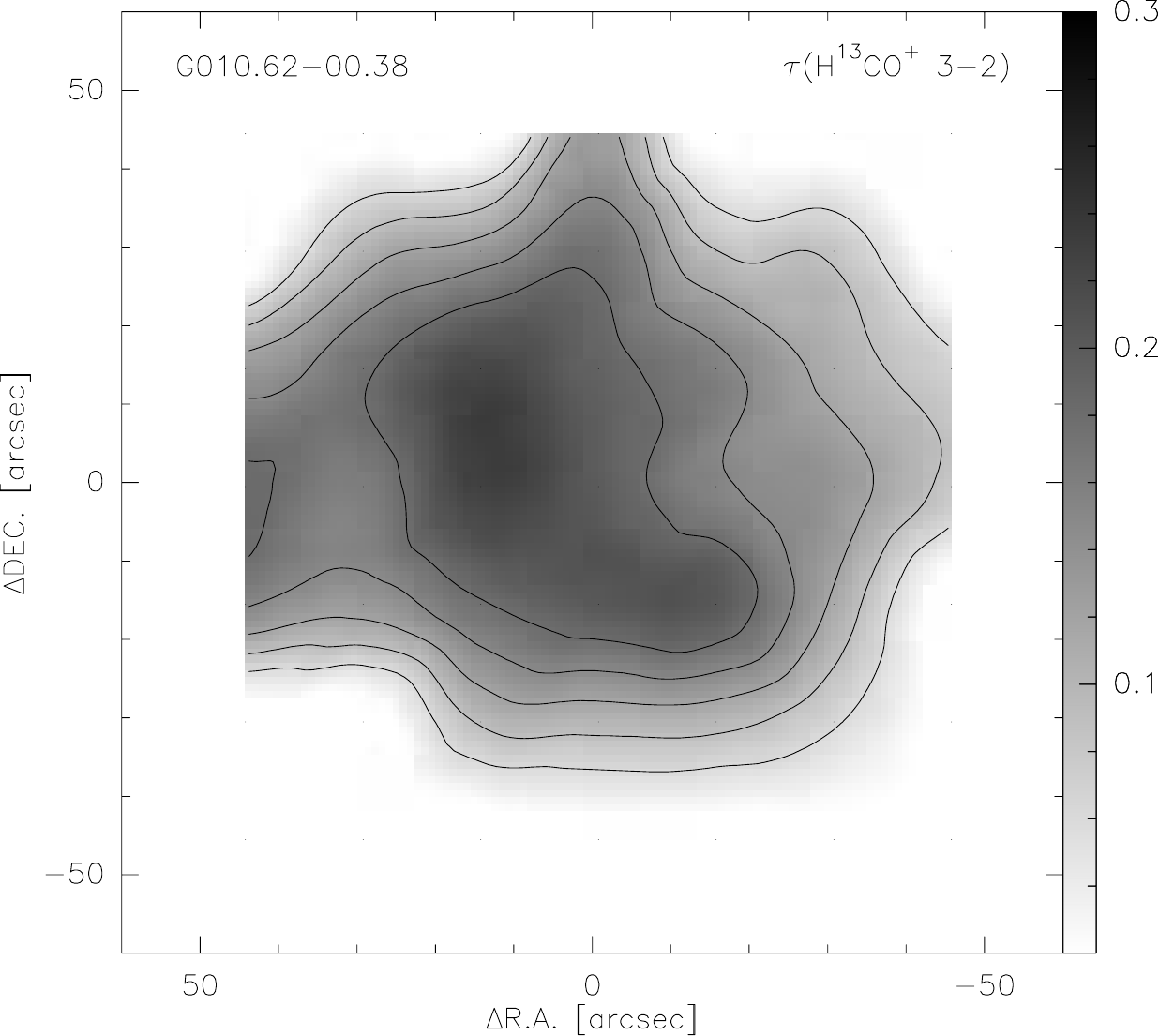}
 \caption{The data reduction results of G010.62-00.38. 
               {\it Top left:} The velocity integrated intensity maps of HCN and H$^{13}$CN 3-2. 
               The mapping size of HCN 3-2 is 2$'\times2'$, while it is 1.5$'\times1.5'$ for H$^{13}$CN 3-2, with a beam size of $\sim$ 27.8$''$.
               The grey scale and the black contour with levels starting from 8 K km s$^{-1}$ in step of 10 K km s$^{-1}$ show the observed HCN 3-2. 
               The red contour with levels starting from 0.6 K km s$^{-1}$ in step of 3 K km s$^{-1}$ represents H$^{13}$CN 3-2.
               {\it Top right:} The spatially resolved $\tau(\textrm{H}^{13}\textrm{CN})$ of G010.62-00.38 is demonstrated by black contour with levels 
               starting from 0.03 in step of 0.04. 
                {\it Bottom left:} The velocity integrated intensity maps of HCO$^+$ and H$^{13}$CO$^+$ 3-2. 
               The mapping size of HCO$^+$ 3-2 is 2$'\times2'$, while it is 1.5$'\times1.5'$ for H$^{13}$CO$^+$ 3-2, with a beam size of $\sim$ 27.8$''$.
               The grey scale and the black contour with levels starting from 6 K km s$^{-1}$ in step of 15 K km s$^{-1}$ show the observed HCO$^+$ 3-2. 
               The red contour with levels starting from 0.85 K km s$^{-1}$ in step of 4 K km s$^{-1}$ represents H$^{13}$CO$^+$ 3-2.
                {\it Bottom right:} The spatially resolved $\tau(\textrm{H}^{13}\textrm{CO$^+$})$ of G010.62-00.38 is demonstrated by black contour with levels 
               starting from 0.06 in step of 0.03. 
                }       
 \label{fig:g01062}
\end{figure*}


 \begin{figure*} 
    \centering
  \includegraphics[width=3.05in]{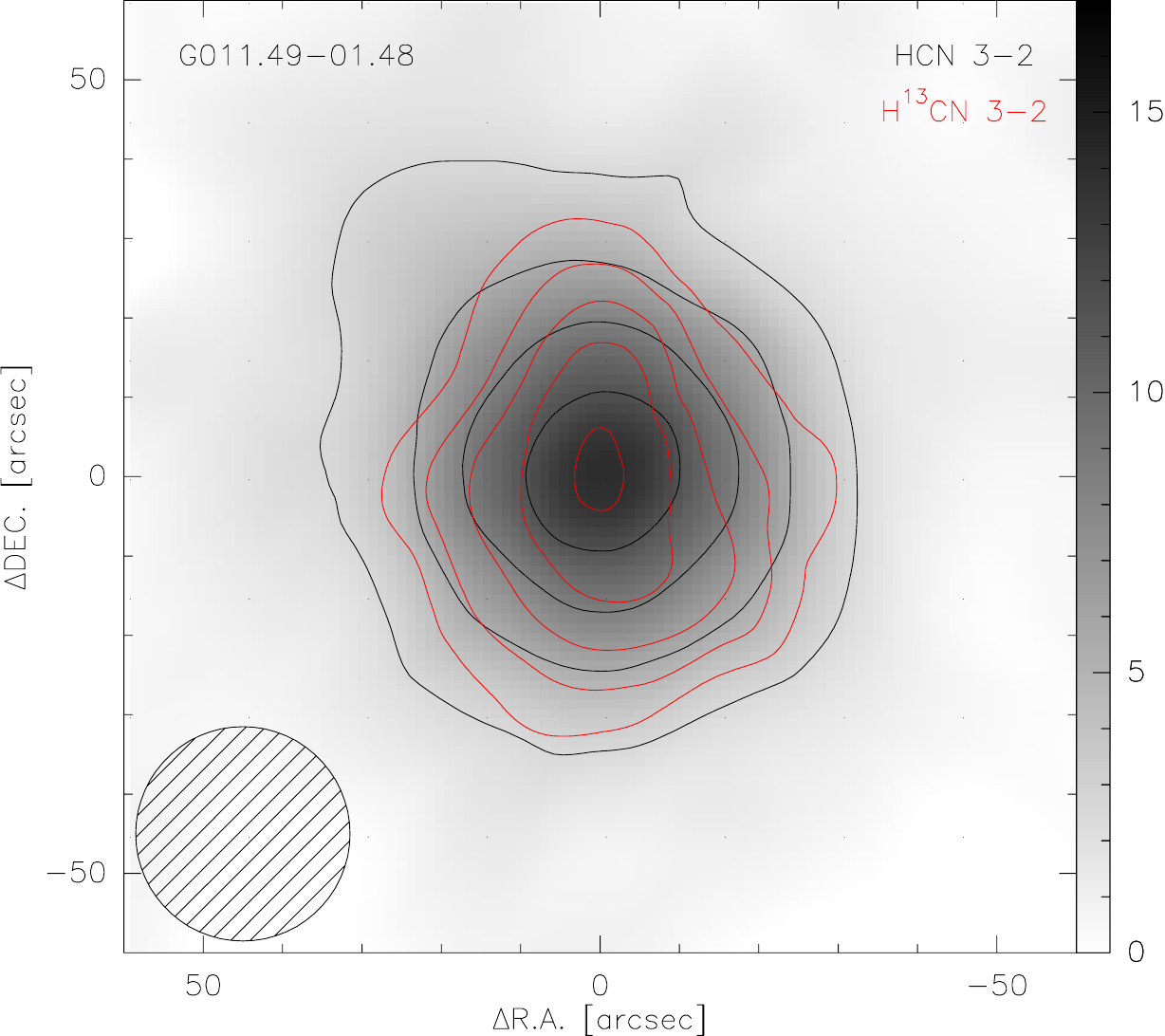}
   \includegraphics[width=3.03in]{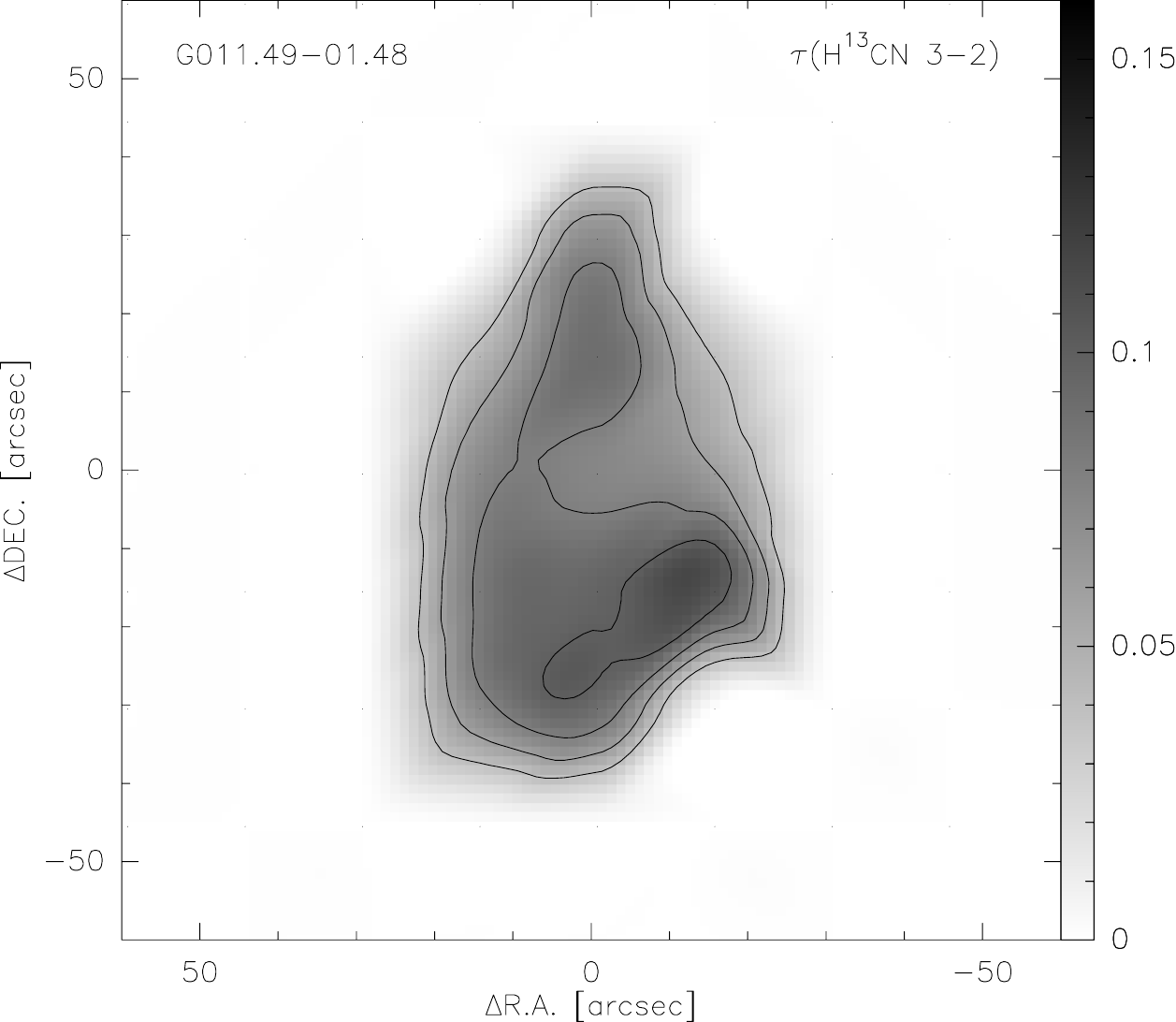}
       \includegraphics[width=3.05in]{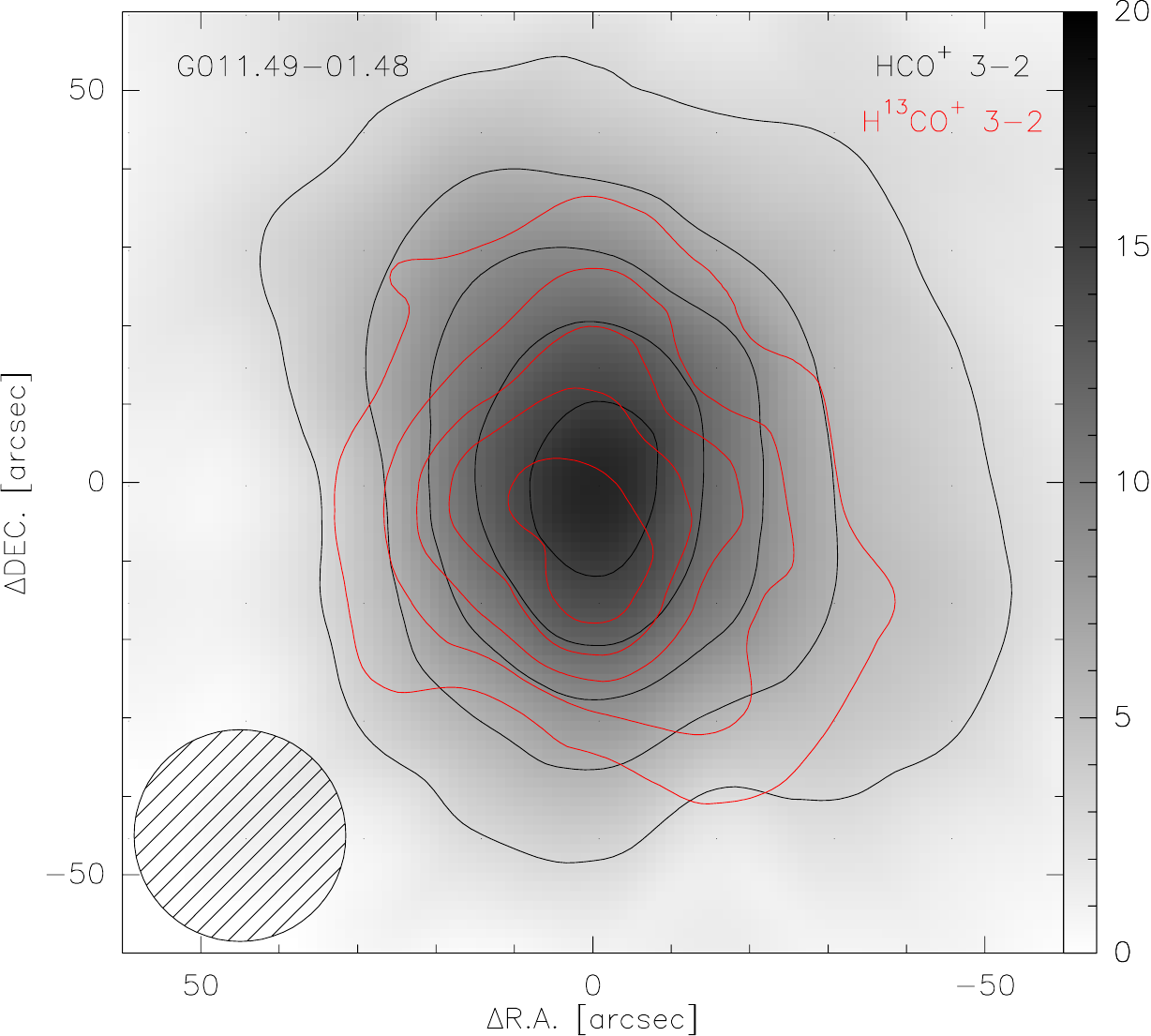}
       \includegraphics[width=3.08in]{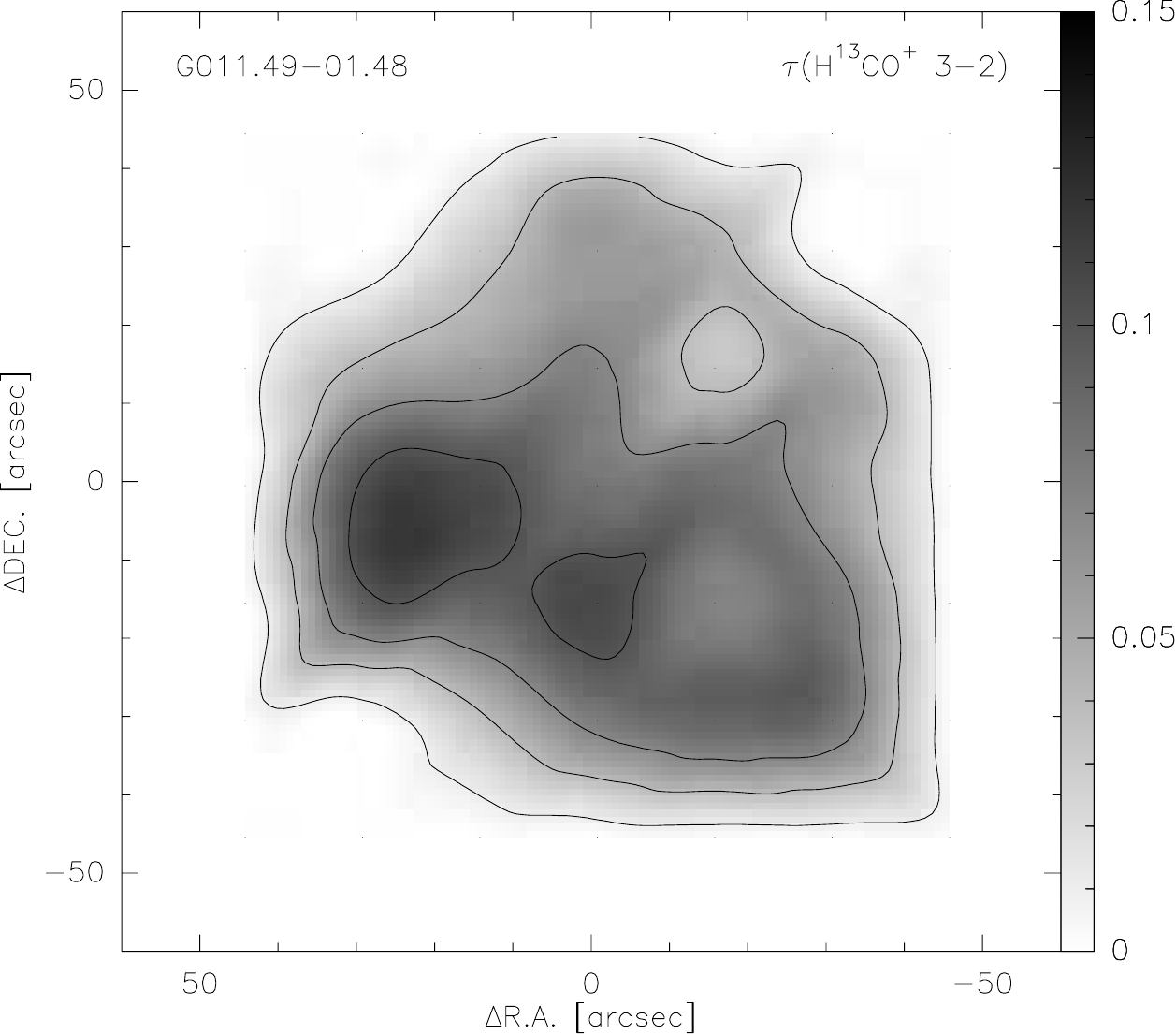}
 \caption{The data reduction results of G011.49-01.48. 
               {\it Top left:} The velocity integrated intensity maps of HCN and H$^{13}$CN 3-2. 
               The mapping size of HCN 3-2 is 2$'\times2'$, while it is 1.5$'\times1.5'$ for H$^{13}$CN 3-2, with a beam size of $\sim$ 27.8$''$.
               The grey scale and the black contour with levels starting from 2.5 K km s$^{-1}$ in step of 3 K km s$^{-1}$ show the observed HCN 3-2. 
               The red contour with levels starting from 0.25 K km s$^{-1}$ in step of 0.2 K km s$^{-1}$ represents H$^{13}$CN 3-2.
               {\it Top right:} The spatially resolved $\tau(\textrm{H}^{13}\textrm{CN})$ of G011.49-01.48 is demonstrated by black contour with levels 
               starting from 0.04 in step of 0.02. 
                {\it Bottom left:} The velocity integrated intensity maps of HCO$^+$ and H$^{13}$CO$^+$ 3-2. 
               The mapping size of HCO$^+$ 3-2 is 2$'\times2'$, while it is 1.5$'\times1.5'$ for H$^{13}$CO$^+$ 3-2, with a beam size of $\sim$ 27.8$''$.
               The grey scale and the black contour with levels starting from 3 K km s$^{-1}$ in step of 3 K km s$^{-1}$ show the observed HCO$^+$ 3-2. 
               The red contour with levels starting from 0.35 K km s$^{-1}$ in step of 0.25 K km s$^{-1}$ represents H$^{13}$CO$^+$ 3-2.
                {\it Bottom right:} The spatially resolved $\tau(\textrm{H}^{13}\textrm{CO$^+$})$ of G011.49-01.48 is demonstrated by black contour with levels 
               starting from 0.01 in step of 0.03. 
                }       
 \label{fig:g01149}
\end{figure*}


 \begin{figure*} 
    \centering
  \includegraphics[width=3.05in]{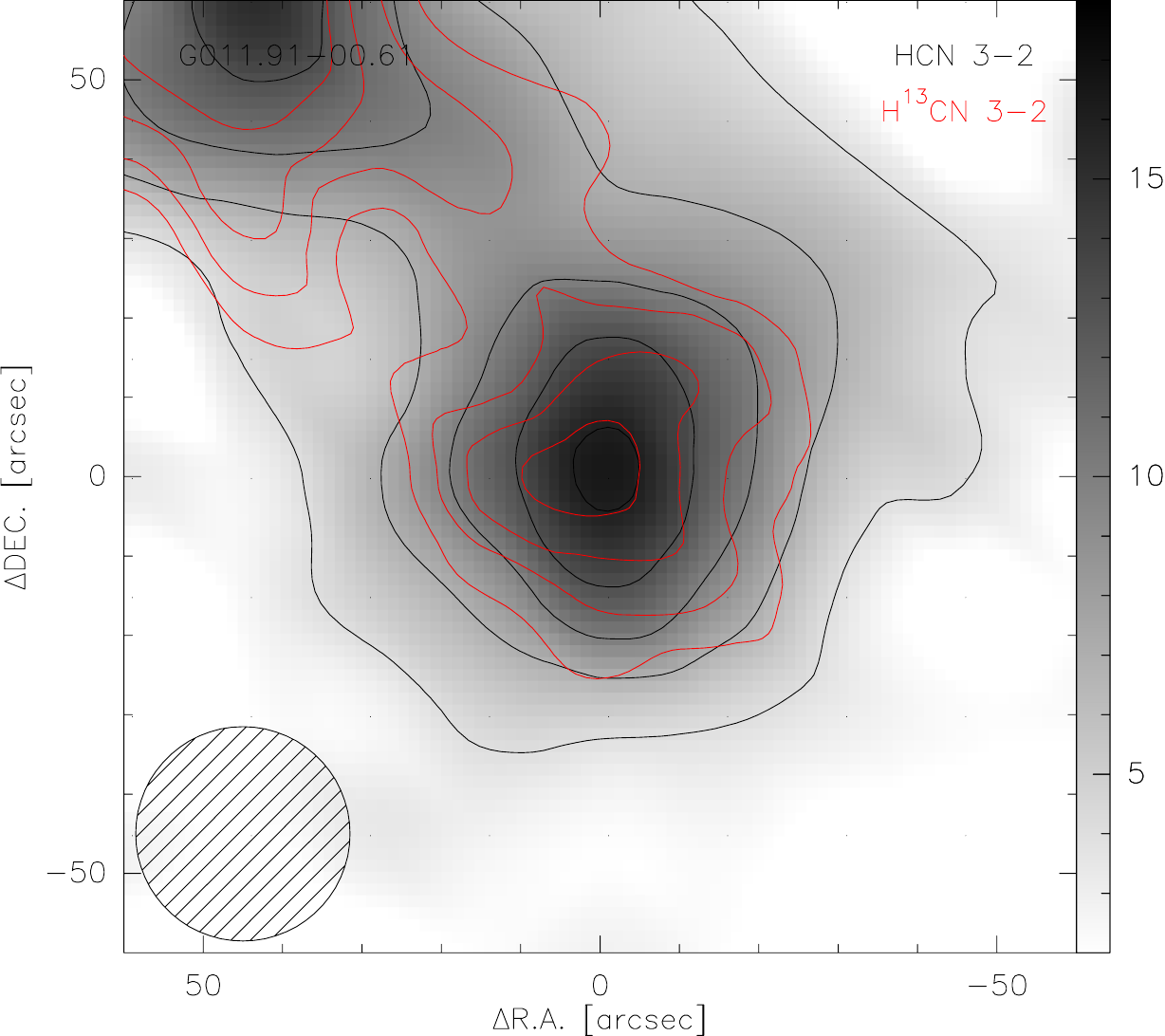}
   \includegraphics[width=3.03in]{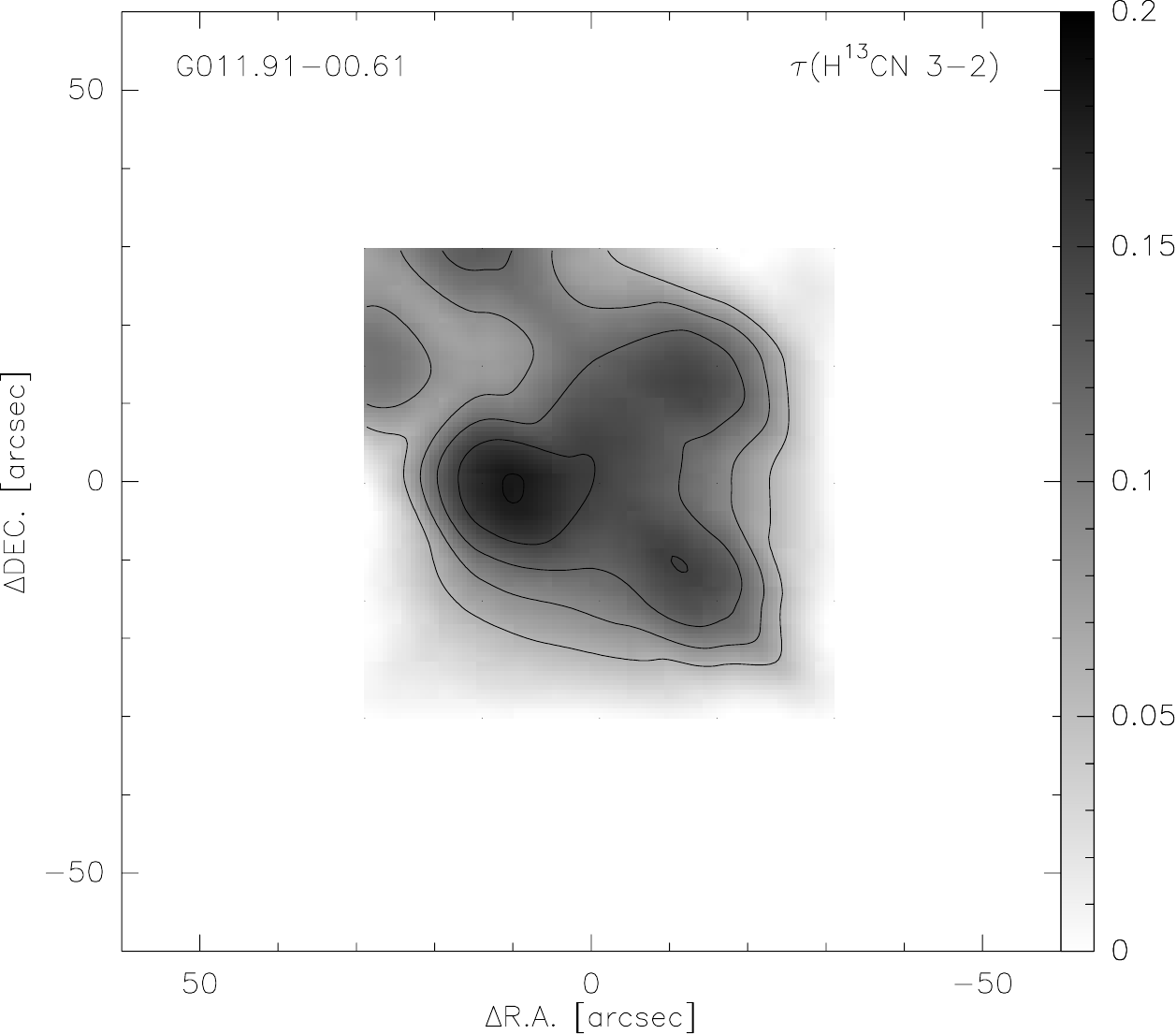}
       \includegraphics[width=3.05in]{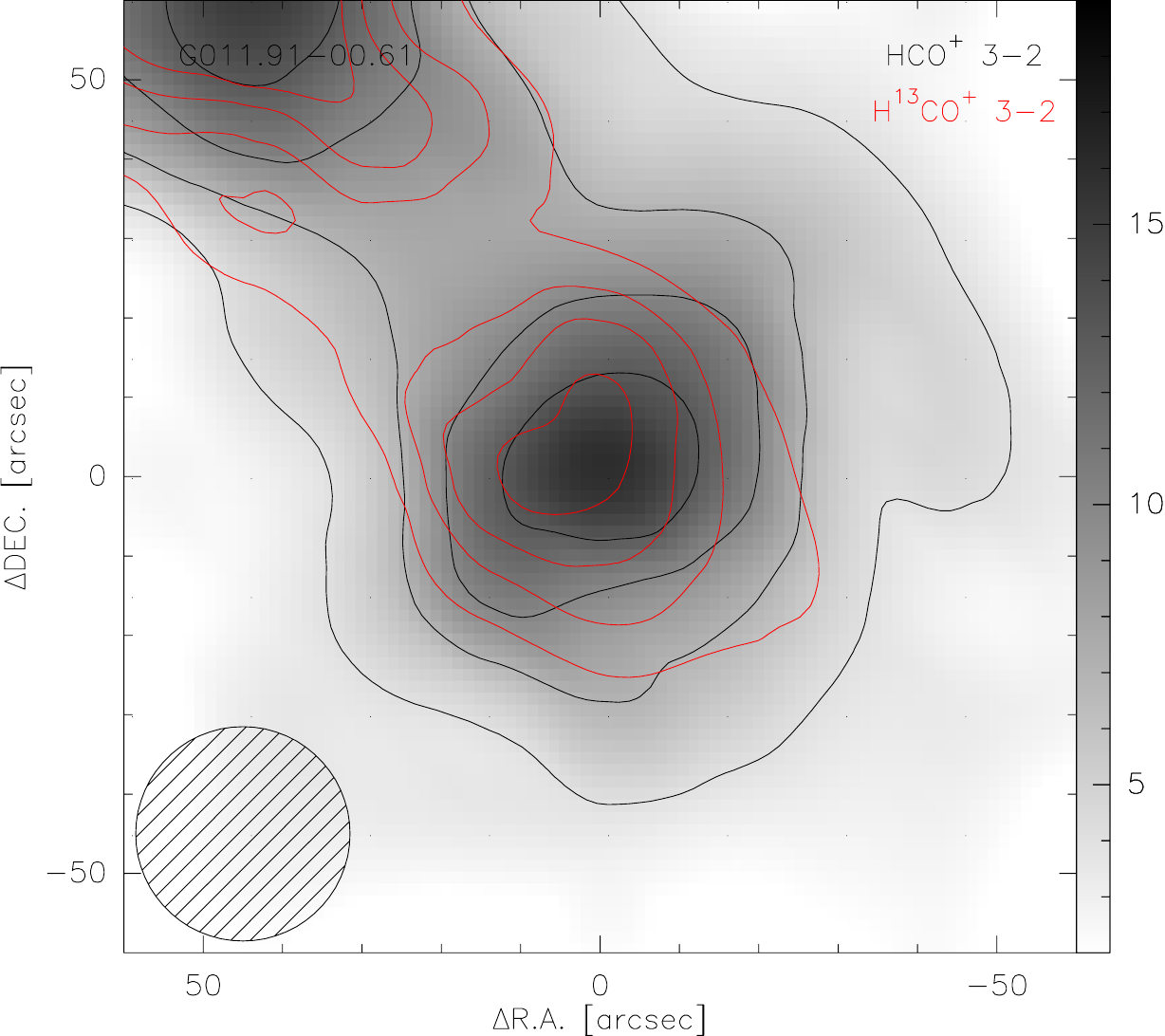}
       \includegraphics[width=3.08in]{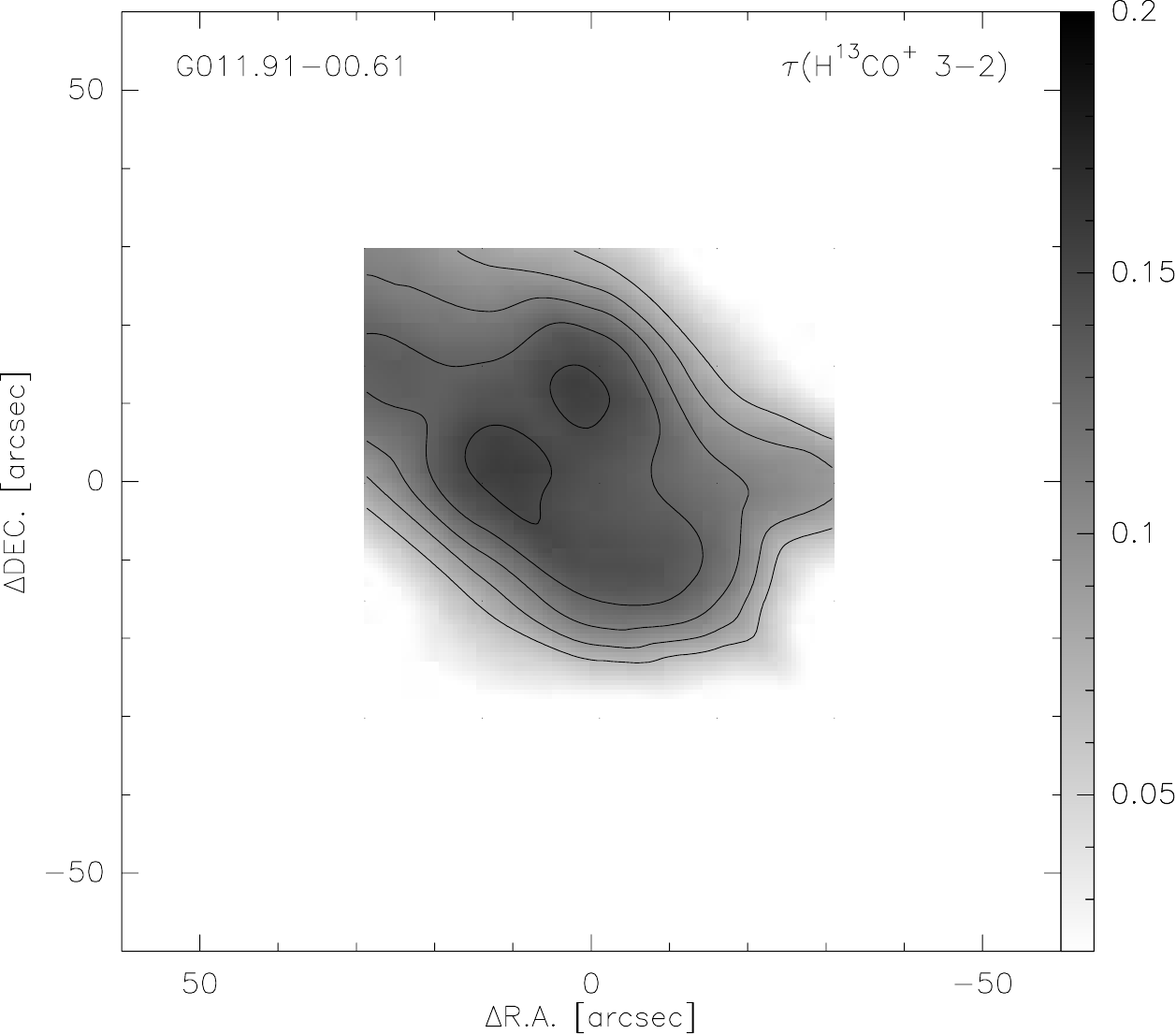}
 \caption{The data reduction results of G011.91-00.61, which is located in the center of the image. Another obvious core is found in the northeast region, 
               which is not fully covered in the observation and also not adopted in the $\tau$ calculation.  
               {\it Top left:} The velocity integrated intensity maps of HCN and H$^{13}$CN 3-2. 
               The mapping size of HCN 3-2 is 2$'\times2'$, while it is 1.5$'\times1.5'$ for H$^{13}$CN 3-2, with a beam size of $\sim$ 27.8$''$.
               The grey scale and the black contour with levels starting from 4 K km s$^{-1}$ in step of 3 K km s$^{-1}$ show the observed HCN 3-2. 
               The red contour with levels starting from 0.6 K km s$^{-1}$ in step of 0.5 K km s$^{-1}$ represents H$^{13}$CN 3-2.
               {\it Top right:} The spatially resolved $\tau(\textrm{H}^{13}\textrm{CN})$ of G011.91-00.61 is demonstrated by black contour with levels 
               starting from 0.06 in step of 0.03. 
                {\it Bottom left:} The velocity integrated intensity maps of HCO$^+$ and H$^{13}$CO$^+$ 3-2. 
               The mapping size of HCO$^+$ 3-2 is 2$'\times2'$, while it is 1.5$'\times1.5'$ for H$^{13}$CO$^+$ 3-2, with a beam size of $\sim$ 27.8$''$.
               The grey scale and the black contour with levels starting from 4 K km s$^{-1}$ in step of 3 K km s$^{-1}$ show the observed HCO$^+$ 3-2. 
               The red contour with levels starting from 0.6 K km s$^{-1}$ in step of 0.4 K km s$^{-1}$ represents H$^{13}$CO$^+$ 3-2.
                {\it Bottom right:} The spatially resolved $\tau(\textrm{H}^{13}\textrm{CO$^+$})$ of G011.91-00.61 is demonstrated by black contour with levels 
               starting from 0.07 in step of 0.02. 
                }       
 \label{fig:g01191}
\end{figure*}


 \begin{figure*} 
    \centering
  \includegraphics[width=3.05in]{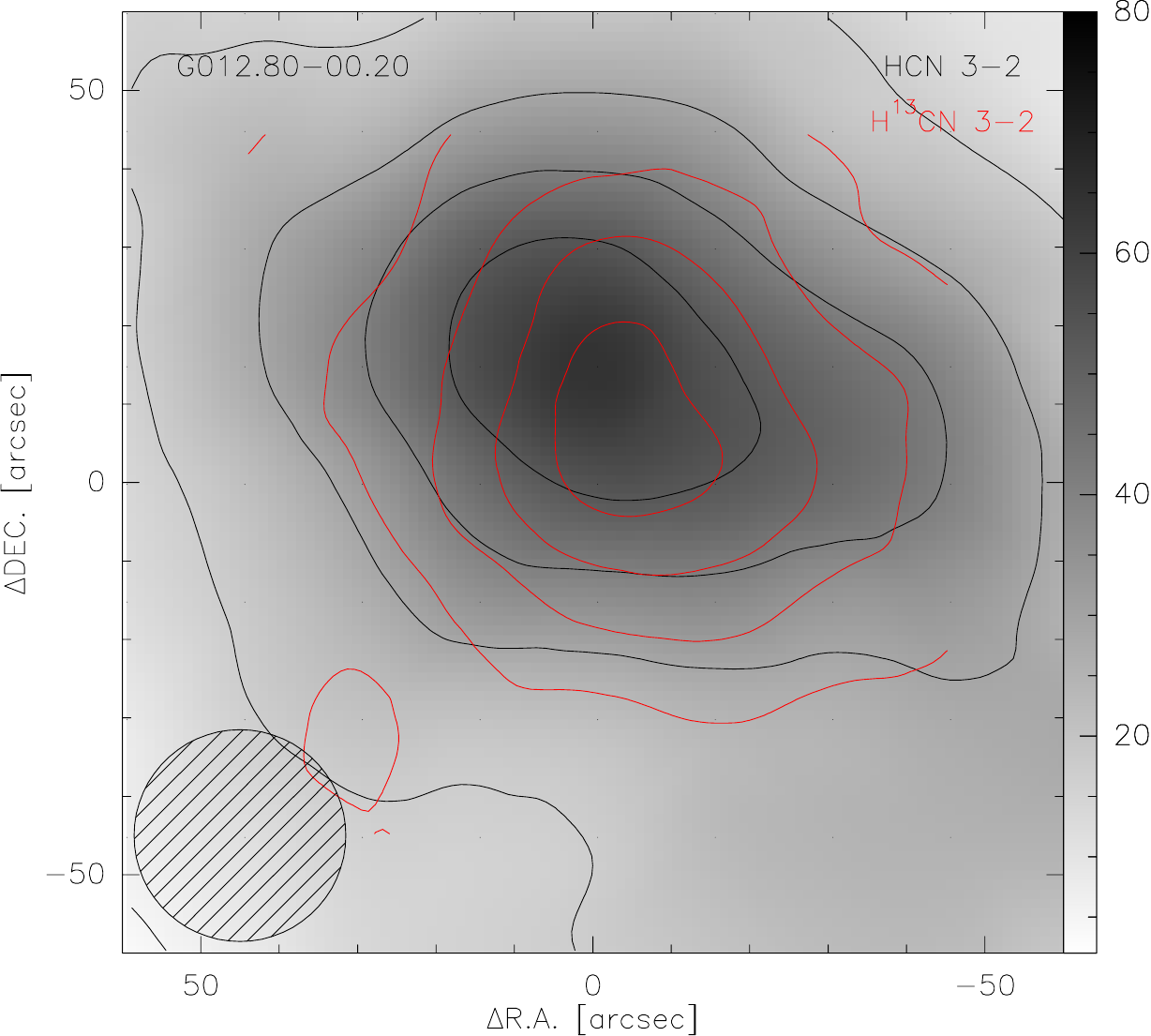}
   \includegraphics[width=3.03in]{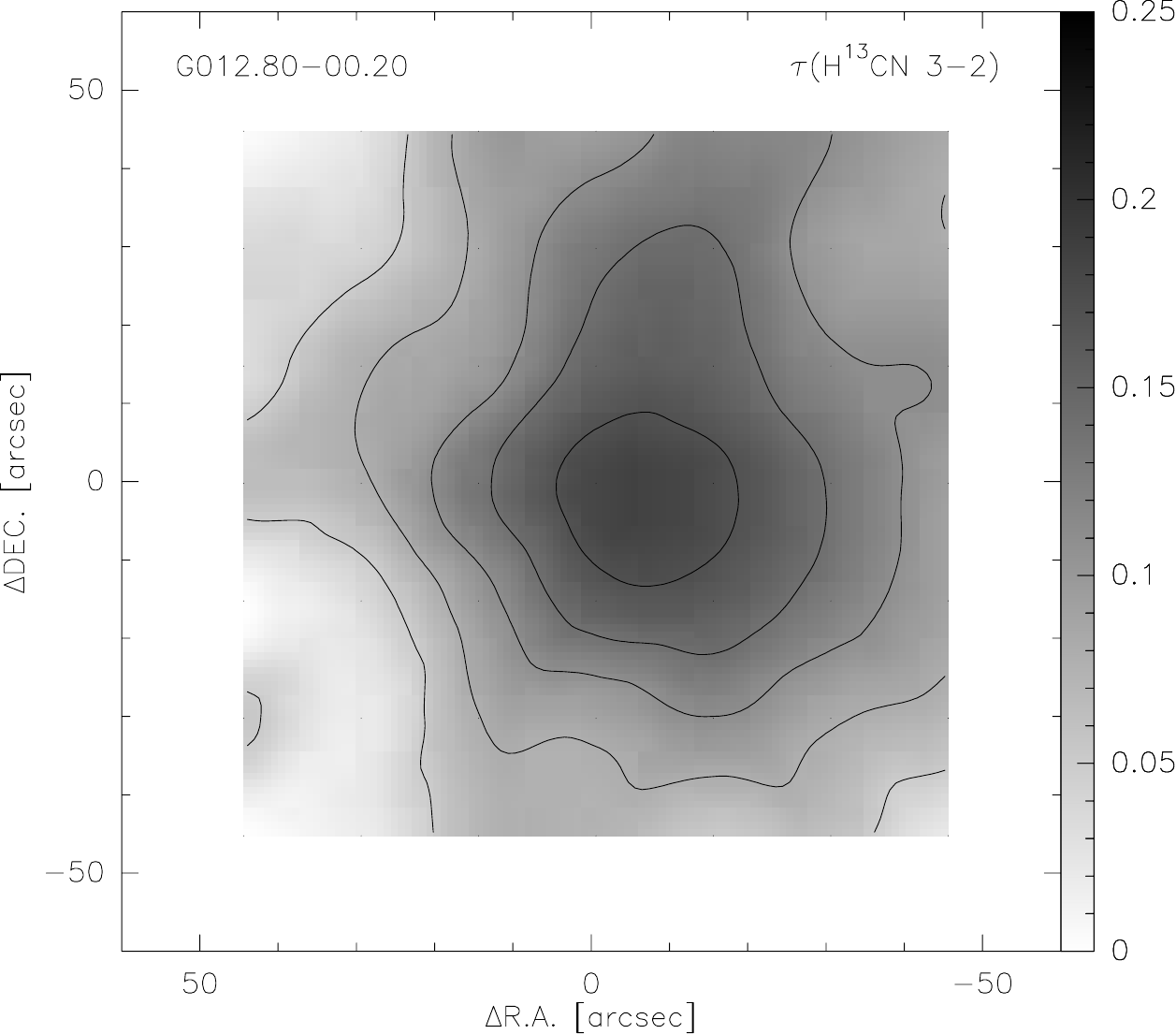}
       \includegraphics[width=3.05in]{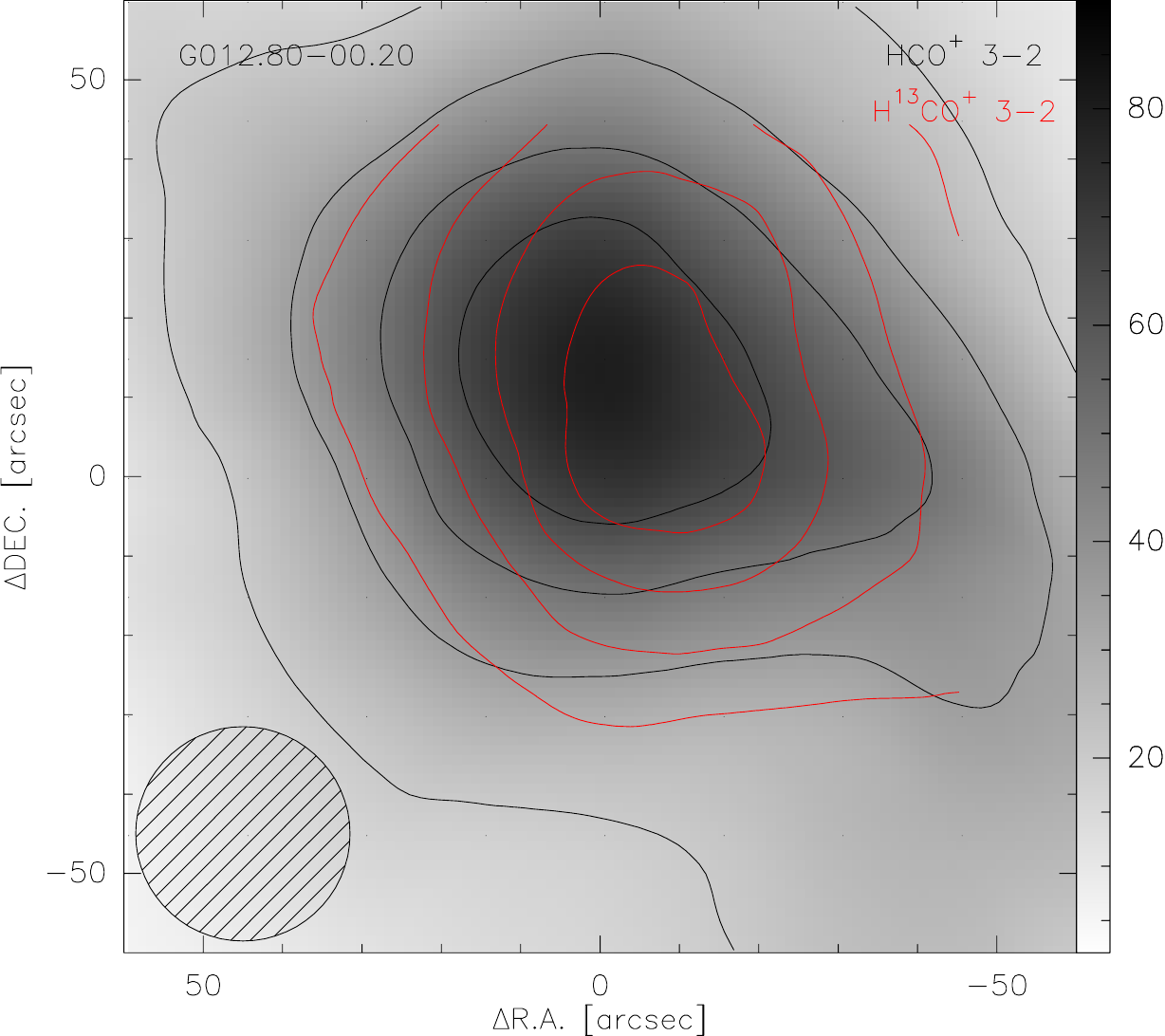}
       \includegraphics[width=3.08in]{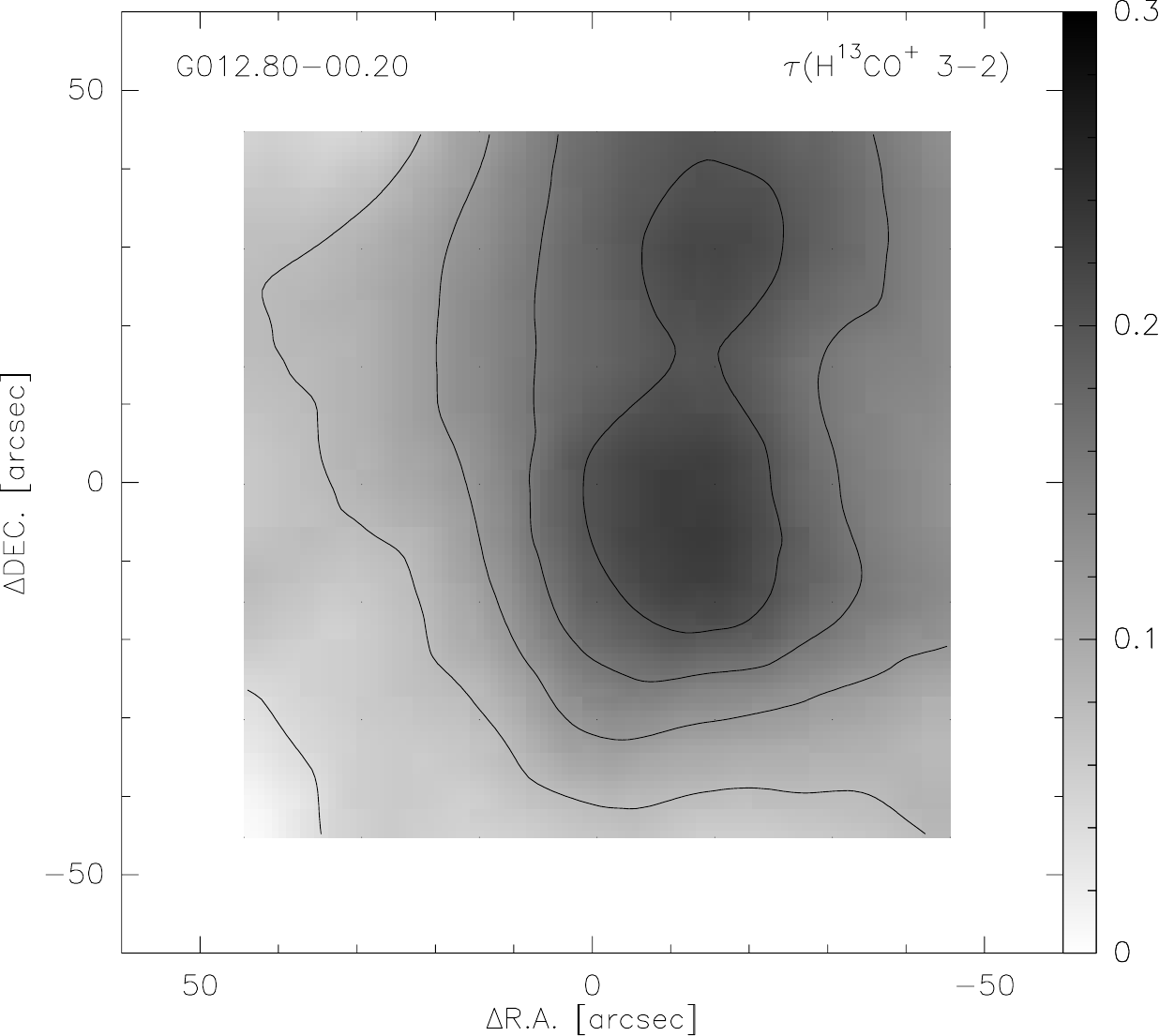}
 \caption{The data reduction results of G012.80-00.20. 
               {\it Top left:} The velocity integrated intensity maps of HCN and H$^{13}$CN 3-2. 
               The mapping size of HCN 3-2 is 2$'\times2'$, while it is 1.5$'\times1.5'$ for H$^{13}$CN 3-2, with a beam size of $\sim$ 27.8$''$.
               The grey scale and the black contour with levels starting from 5 K km s$^{-1}$ in step of 12 K km s$^{-1}$ show the observed HCN 3-2. 
               The red contour with levels starting from 0.5 K km s$^{-1}$ in step of 2 K km s$^{-1}$ represents H$^{13}$CN 3-2.
               {\it Top right:} The spatially resolved $\tau(\textrm{H}^{13}\textrm{CN})$ of G012.80-00.20 is demonstrated by black contour with levels 
               starting from 0.05 in step of 0.03. 
                {\it Bottom left:} The velocity integrated intensity maps of HCO$^+$ and H$^{13}$CO$^+$ 3-2. 
               The mapping size of HCO$^+$ 3-2 is 2$'\times2'$, while it is 1.5$'\times1.5'$ for H$^{13}$CO$^+$ 3-2, with a beam size of $\sim$ 27.8$''$.
               The grey scale and the black contour with levels starting from 5 K km s$^{-1}$ in step of 15 K km s$^{-1}$ show the observed HCO$^+$ 3-2. 
               The red contour with levels starting from 0.3 K km s$^{-1}$ in step of 3 K km s$^{-1}$ represents H$^{13}$CO$^+$ 3-2.
                {\it Bottom right:} The spatially resolved $\tau(\textrm{H}^{13}\textrm{CO$^+$})$ of G012.80-00.20 is demonstrated by black contour with levels 
               starting from 0.04 in step of 0.04. 
                }       
 \label{fig:g01280}
\end{figure*}


 \begin{figure*} 
    \centering
  \includegraphics[width=3.05in]{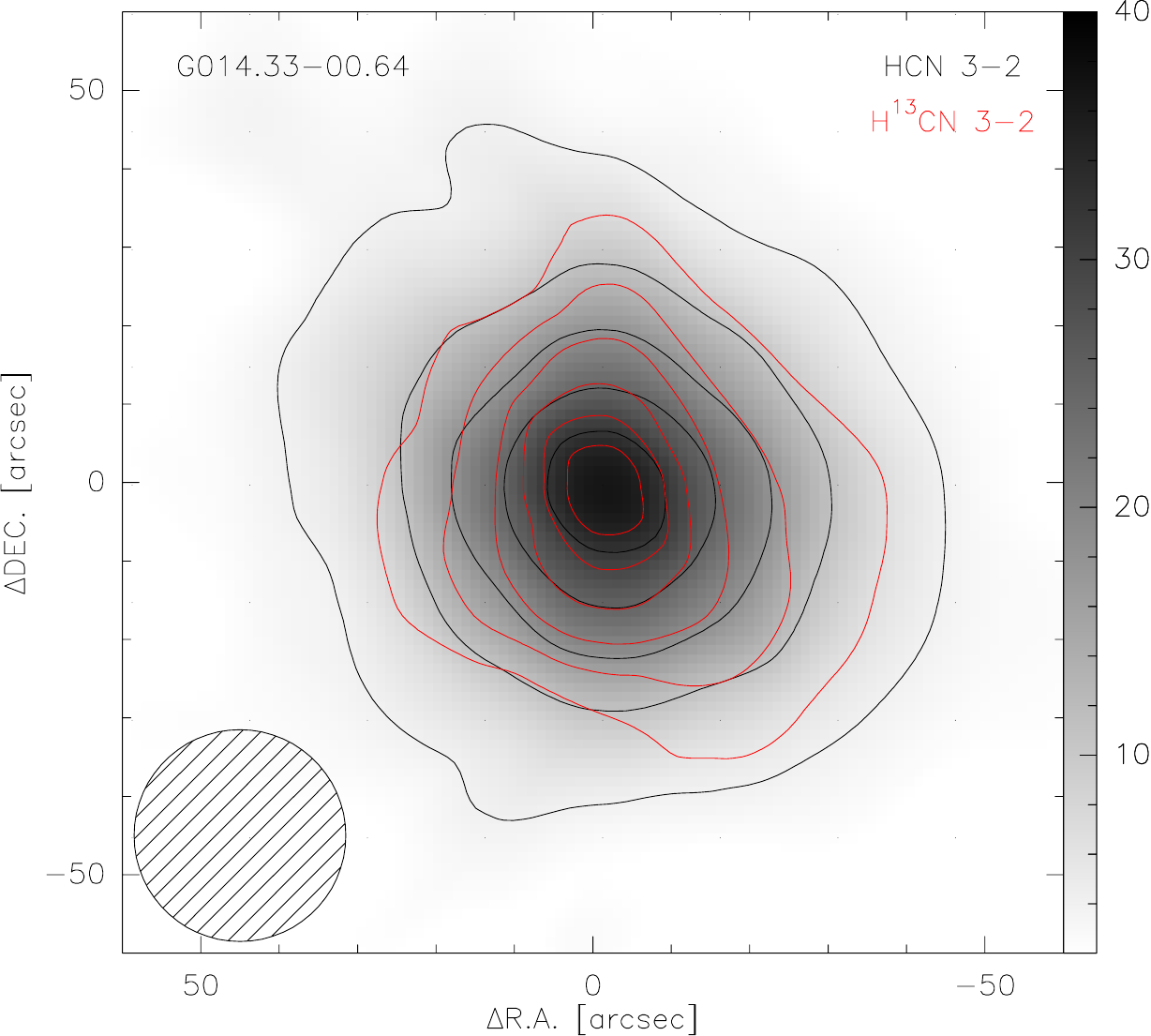}
   \includegraphics[width=3.03in]{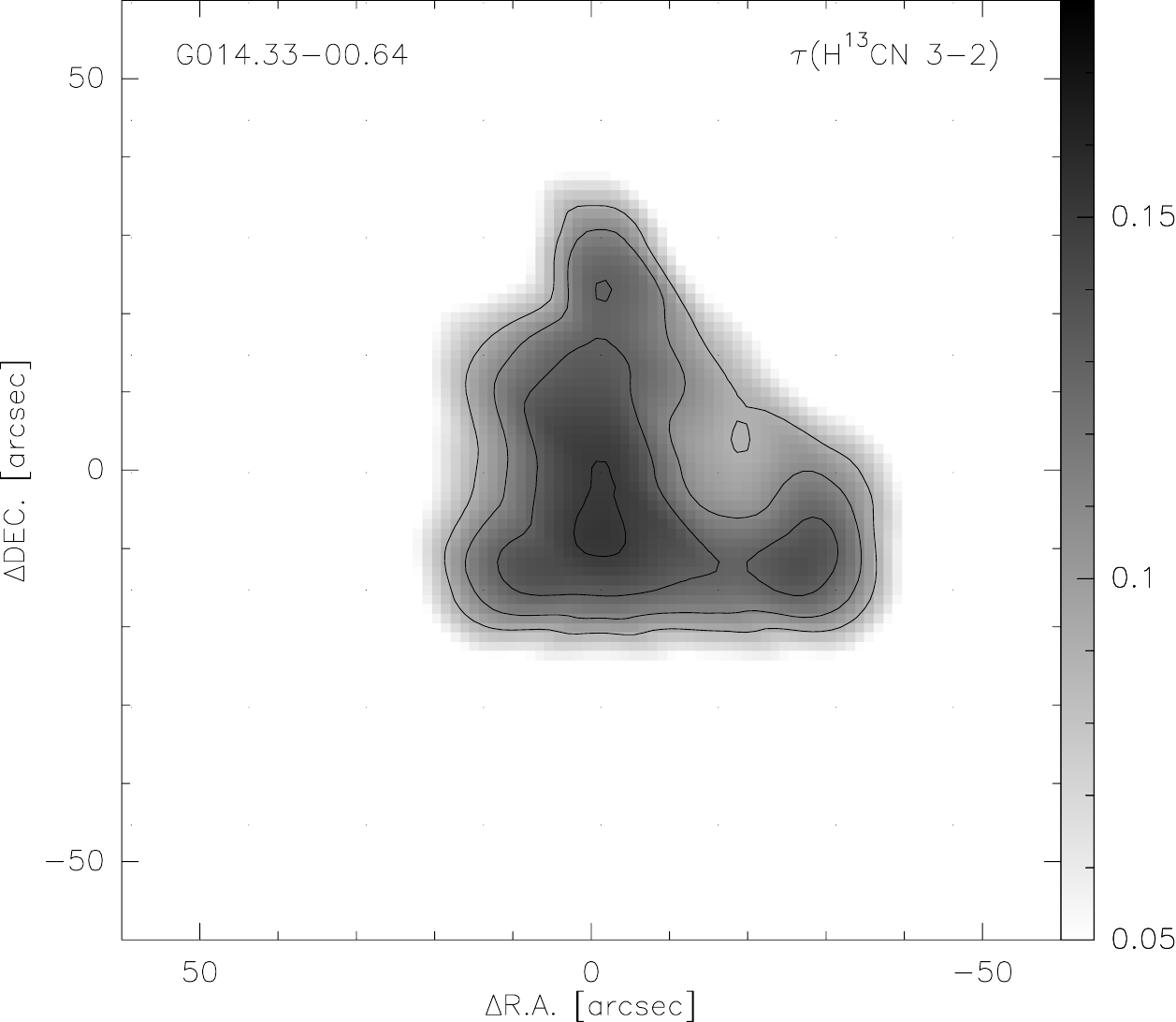}
       \includegraphics[width=3.05in]{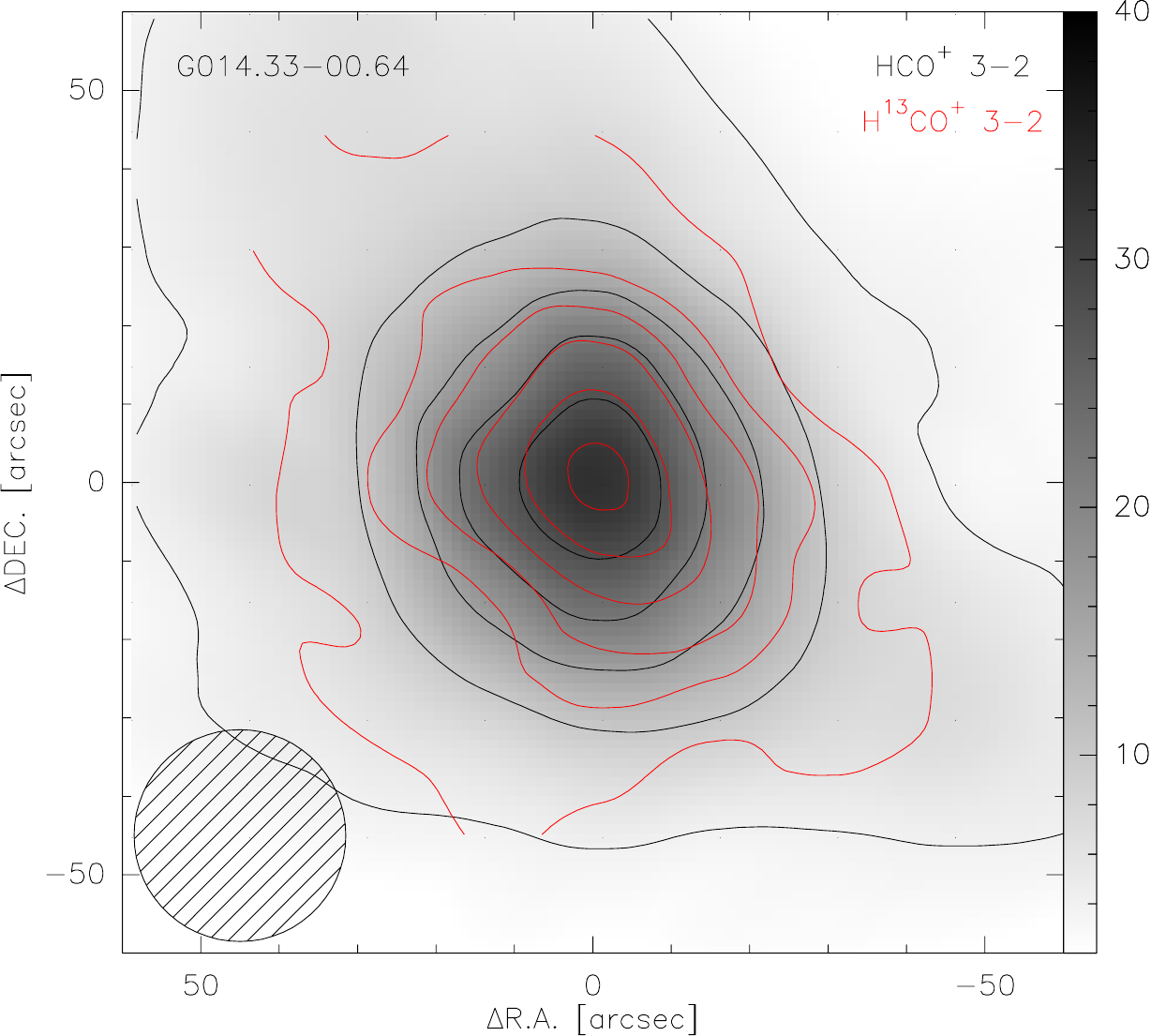}
       \includegraphics[width=3.08in]{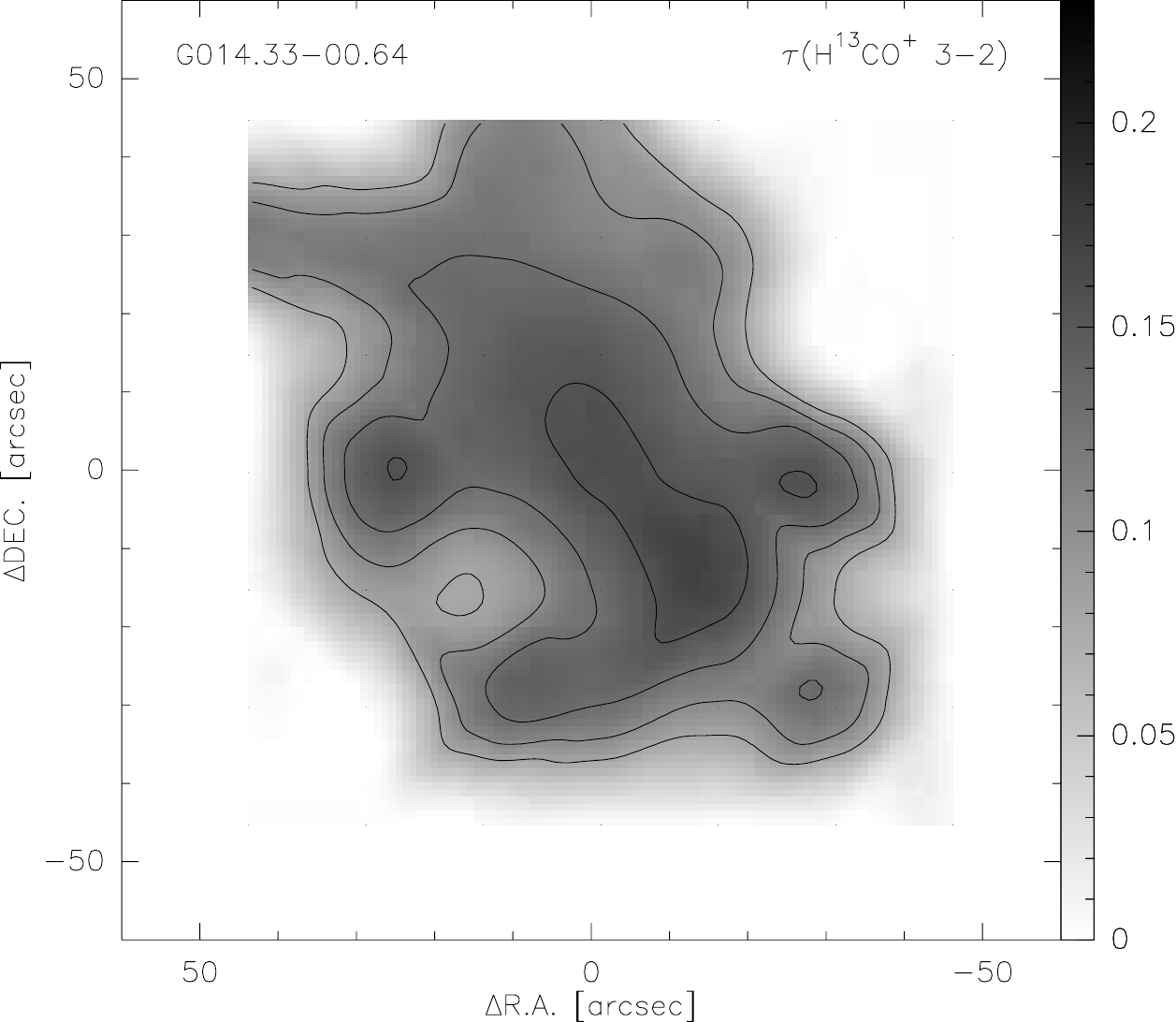}
 \caption{The data reduction results of G014.33-00.64. 
               {\it Top left:} The velocity integrated intensity maps of HCN and H$^{13}$CN 3-2. 
               The mapping size of HCN 3-2 is 2$'\times2'$, while it is 1.5$'\times1.5'$ for H$^{13}$CN 3-2, with a beam size of $\sim$ 27.8$''$.
               The grey scale and the black contour with levels starting from 4 K km s$^{-1}$ in step of 7 K km s$^{-1}$ show the observed HCN 3-2. 
               The red contour with levels starting from 0.7 K km s$^{-1}$ in step of 0.8 K km s$^{-1}$ represents H$^{13}$CN 3-2.
               {\it Top right:} The spatially resolved $\tau(\textrm{H}^{13}\textrm{CN})$ of G014.33-00.64 is demonstrated by black contour with levels 
               starting from 0.09 in step of 0.02. 
                {\it Bottom left:} The velocity integrated intensity maps of HCO$^+$ and H$^{13}$CO$^+$ 3-2. 
               The mapping size of HCO$^+$ 3-2 is 2$'\times2'$, while it is 1.5$'\times1.5'$ for H$^{13}$CO$^+$ 3-2, with a beam size of $\sim$ 27.8$''$.
               The grey scale and the black contour with levels starting from 4 K km s$^{-1}$ in step of 6 K km s$^{-1}$ show the observed HCO$^+$ 3-2. 
               The red contour with levels starting from 0.6 K km s$^{-1}$ in step of 0.8 K km s$^{-1}$ represents H$^{13}$CO$^+$ 3-2.
                {\it Bottom right:} The spatially resolved $\tau(\textrm{H}^{13}\textrm{CO$^+$})$ of G014.33-00.64 is demonstrated by black contour with levels 
               starting from 0.08 in step of 0.03. 
                }       
 \label{fig:g01433}
\end{figure*}


 \begin{figure*} 
    \centering
  \includegraphics[width=3.05in]{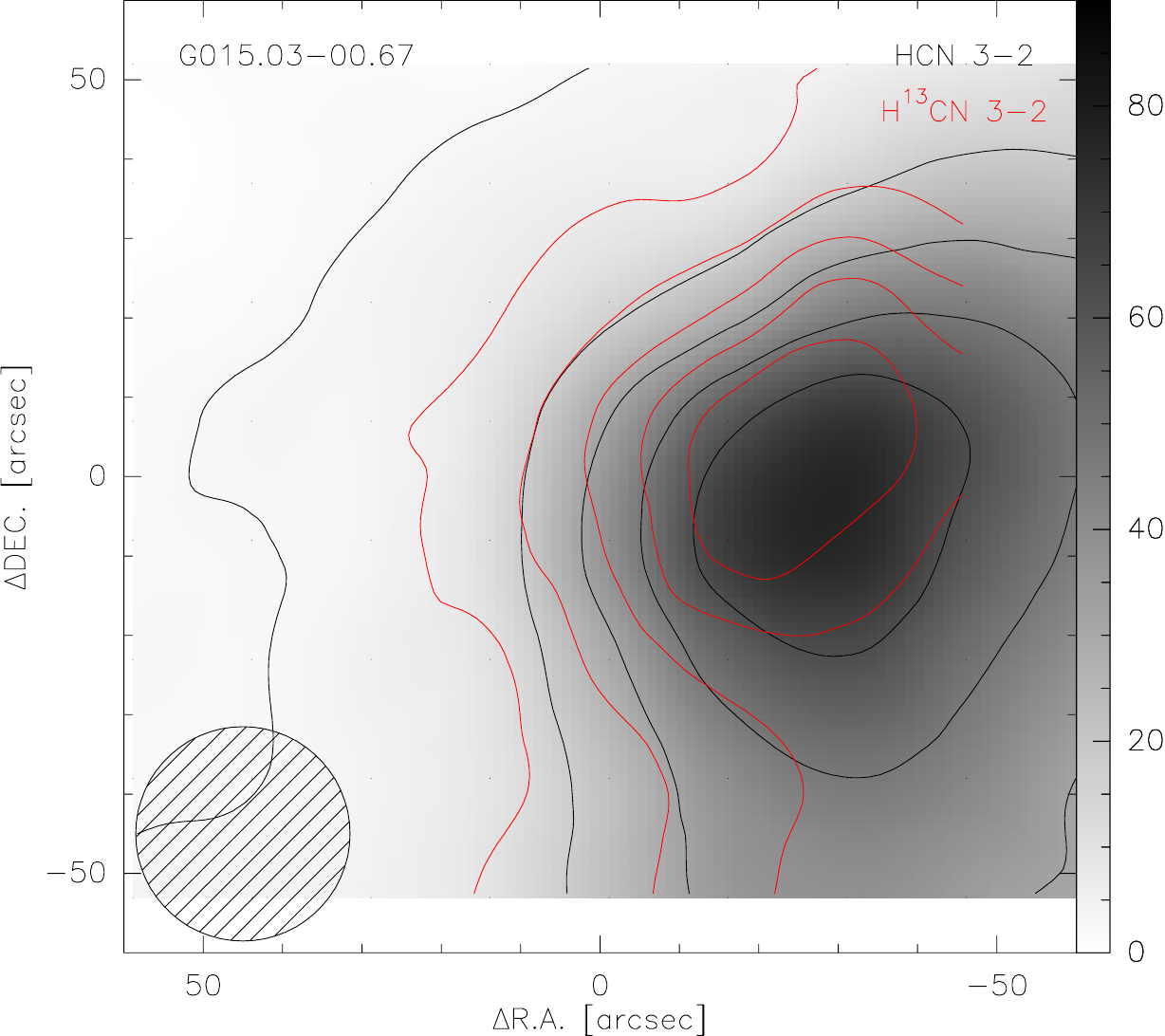}
   \includegraphics[width=3.03in]{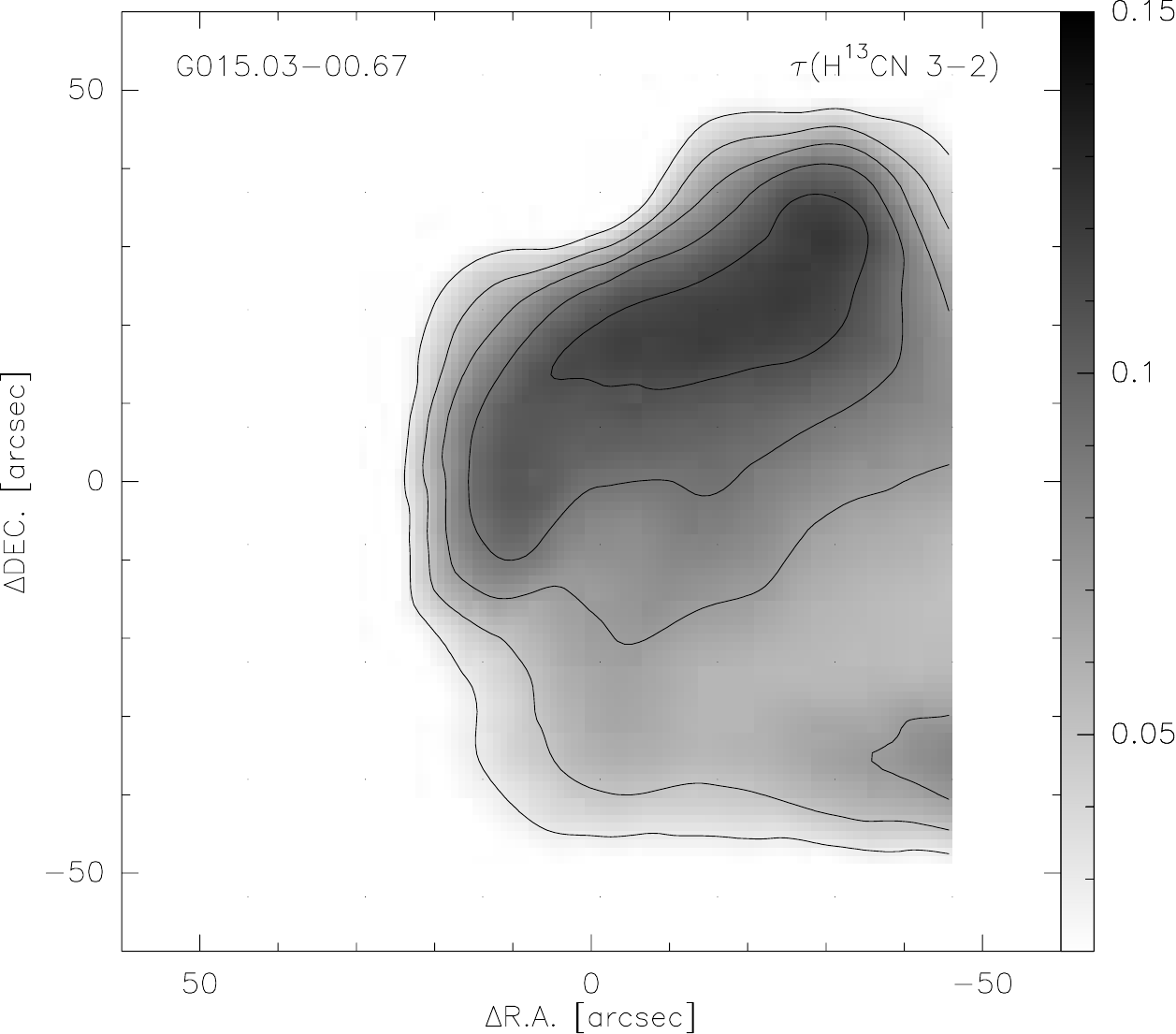}
       \includegraphics[width=3.05in]{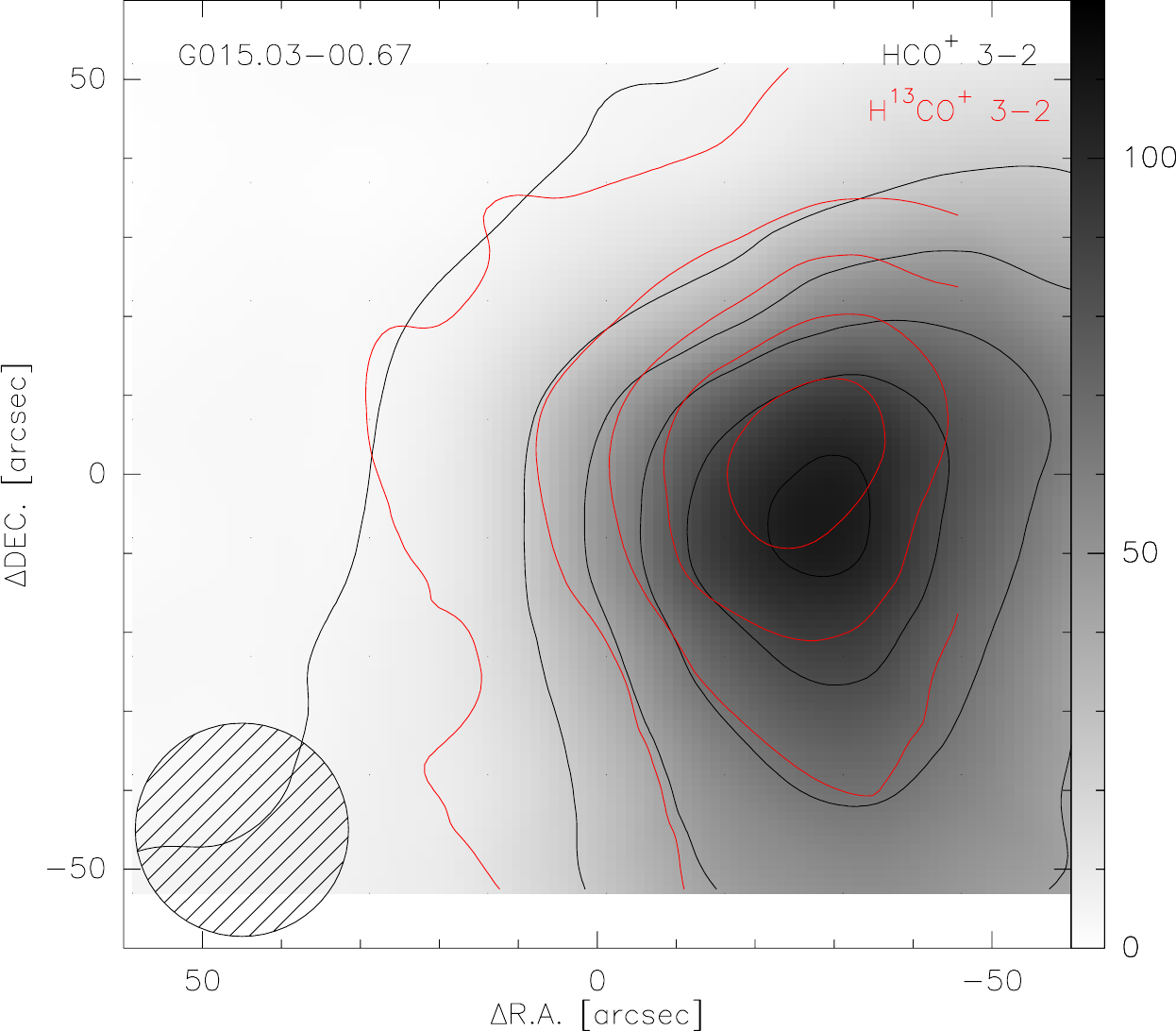}
       \includegraphics[width=3.08in]{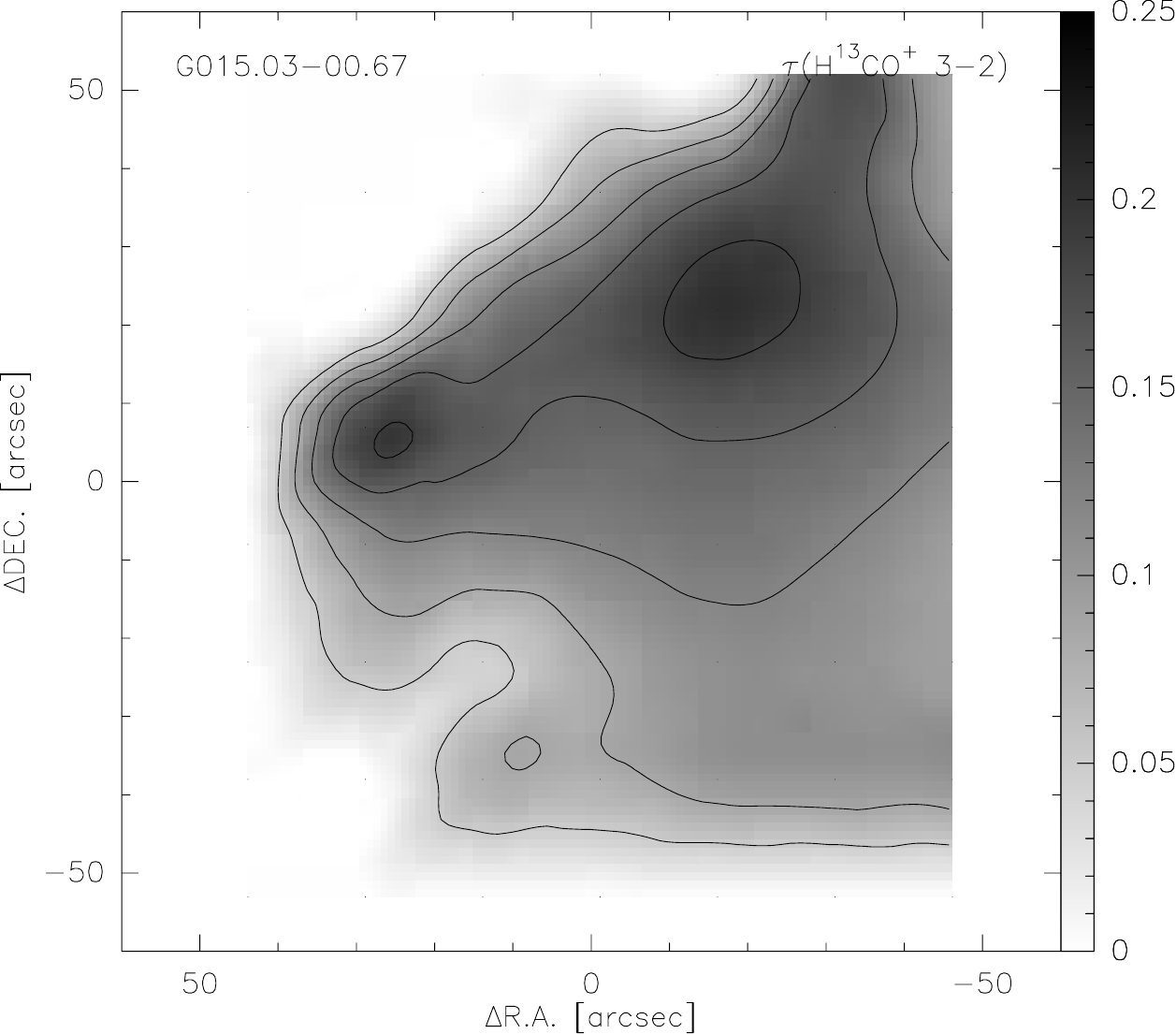}
 \caption{The data reduction results of G015.03-00.67. 
               {\it Top left:} The velocity integrated intensity maps of HCN and H$^{13}$CN 3-2. 
               The mapping size of HCN 3-2 is 2$'\times2'$, while it is 1.5$'\times1.5'$ for H$^{13}$CN 3-2, with a beam size of $\sim$ 27.8$''$.
               The grey scale and the black contour with levels starting from 2.5 K km s$^{-1}$ in step of 15 K km s$^{-1}$ show the observed HCN 3-2. 
               The red contour with levels starting from 0.5 K km s$^{-1}$ in step of 1.2 K km s$^{-1}$ represents H$^{13}$CN 3-2.
               {\it Top right:} The spatially resolved $\tau(\textrm{H}^{13}\textrm{CN})$ of G015.03-00.67 is demonstrated by black contour with levels 
               starting from 0.03 in step of 0.02. 
                {\it Bottom left:} The velocity integrated intensity maps of HCO$^+$ and H$^{13}$CO$^+$ 3-2. 
               The mapping size of HCO$^+$ 3-2 is 2$'\times2'$, while it is 1.5$'\times1.5'$ for H$^{13}$CO$^+$ 3-2, with a beam size of $\sim$ 27.8$''$.
               The grey scale and the black contour with levels starting from 4 K km s$^{-1}$ in step of 20 K km s$^{-1}$ show the observed HCO$^+$ 3-2. 
               The red contour with levels starting from 0.6 K km s$^{-1}$ in step of 3 K km s$^{-1}$ represents H$^{13}$CO$^+$ 3-2.
                {\it Bottom right:} The spatially resolved $\tau(\textrm{H}^{13}\textrm{CO$^+$})$ of G015.03-00.67 is demonstrated by black contour with levels 
               starting from 0.05 in step of 0.04. 
                }       
 \label{fig:g01503}
\end{figure*}


 \begin{figure*} 
    \centering
  \includegraphics[width=3.05in]{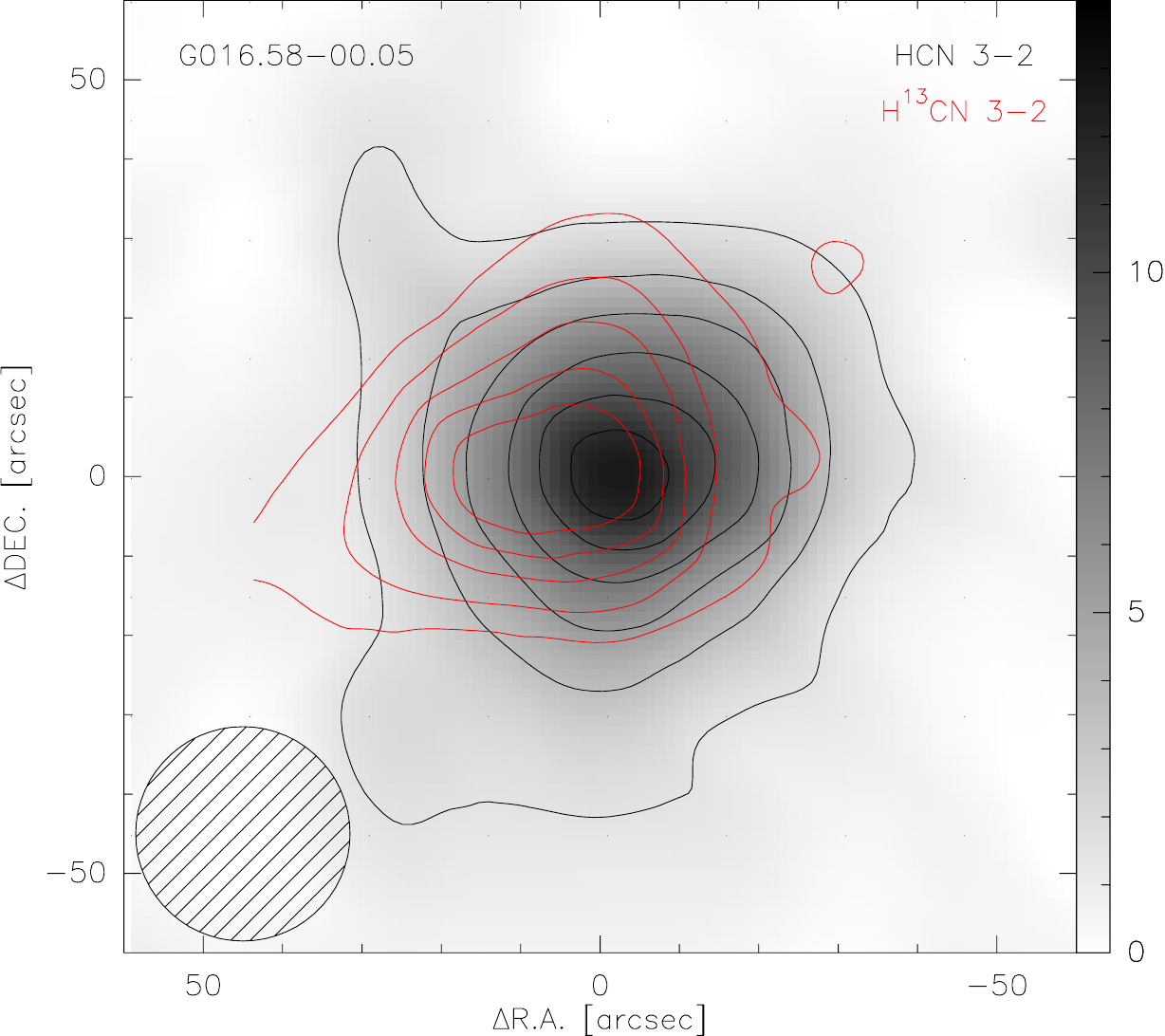}
   \includegraphics[width=3.03in]{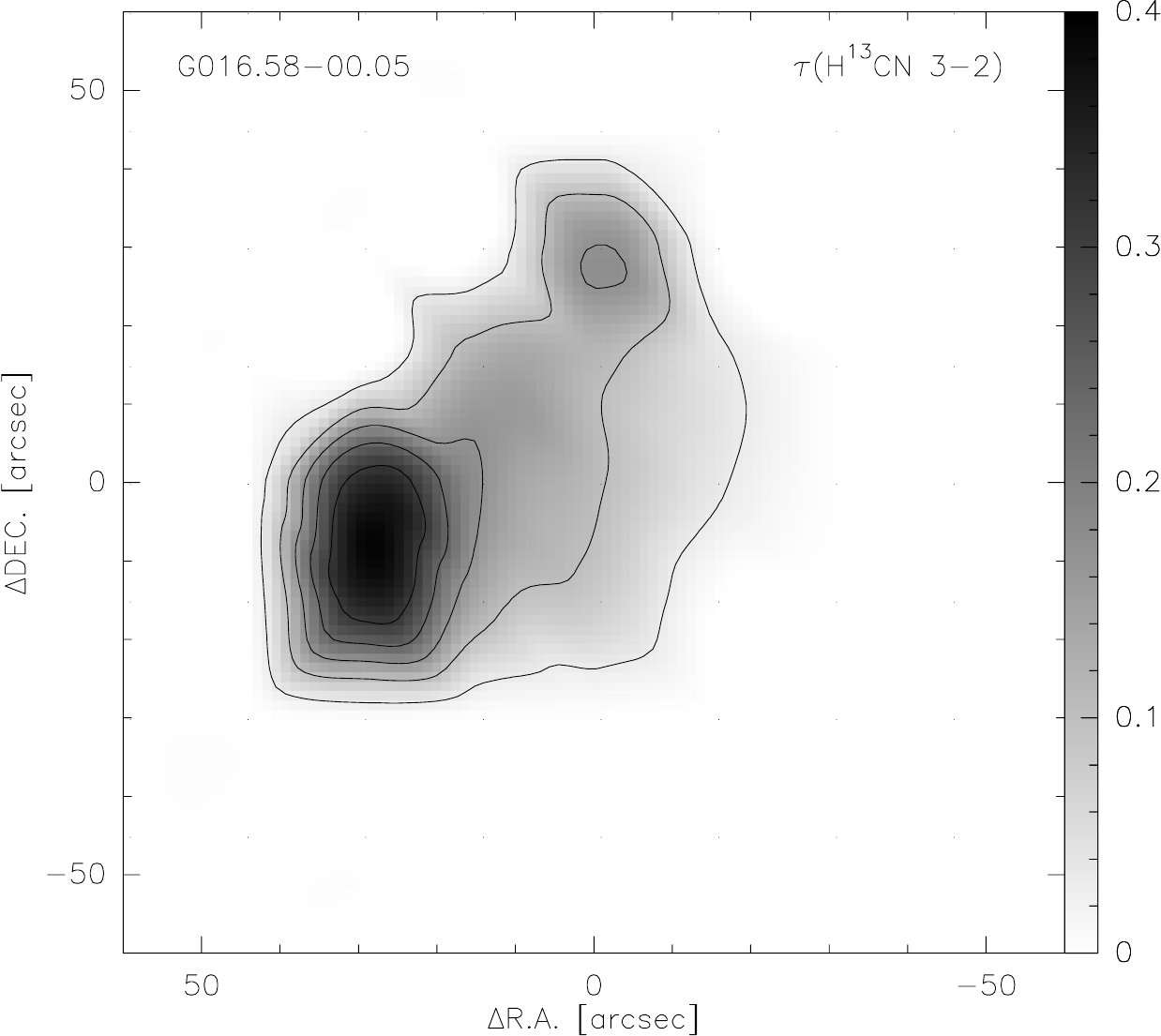}
       \includegraphics[width=3.05in]{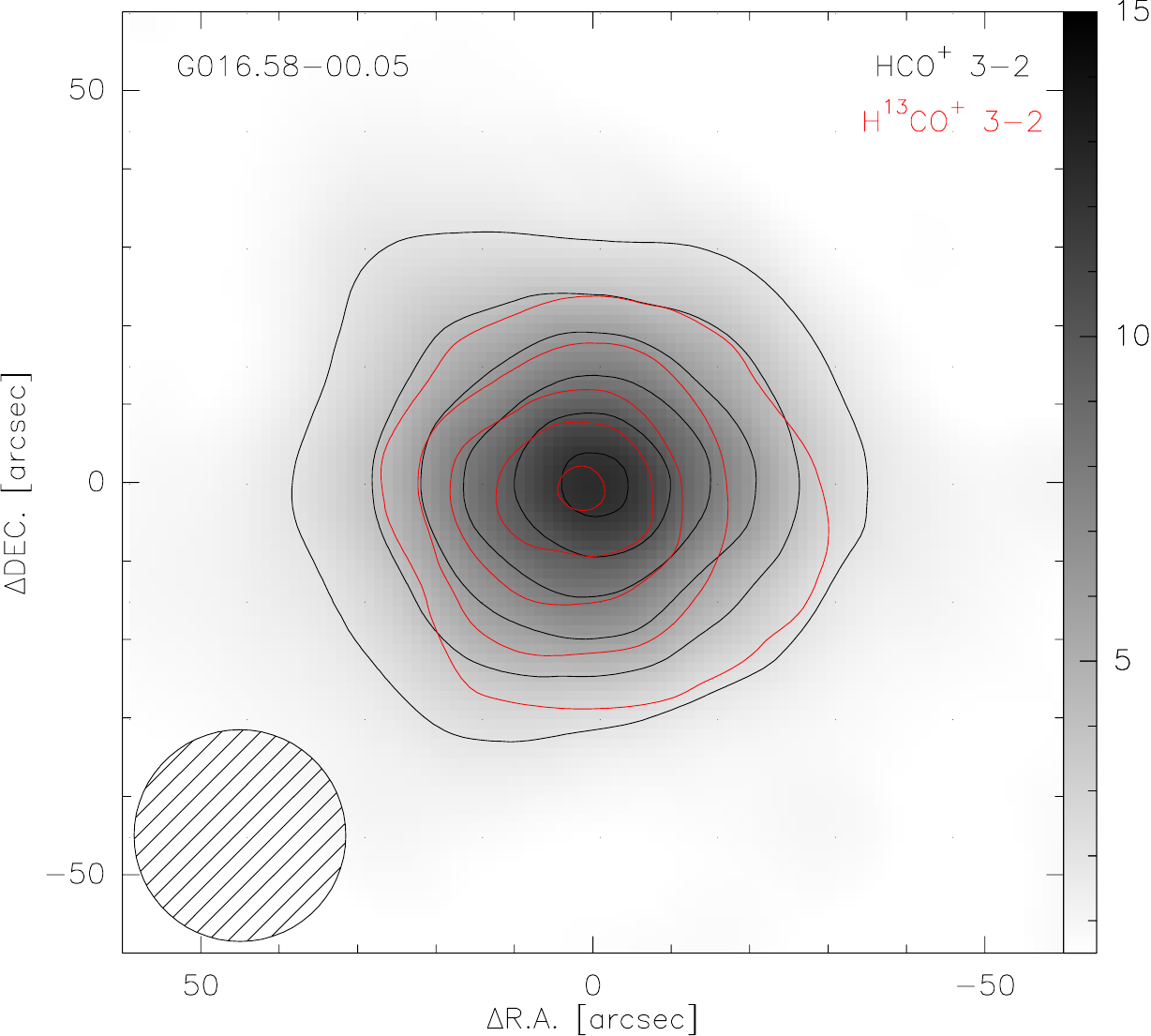}
       \includegraphics[width=3.08in]{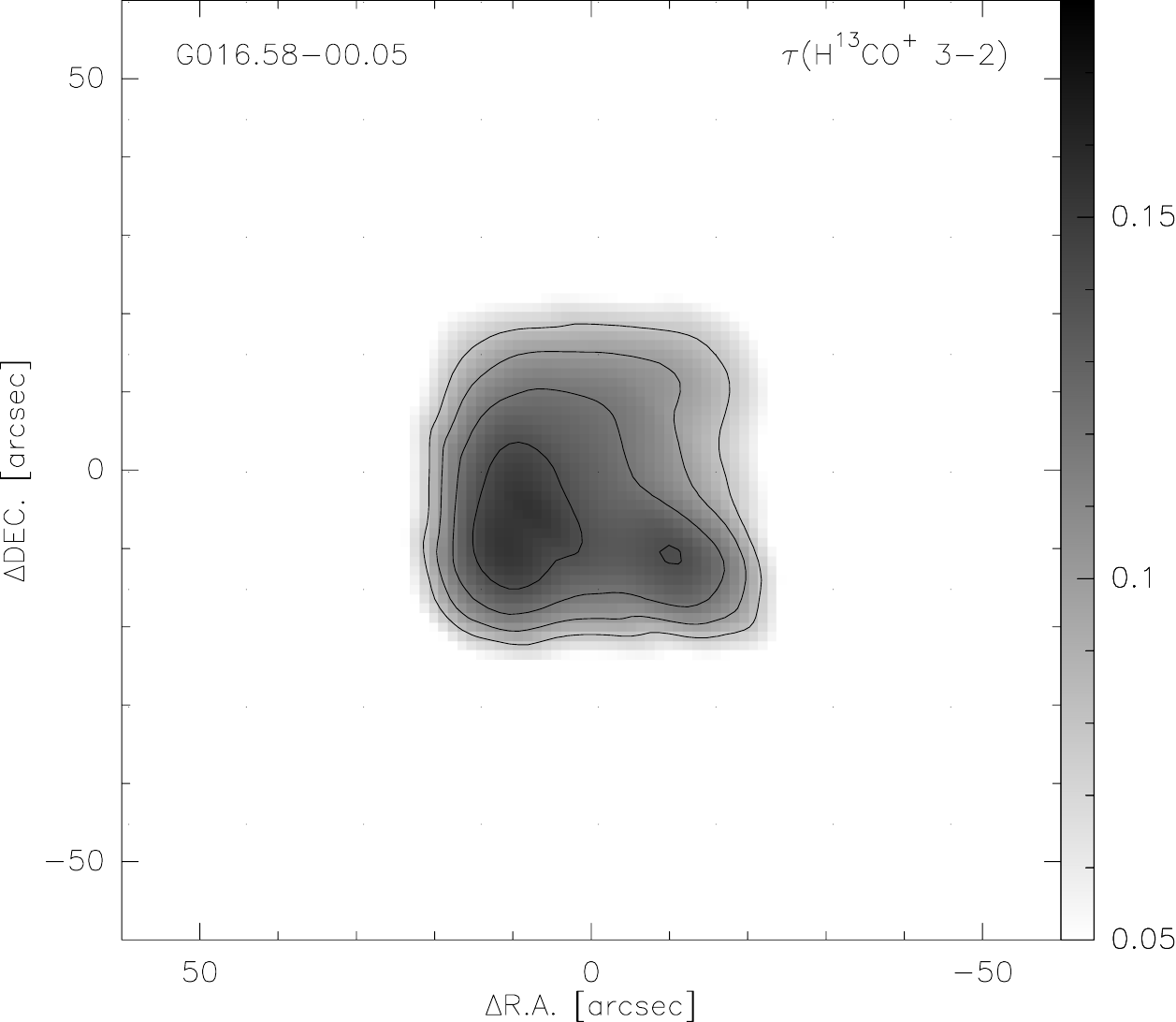}
 \caption{The data reduction results of G016.58-00.05. 
               {\it Top left:} The velocity integrated intensity maps of HCN and H$^{13}$CN 3-2. 
               The mapping size of HCN 3-2 is 2$'\times2'$, while it is 1.5$'\times1.5'$ for H$^{13}$CN 3-2, with a beam size of $\sim$ 27.8$''$.
               The grey scale and the black contour with levels starting from 2 K km s$^{-1}$ in step of 2 K km s$^{-1}$ show the observed HCN 3-2. 
               The red contour with levels starting from 0.25 K km s$^{-1}$ in step of 0.2 K km s$^{-1}$ represents H$^{13}$CN 3-2.
               {\it Top right:} The spatially resolved $\tau(\textrm{H}^{13}\textrm{CN})$ of G016.58-00.05 is demonstrated by black contour with levels 
               starting from 0.03 in step of 0.07. 
                {\it Bottom left:} The velocity integrated intensity maps of HCO$^+$ and H$^{13}$CO$^+$ 3-2. 
               The mapping size of HCO$^+$ 3-2 is 2$'\times2'$, while it is 1.5$'\times1.5'$ for H$^{13}$CO$^+$ 3-2, with a beam size of $\sim$ 27.8$''$.
               The grey scale and the black contour with levels starting from 2 K km s$^{-1}$ in step of 2 K km s$^{-1}$ show the observed HCO$^+$ 3-2. 
               The red contour with levels starting from 0.3 K km s$^{-1}$ in step of 0.3 K km s$^{-1}$ represents H$^{13}$CO$^+$ 3-2.
                {\it Bottom right:} The spatially resolved $\tau(\textrm{H}^{13}\textrm{CO$^+$})$ of G016.58-00.05 is demonstrated by black contour with levels 
               starting from 0.08 in step of 0.02. 
                }       
 \label{fig:g01658}
\end{figure*}


 \begin{figure*} 
    \centering
  \includegraphics[width=3.05in]{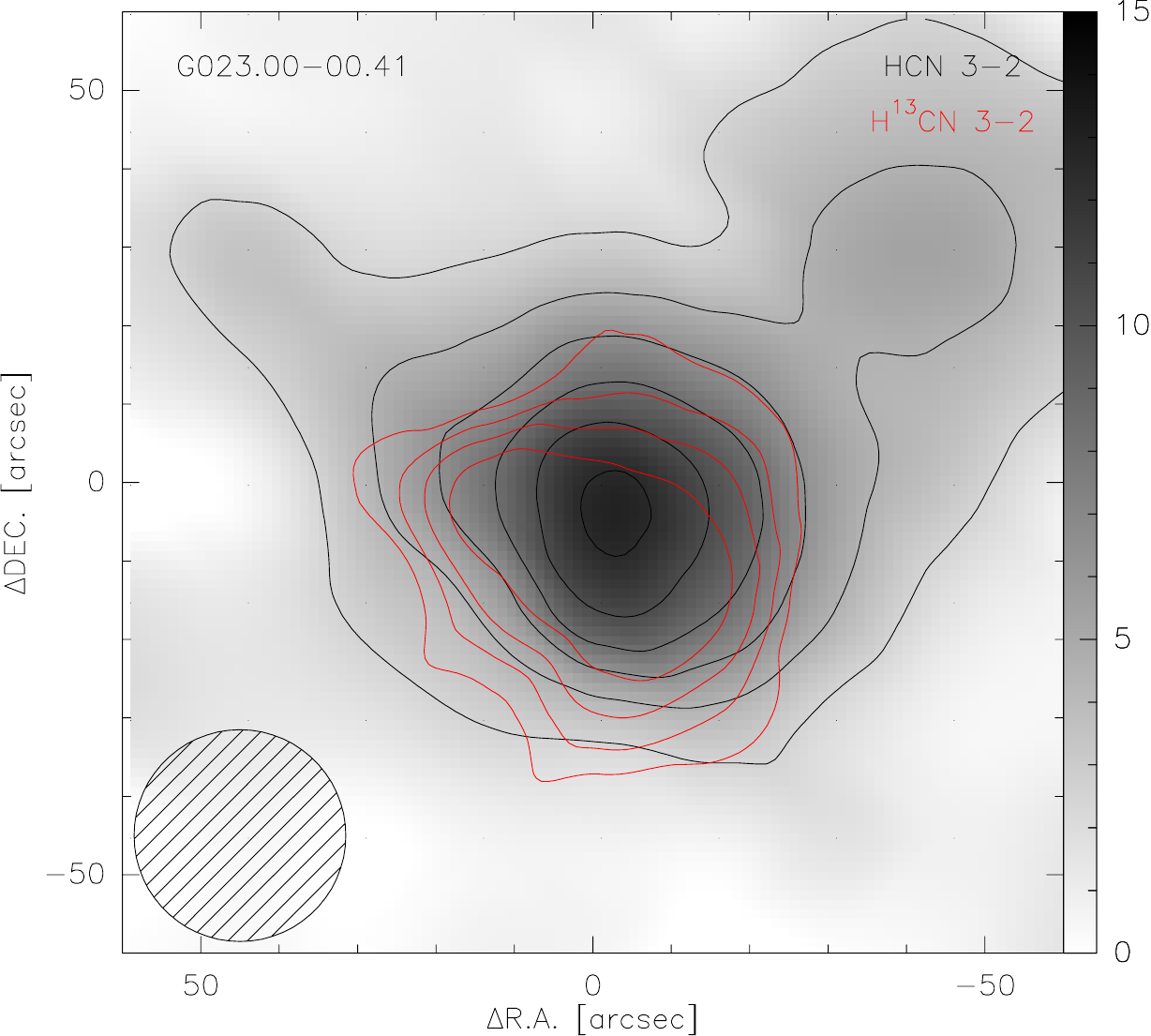}
   \includegraphics[width=3.03in]{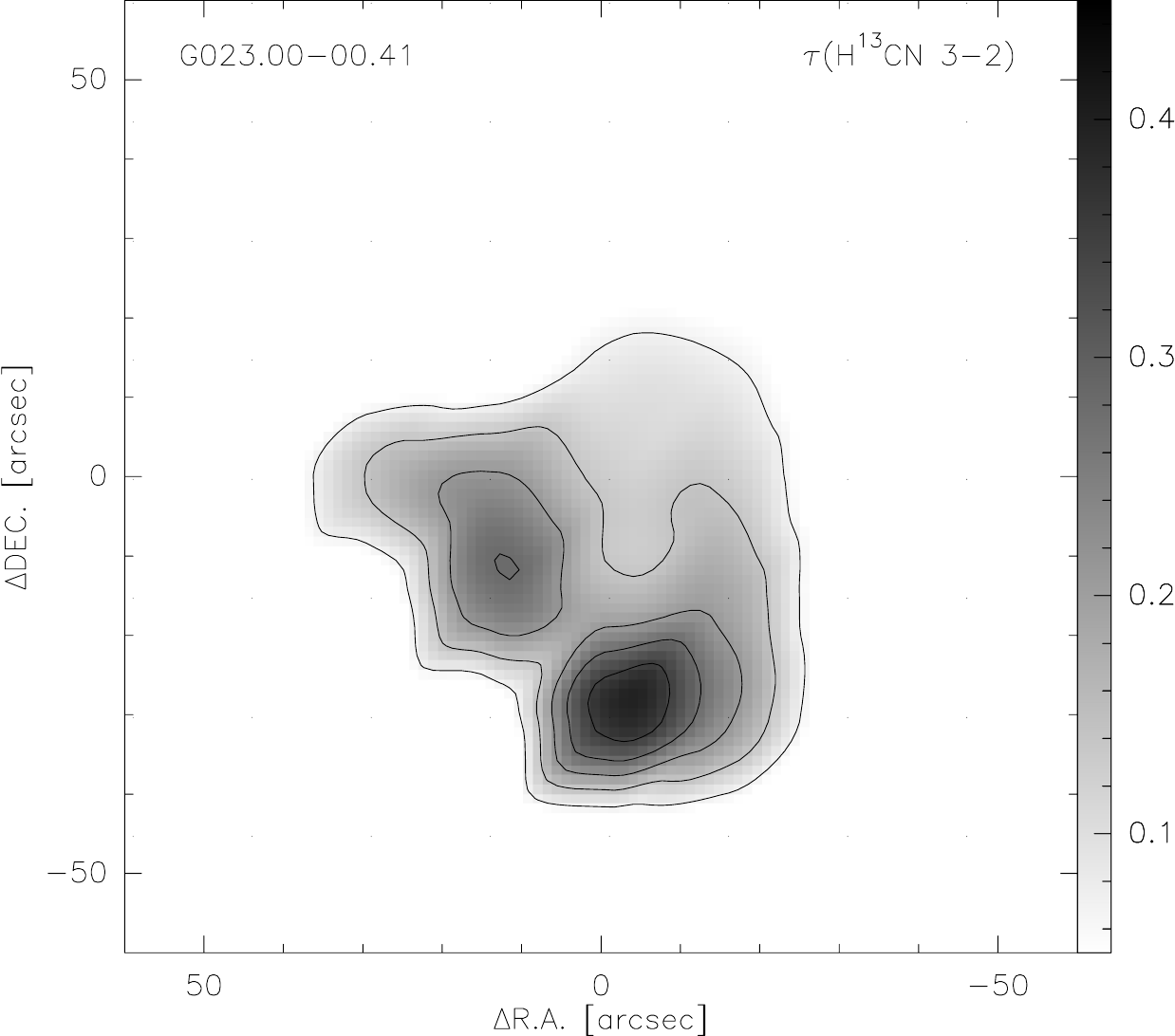}
       \includegraphics[width=3.05in]{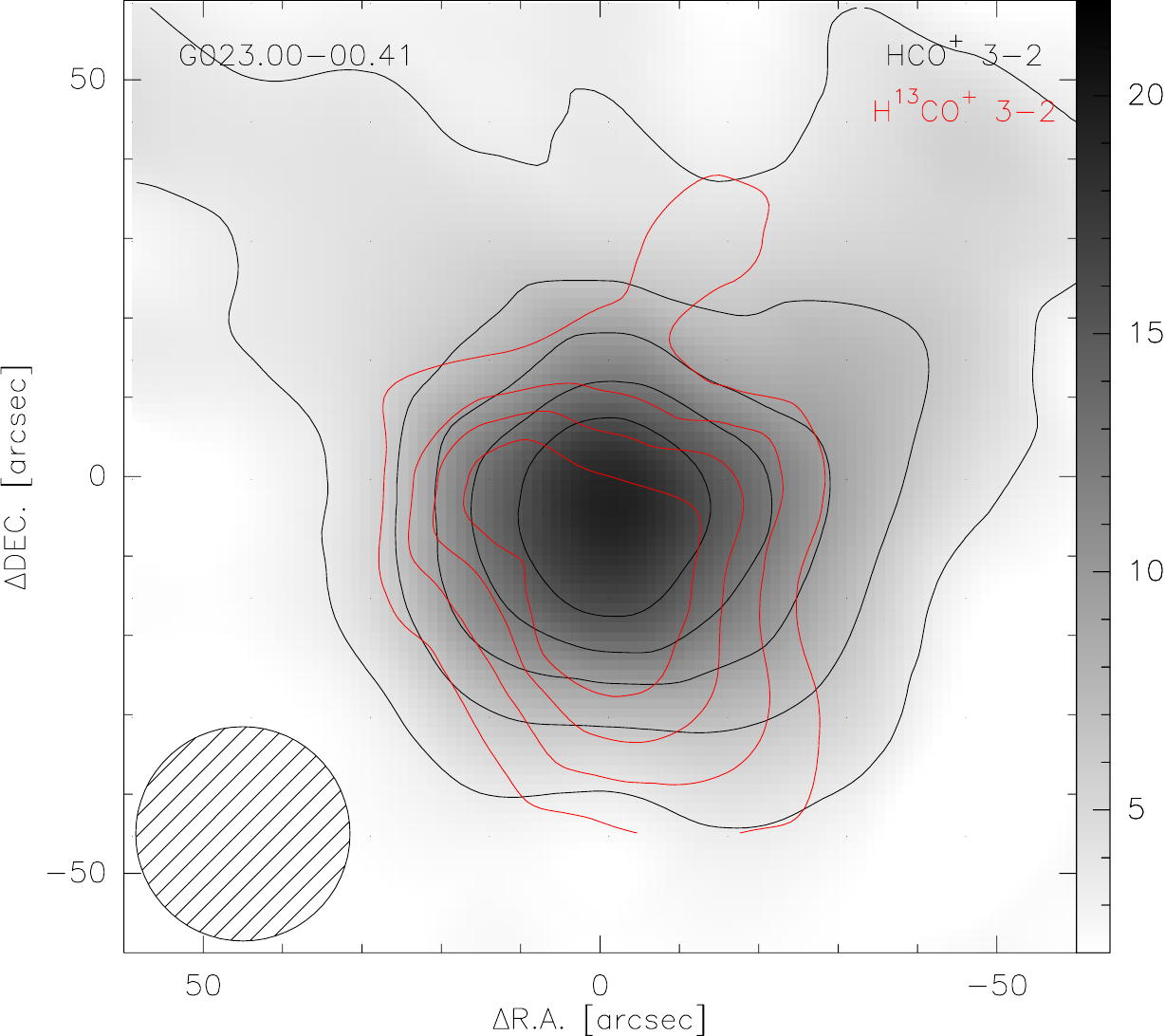}
       \includegraphics[width=3.08in]{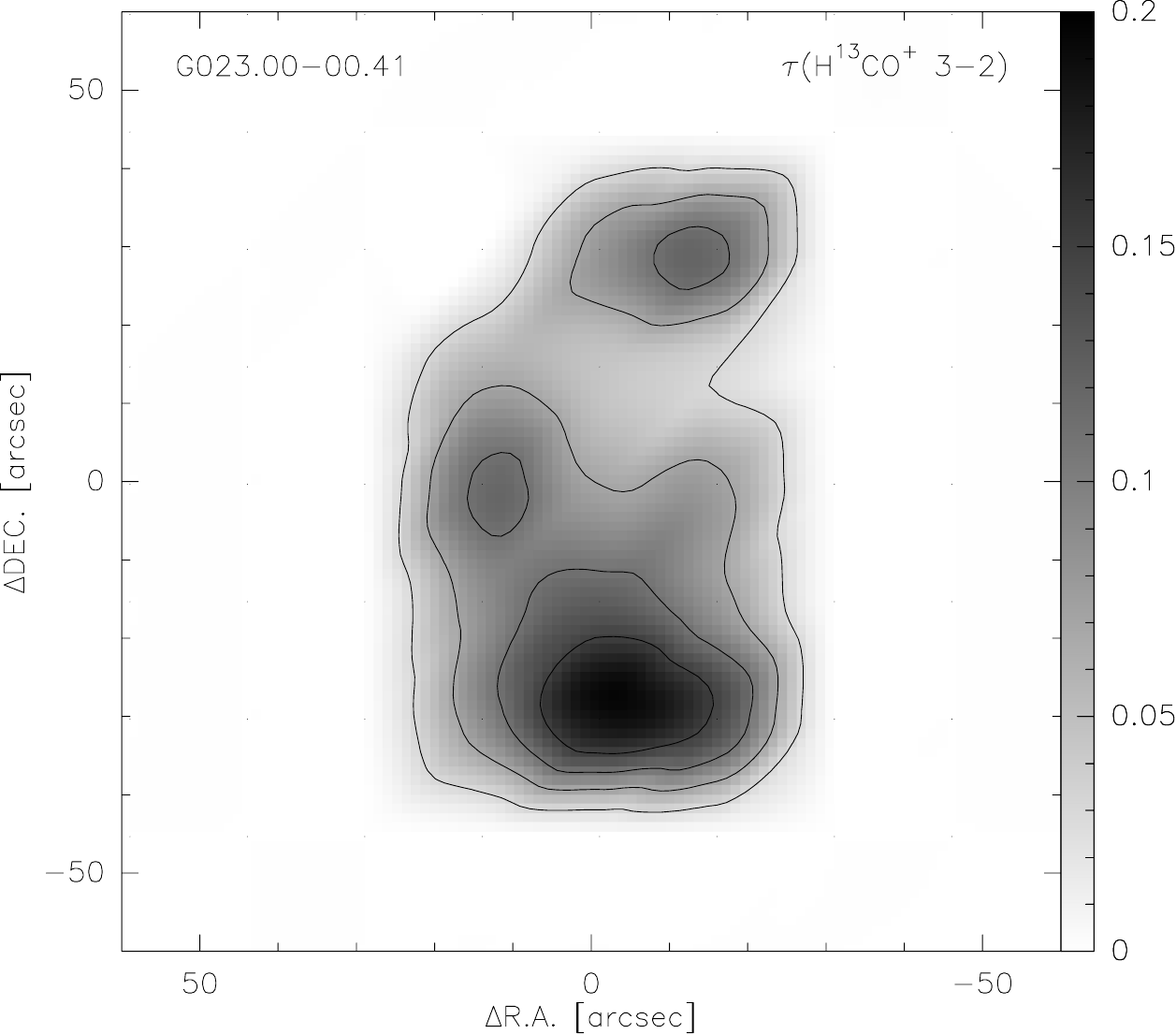}
 \caption{The data reduction results of G023.00-00.41. 
               {\it Top left:} The velocity integrated intensity maps of HCN and H$^{13}$CN 3-2. 
               The mapping size of HCN 3-2 is 2$'\times2'$, while it is 1.5$'\times1.5'$ for H$^{13}$CN 3-2, with a beam size of $\sim$ 27.8$''$.
               The grey scale and the black contour with levels starting from 2 K km s$^{-1}$ in step of 2 K km s$^{-1}$ show the observed HCN 3-2. 
               The red contour with levels starting from 0.3 K km s$^{-1}$ in step of 0.2 K km s$^{-1}$ represents H$^{13}$CN 3-2.
               {\it Top right:} The spatially resolved $\tau(\textrm{H}^{13}\textrm{CN})$ of G023.00-00.41 is demonstrated by black contour with levels 
               starting from 0.07 in step of 0.07. 
                {\it Bottom left:} The velocity integrated intensity maps of HCO$^+$ and H$^{13}$CO$^+$ 3-2. 
               The mapping size of HCO$^+$ 3-2 is 2$'\times2'$, while it is 1.5$'\times1.5'$ for H$^{13}$CO$^+$ 3-2, with a beam size of $\sim$ 27.8$''$.
               The grey scale and the black contour with levels starting from 3.5 K km s$^{-1}$ in step of 3 K km s$^{-1}$ show the observed HCO$^+$ 3-2. 
               The red contour with levels starting from 0.4 K km s$^{-1}$ in step of 0.3 K km s$^{-1}$ represents H$^{13}$CO$^+$ 3-2.
                {\it Bottom right:} The spatially resolved $\tau(\textrm{H}^{13}\textrm{CO$^+$})$ of G023.00-00.41 is demonstrated by black contour with levels 
               starting from 0.03 in step of 0.04. 
                }       
 \label{fig:g02300}
\end{figure*}


 \begin{figure*} 
    \centering
  \includegraphics[width=3.05in]{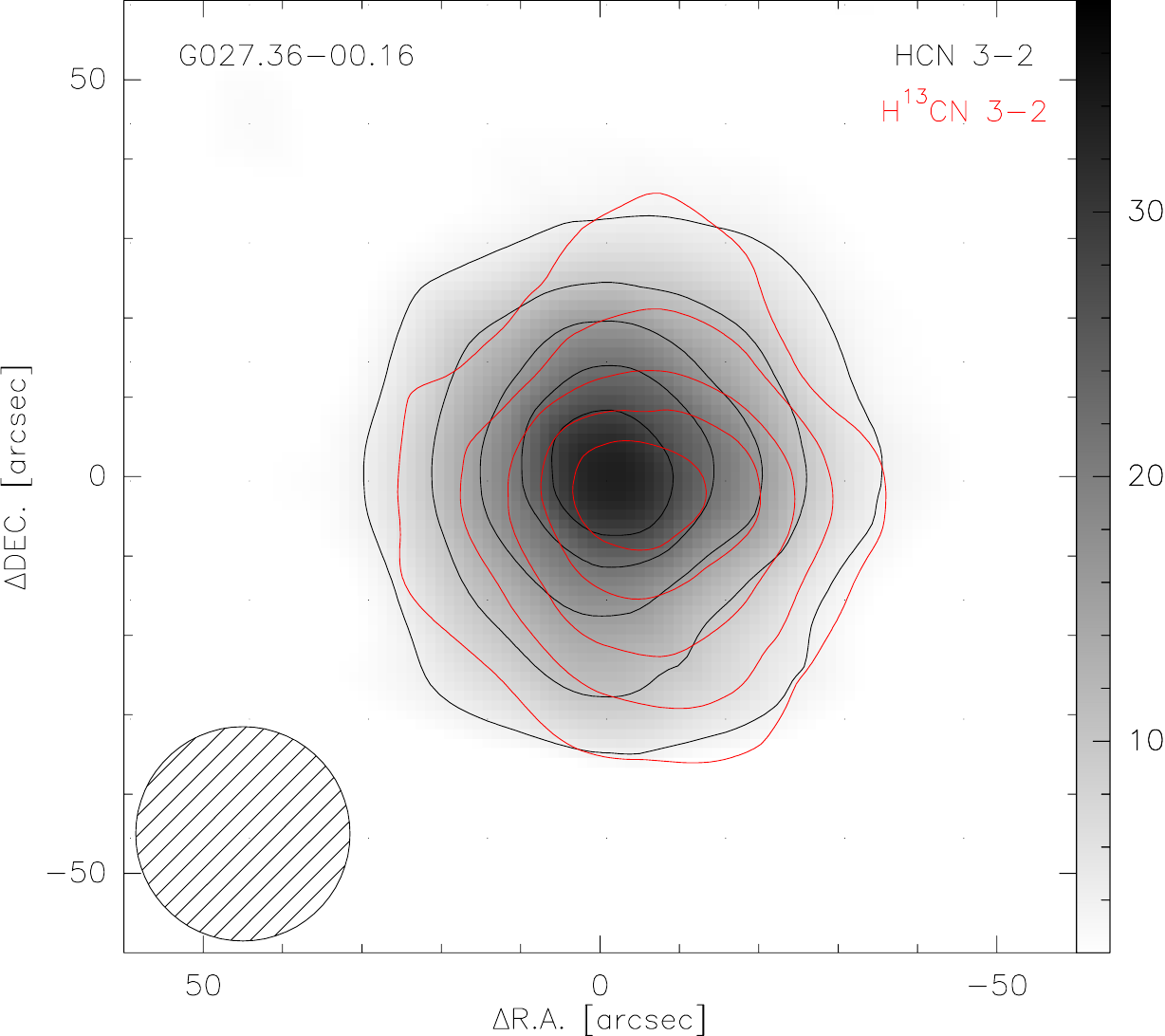}
   \includegraphics[width=3.03in]{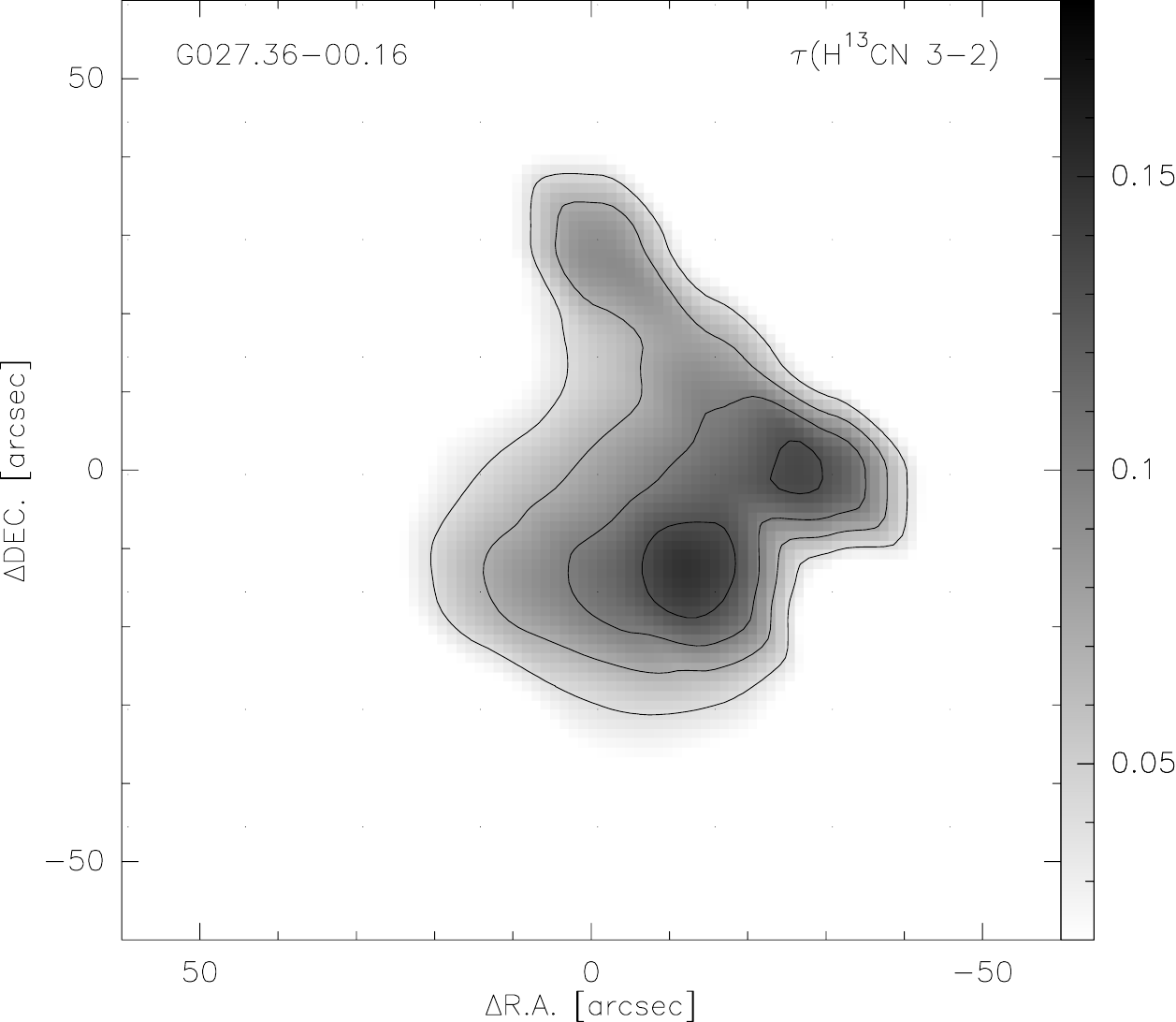}
       \includegraphics[width=3.05in]{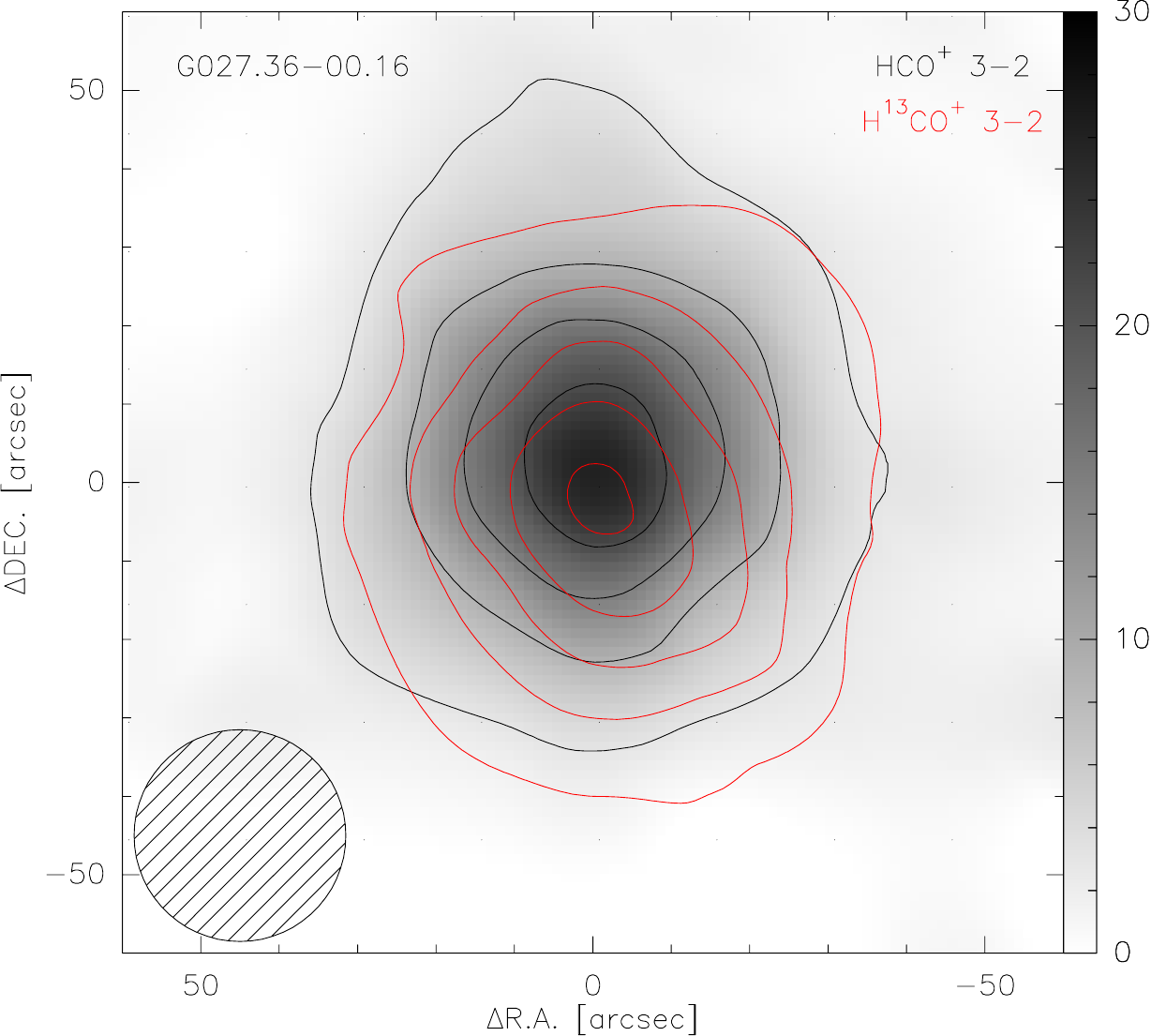}
       \includegraphics[width=3.08in]{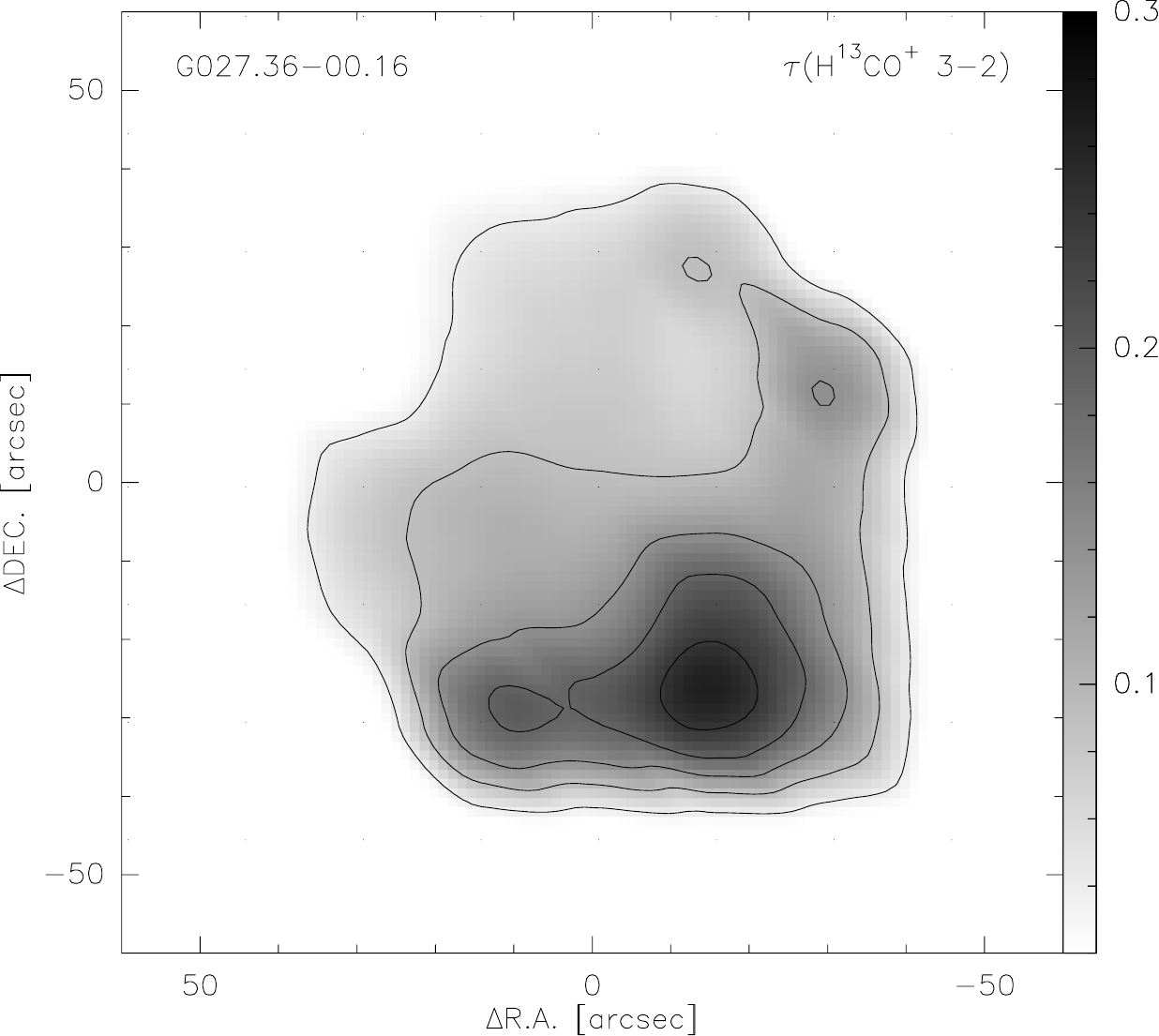}
 \caption{The data reduction results of G027.36-00.16. 
               {\it Top left:} The velocity integrated intensity maps of HCN and H$^{13}$CN 3-2. 
               The mapping size of HCN 3-2 is 2$'\times2'$, while it is 1.5$'\times1.5'$ for H$^{13}$CN 3-2, with a beam size of $\sim$ 27.8$''$.
               The grey scale and the black contour with levels starting from 4 K km s$^{-1}$ in step of 6 K km s$^{-1}$ show the observed HCN 3-2. 
               The red contour with levels starting from 0.4 K km s$^{-1}$ in step of 0.5 K km s$^{-1}$ represents H$^{13}$CN 3-2.
               {\it Top right:} The spatially resolved $\tau(\textrm{H}^{13}\textrm{CN})$ of G027.36-00.16 is demonstrated by black contour with levels 
               starting from 0.04 in step of 0.03. 
                {\it Bottom left:} The velocity integrated intensity maps of HCO$^+$ and H$^{13}$CO$^+$ 3-2. 
               The mapping size of HCO$^+$ 3-2 is 2$'\times2'$, while it is 1.5$'\times1.5'$ for H$^{13}$CO$^+$ 3-2, with a beam size of $\sim$ 27.8$''$.
               The grey scale and the black contour with levels starting from 3 K km s$^{-1}$ in step of 6 K km s$^{-1}$ show the observed HCO$^+$ 3-2. 
               The red contour with levels starting from 0.3 K km s$^{-1}$ in step of 0.5 K km s$^{-1}$ represents H$^{13}$CO$^+$ 3-2.
                {\it Bottom right:} The spatially resolved $\tau(\textrm{H}^{13}\textrm{CO$^+$})$ of G027.36-00.16 is demonstrated by black contour with levels 
               starting from 0.04 in step of 0.05. 
                }       
 \label{fig:g02736}
\end{figure*}


 \begin{figure*} 
    \centering
  \includegraphics[width=3.05in]{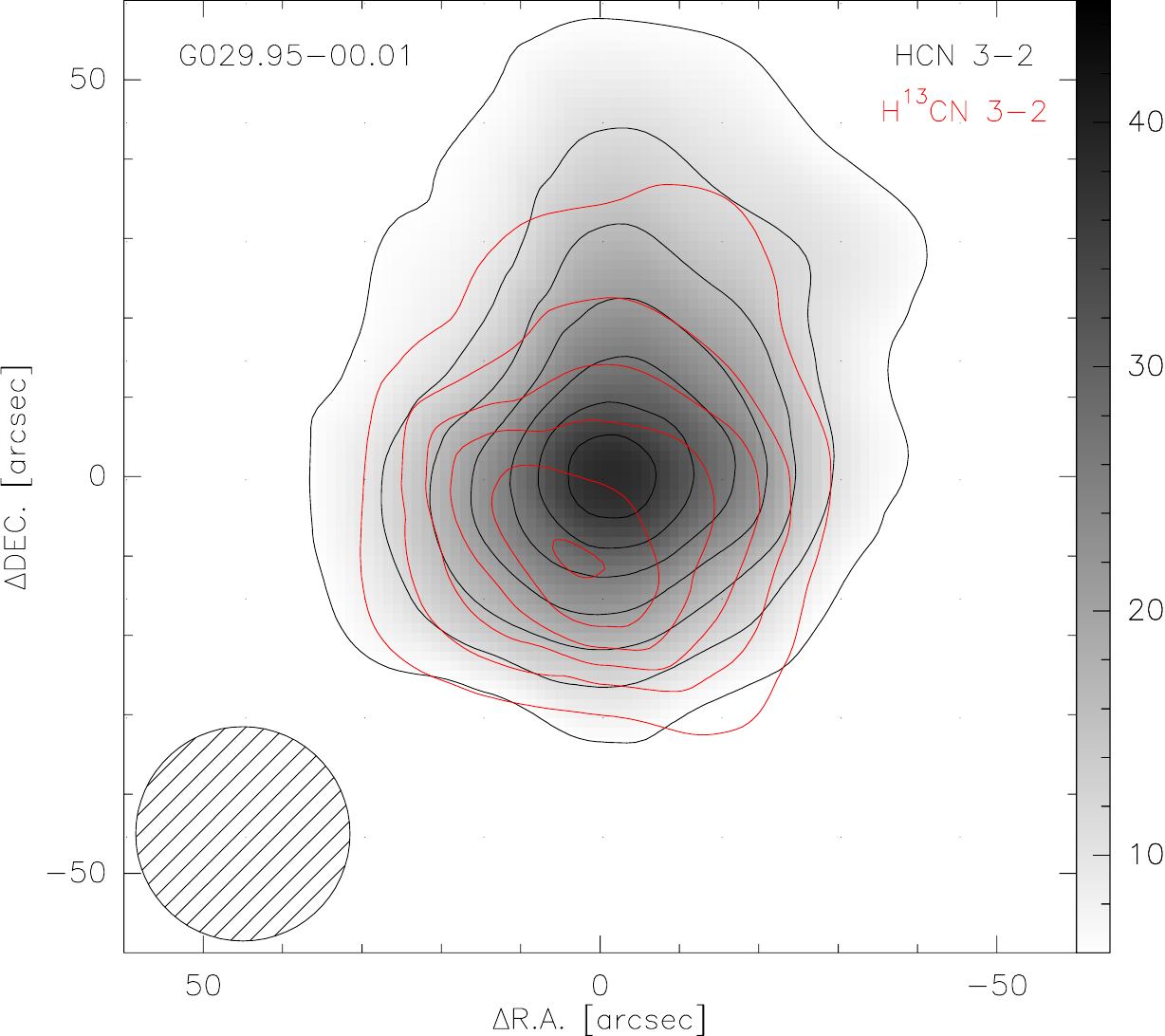}
   \includegraphics[width=3.03in]{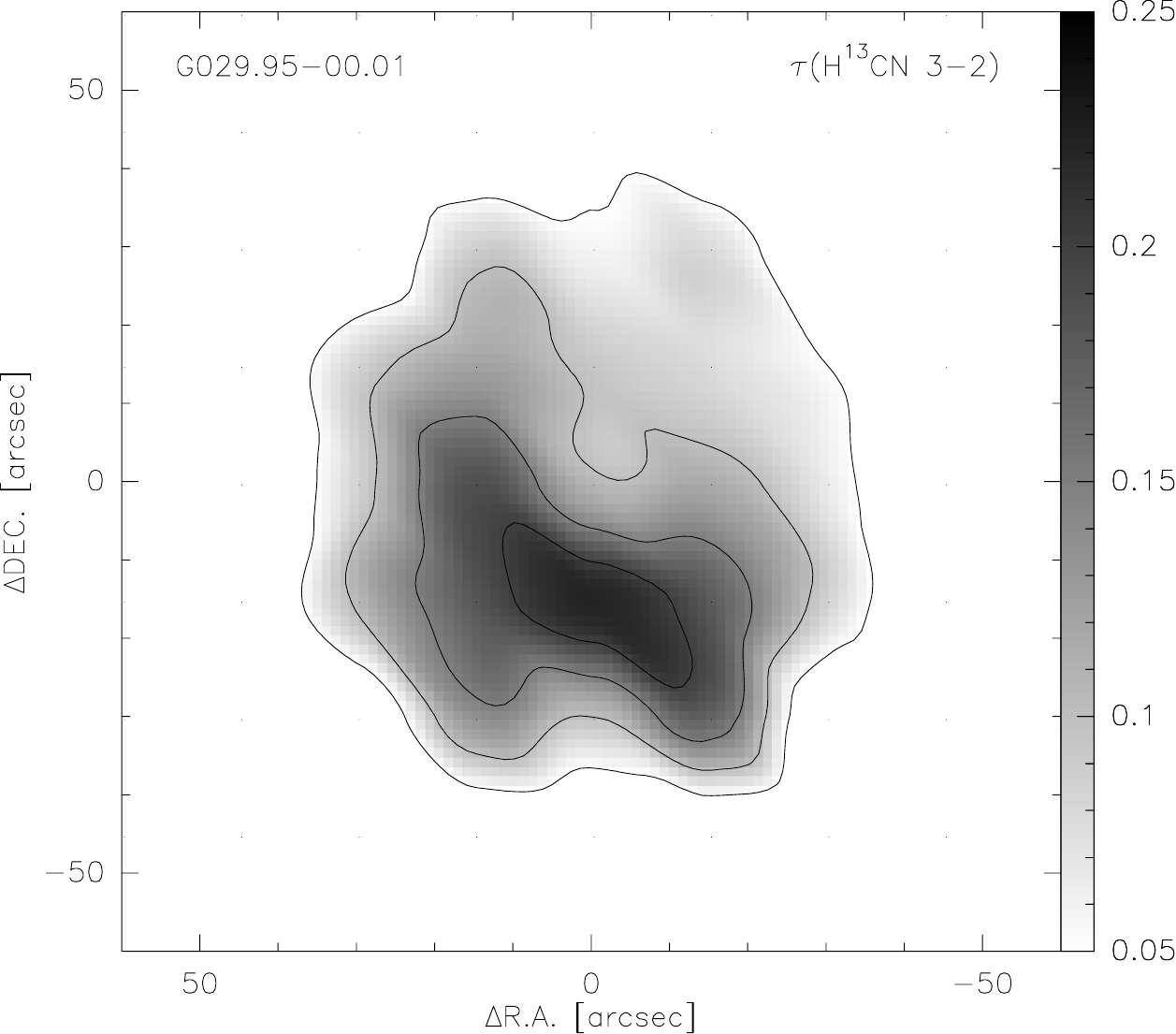}
       \includegraphics[width=3.05in]{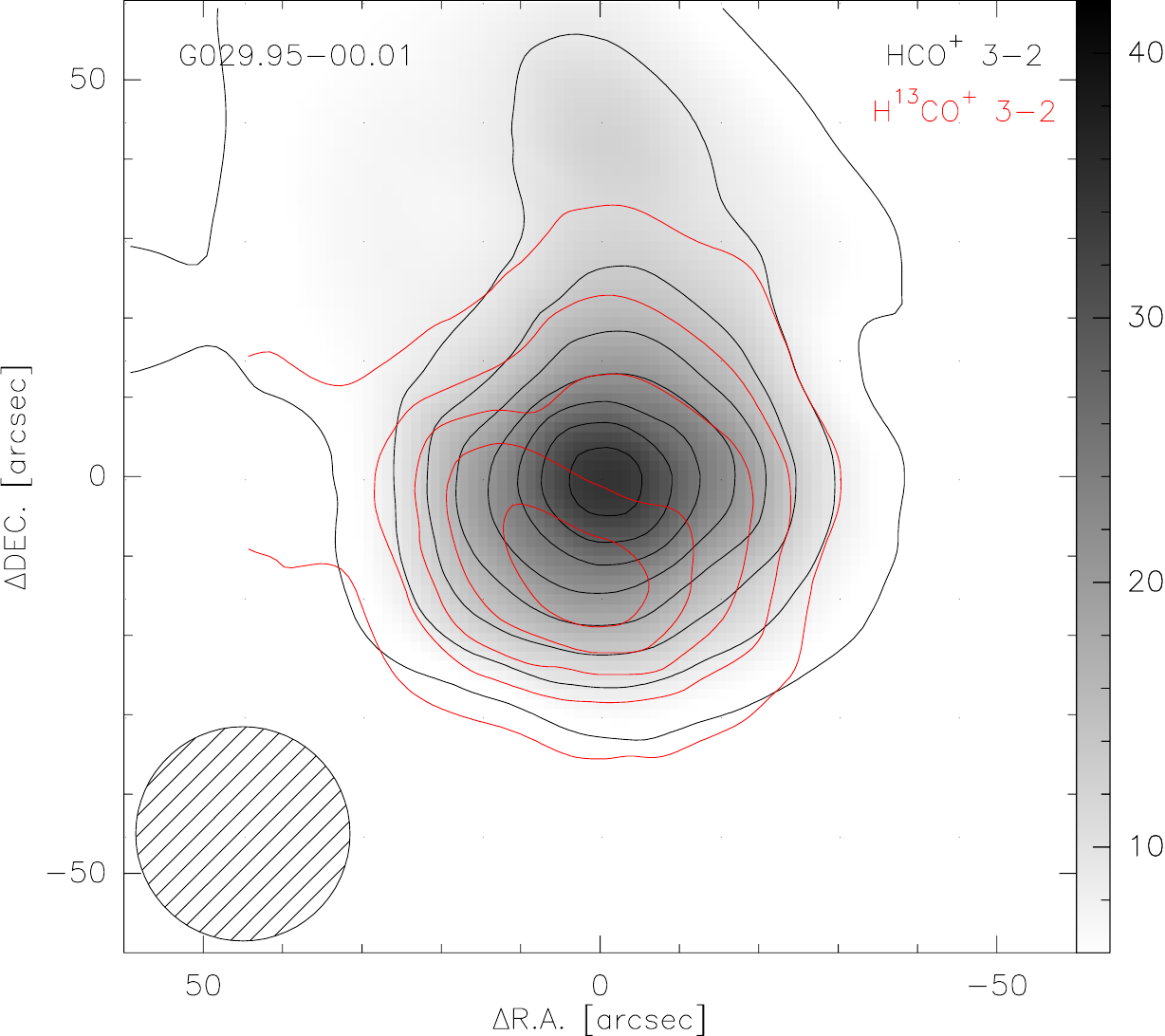}
       \includegraphics[width=3.08in]{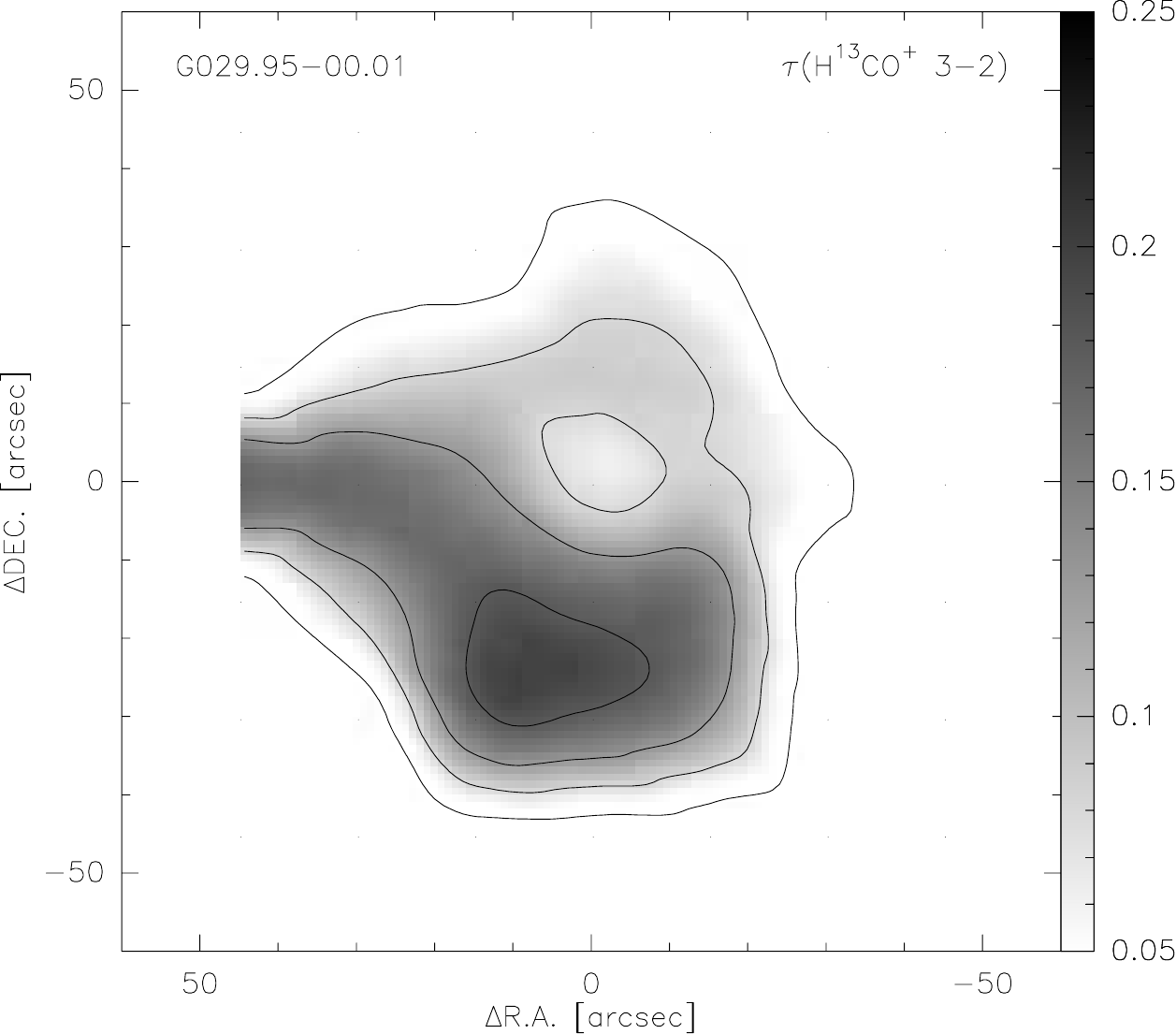}
 \caption{The data reduction results of G029.95-00.01. 
               {\it Top left:} The velocity integrated intensity maps of HCN and H$^{13}$CN 3-2. 
               The mapping size of HCN 3-2 is 2$'\times2'$, while it is 1.5$'\times1.5'$ for H$^{13}$CN 3-2, with a beam size of $\sim$ 27.8$''$.
               The grey scale and the black contour with levels starting from 6 K km s$^{-1}$ in step of 5 K km s$^{-1}$ show the observed HCN 3-2. 
               The red contour with levels starting from 0.8 K km s$^{-1}$ in step of 0.8 K km s$^{-1}$ represents H$^{13}$CN 3-2.
               {\it Top right:} The spatially resolved $\tau(\textrm{H}^{13}\textrm{CN})$ of G029.95-00.01 is demonstrated by black contour with levels 
               starting from 0.05 in step of 0.05. 
                {\it Bottom left:} The velocity integrated intensity maps of HCO$^+$ and H$^{13}$CO$^+$ 3-2. 
               The mapping size of HCO$^+$ 3-2 is 2$'\times2'$, while it is 1.5$'\times1.5'$ for H$^{13}$CO$^+$ 3-2, with a beam size of $\sim$ 27.8$''$.
               The grey scale and the black contour with levels starting from 5 K km s$^{-1}$ in step of 4 K km s$^{-1}$ show the observed HCO$^+$ 3-2. 
               The red contour with levels starting from 0.5 K km s$^{-1}$ in step of 0.6 K km s$^{-1}$ represents H$^{13}$CO$^+$ 3-2.
                {\it Bottom right:} The spatially resolved $\tau(\textrm{H}^{13}\textrm{CO$^+$})$ of G029.95-00.01 is demonstrated by black contour with levels 
               starting from 0.03 in step of 0.05. 
                }       
 \label{fig:g002995}
\end{figure*}


 \begin{figure*} 
    \centering
  \includegraphics[width=3.05in]{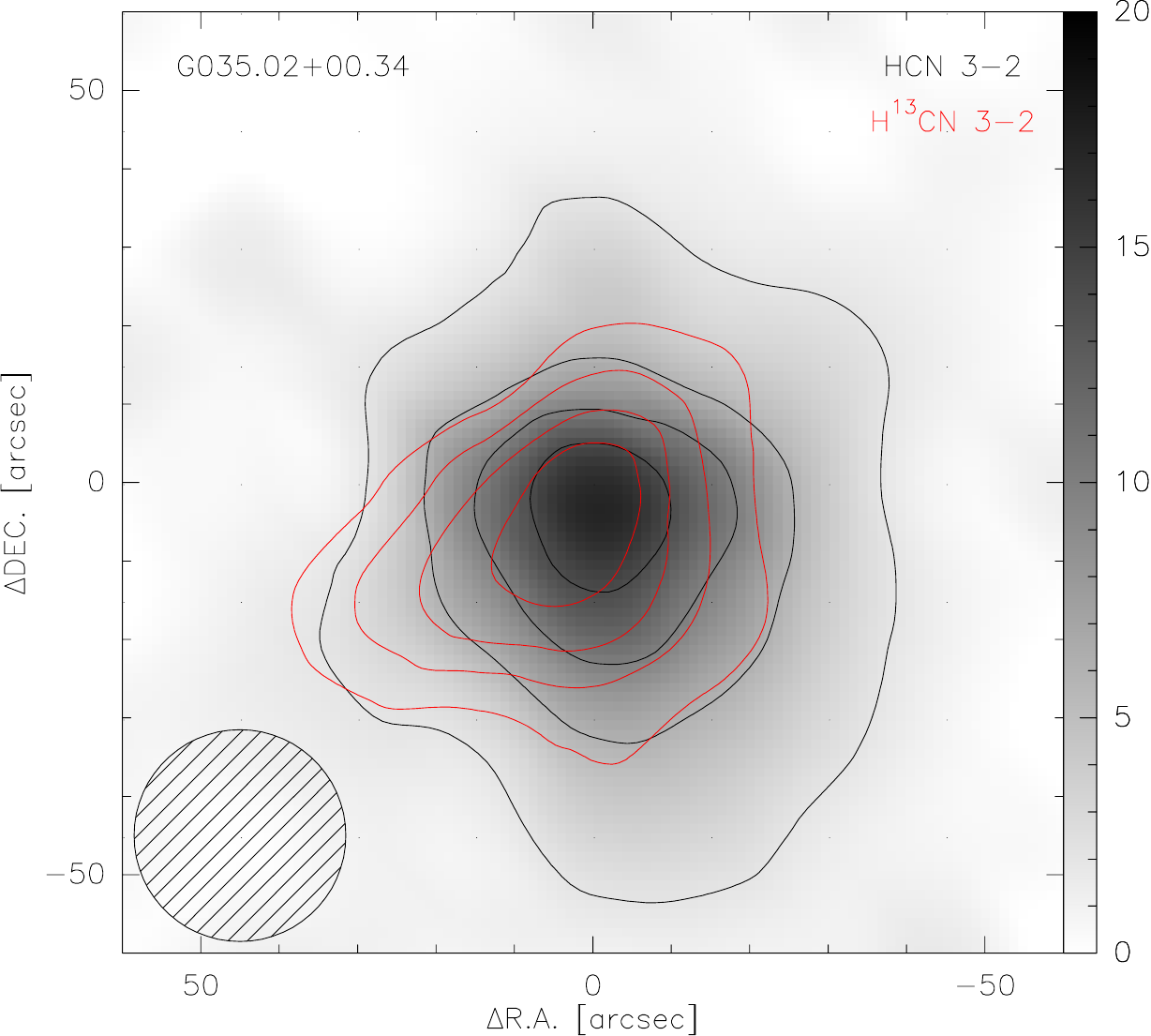}
   \includegraphics[width=3.03in]{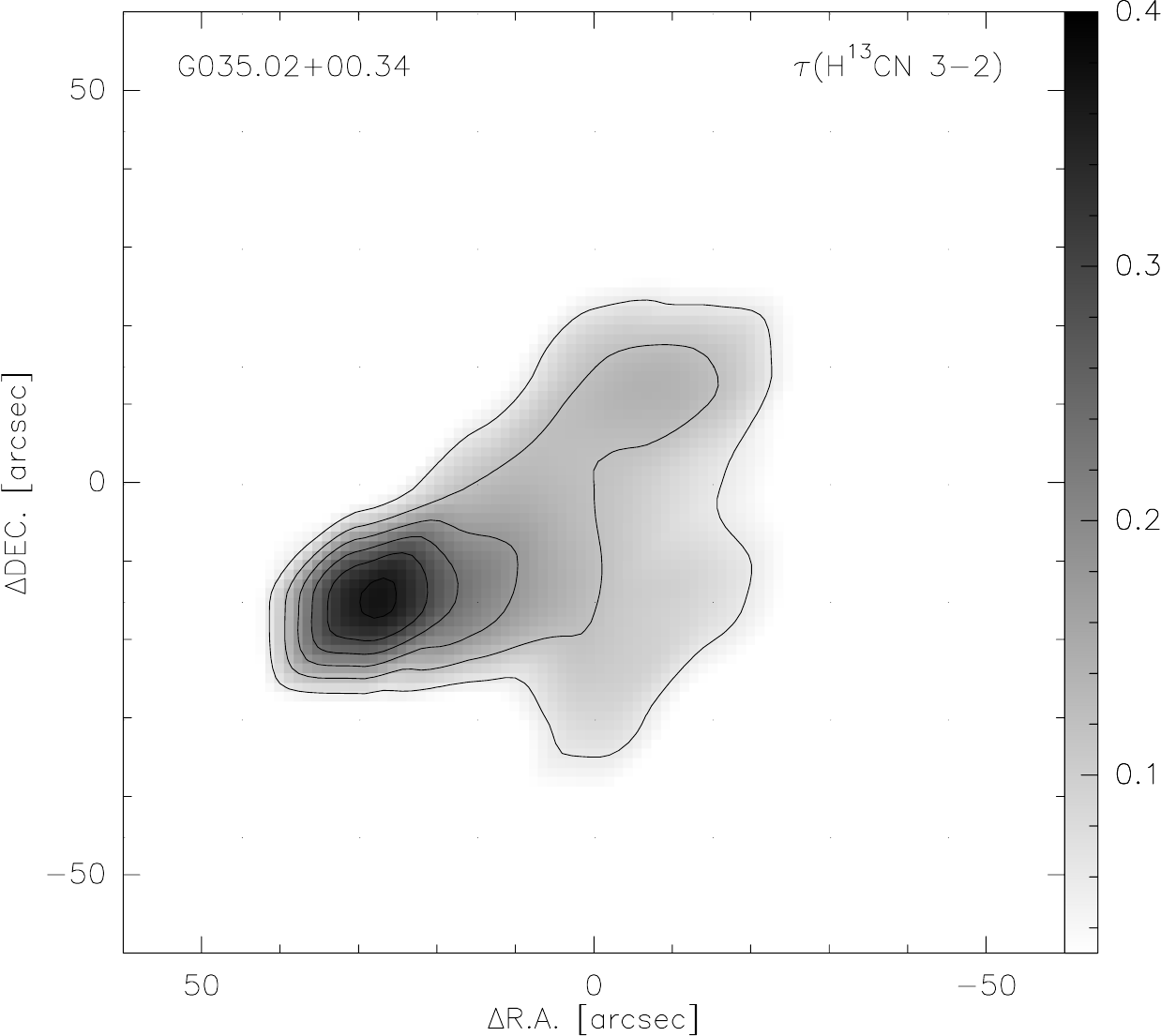}
       \includegraphics[width=3.05in]{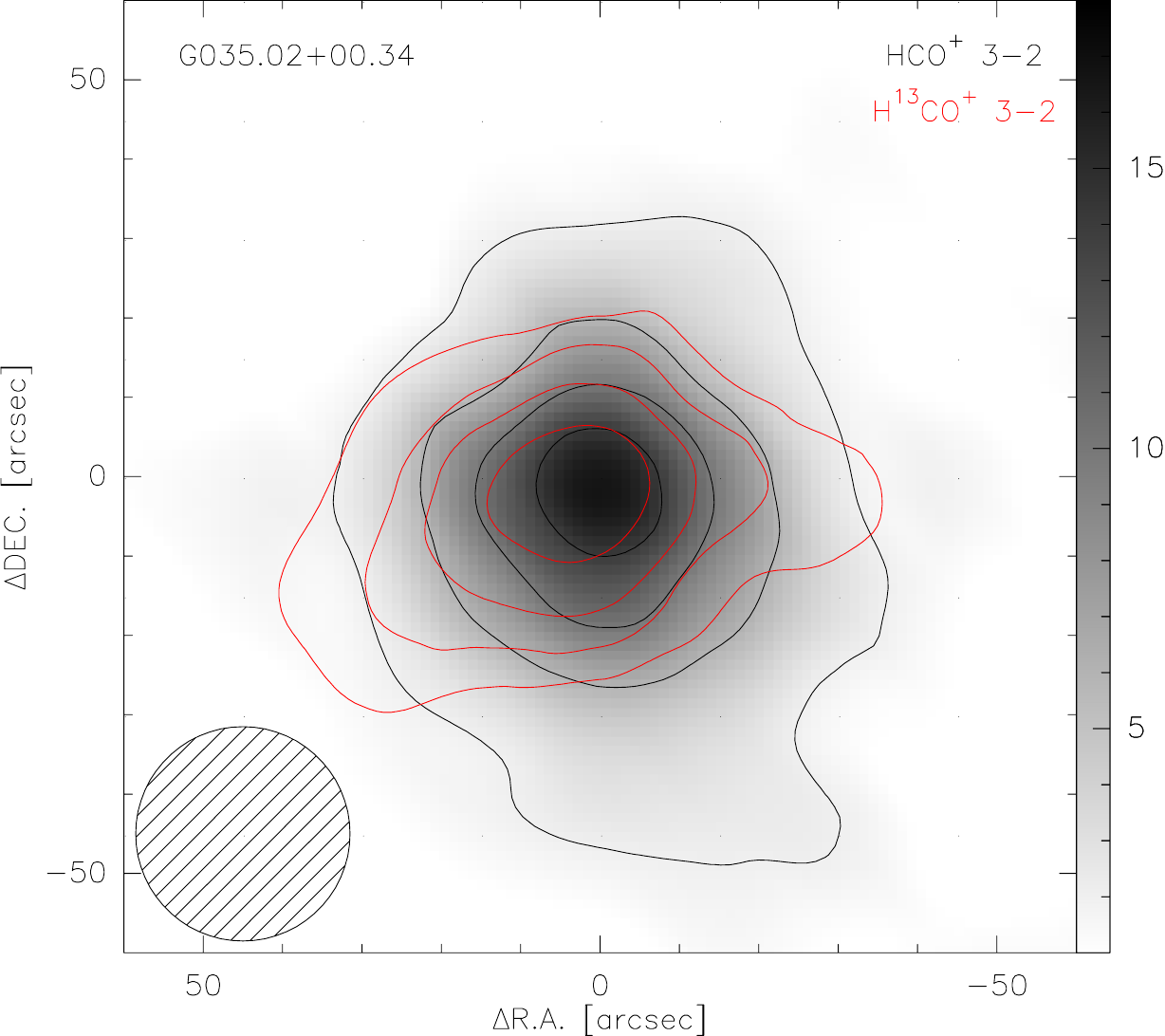}
       \includegraphics[width=3.08in]{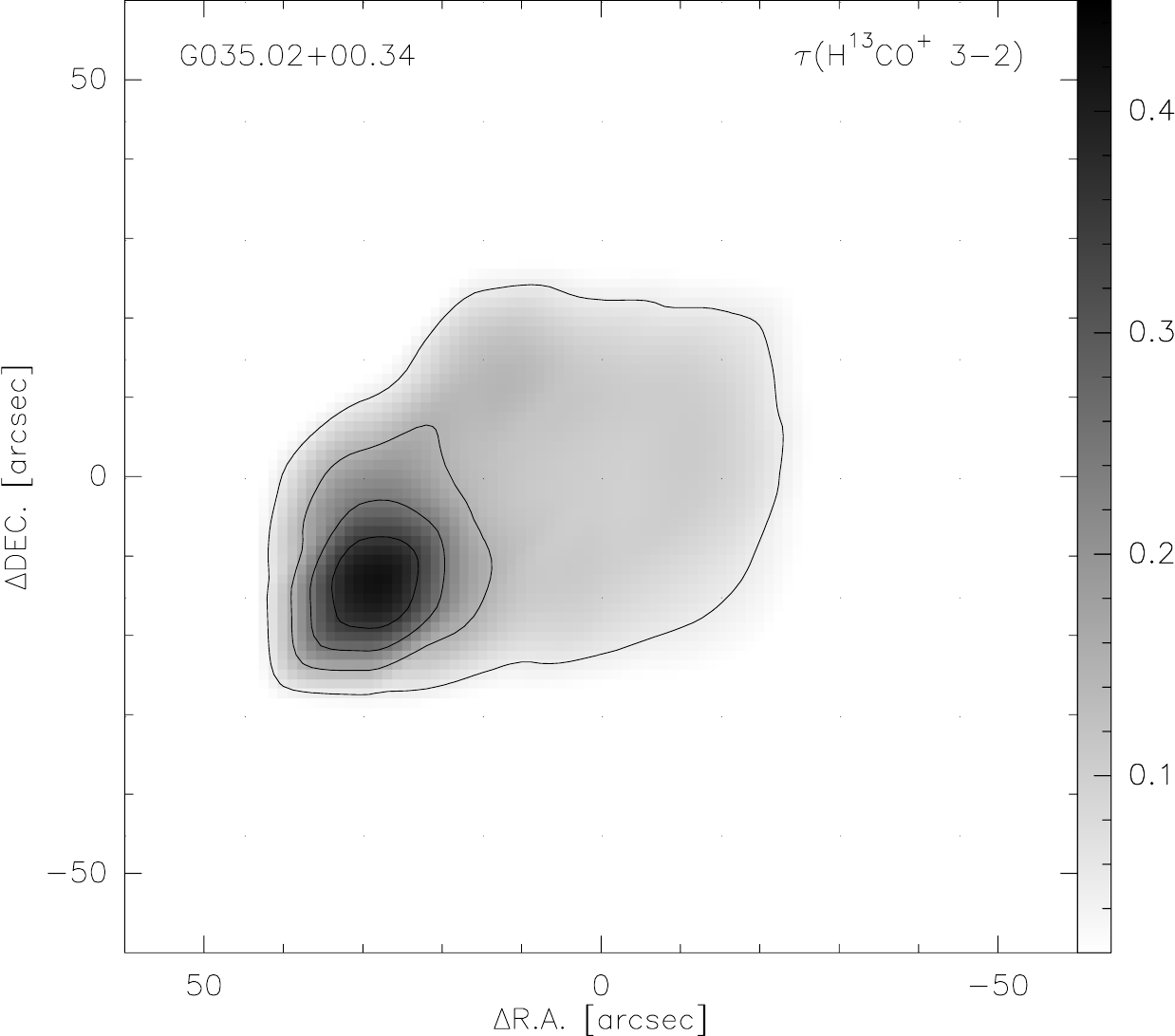}
 \caption{The data reduction results of G035.02+00.34. 
               {\it Top left:} The velocity integrated intensity maps of HCN and H$^{13}$CN 3-2. 
               The mapping size of HCN 3-2 is 2$'\times2'$, while it is 1.5$'\times1.5'$ for H$^{13}$CN 3-2, with a beam size of $\sim$ 27.8$''$.
               The grey scale and the black contour with levels starting from 2 K km s$^{-1}$ in step of 4 K km s$^{-1}$ show the observed HCN 3-2. 
               The red contour with levels starting from 0.4 K km s$^{-1}$ in step of 0.4 K km s$^{-1}$ represents H$^{13}$CN 3-2.
               {\it Top right:} The spatially resolved $\tau(\textrm{H}^{13}\textrm{CN})$ of G035.02+00.34 is demonstrated by black contour with levels 
               starting from 0.07 in step of 0.07. 
                {\it Bottom left:} The velocity integrated intensity maps of HCO$^+$ and H$^{13}$CO$^+$ 3-2. 
               The mapping size of HCO$^+$ 3-2 is 2$'\times2'$, while it is 1.5$'\times1.5'$ for H$^{13}$CO$^+$ 3-2, with a beam size of $\sim$ 27.8$''$.
               The grey scale and the black contour with levels starting from 2 K km s$^{-1}$ in step of 4 K km s$^{-1}$ show the observed HCO$^+$ 3-2. 
               The red contour with levels starting from 0.4 K km s$^{-1}$ in step of 0.3 K km s$^{-1}$ represents H$^{13}$CO$^+$ 3-2.
                {\it Bottom right:} The spatially resolved $\tau(\textrm{H}^{13}\textrm{CO$^+$})$ of G035.02+00.34 is demonstrated by black contour with levels 
               starting from 0.05 in step of 0.1. 
                }       
 \label{fig:g03502}
\end{figure*}


 \begin{figure*} 
    \centering
  \includegraphics[width=3.05in]{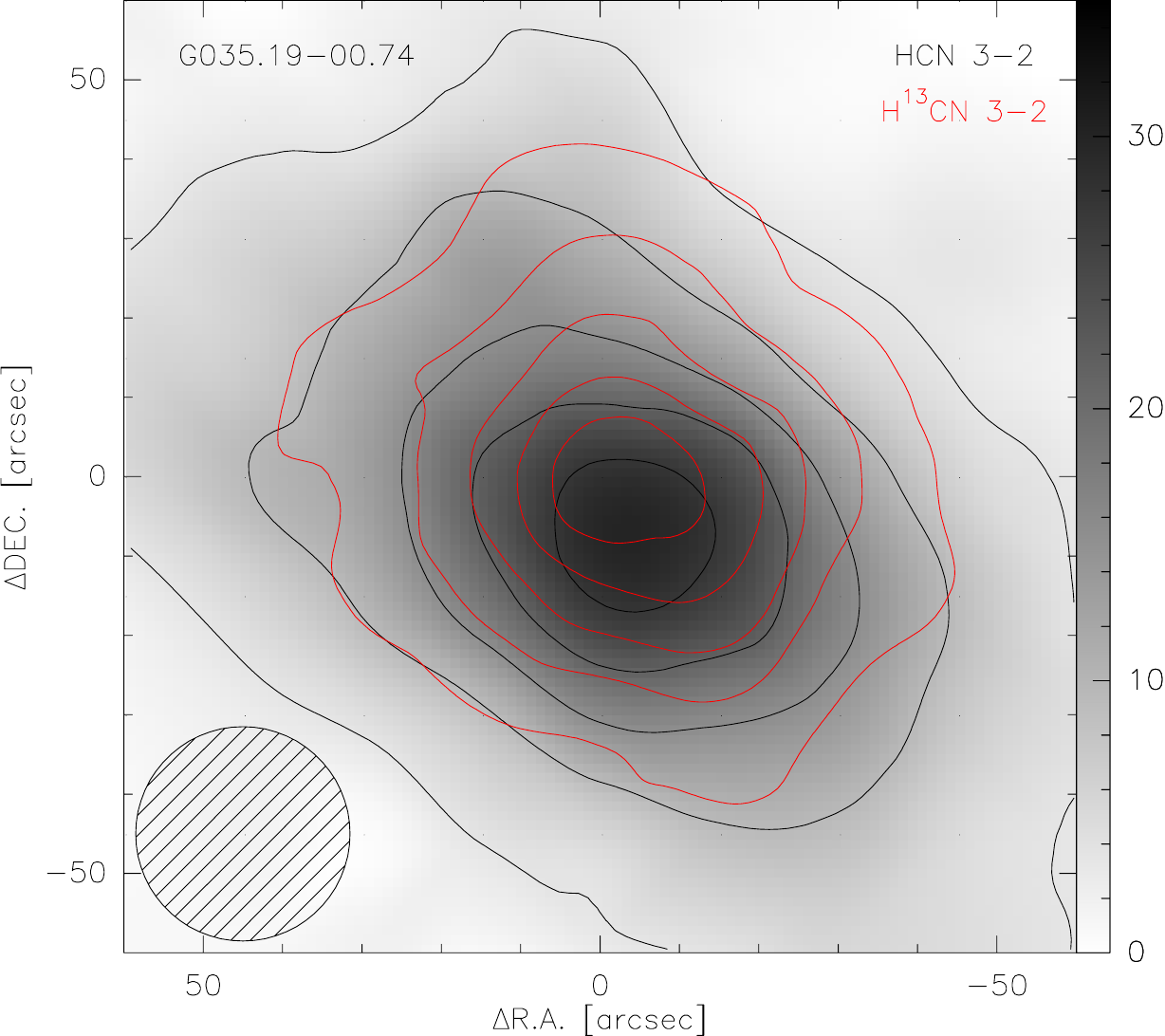}
   \includegraphics[width=3.03in]{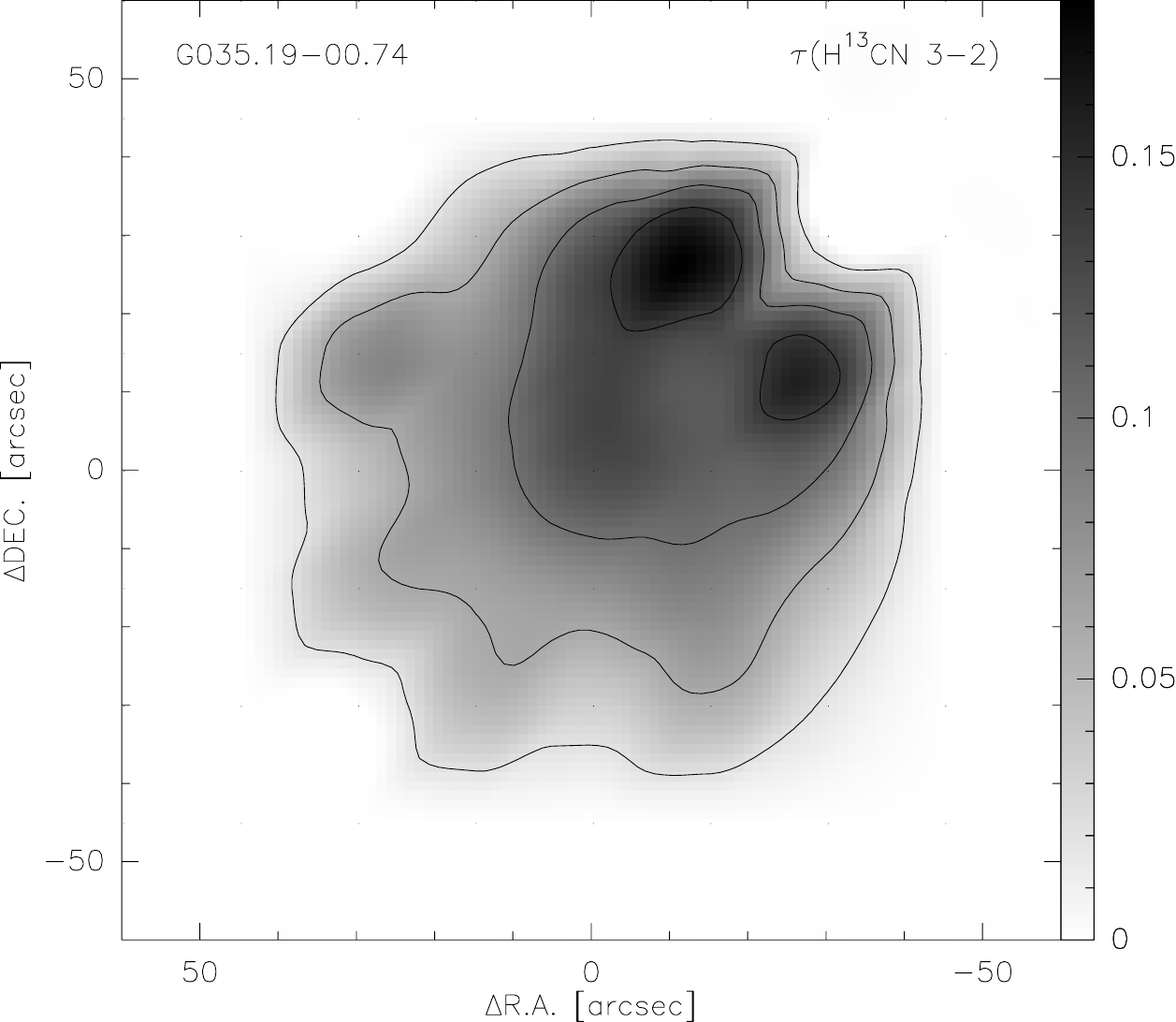}
       \includegraphics[width=3.05in]{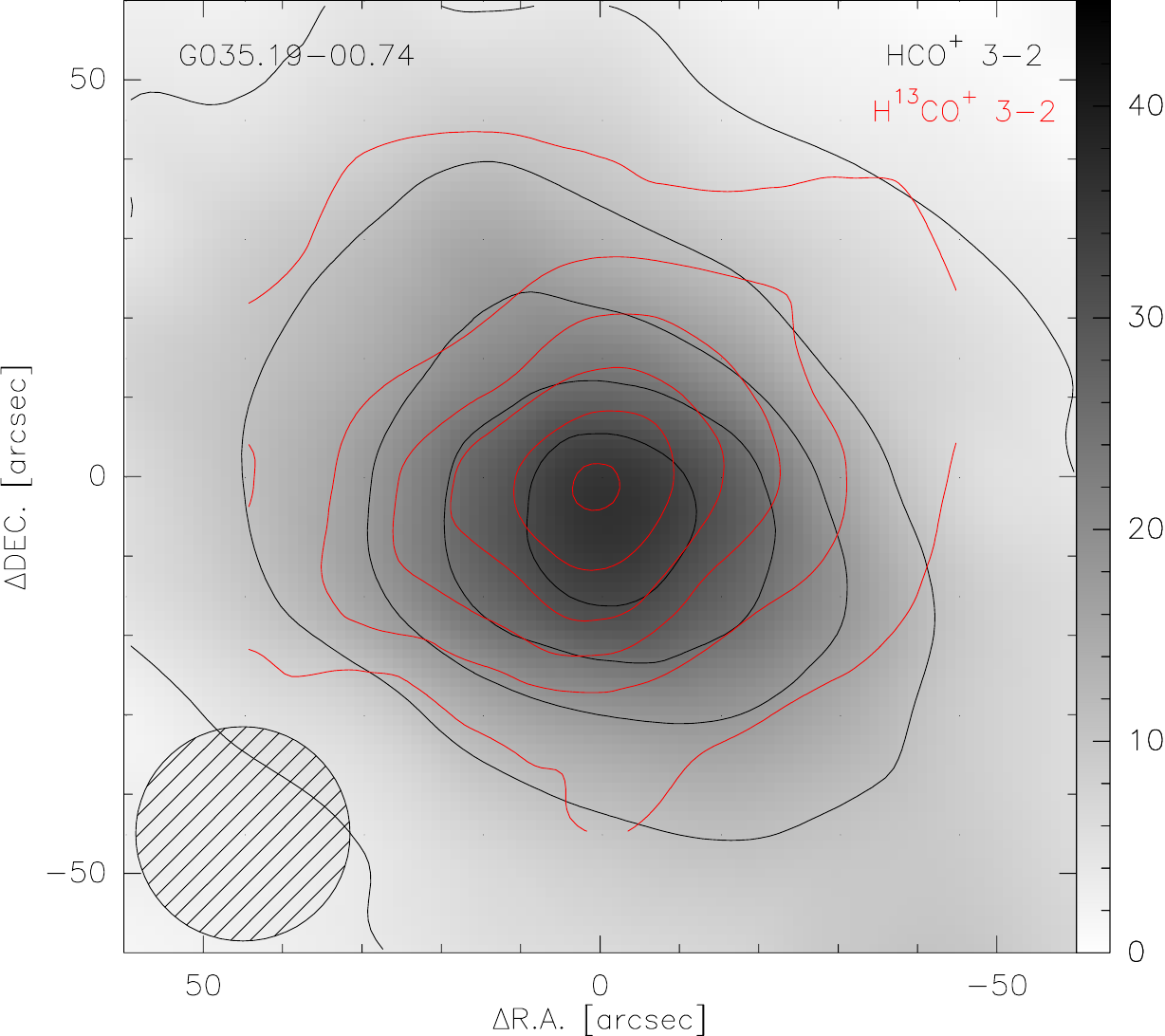}
       \includegraphics[width=3.08in]{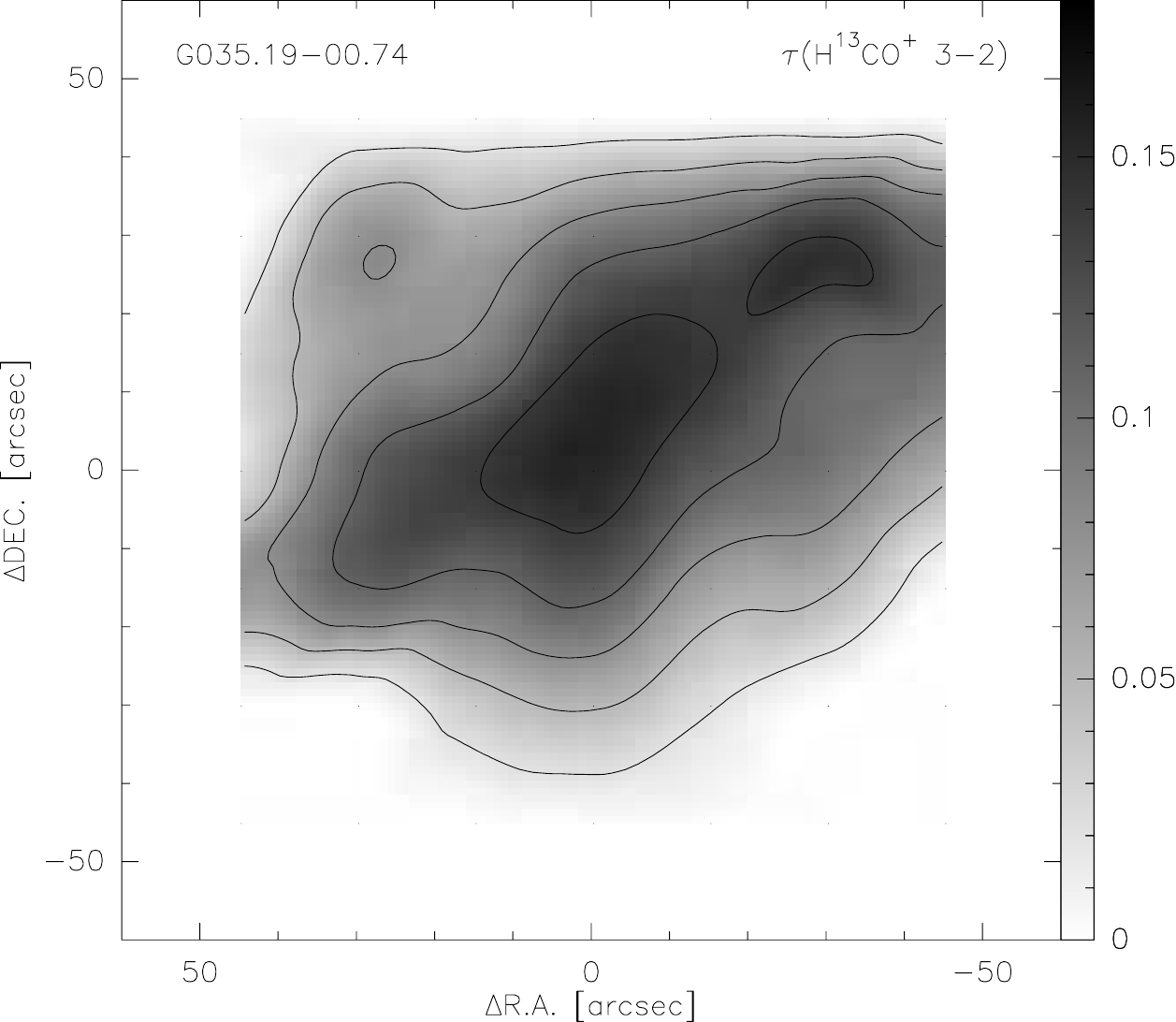}
 \caption{The data reduction results of G035.19-00.74. 
               {\it Top left:} The velocity integrated intensity maps of HCN and H$^{13}$CN 3-2. 
               The mapping size of HCN 3-2 is 2$'\times2'$, while it is 1.5$'\times1.5'$ for H$^{13}$CN 3-2, with a beam size of $\sim$ 27.8$''$.
               The grey scale and the black contour with levels starting from 4 K km s$^{-1}$ in step of 6 K km s$^{-1}$ show the observed HCN 3-2. 
               The red contour with levels starting from 0.5 K km s$^{-1}$ in step of 0.6 K km s$^{-1}$ represents H$^{13}$CN 3-2.
               {\it Top right:} The spatially resolved $\tau(\textrm{H}^{13}\textrm{CN})$ of G035.19-00.74 is demonstrated by black contour with levels 
               starting from 0.02 in step of 0.04. 
                {\it Bottom left:} The velocity integrated intensity maps of HCO$^+$ and H$^{13}$CO$^+$ 3-2. 
               The mapping size of HCO$^+$ 3-2 is 2$'\times2'$, while it is 1.5$'\times1.5'$ for H$^{13}$CO$^+$ 3-2, with a beam size of $\sim$ 27.8$''$.
               The grey scale and the black contour with levels starting from 4 K km s$^{-1}$ in step of 7 K km s$^{-1}$ show the observed HCO$^+$ 3-2. 
               The red contour with levels starting from 0.5 K km s$^{-1}$ in step of 0.9 K km s$^{-1}$ represents H$^{13}$CO$^+$ 3-2.
                {\it Bottom right:} The spatially resolved $\tau(\textrm{H}^{13}\textrm{CO$^+$})$ of G035.19-00.74 is demonstrated by black contour with levels 
               starting from 0.02 in step of 0.03. 
                }       
 \label{fig:g03519}
\end{figure*}


 \begin{figure*} 
    \centering
  \includegraphics[width=3.05in]{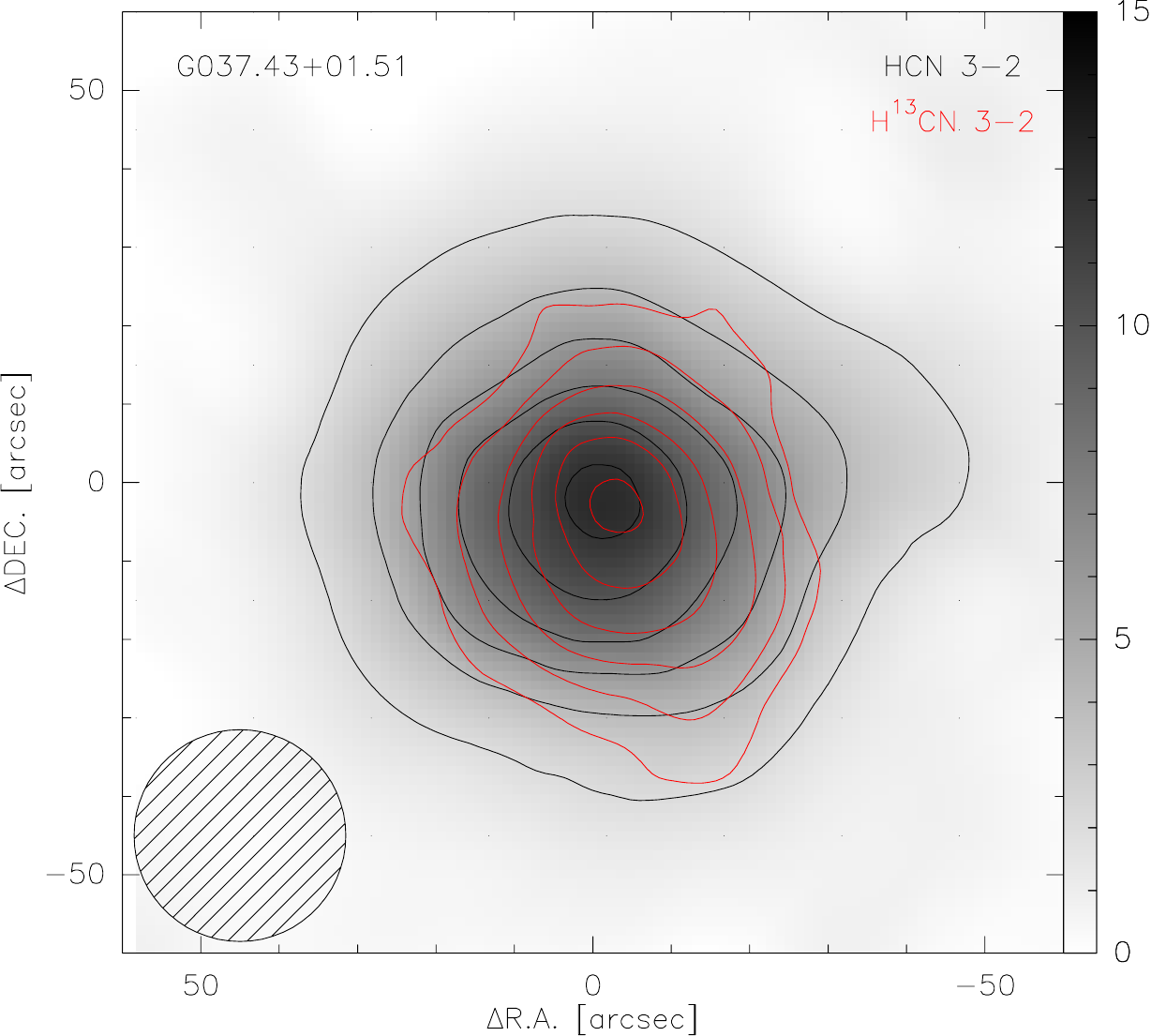}
   \includegraphics[width=3.03in]{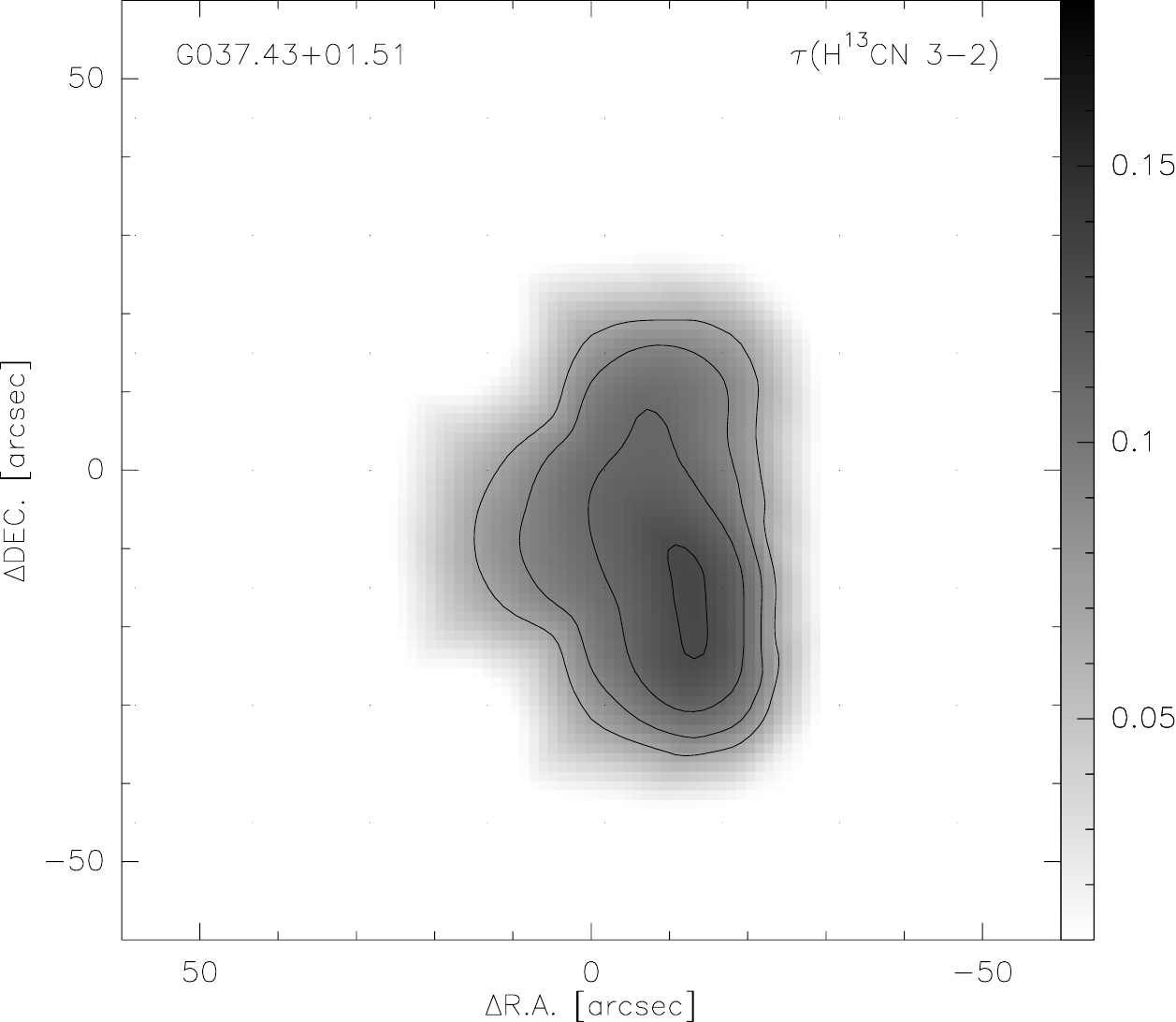}
       \includegraphics[width=3.05in]{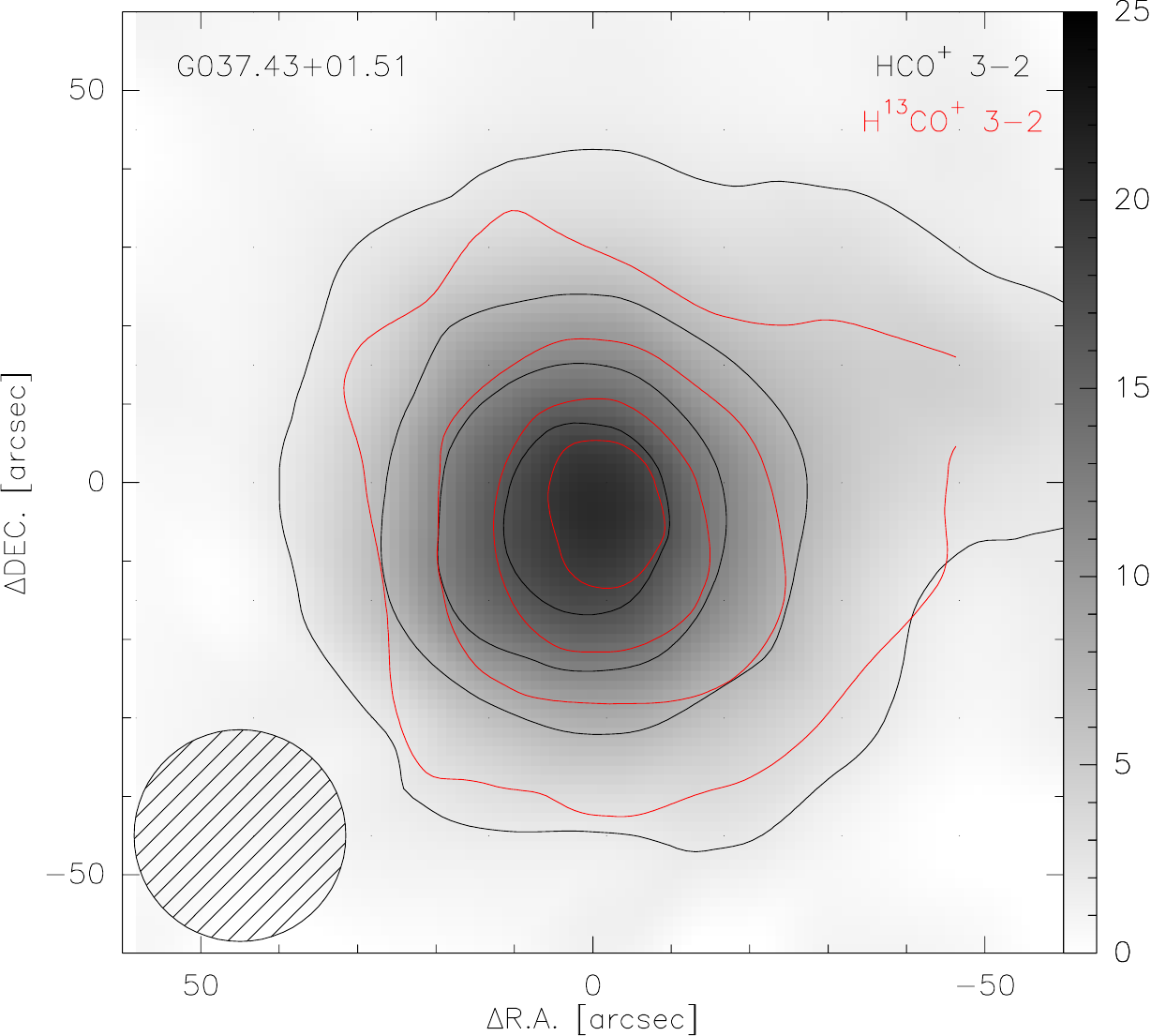}
       \includegraphics[width=3.08in]{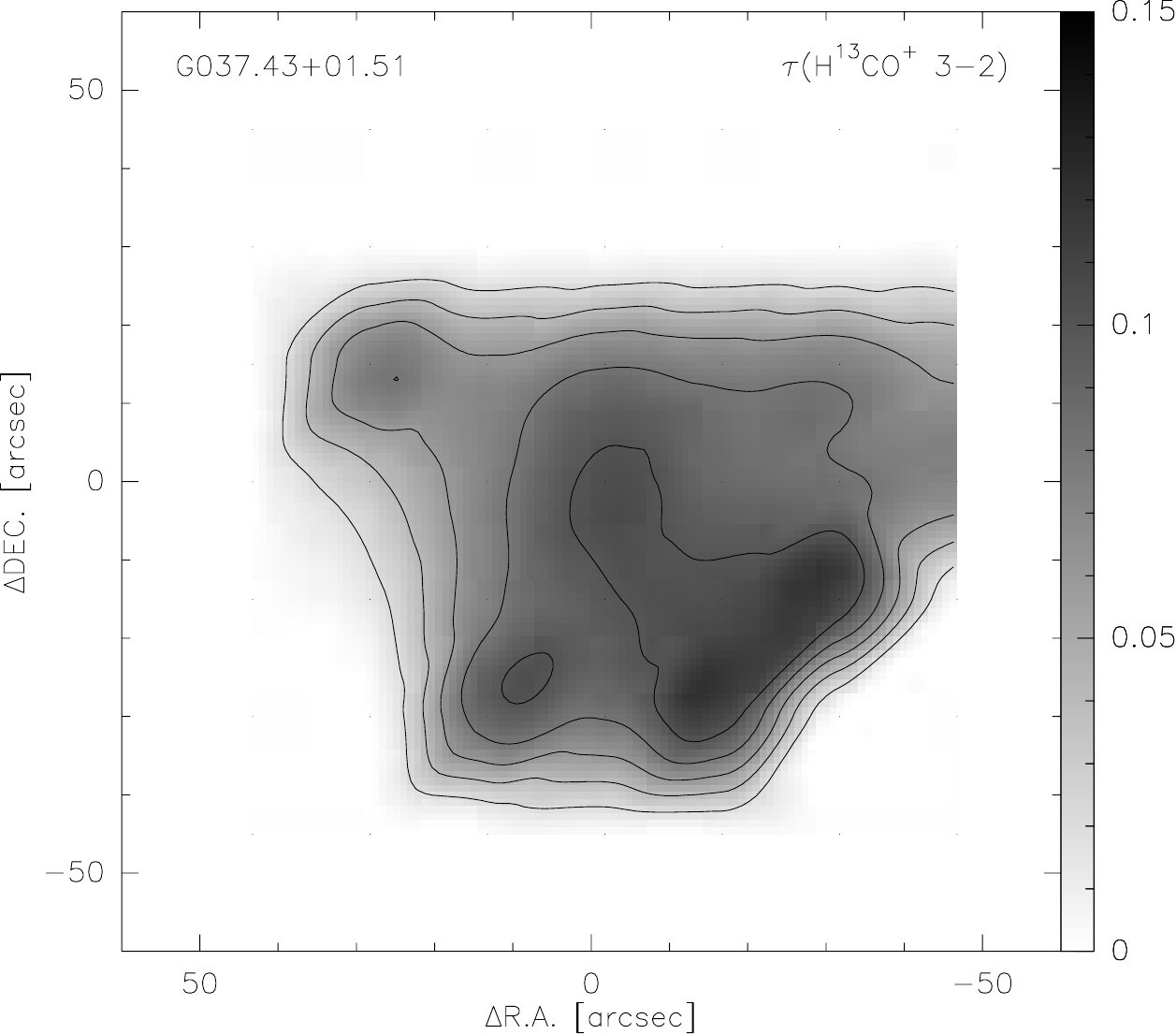}
 \caption{The data reduction results of G037.43+01.51. 
               {\it Top left:} The velocity integrated intensity maps of HCN and H$^{13}$CN 3-2. 
               The mapping size of HCN 3-2 is 2$'\times2'$, while it is 1.5$'\times1.5'$ for H$^{13}$CN 3-2, with a beam size of $\sim$ 27.8$''$.
               The grey scale and the black contour with levels starting from 2 K km s$^{-1}$ in step of 2 K km s$^{-1}$ show the observed HCN 3-2. 
               The red contour with levels starting from 0.3 K km s$^{-1}$ in step of 0.2 K km s$^{-1}$ represents H$^{13}$CN 3-2.
               {\it Top right:} The spatially resolved $\tau(\textrm{H}^{13}\textrm{CN})$ of G037.43+01.51 is demonstrated by black contour with levels 
               starting from 0.07 in step of 0.02. 
                {\it Bottom left:} The velocity integrated intensity maps of HCO$^+$ and H$^{13}$CO$^+$ 3-2. 
               The mapping size of HCO$^+$ 3-2 is 2$'\times2'$, while it is 1.5$'\times1.5'$ for H$^{13}$CO$^+$ 3-2, with a beam size of $\sim$ 27.8$''$.
               The grey scale and the black contour with levels starting from 1.5 K km s$^{-1}$ in step of 5 K km s$^{-1}$ show the observed HCO$^+$ 3-2. 
               The red contour with levels starting from 0.3 K km s$^{-1}$ in step of 0.5 K km s$^{-1}$ represents H$^{13}$CO$^+$ 3-2.
                {\it Bottom right:} The spatially resolved $\tau(\textrm{H}^{13}\textrm{CO$^+$})$ of G037.43+01.51 is demonstrated by black contour with levels 
               starting from 0.02 in step of 0.02. 
                }       
 \label{fig:g03743}
\end{figure*}


 \begin{figure*} 
    \centering
  \includegraphics[width=3.05in]{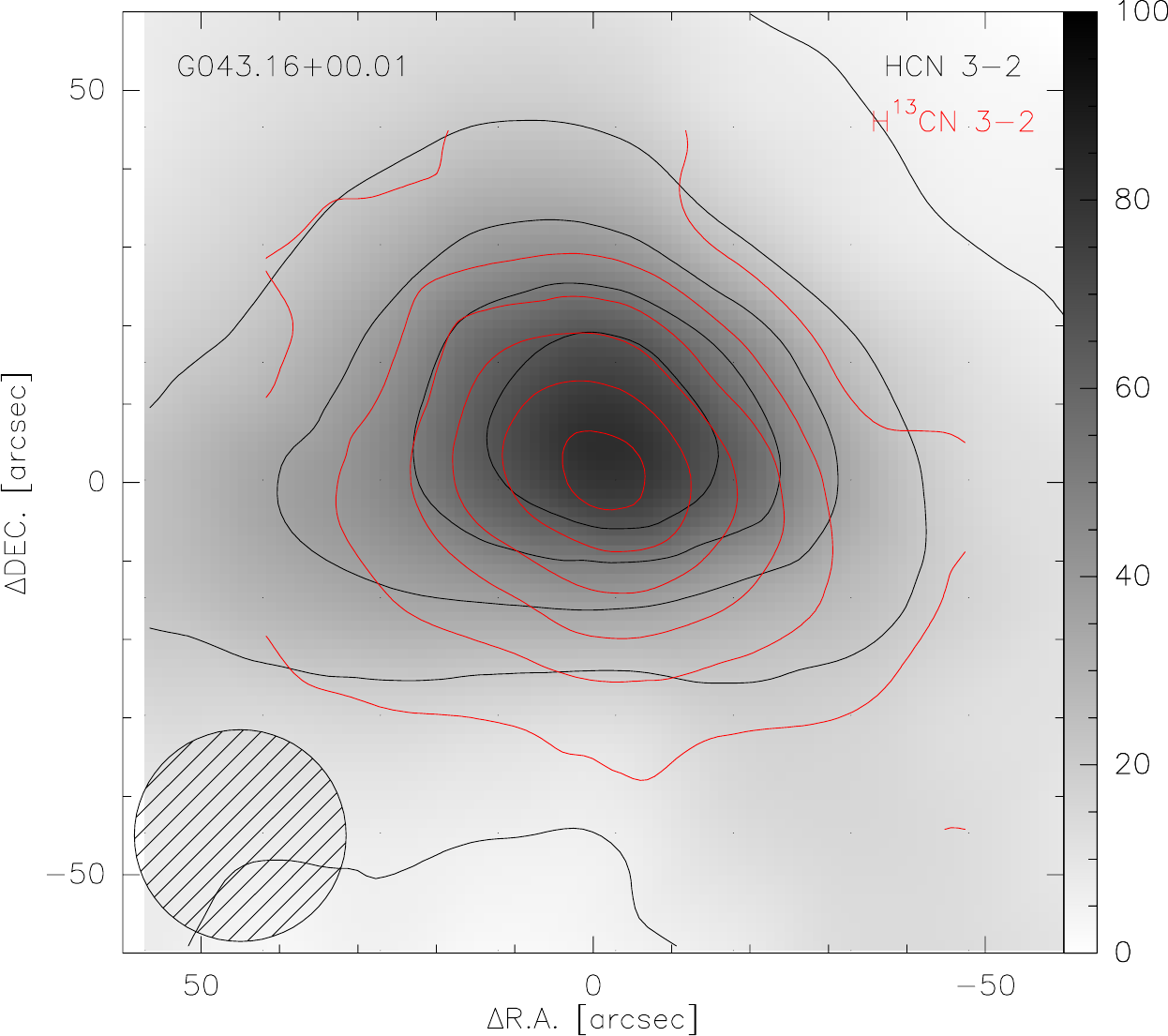}
   \includegraphics[width=3.03in]{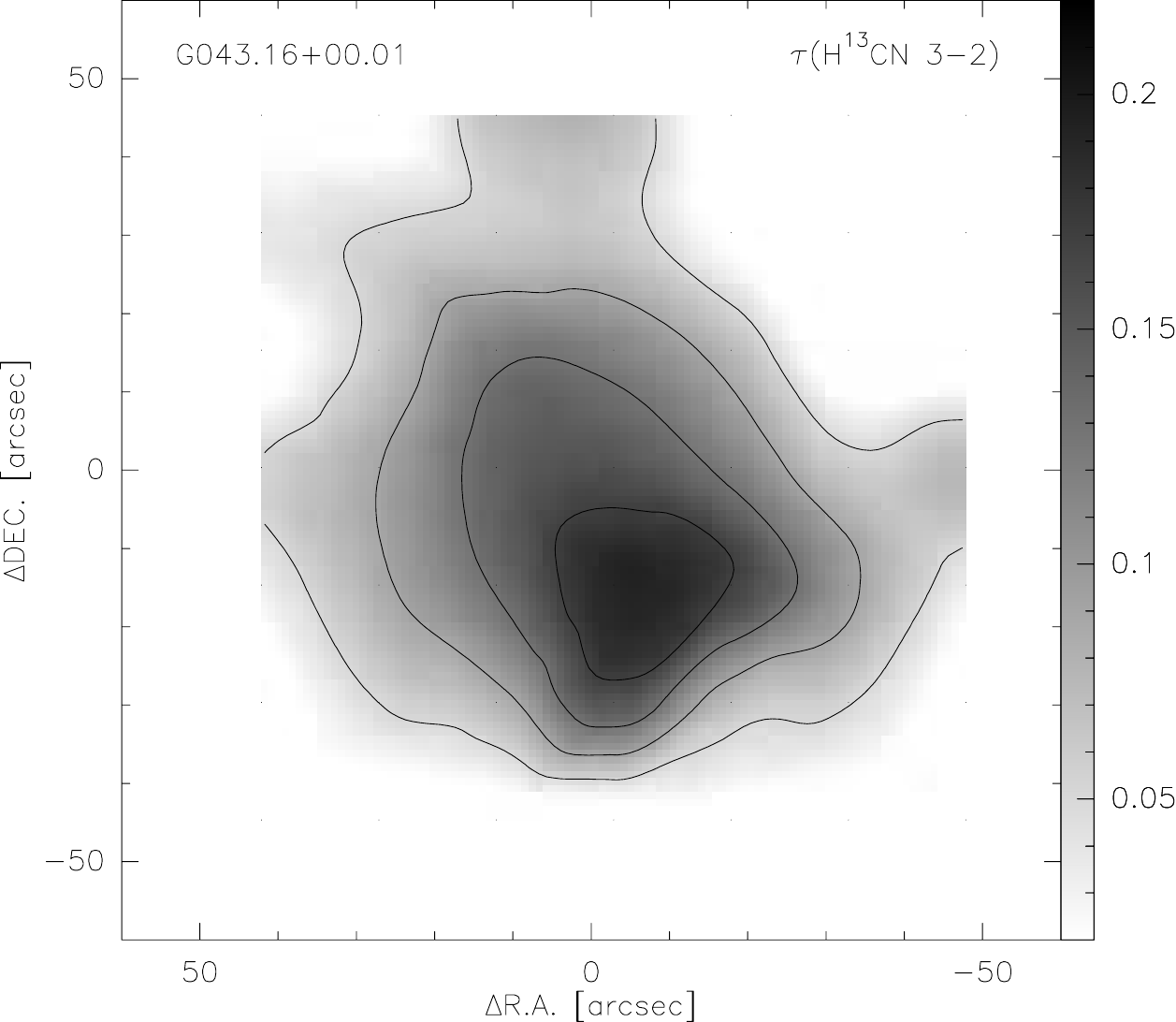}
       \includegraphics[width=3.05in]{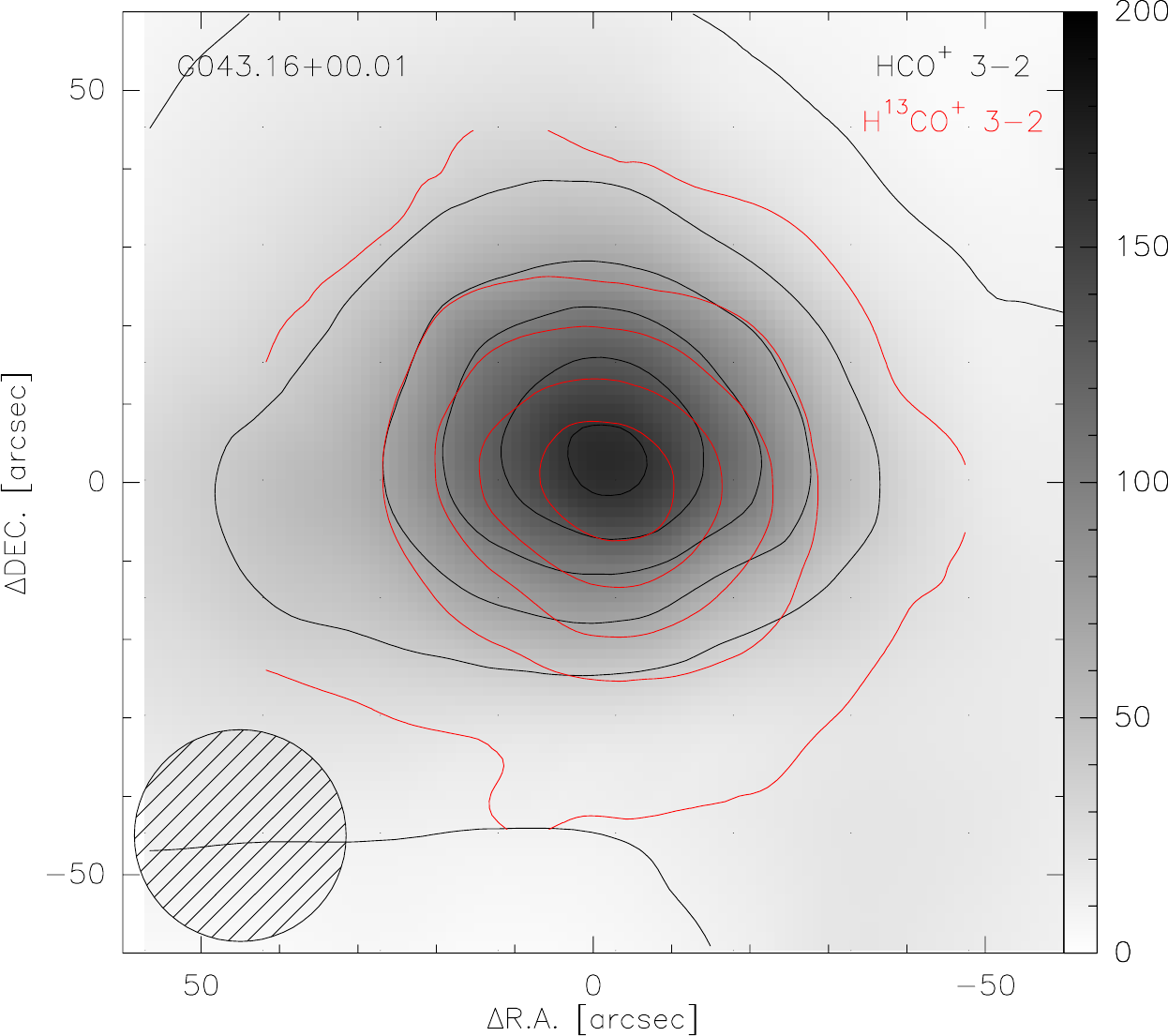}
       \includegraphics[width=3.08in]{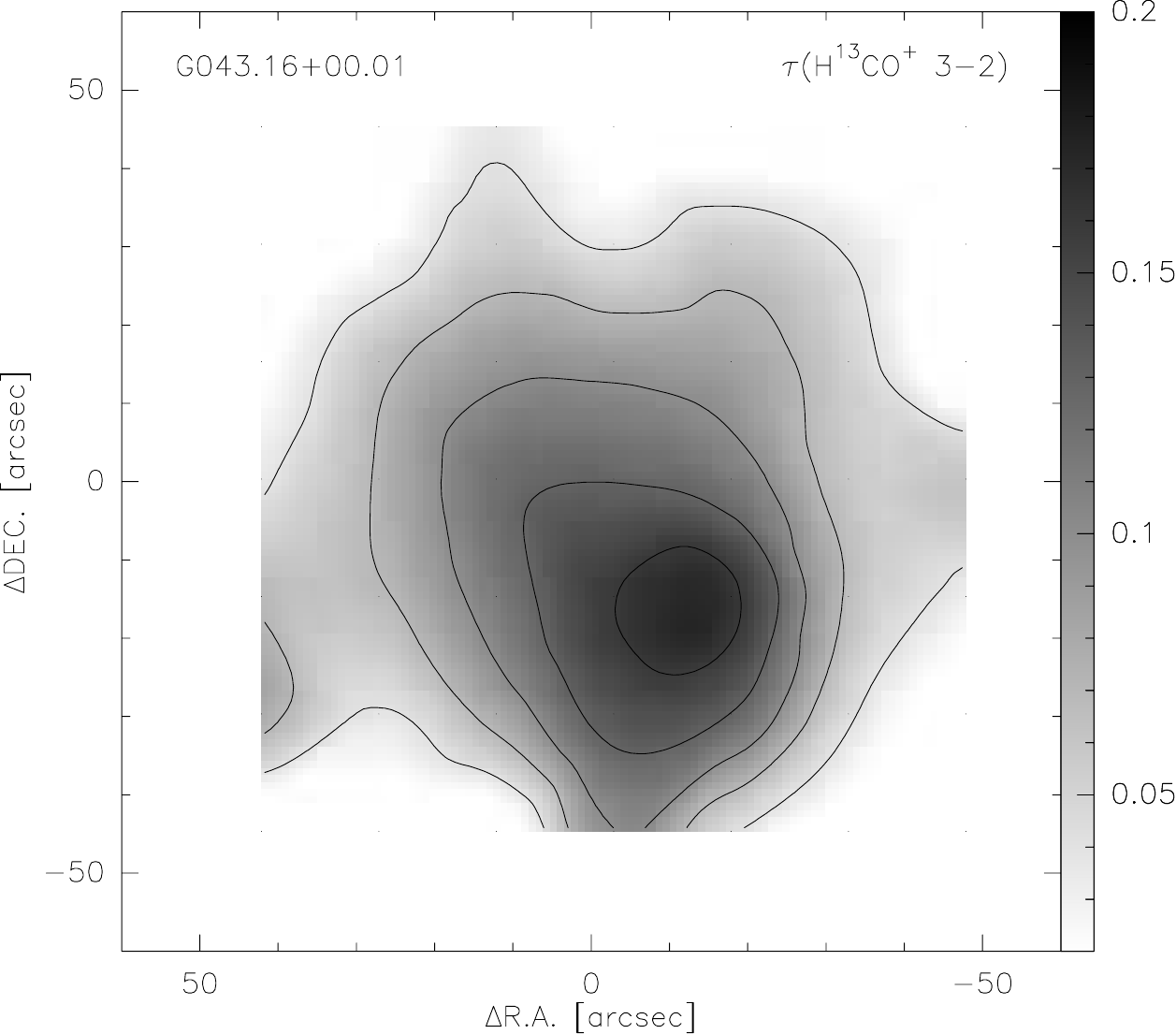}
 \caption{The data reduction results of G043.16+00.01. 
               {\it Top left:} The velocity integrated intensity maps of HCN and H$^{13}$CN 3-2. 
               The mapping size of HCN 3-2 is 2$'\times2'$, while it is 1.5$'\times1.5'$ for H$^{13}$CN 3-2, with a beam size of $\sim$ 27.8$''$.
               The grey scale and the black contour with levels starting from 6 K km s$^{-1}$ in step of 15 K km s$^{-1}$ show the observed HCN 3-2. 
               The red contour with levels starting from 1 K km s$^{-1}$ in step of 2 K km s$^{-1}$ represents H$^{13}$CN 3-2.
               {\it Top right:} The spatially resolved $\tau(\textrm{H}^{13}\textrm{CN})$ of G043.16+00.01 is demonstrated by black contour with levels 
               starting from 0.05 in step of 0.04. 
                {\it Bottom left:} The velocity integrated intensity maps of HCO$^+$ and H$^{13}$CO$^+$ 3-2. 
               The mapping size of HCO$^+$ 3-2 is 2$'\times2'$, while it is 1.5$'\times1.5'$ for H$^{13}$CO$^+$ 3-2, with a beam size of $\sim$ 27.8$''$.
               The grey scale and the black contour with levels starting from 10 K km s$^{-1}$ in step of 30 K km s$^{-1}$ show the observed HCO$^+$ 3-2. 
               The red contour with levels starting from 1 K km s$^{-1}$ in step of 4 K km s$^{-1}$ represents H$^{13}$CO$^+$ 3-2.
                {\it Bottom right:} The spatially resolved $\tau(\textrm{H}^{13}\textrm{CO$^+$})$ of G043.16+00.01 is demonstrated by black contour with levels 
               starting from 0.04 in step of 0.03. 
                }       
 \label{fig:g04316}
\end{figure*}


 \begin{figure*} 
    \centering
  \includegraphics[width=3.05in]{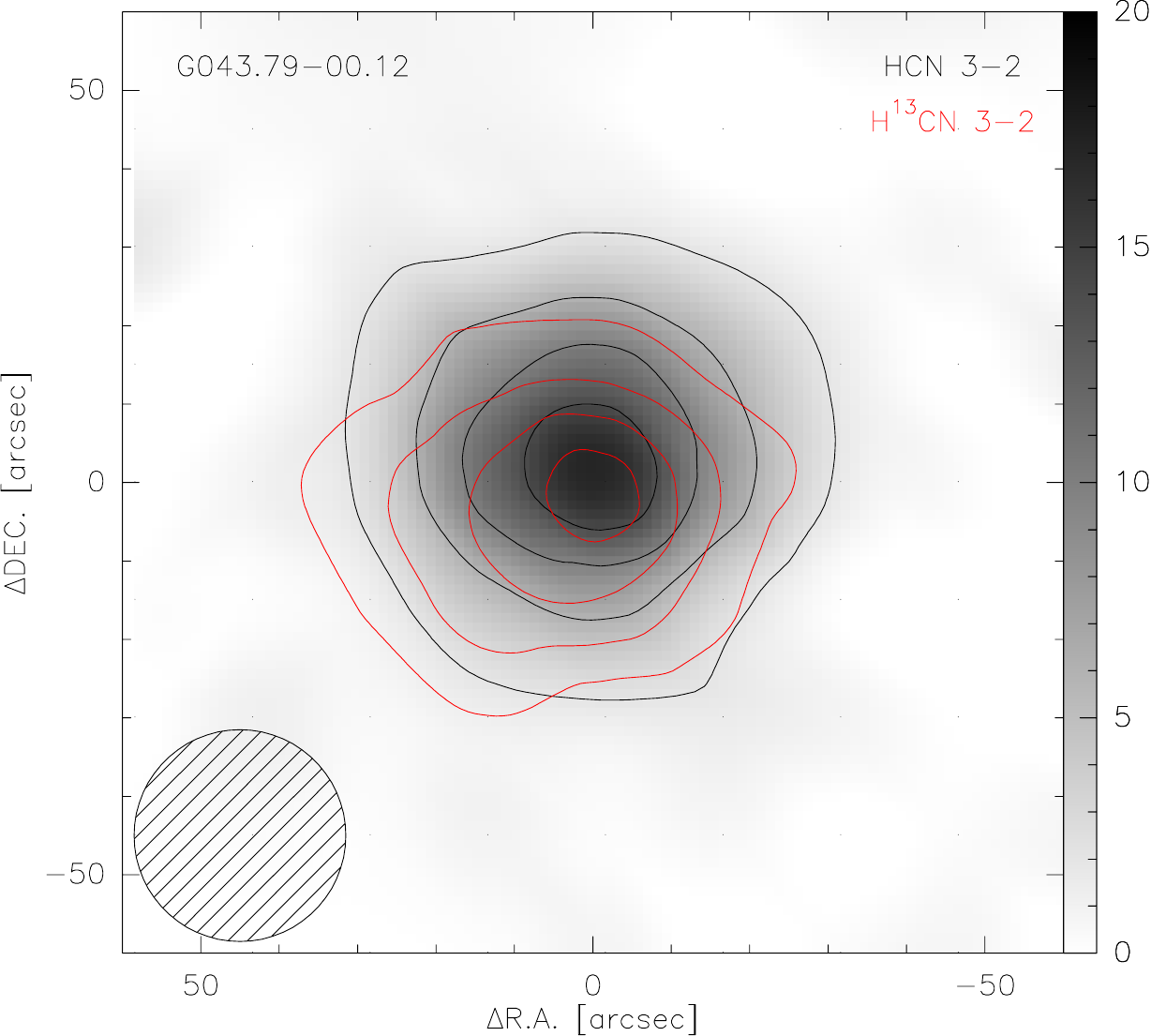}
   \includegraphics[width=3.03in]{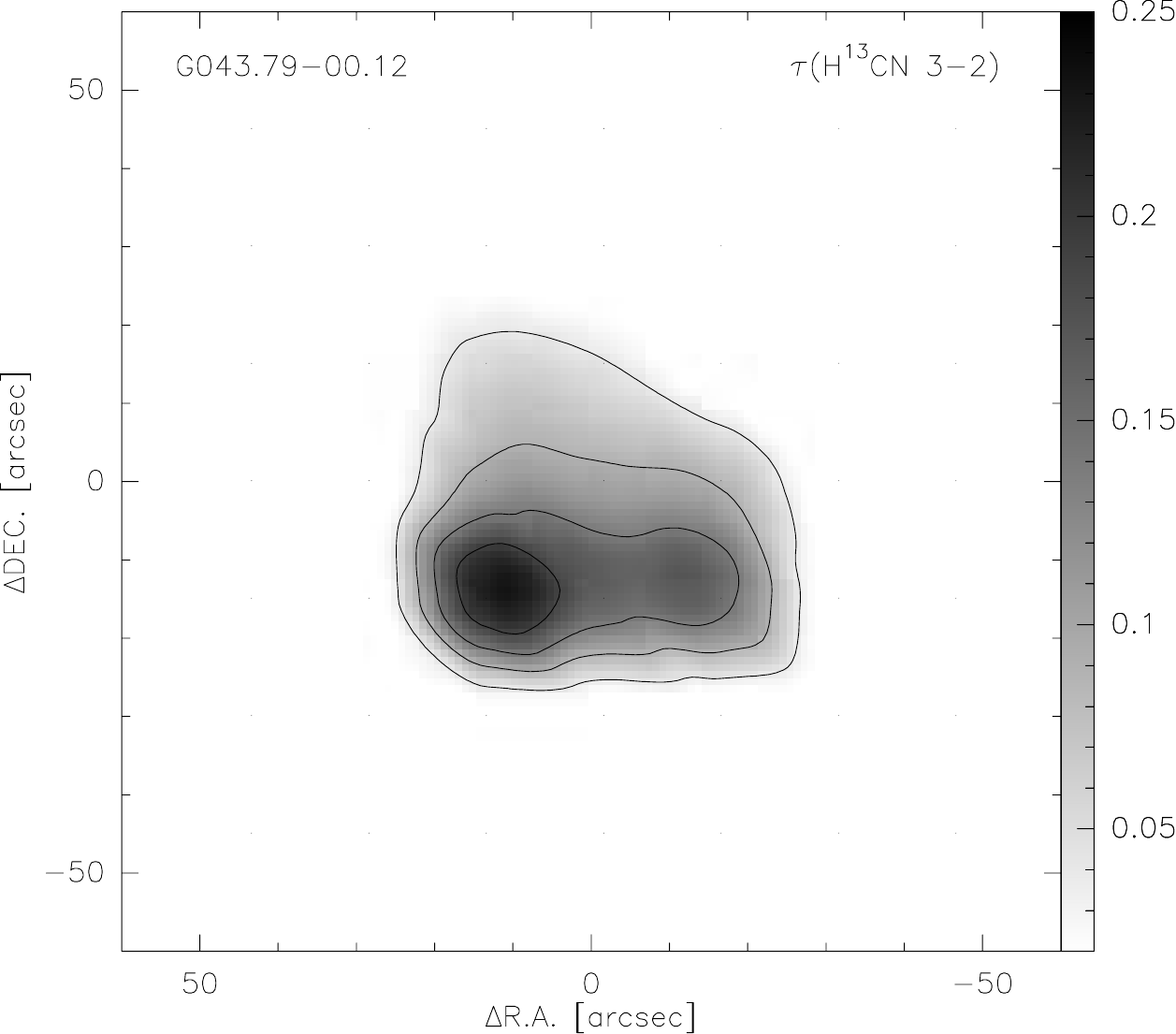}
       \includegraphics[width=3.05in]{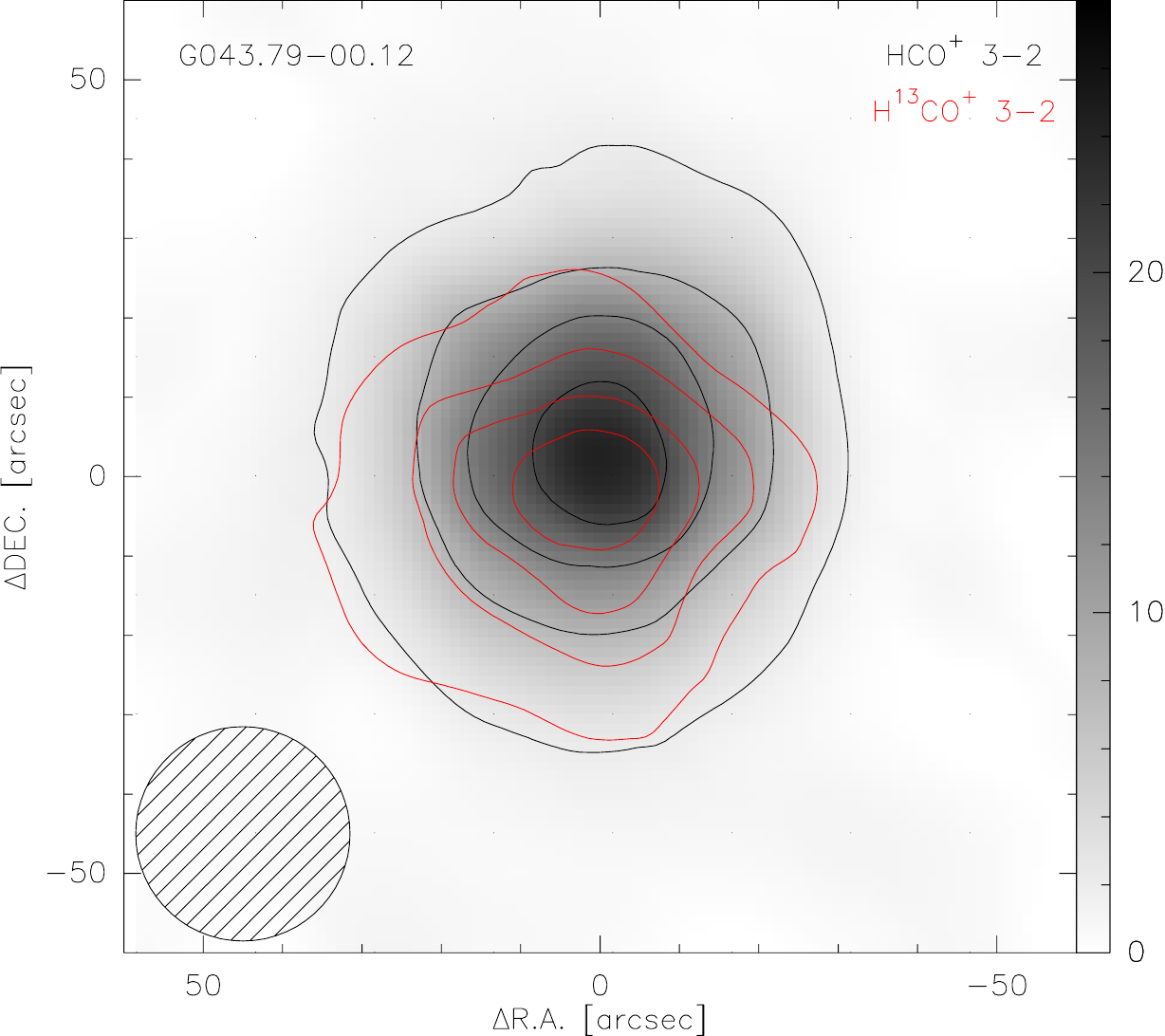}
       \includegraphics[width=3.08in]{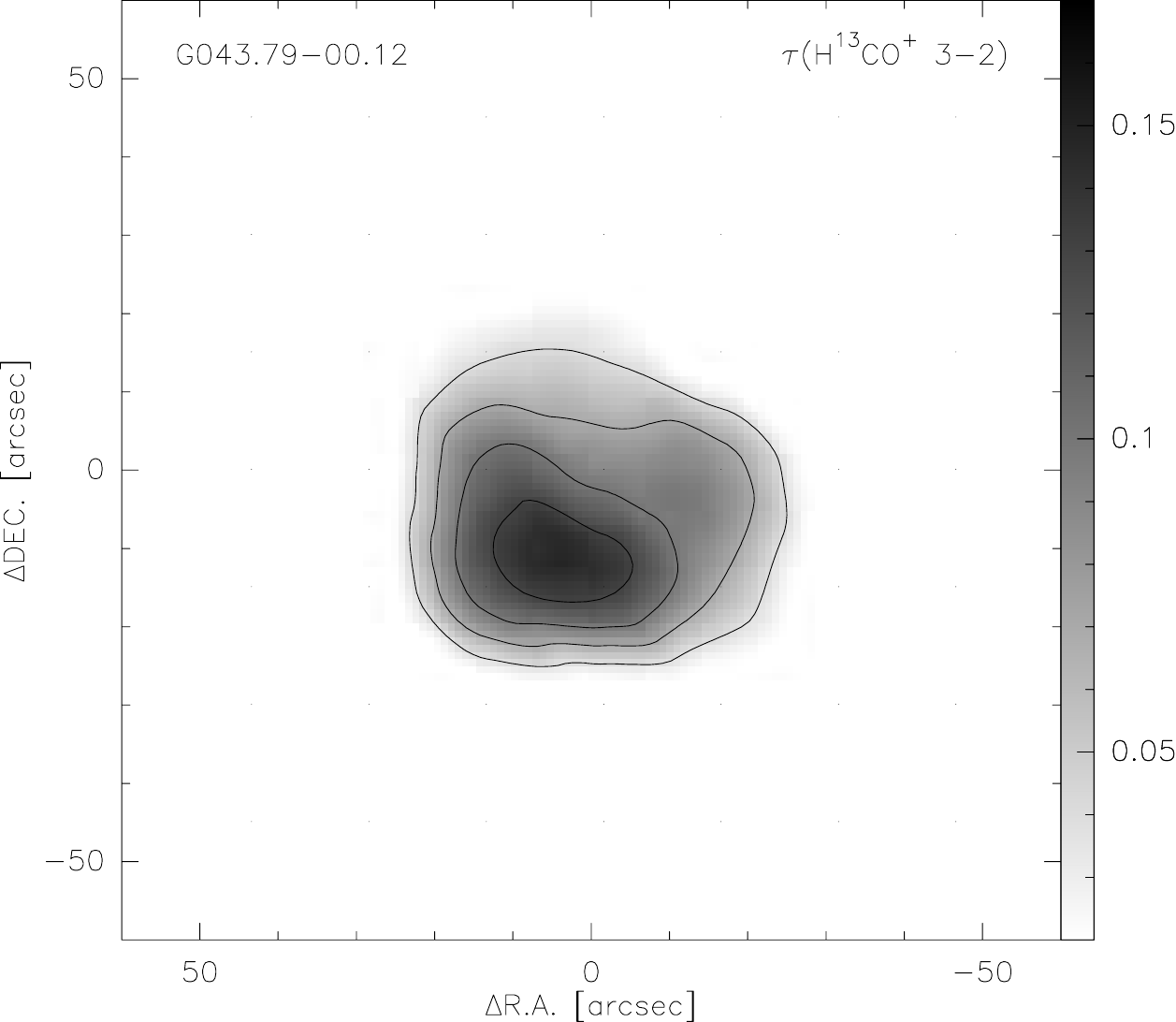}
 \caption{The data reduction results of G043.79-00.12. 
               {\it Top left:} The velocity integrated intensity maps of HCN and H$^{13}$CN 3-2. 
               The mapping size of HCN 3-2 is 2$'\times2'$, while it is 1.5$'\times1.5'$ for H$^{13}$CN 3-2, with a beam size of $\sim$ 27.8$''$.
               The grey scale and the black contour with levels starting from 2 K km s$^{-1}$ in step of 4 K km s$^{-1}$ show the observed HCN 3-2. 
               The red contour with levels starting from 0.3 K km s$^{-1}$ in step of 0.4 K km s$^{-1}$ represents H$^{13}$CN 3-2.
               {\it Top right:} The spatially resolved $\tau(\textrm{H}^{13}\textrm{CN})$ of G043.79-00.12 is demonstrated by black contour with levels 
               starting from 0.04 in step of 0.05. 
                {\it Bottom left:} The velocity integrated intensity maps of HCO$^+$ and H$^{13}$CO$^+$ 3-2. 
               The mapping size of HCO$^+$ 3-2 is 2$'\times2'$, while it is 1.5$'\times1.5'$ for H$^{13}$CO$^+$ 3-2, with a beam size of $\sim$ 27.8$''$.
               The grey scale and the black contour with levels starting from 2 K km s$^{-1}$ in step of 6 K km s$^{-1}$ show the observed HCO$^+$ 3-2. 
               The red contour with levels starting from 0.3 K km s$^{-1}$ in step of 0.5 K km s$^{-1}$ represents H$^{13}$CO$^+$ 3-2.
                {\it Bottom right:} The spatially resolved $\tau(\textrm{H}^{13}\textrm{CO$^+$})$ of G043.79-00.12 is demonstrated by black contour with levels 
               starting from 0.04 in step of 0.03. 
                }       
 \label{fig:g04379}
\end{figure*}


 \begin{figure*} 
    \centering
  \includegraphics[width=3.05in]{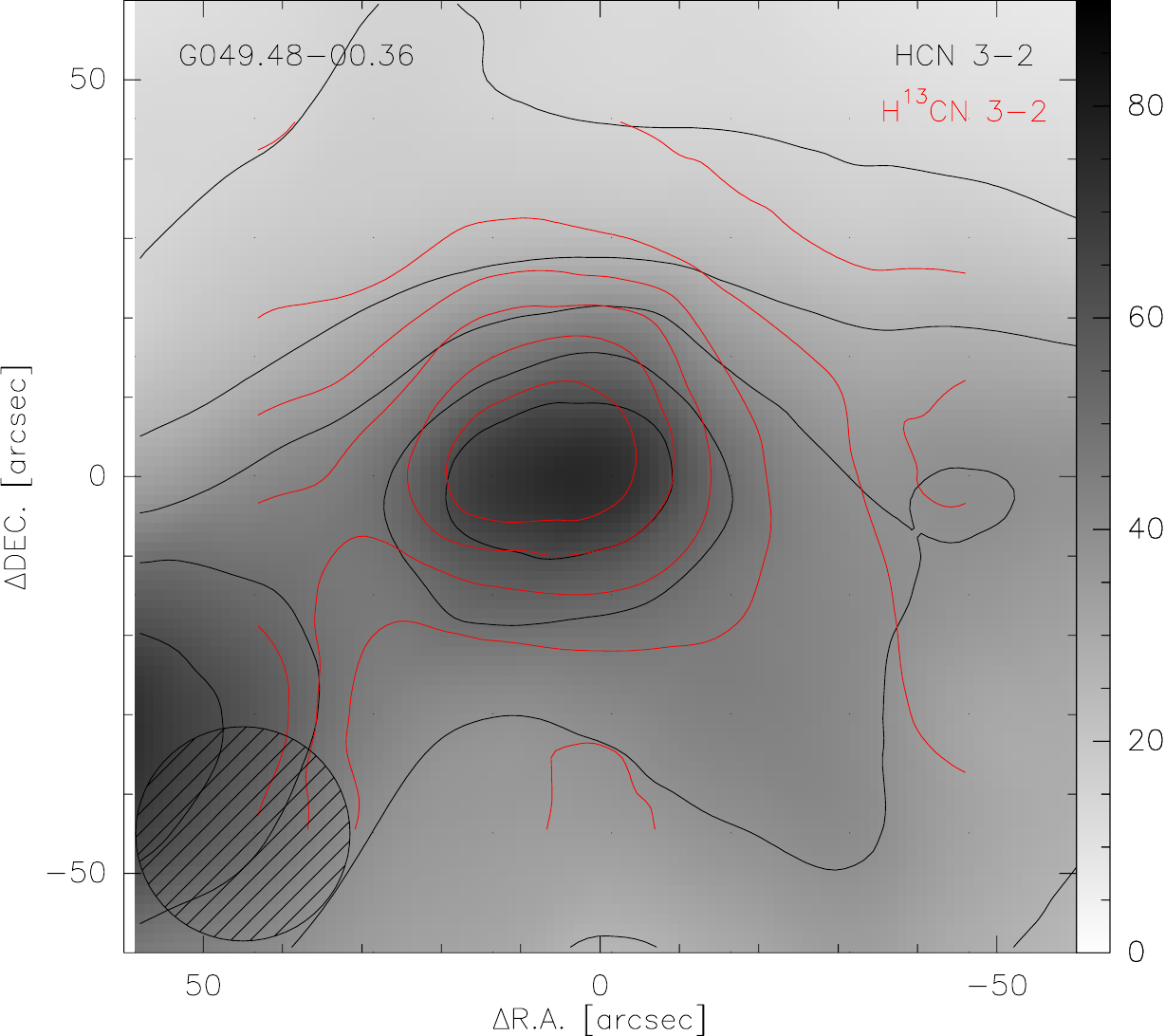}
   \includegraphics[width=3.03in]{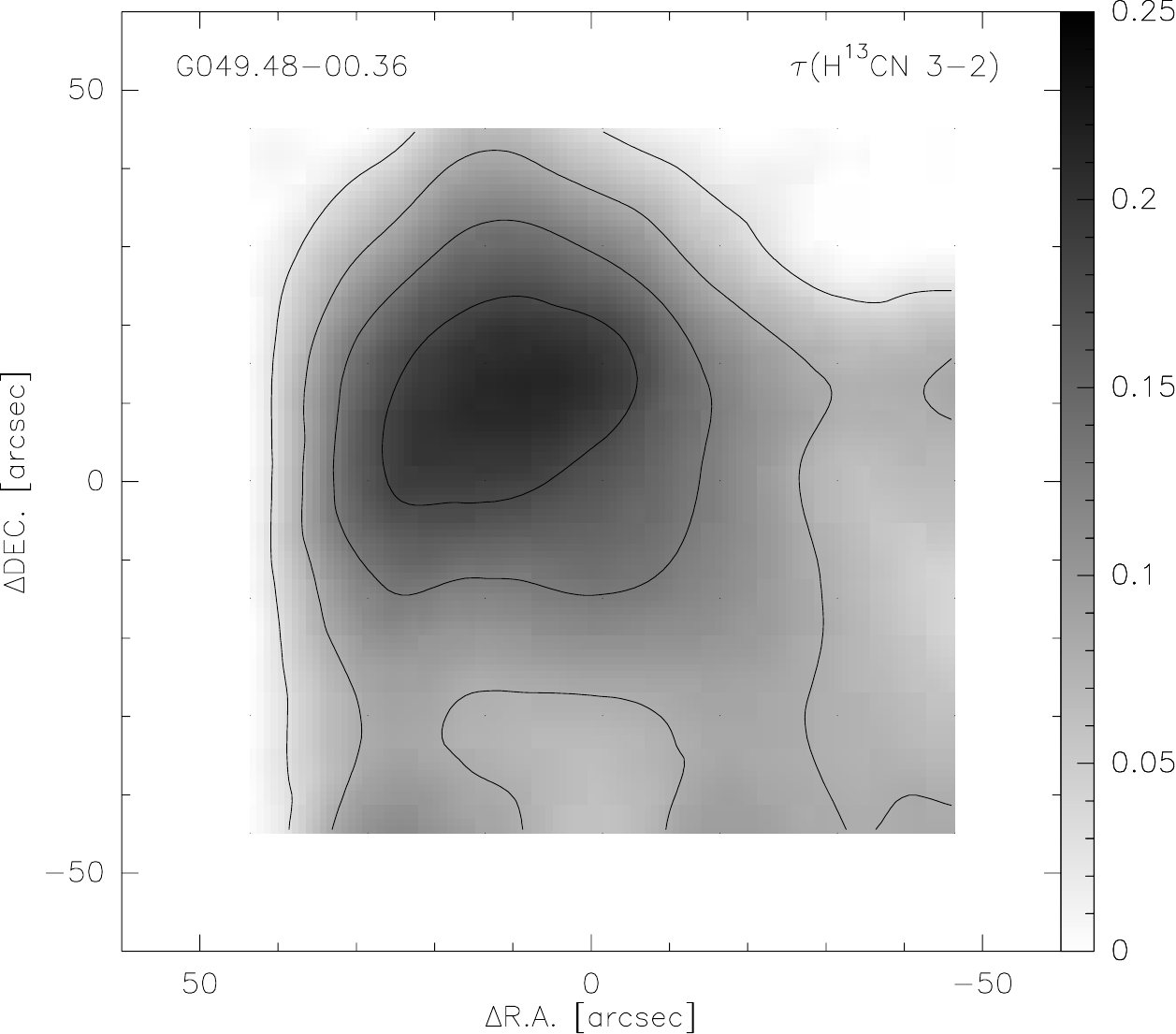}
       \includegraphics[width=3.05in]{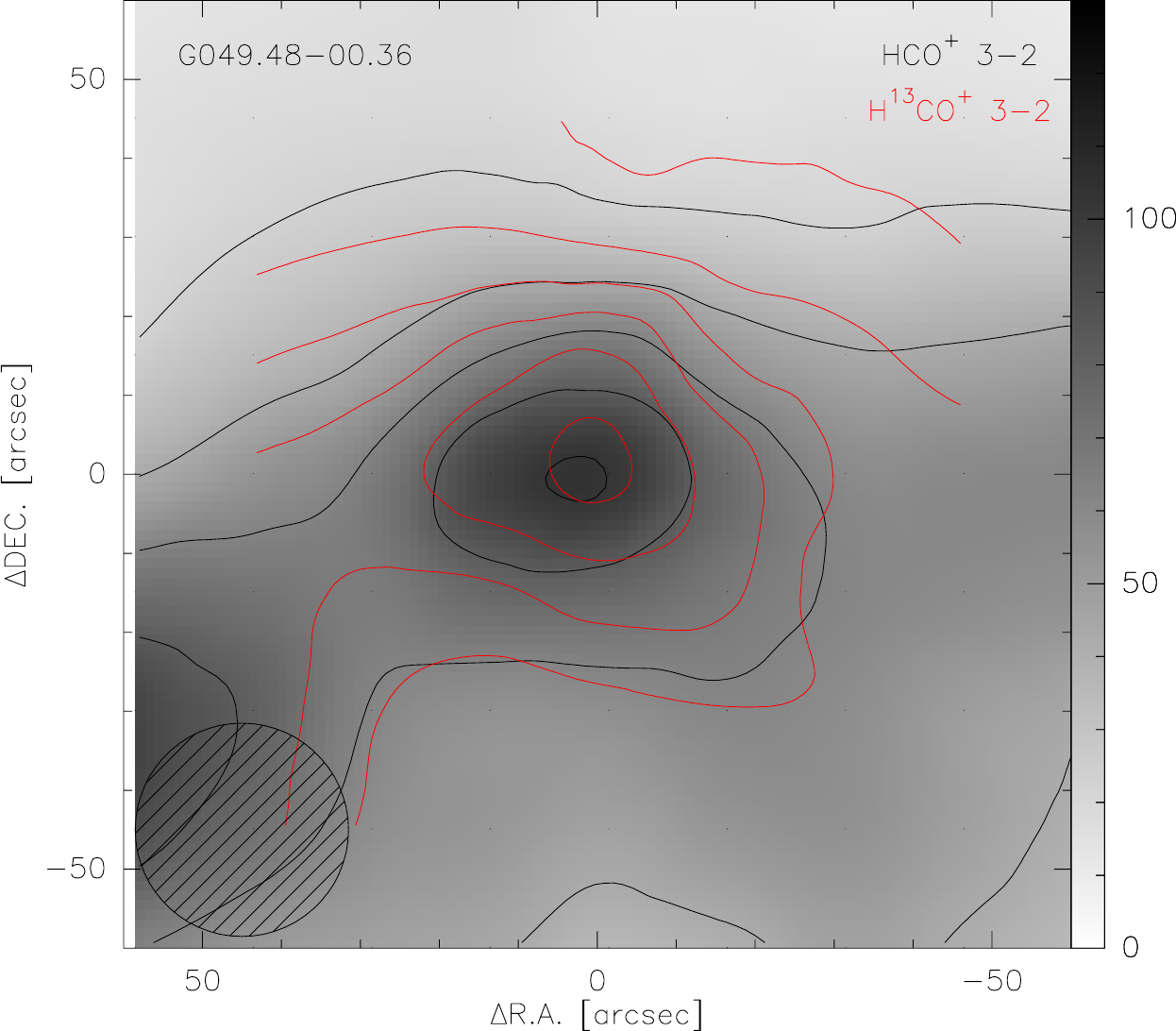}
       \includegraphics[width=3.08in]{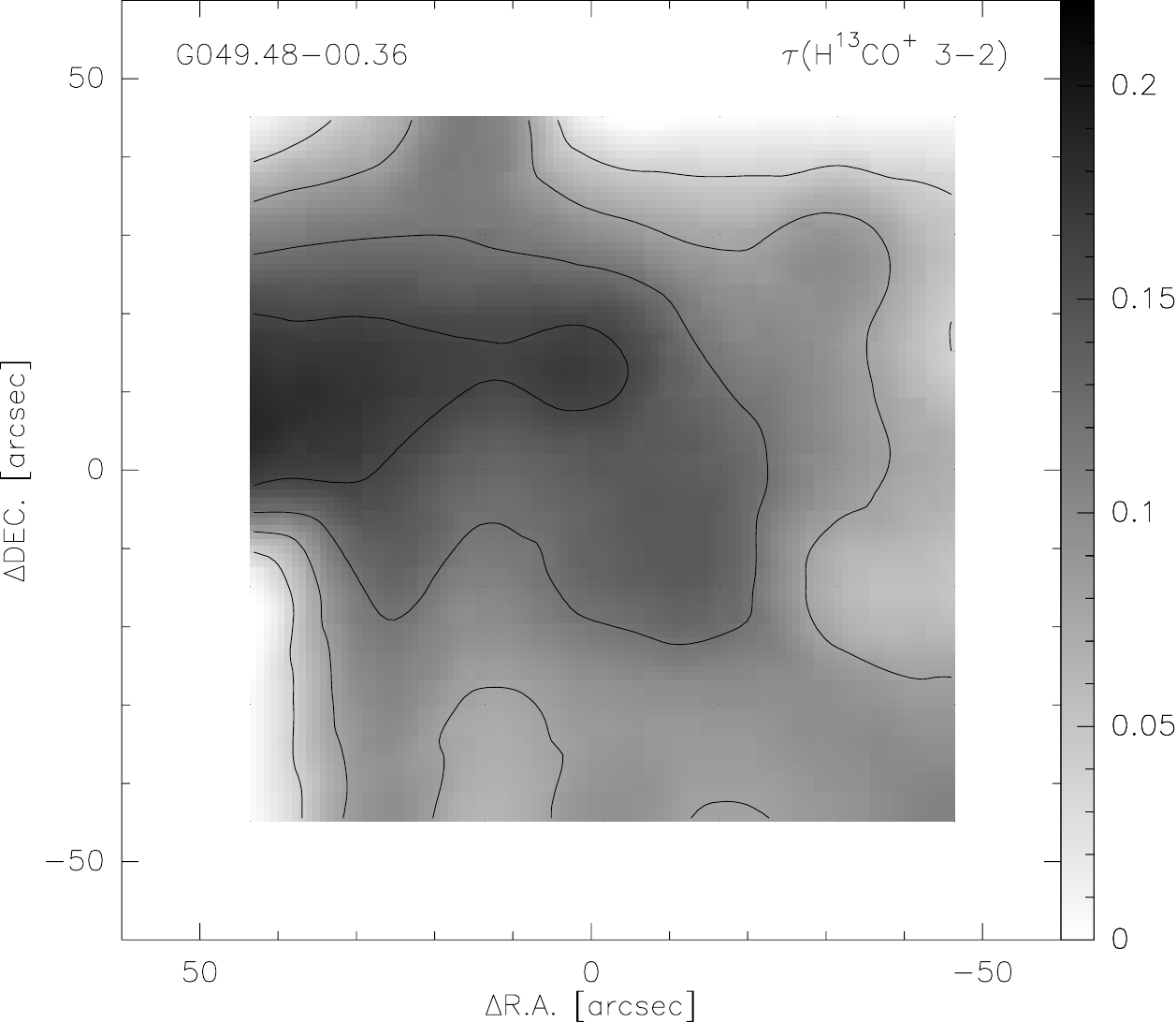}
 \caption{The data reduction results of G049.48-00.36. 
               {\it Top left:} The velocity integrated intensity maps of HCN and H$^{13}$CN 3-2. 
               The mapping size of HCN 3-2 is 2$'\times2'$, while it is 1.5$'\times1.5'$ for H$^{13}$CN 3-2, with a beam size of $\sim$ 27.8$''$.
               The grey scale and the black contour with levels starting from 3 K km s$^{-1}$ in step of 12 K km s$^{-1}$ show the observed HCN 3-2. 
               The red contour with levels starting from 0.6 K km s$^{-1}$ in step of 2 K km s$^{-1}$ represents H$^{13}$CN 3-2.
               {\it Top right:} The spatially resolved $\tau(\textrm{H}^{13}\textrm{CN})$ of G049.48-00.36 is demonstrated by black contour with levels 
               starting from 0.03 in step of 0.05. 
                {\it Bottom left:} The velocity integrated intensity maps of HCO$^+$ and H$^{13}$CO$^+$ 3-2. 
               The mapping size of HCO$^+$ 3-2 is 2$'\times2'$, while it is 1.5$'\times1.5'$ for H$^{13}$CO$^+$ 3-2, with a beam size of $\sim$ 27.8$''$.
               The grey scale and the black contour with levels starting from 3 K km s$^{-1}$ in step of 20 K km s$^{-1}$ show the observed HCO$^+$ 3-2. 
               The red contour with levels starting from 0.7 K km s$^{-1}$ in step of 2.5 K km s$^{-1}$ represents H$^{13}$CO$^+$ 3-2.
                {\it Bottom right:} The spatially resolved $\tau(\textrm{H}^{13}\textrm{CO$^+$})$ of G049.48-00.36 is demonstrated by black contour with levels 
               starting from 0.04 in step of 0.04. 
                }       
 \label{fig:g049480036}
\end{figure*}


 \begin{figure*} 
    \centering
  \includegraphics[width=3.05in]{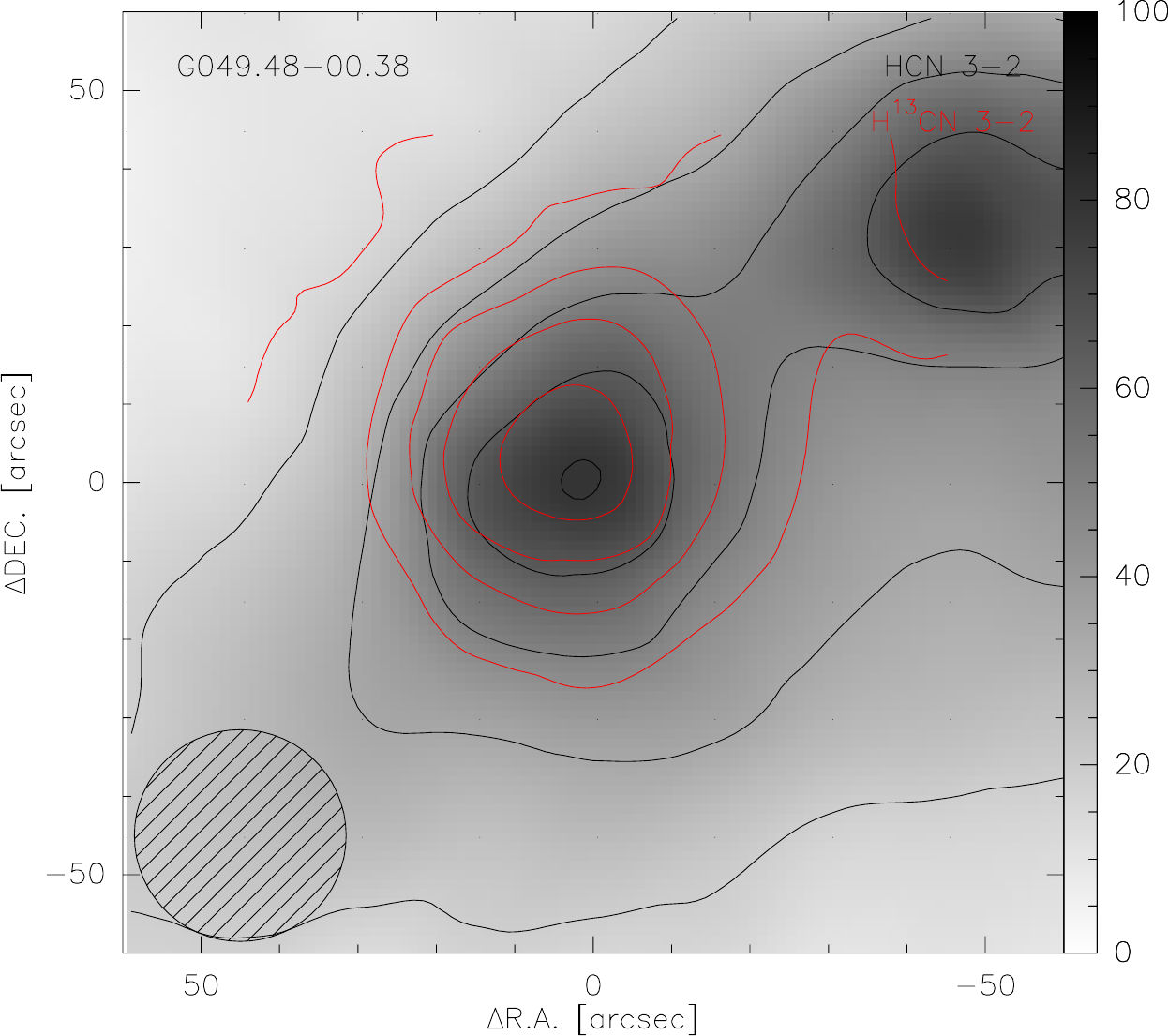}
   \includegraphics[width=3.03in]{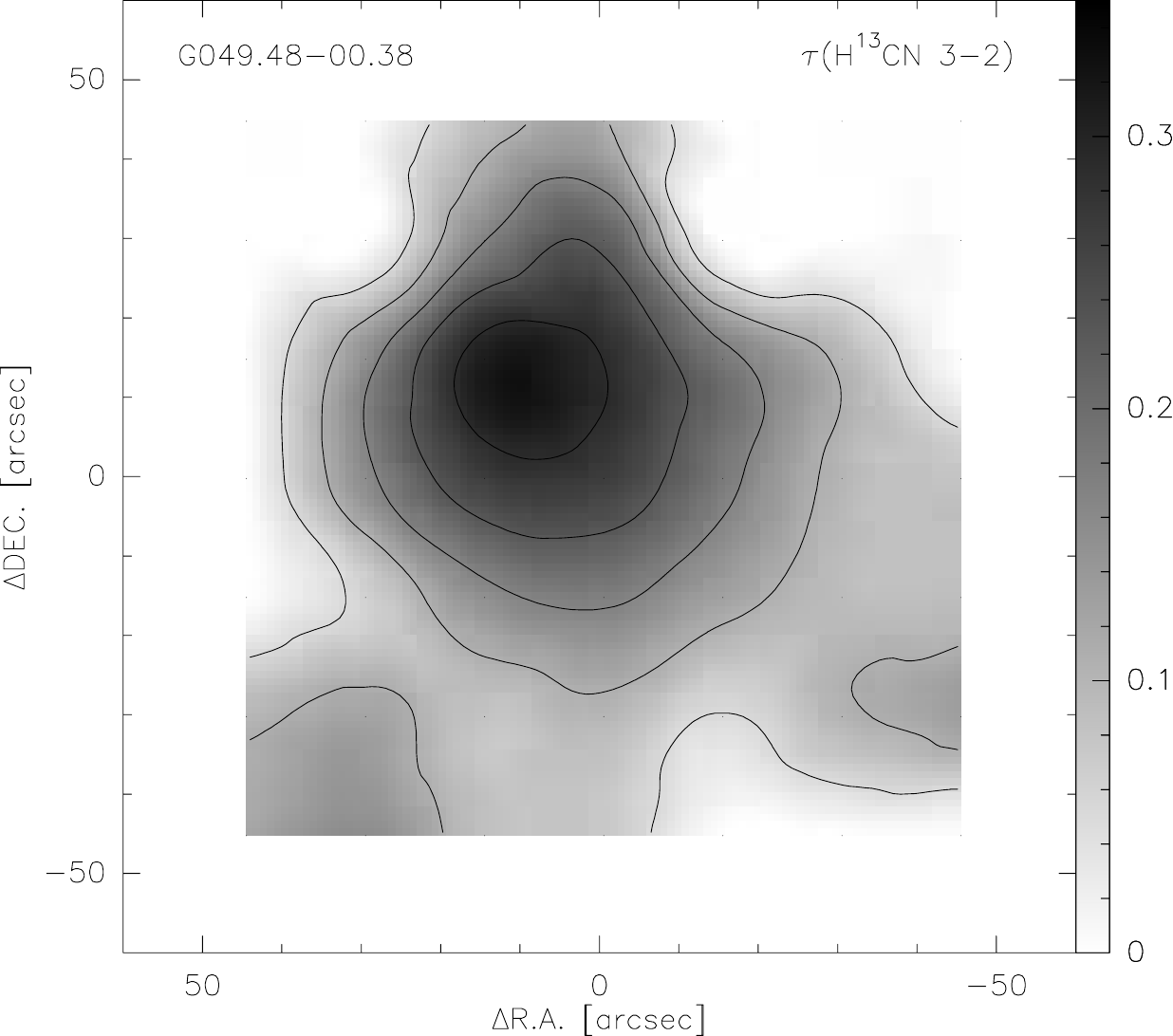}
       \includegraphics[width=3.05in]{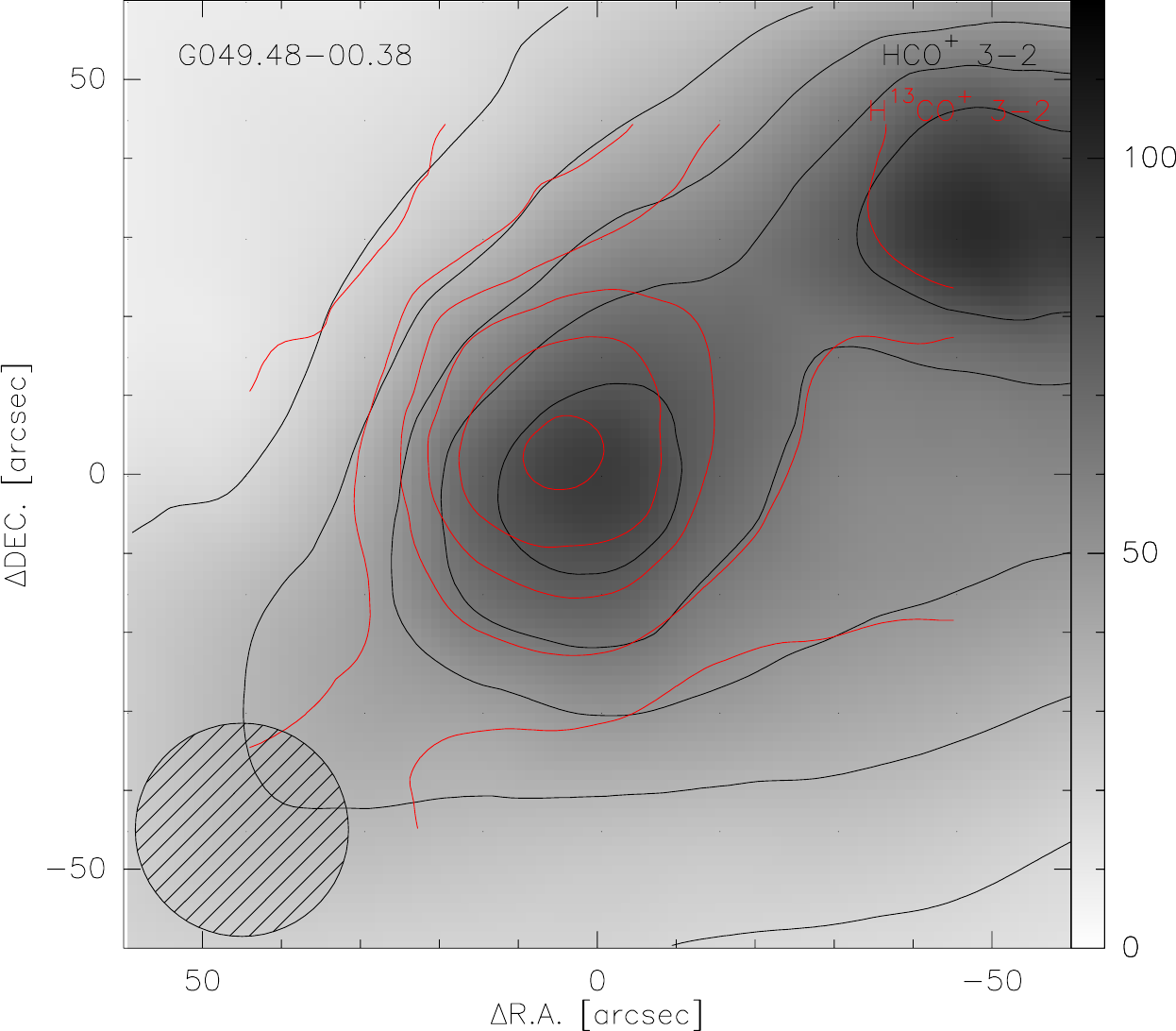}
       \includegraphics[width=3.08in]{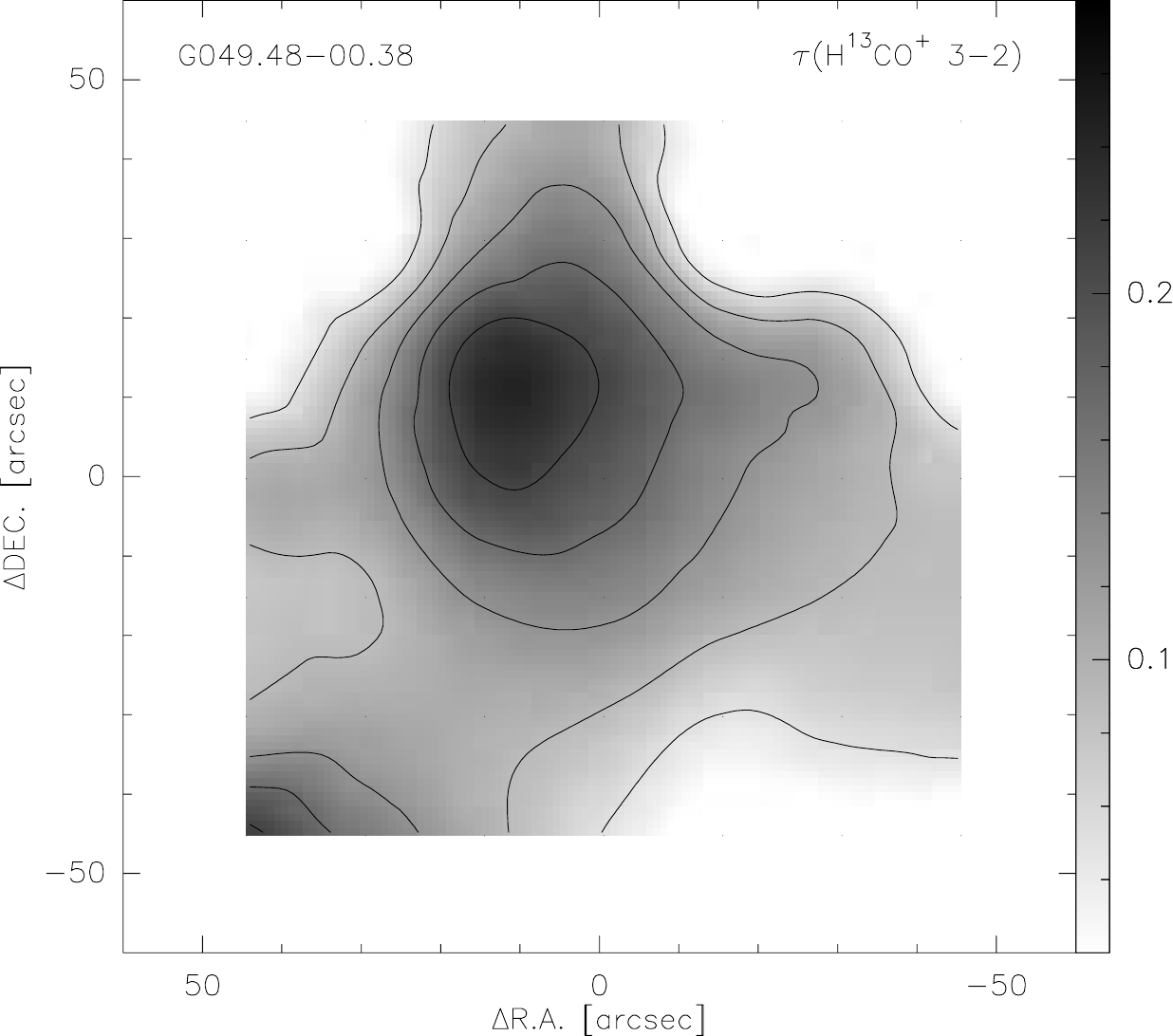}
 \caption{The data reduction results of G049.48-00.38. 
               {\it Top left:} The velocity integrated intensity maps of HCN and H$^{13}$CN 3-2. 
               The mapping size of HCN 3-2 is 2$'\times2'$, while it is 1.5$'\times1.5'$ for H$^{13}$CN 3-2, with a beam size of $\sim$ 27.8$''$.
               The grey scale and the black contour with levels starting from 4 K km s$^{-1}$ in step of 15 K km s$^{-1}$ show the observed HCN 3-2. 
               The red contour with levels starting from 0.9 K km s$^{-1}$ in step of 4 K km s$^{-1}$ represents H$^{13}$CN 3-2.
               {\it Top right:} The spatially resolved $\tau(\textrm{H}^{13}\textrm{CN})$ of G049.48-00.38 is demonstrated by black contour with levels 
               starting from 0.05 in step of 0.06. 
                {\it Bottom left:} The velocity integrated intensity maps of HCO$^+$ and H$^{13}$CO$^+$ 3-2. 
               The mapping size of HCO$^+$ 3-2 is 2$'\times2'$, while it is 1.5$'\times1.5'$ for H$^{13}$CO$^+$ 3-2, with a beam size of $\sim$ 27.8$''$.
               The grey scale and the black contour with levels starting from 4 K km s$^{-1}$ in step of 15 K km s$^{-1}$ show the observed HCO$^+$ 3-2. 
               The red contour with levels starting from 0.9 K km s$^{-1}$ in step of 3 K km s$^{-1}$ represents H$^{13}$CO$^+$ 3-2.
                {\it Bottom right:} The spatially resolved $\tau(\textrm{H}^{13}\textrm{CO$^+$})$ of G049.48-00.38 is demonstrated by black contour with levels 
               starting from 0.05 in step of 0.04. 
                }       
 \label{fig:g049480038}
\end{figure*}


 \begin{figure*} 
    \centering
  \includegraphics[width=3.05in]{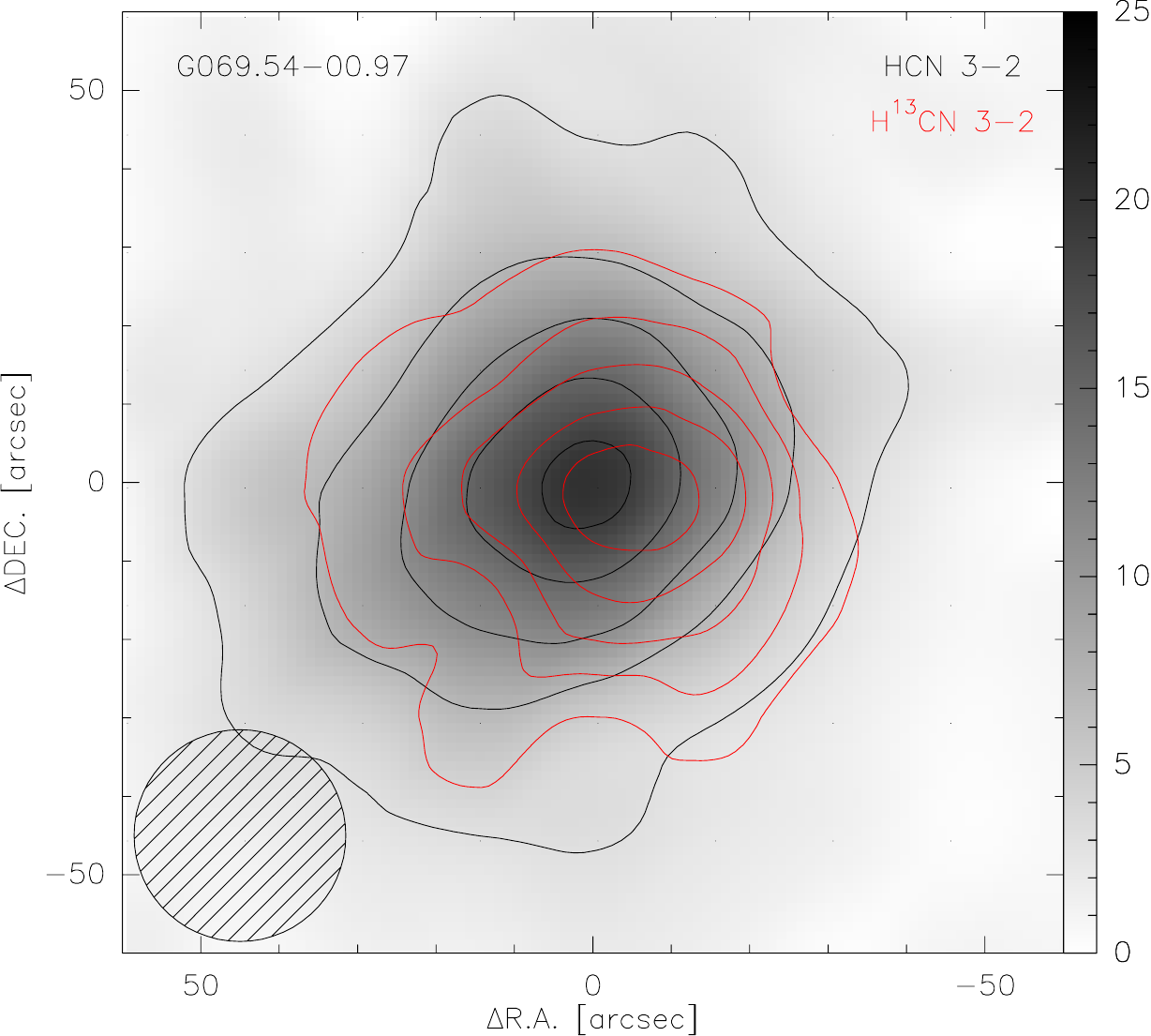}
   \includegraphics[width=3.03in]{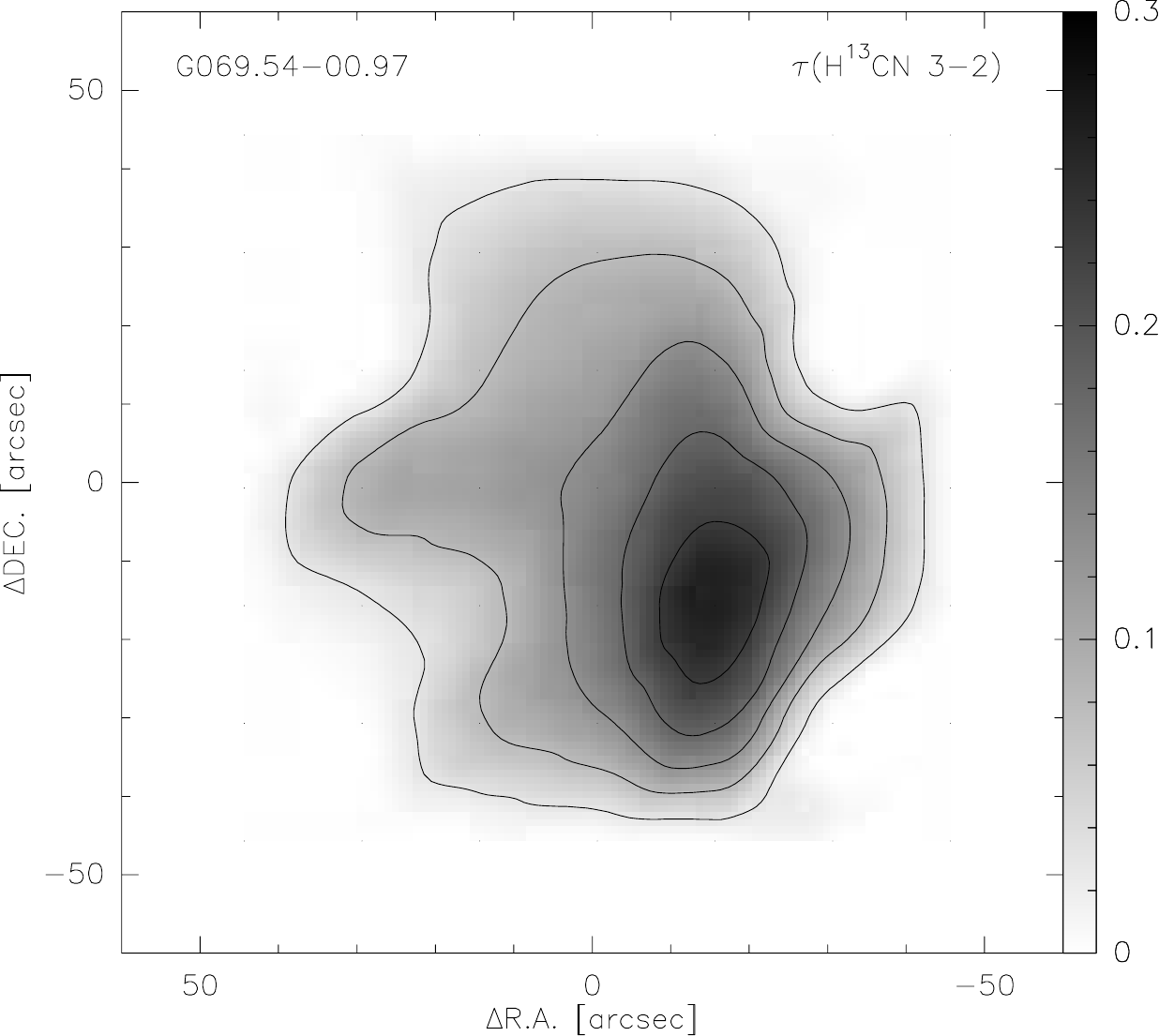}
       \includegraphics[width=3.05in]{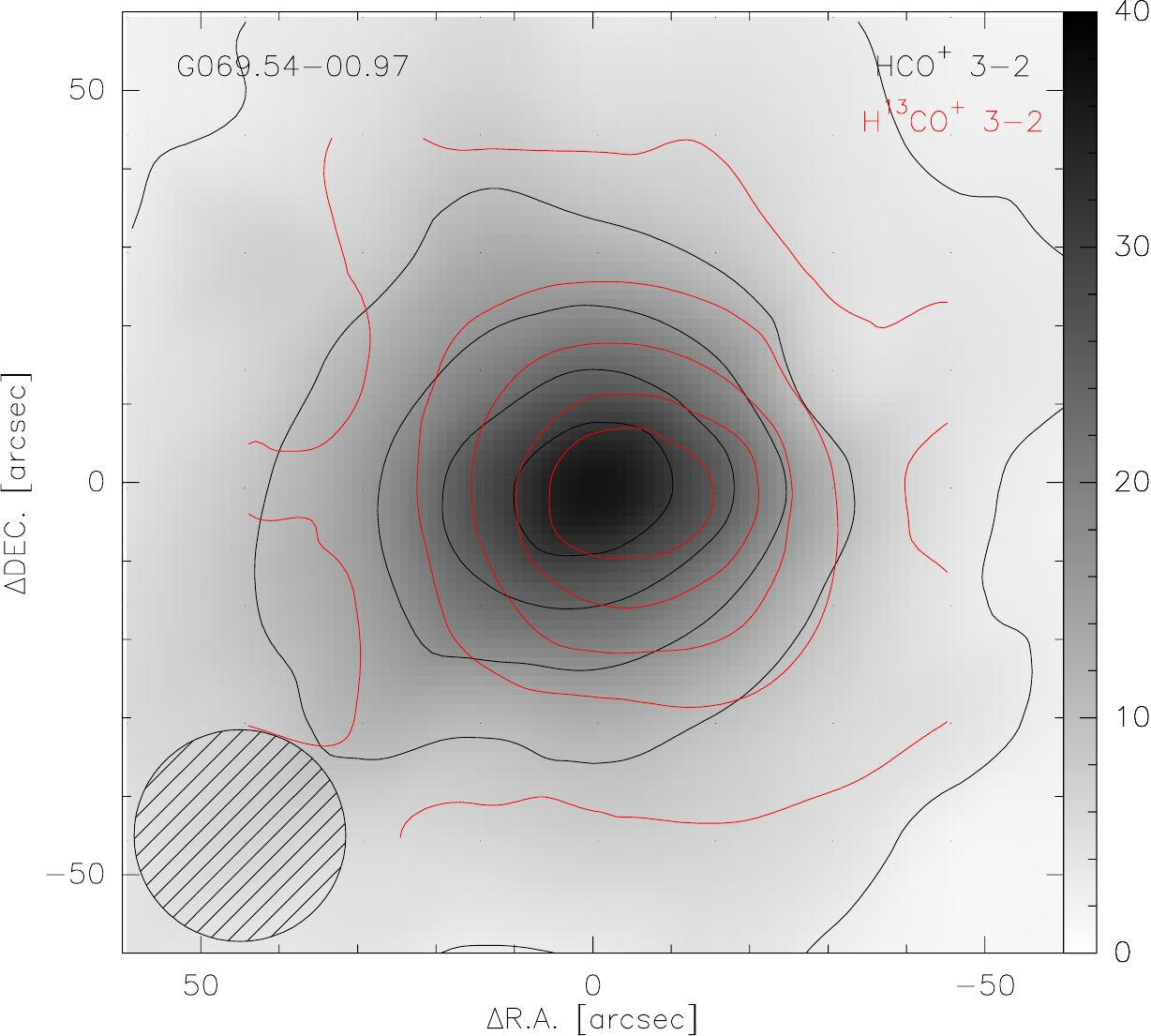}
       \includegraphics[width=3.08in]{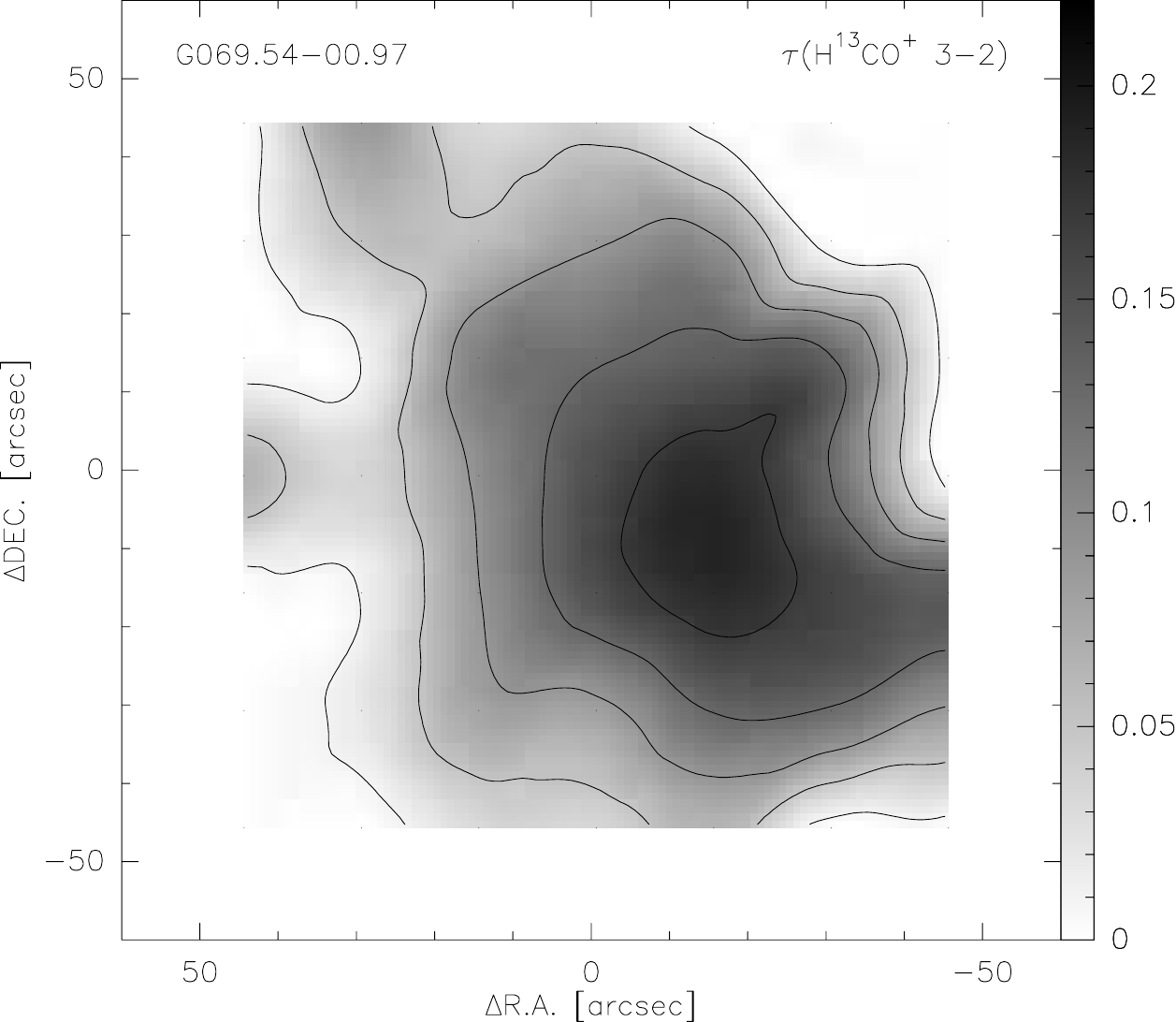}
 \caption{The data reduction results of G069.54-00.97. 
               {\it Top left:} The velocity integrated intensity maps of HCN and H$^{13}$CN 3-2. 
               The mapping size of HCN 3-2 is 2$'\times2'$, while it is 1.5$'\times1.5'$ for H$^{13}$CN 3-2, with a beam size of $\sim$ 27.8$''$.
               The grey scale and the black contour with levels starting from 3 K km s$^{-1}$ in step of 4 K km s$^{-1}$ show the observed HCN 3-2. 
               The red contour with levels starting from 0.5 K km s$^{-1}$ in step of 0.5 K km s$^{-1}$ represents H$^{13}$CN 3-2.
               {\it Top right:} The spatially resolved $\tau(\textrm{H}^{13}\textrm{CN})$ of G069.54-00.97 is demonstrated by black contour with levels 
               starting from 0.03 in step of 0.05. 
                {\it Bottom left:} The velocity integrated intensity maps of HCO$^+$ and H$^{13}$CO$^+$ 3-2. 
               The mapping size of HCO$^+$ 3-2 is 2$'\times2'$, while it is 1.5$'\times1.5'$ for H$^{13}$CO$^+$ 3-2, with a beam size of $\sim$ 27.8$''$.
               The grey scale and the black contour with levels starting from 3 K km s$^{-1}$ in step of 7 K km s$^{-1}$ show the observed HCO$^+$ 3-2. 
               The red contour with levels starting from 0.4 K km s$^{-1}$ in step of 1 K km s$^{-1}$ represents H$^{13}$CO$^+$ 3-2.
                {\it Bottom right:} The spatially resolved $\tau(\textrm{H}^{13}\textrm{CO$^+$})$ of G069.54-00.97 is demonstrated by black contour with levels 
               starting from 0.01 in step of 0.04. 
                }       
 \label{fig:g06954}
\end{figure*}


 \begin{figure*} 
    \centering
  \includegraphics[width=3.05in]{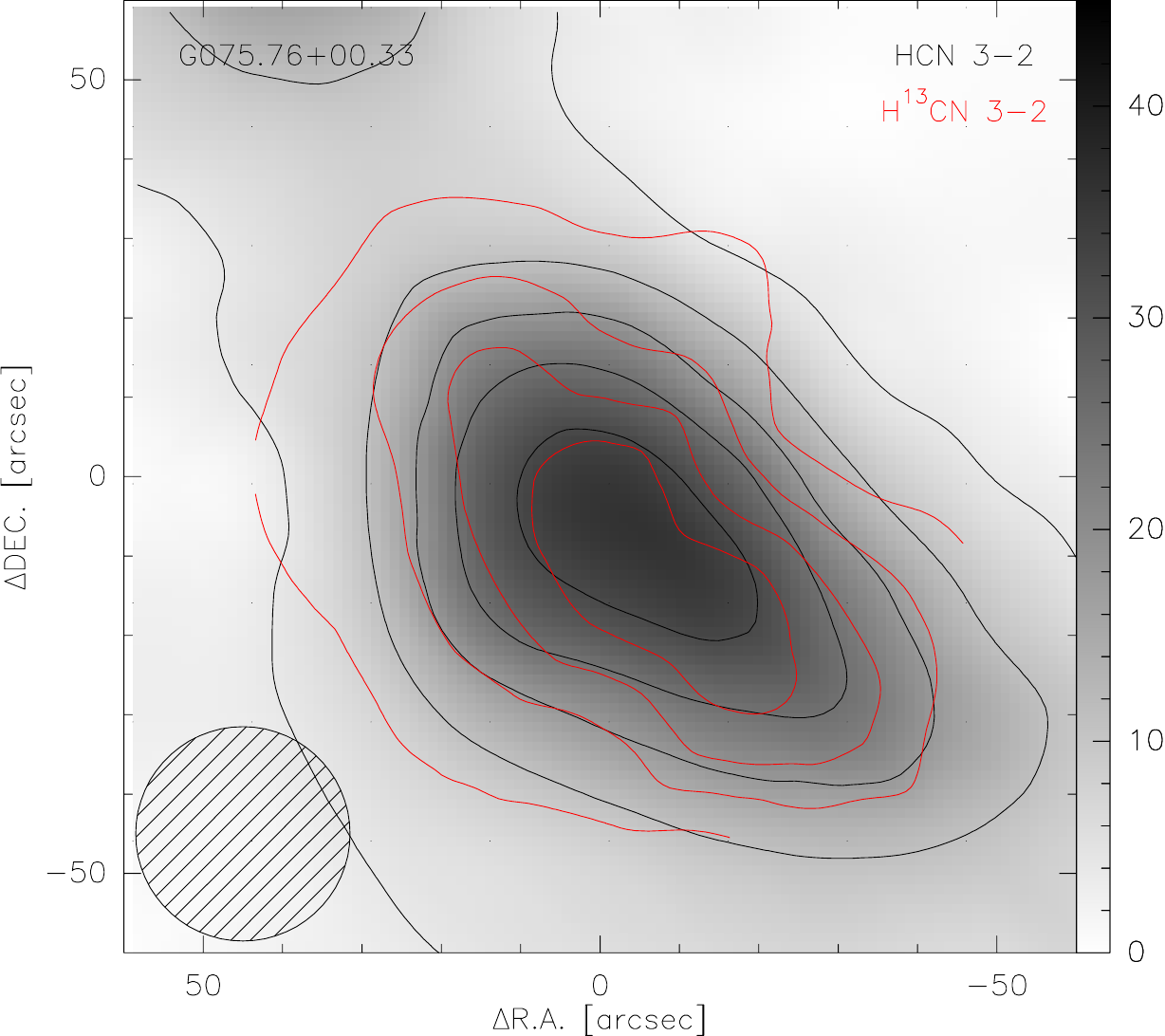}
   \includegraphics[width=3.03in]{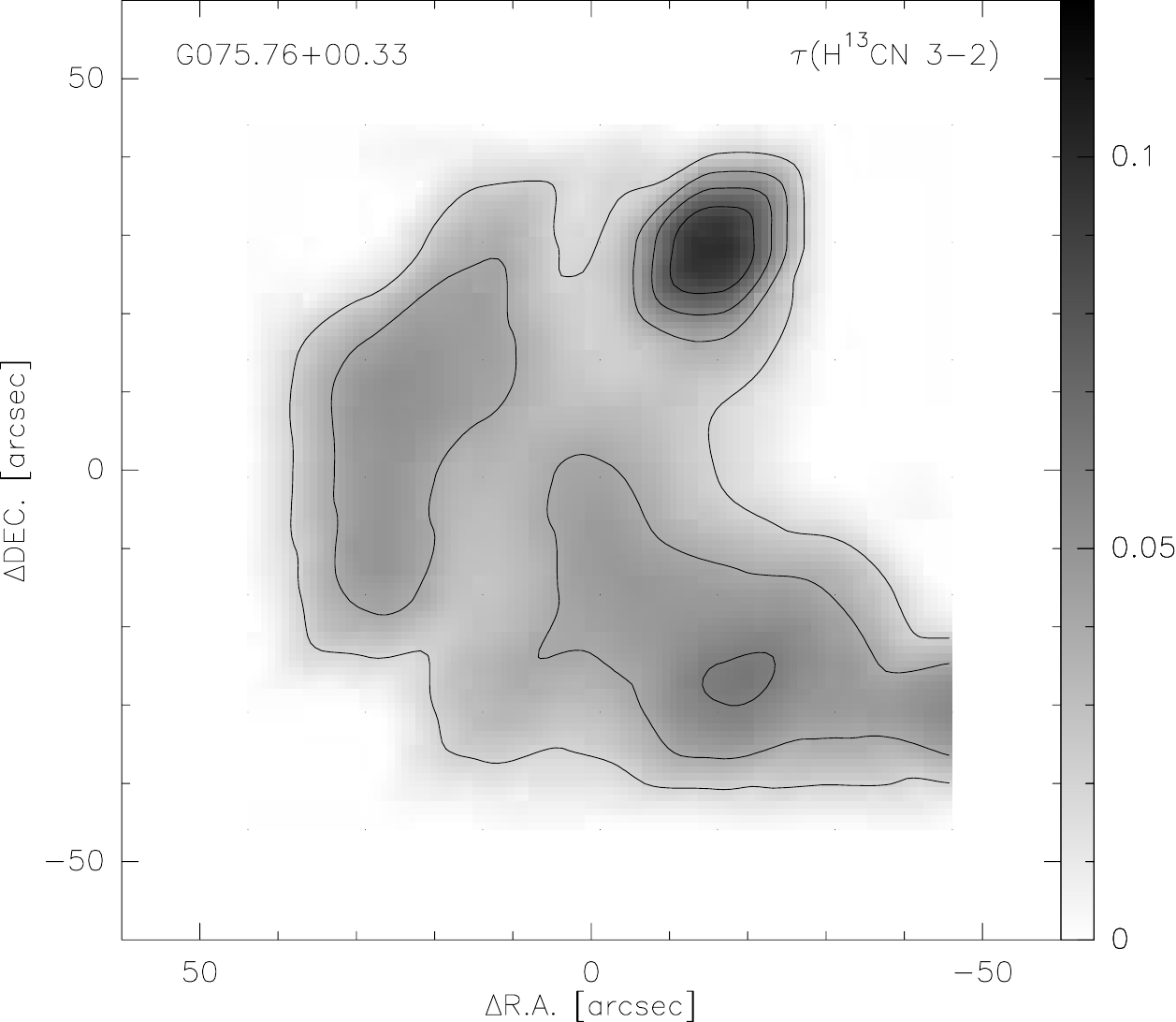}
       \includegraphics[width=3.05in]{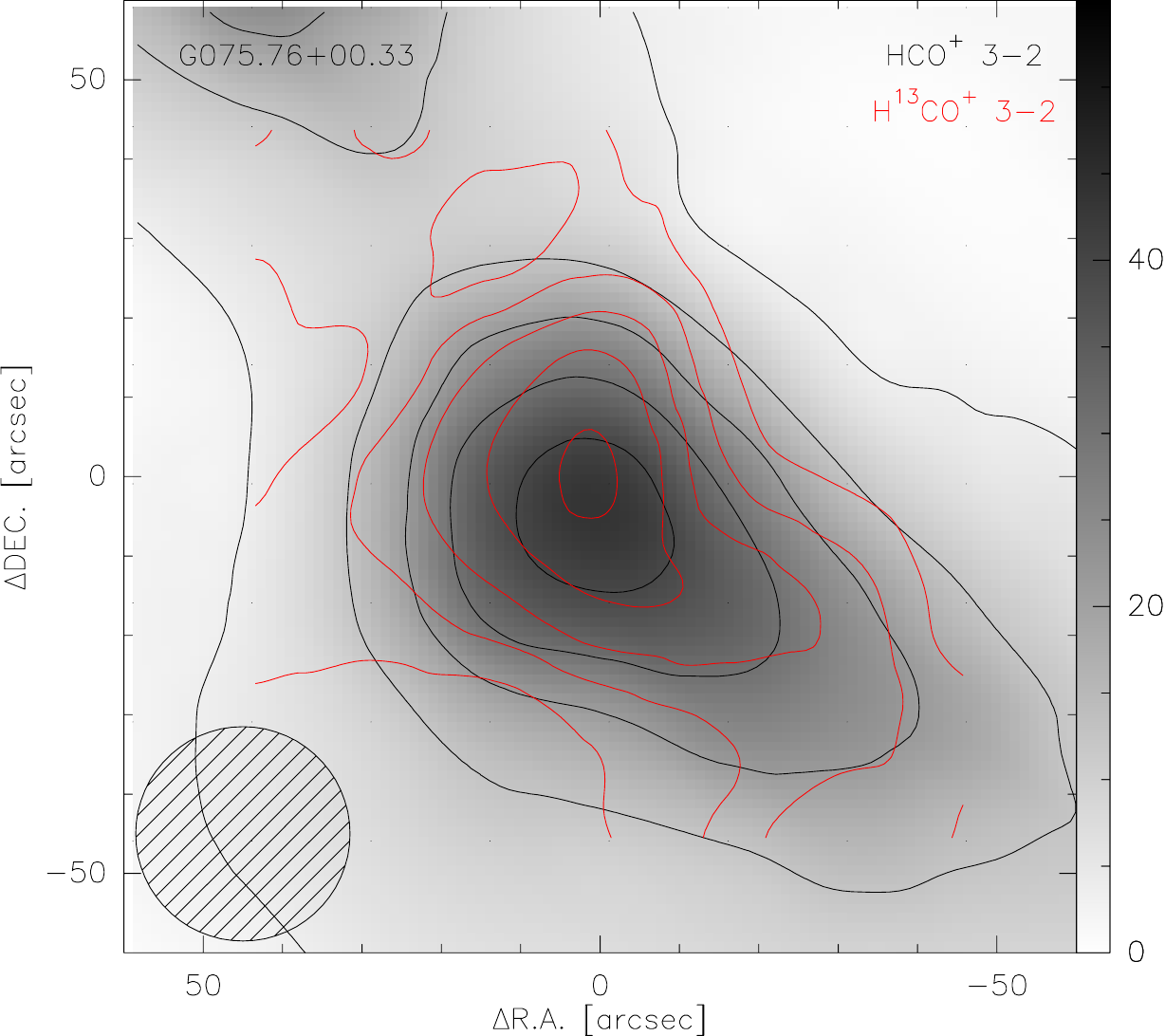}
       \includegraphics[width=3.08in]{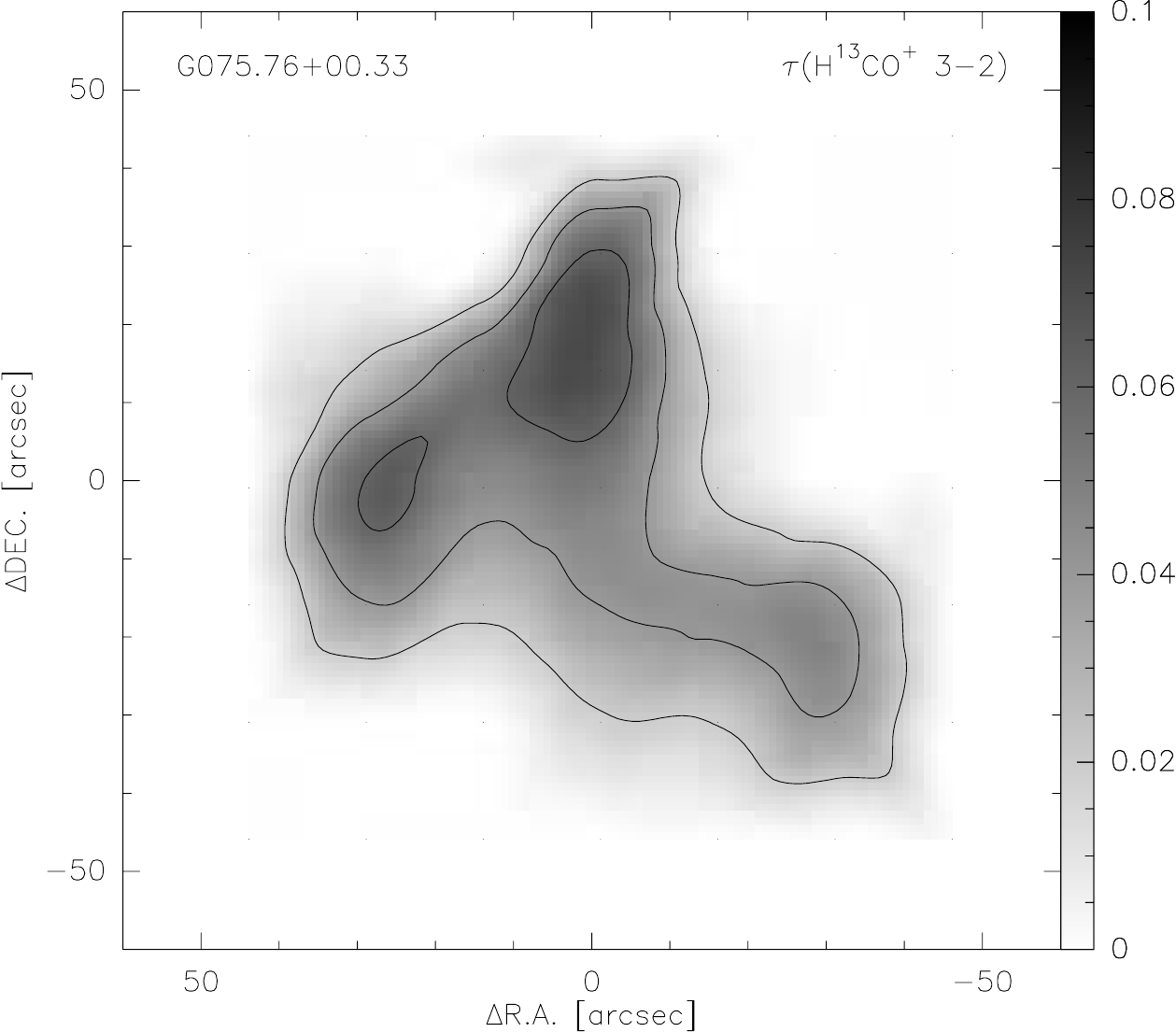}
 \caption{The data reduction results of G075.76+00.33. 
               {\it Top left:} The velocity integrated intensity maps of HCN and H$^{13}$CN 3-2. 
               The mapping size of HCN 3-2 is 2$'\times2'$, while it is 1.5$'\times1.5'$ for H$^{13}$CN 3-2, with a beam size of $\sim$ 27.8$''$.
               The grey scale and the black contour with levels starting from 4 K km s$^{-1}$ in step of 7 K km s$^{-1}$ show the observed HCN 3-2. 
               The red contour with levels starting from 0.4 K km s$^{-1}$ in step of 0.4 K km s$^{-1}$ represents H$^{13}$CN 3-2.
               {\it Top right:} The spatially resolved $\tau(\textrm{H}^{13}\textrm{CN})$ of G075.76+00.33 is demonstrated by black contour with levels 
               starting from 0.02 in step of 0.02. 
                {\it Bottom left:} The velocity integrated intensity maps of HCO$^+$ and H$^{13}$CO$^+$ 3-2. 
               The mapping size of HCO$^+$ 3-2 is 2$'\times2'$, while it is 1.5$'\times1.5'$ for H$^{13}$CO$^+$ 3-2, with a beam size of $\sim$ 27.8$''$.
               The grey scale and the black contour with levels starting from 4 K km s$^{-1}$ in step of 9 K km s$^{-1}$ show the observed HCO$^+$ 3-2. 
               The red contour with levels starting from 0.4 K km s$^{-1}$ in step of 0.5 K km s$^{-1}$ represents H$^{13}$CO$^+$ 3-2.
                {\it Bottom right:} The spatially resolved $\tau(\textrm{H}^{13}\textrm{CO$^+$})$ of G075.76+00.33 is demonstrated by black contour with levels 
               starting from 0.02 in step of 0.02. 
                }       
 \label{fig:g07576}
\end{figure*}


 \begin{figure*} 
    \centering
  \includegraphics[width=3.05in]{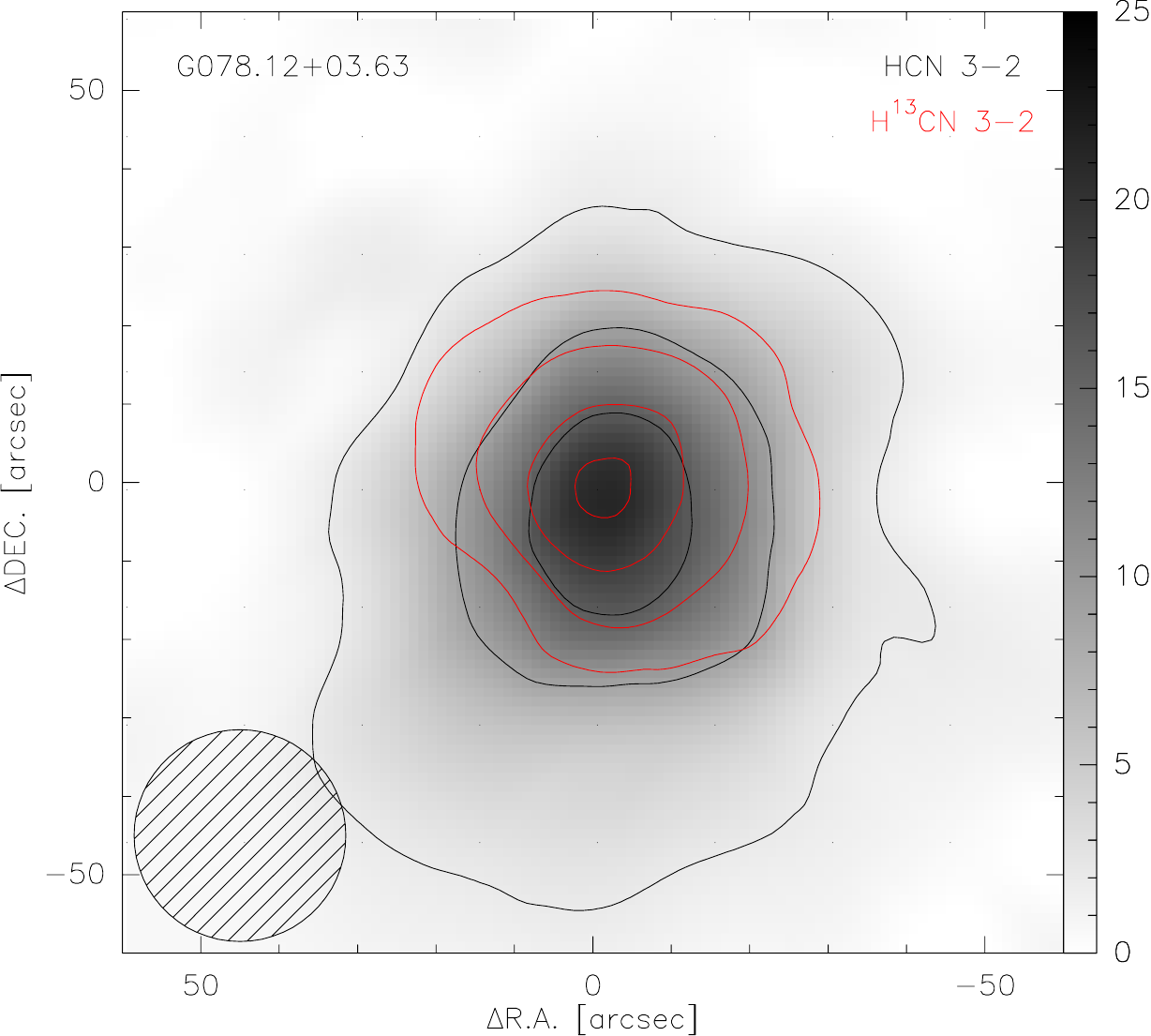}
   \includegraphics[width=3.03in]{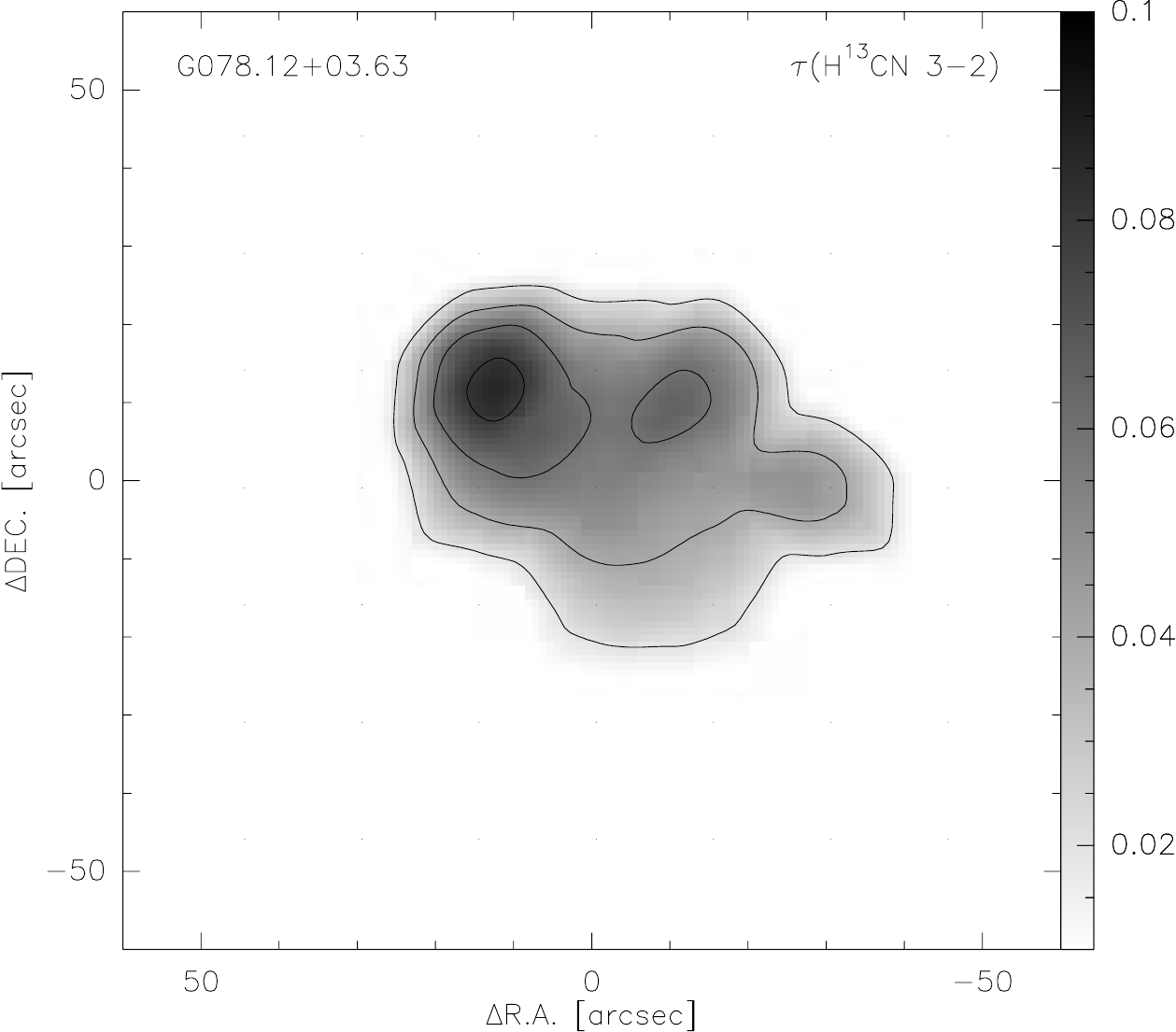}
       \includegraphics[width=3.05in]{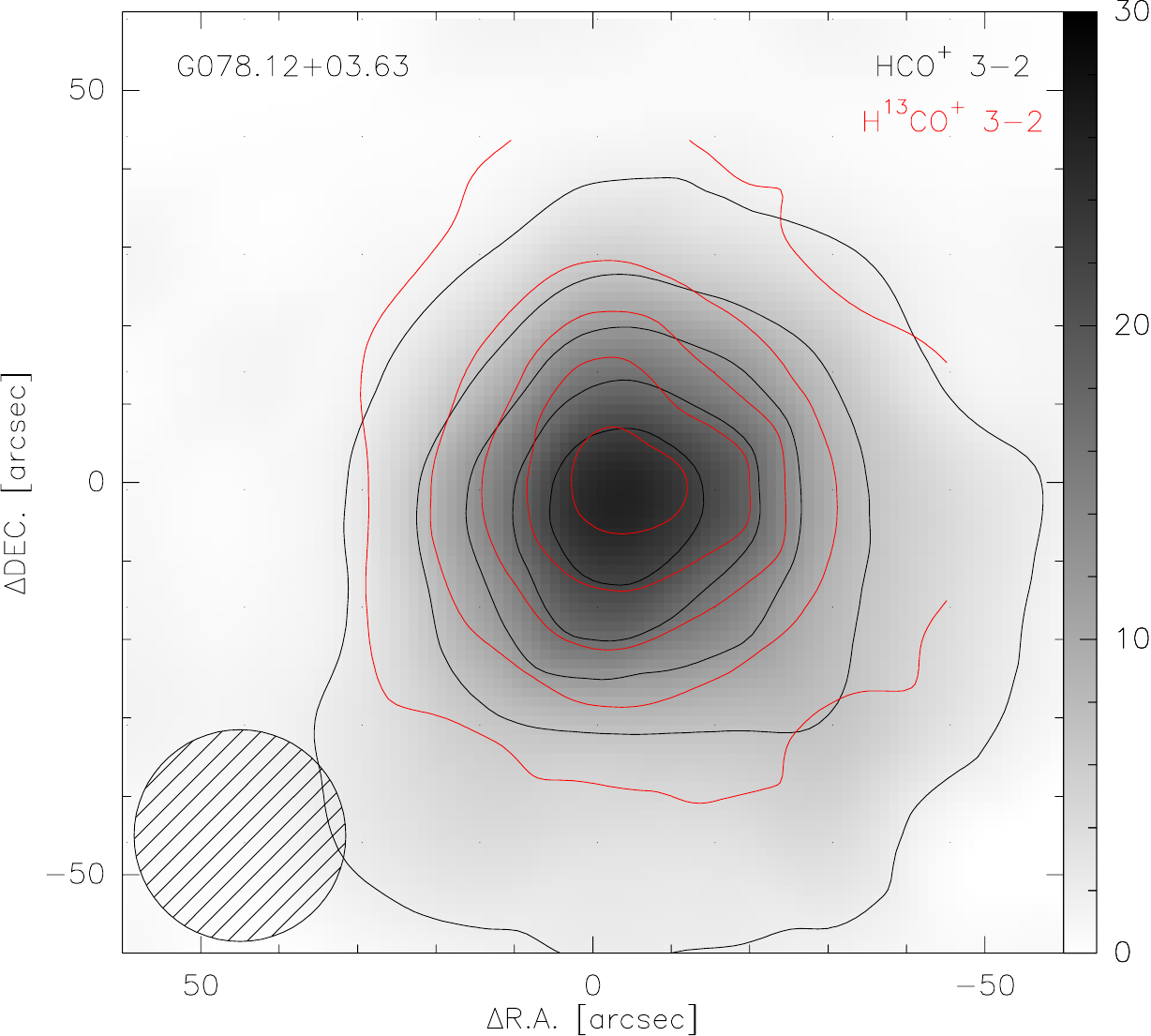}
       \includegraphics[width=3.08in]{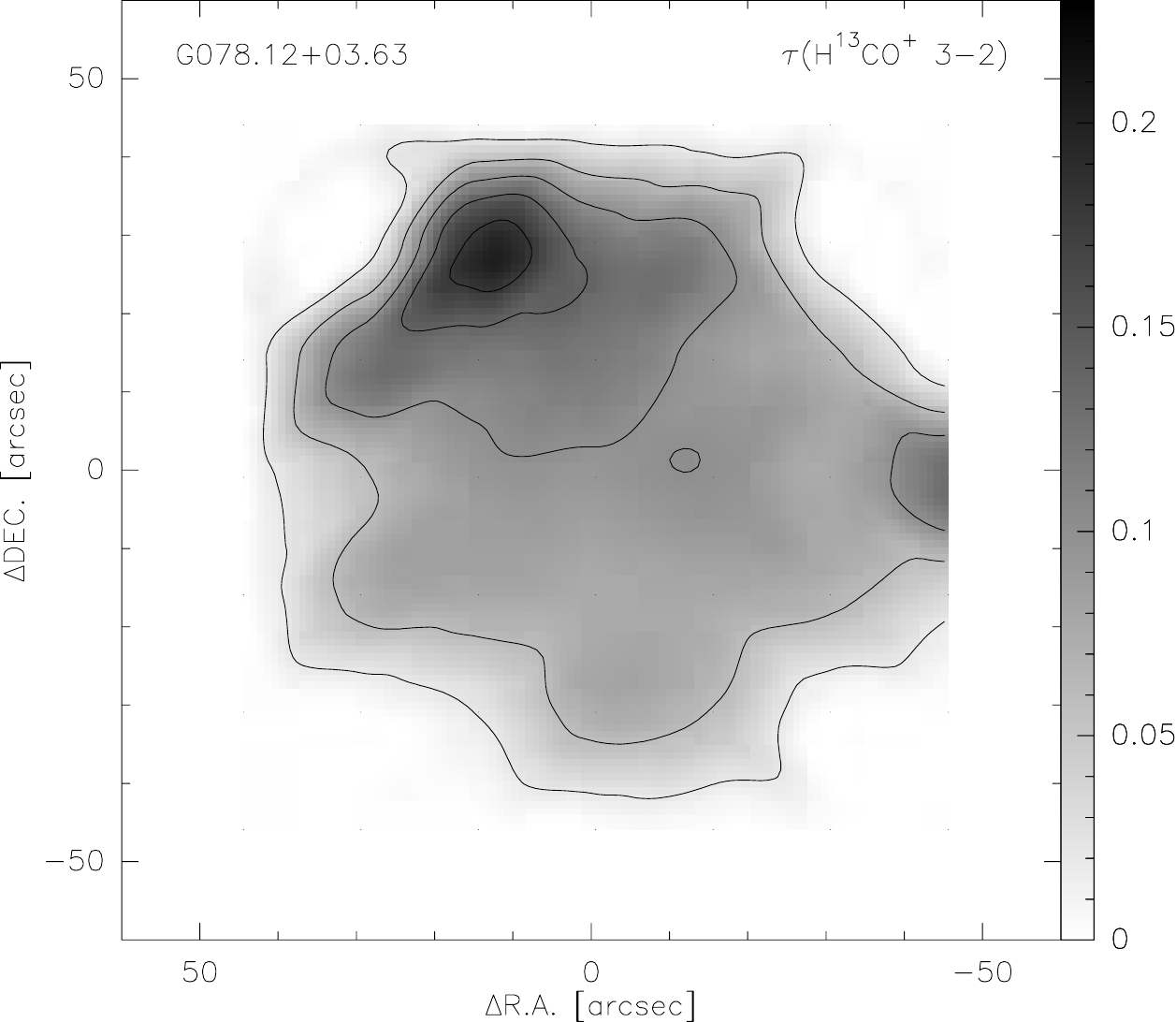}
 \caption{The data reduction results of G078.12+03.63. 
               {\it Top left:} The velocity integrated intensity maps of HCN and H$^{13}$CN 3-2. 
               The mapping size of HCN 3-2 is 2$'\times2'$, while it is 1.5$'\times1.5'$ for H$^{13}$CN 3-2, with a beam size of $\sim$ 27.8$''$.
               The grey scale and the black contour with levels starting from 2 K km s$^{-1}$ in step of 7 K km s$^{-1}$ show the observed HCN 3-2. 
               The red contour with levels starting from 0.3 K km s$^{-1}$ in step of 0.3 K km s$^{-1}$ represents H$^{13}$CN 3-2.
               {\it Top right:} The spatially resolved $\tau(\textrm{H}^{13}\textrm{CN})$ of G078.12+03.63 is demonstrated by black contour with levels 
               starting from 0.02 in step of 0.02. 
                {\it Bottom left:} The velocity integrated intensity maps of HCO$^+$ and H$^{13}$CO$^+$ 3-2. 
               The mapping size of HCO$^+$ 3-2 is 2$'\times2'$, while it is 1.5$'\times1.5'$ for H$^{13}$CO$^+$ 3-2, with a beam size of $\sim$ 27.8$''$.
               The grey scale and the black contour with levels starting from 2 K km s$^{-1}$ in step of 5 K km s$^{-1}$ show the observed HCO$^+$ 3-2. 
               The red contour with levels starting from 0.2 K km s$^{-1}$ in step of 0.5 K km s$^{-1}$ represents H$^{13}$CO$^+$ 3-2.
                {\it Bottom right:} The spatially resolved $\tau(\textrm{H}^{13}\textrm{CO$^+$})$ of G078.12+03.63 is demonstrated by black contour with levels 
               starting from 0.02 in step of 0.04. 
                }       
 \label{fig:g07812}
\end{figure*}


 \begin{figure*} 
    \centering
  \includegraphics[width=3.05in]{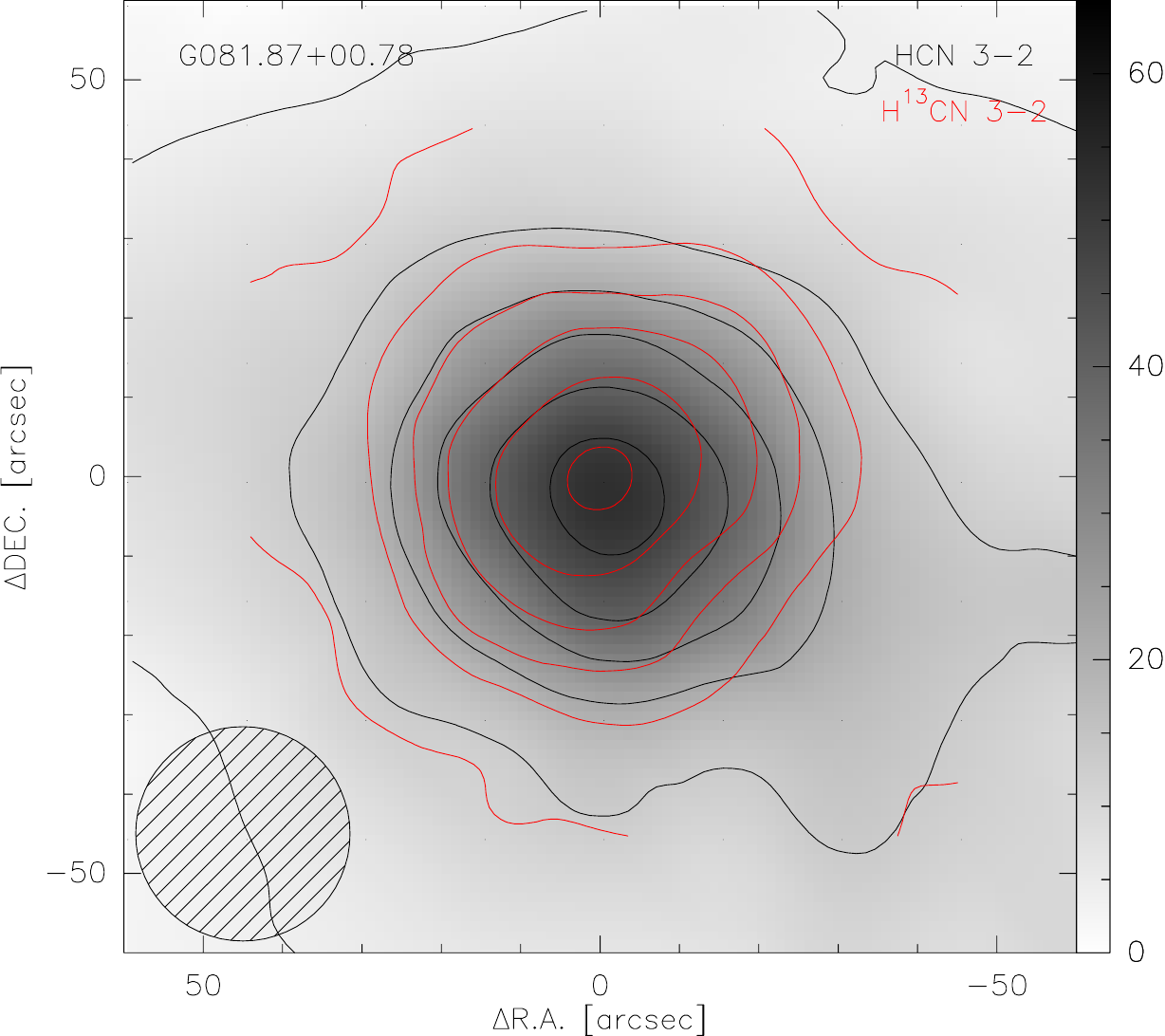}
   \includegraphics[width=3.03in]{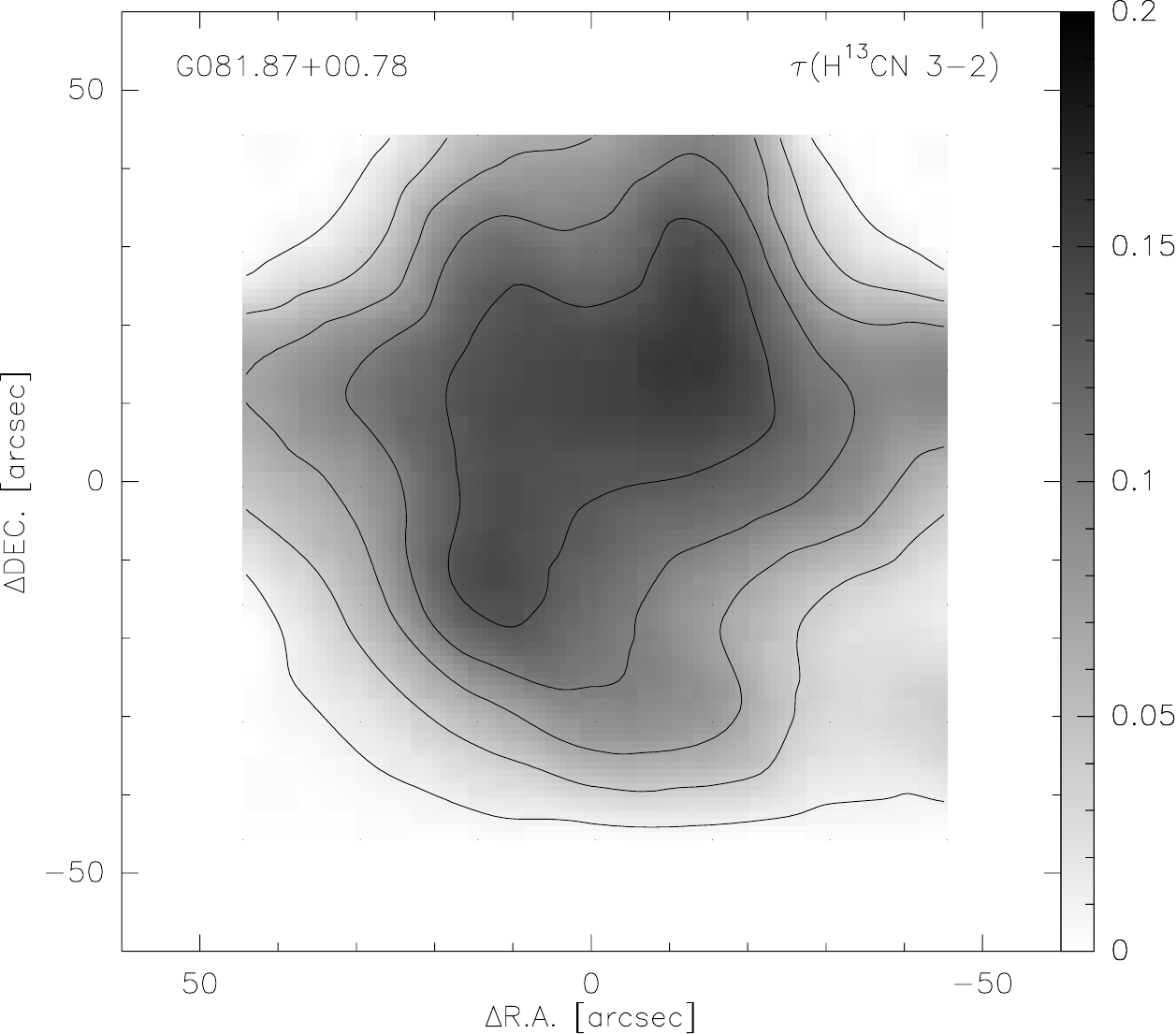}
       \includegraphics[width=3.05in]{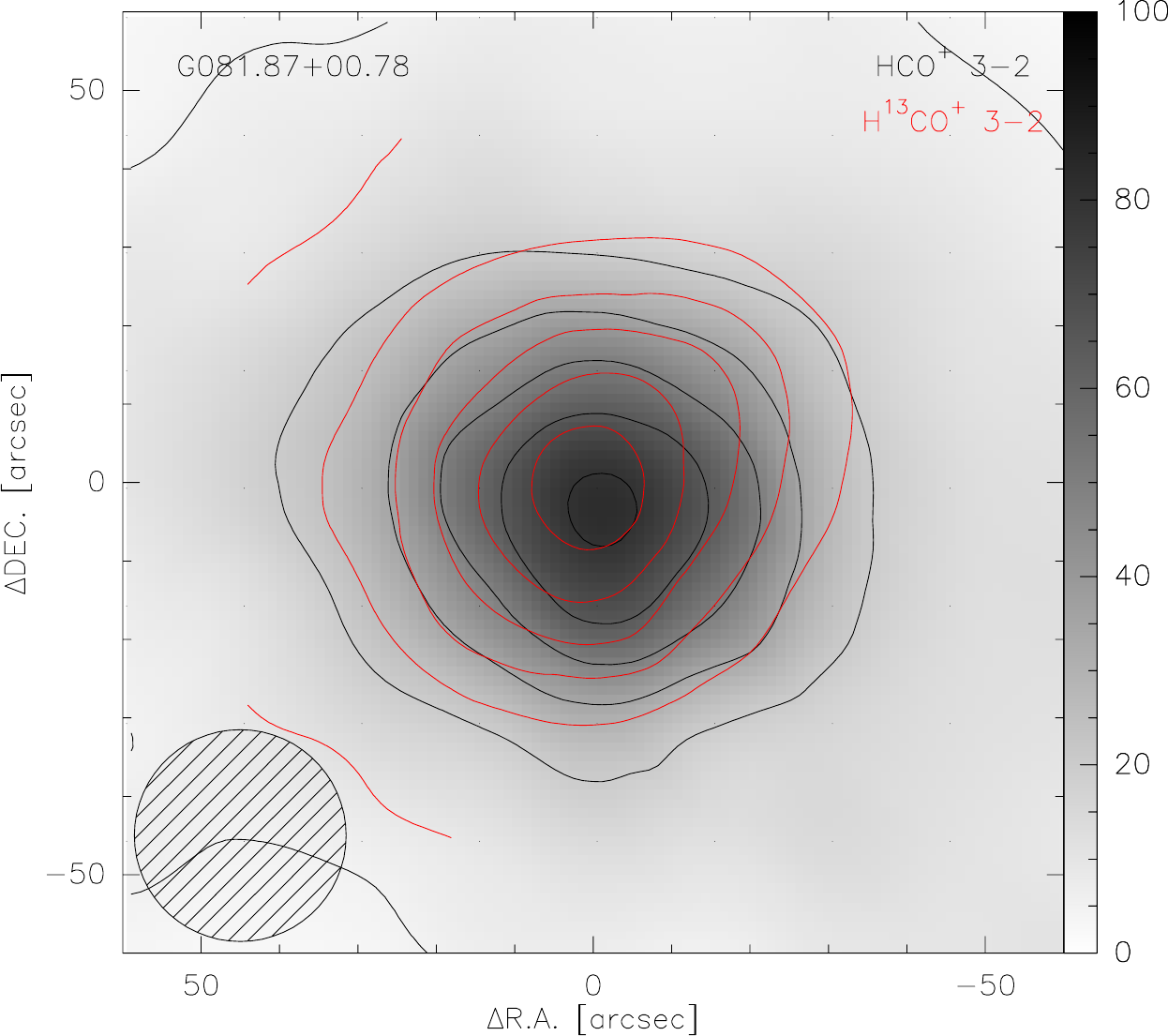}
       \includegraphics[width=3.08in]{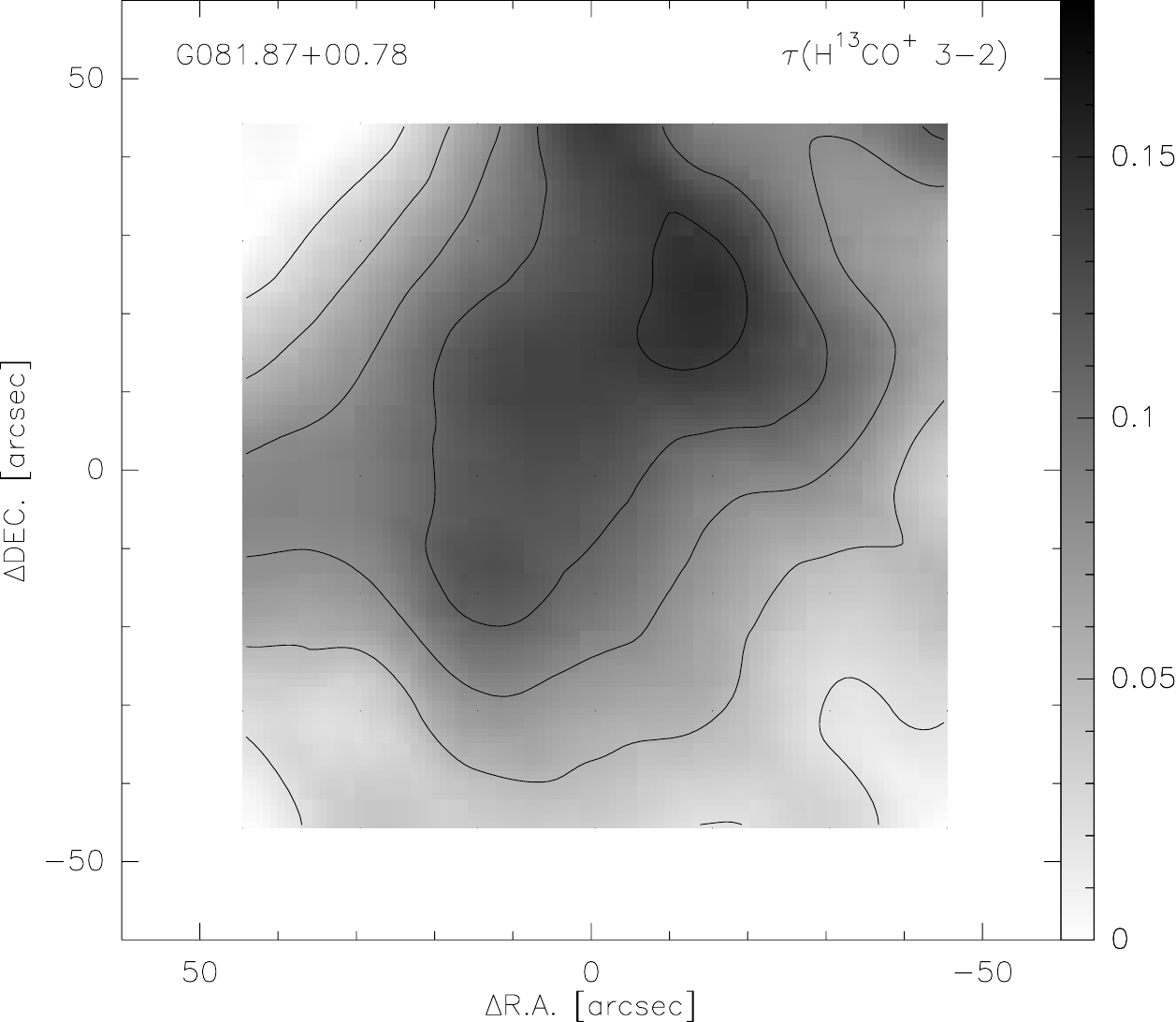}
 \caption{The data reduction results of G081.87+00.78. 
               {\it Top left:} The velocity integrated intensity maps of HCN and H$^{13}$CN 3-2. 
               The mapping size of HCN 3-2 is 2$'\times2'$, while it is 1.5$'\times1.5'$ for H$^{13}$CN 3-2, with a beam size of $\sim$ 27.8$''$.
               The grey scale and the black contour with levels starting from 4 K km s$^{-1}$ in step of 9 K km s$^{-1}$ show the observed HCN 3-2. 
               The red contour with levels starting from 0.4 K km s$^{-1}$ in step of 1.2 K km s$^{-1}$ represents H$^{13}$CN 3-2.
               {\it Top right:} The spatially resolved $\tau(\textrm{H}^{13}\textrm{CN})$ of G081.87+00.78 is demonstrated by black contour with levels 
               starting from 0.01 in step of 0.03. 
                {\it Bottom left:} The velocity integrated intensity maps of HCO$^+$ and H$^{13}$CO$^+$ 3-2. 
               The mapping size of HCO$^+$ 3-2 is 2$'\times2'$, while it is 1.5$'\times1.5'$ for H$^{13}$CO$^+$ 3-2, with a beam size of $\sim$ 27.8$''$.
               The grey scale and the black contour with levels starting from 5 K km s$^{-1}$ in step of 15 K km s$^{-1}$ show the observed HCO$^+$ 3-2. 
               The red contour with levels starting from 0.4 K km s$^{-1}$ in step of 1.6 K km s$^{-1}$ represents H$^{13}$CO$^+$ 3-2.
                {\it Bottom right:} The spatially resolved $\tau(\textrm{H}^{13}\textrm{CO$^+$})$ of G081.87+00.78 is demonstrated by black contour with levels 
               starting from 0.02 in step of 0.03. 
                }       
 \label{fig:g08187}
\end{figure*}


 \begin{figure*} 
    \centering
  \includegraphics[width=3.05in]{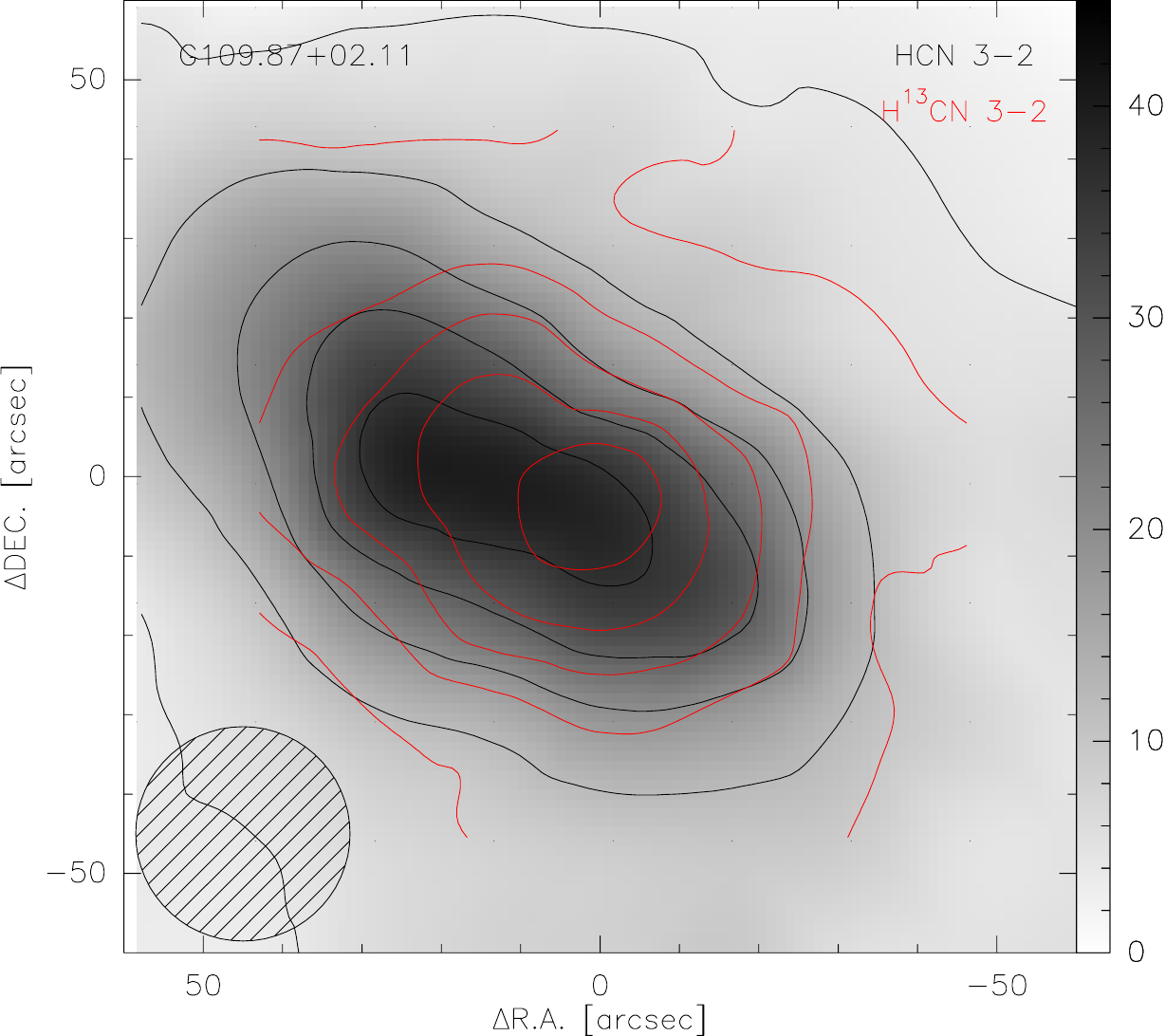}
   \includegraphics[width=3.03in]{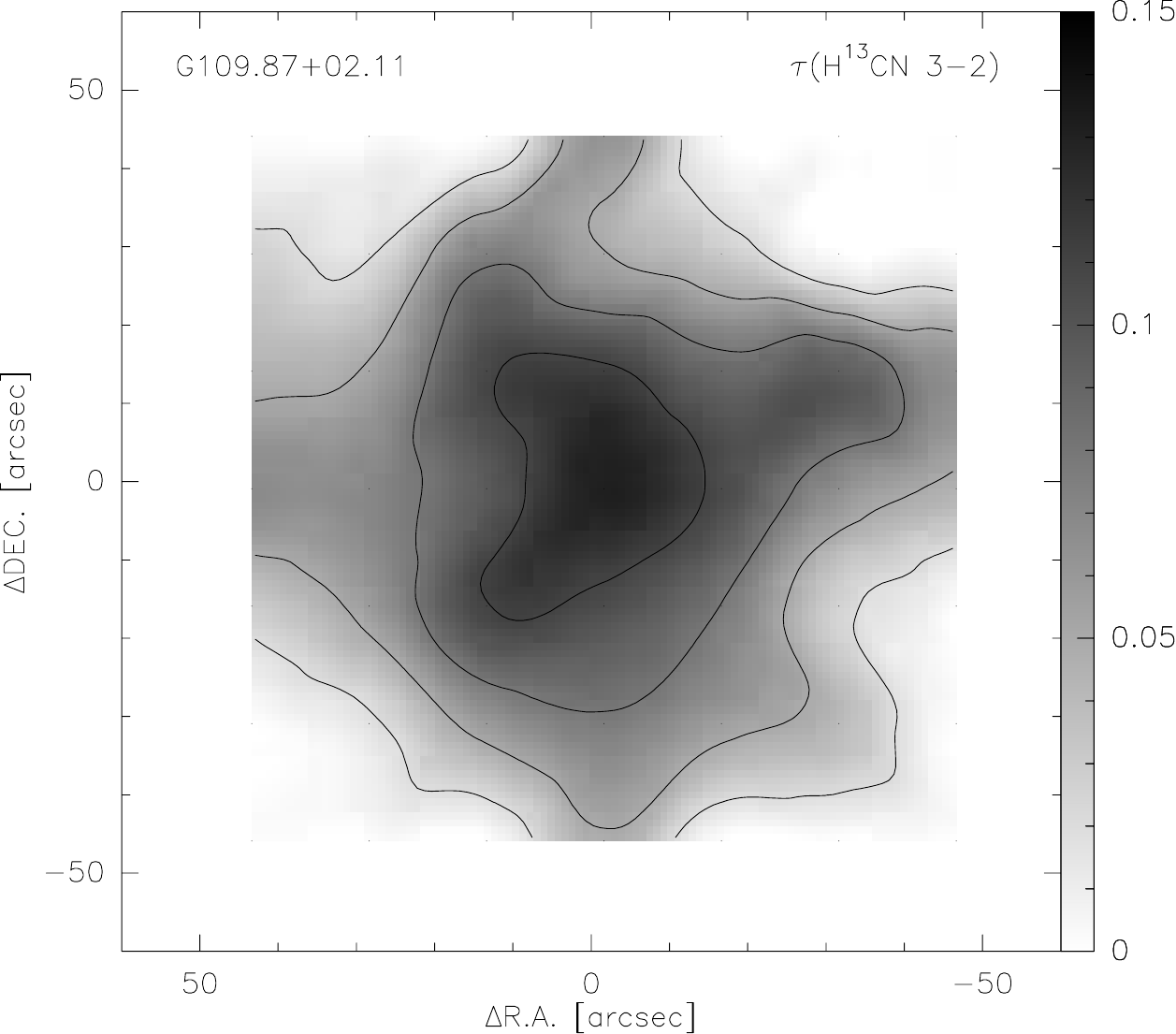}
       \includegraphics[width=3.05in]{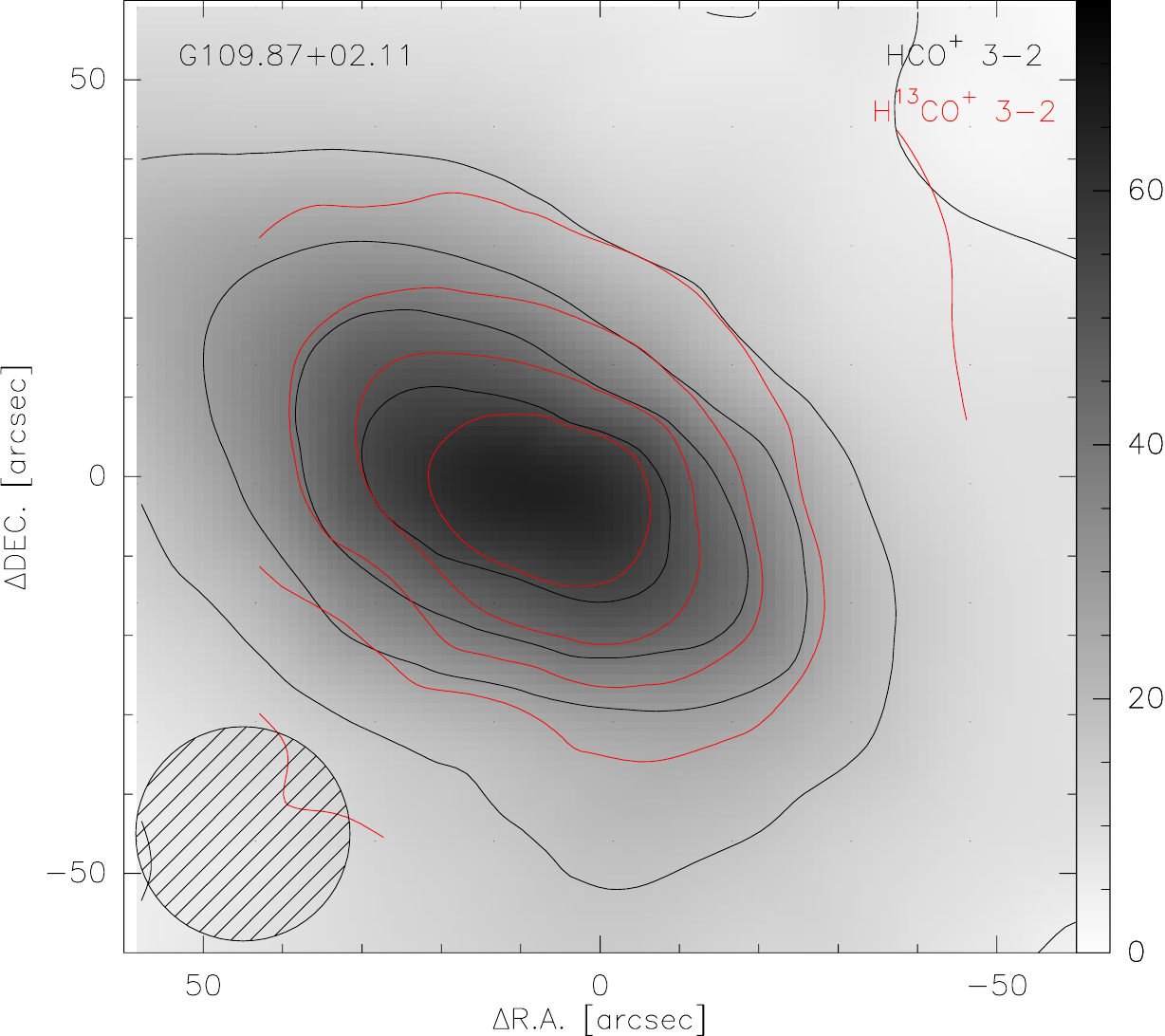}
       \includegraphics[width=3.08in]{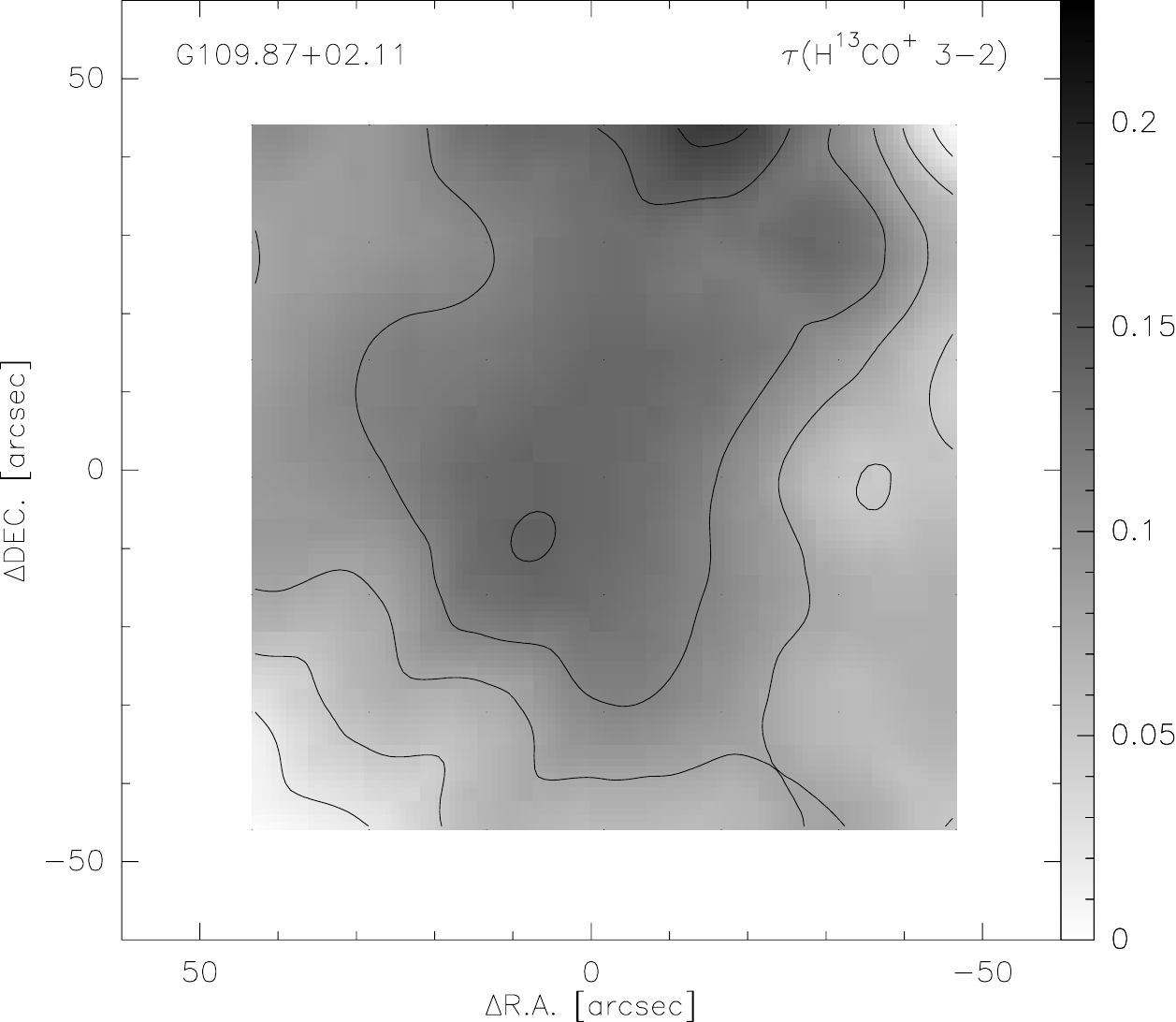}
 \caption{The data reduction results of G109.87+02.11. 
               {\it Top left:} The velocity integrated intensity maps of HCN and H$^{13}$CN 3-2. 
               The mapping size of HCN 3-2 is 2$'\times2'$, while it is 1.5$'\times1.5'$ for H$^{13}$CN 3-2, with a beam size of $\sim$ 27.8$''$.
               The grey scale and the black contour with levels starting from 4 K km s$^{-1}$ in step of 8 K km s$^{-1}$ show the observed HCN 3-2. 
               The red contour with levels starting from 0.4 K km s$^{-1}$ in step of 0.9 K km s$^{-1}$ represents H$^{13}$CN 3-2.
               {\it Top right:} The spatially resolved $\tau(\textrm{H}^{13}\textrm{CN})$ of G109.87+02.11 is demonstrated by black contour with levels 
               starting from 0.02 in step of 0.03. 
                {\it Bottom left:} The velocity integrated intensity maps of HCO$^+$ and H$^{13}$CO$^+$ 3-2. 
               The mapping size of HCO$^+$ 3-2 is 2$'\times2'$, while it is 1.5$'\times1.5'$ for H$^{13}$CO$^+$ 3-2, with a beam size of $\sim$ 27.8$''$.
               The grey scale and the black contour with levels starting from 5 K km s$^{-1}$ in step of 12 K km s$^{-1}$ show the observed HCO$^+$ 3-2. 
               The red contour with levels starting from 0.5 K km s$^{-1}$ in step of 1.6 K km s$^{-1}$ represents H$^{13}$CO$^+$ 3-2.
                {\it Bottom right:} The spatially resolved $\tau(\textrm{H}^{13}\textrm{CO$^+$})$ of G109.87+02.11 is demonstrated by black contour with levels 
               starting from 0.02 in step of 0.04. 
                }       
 \label{fig:g10987}
\end{figure*}


 \begin{figure*} 
    \centering
  \includegraphics[width=3.05in]{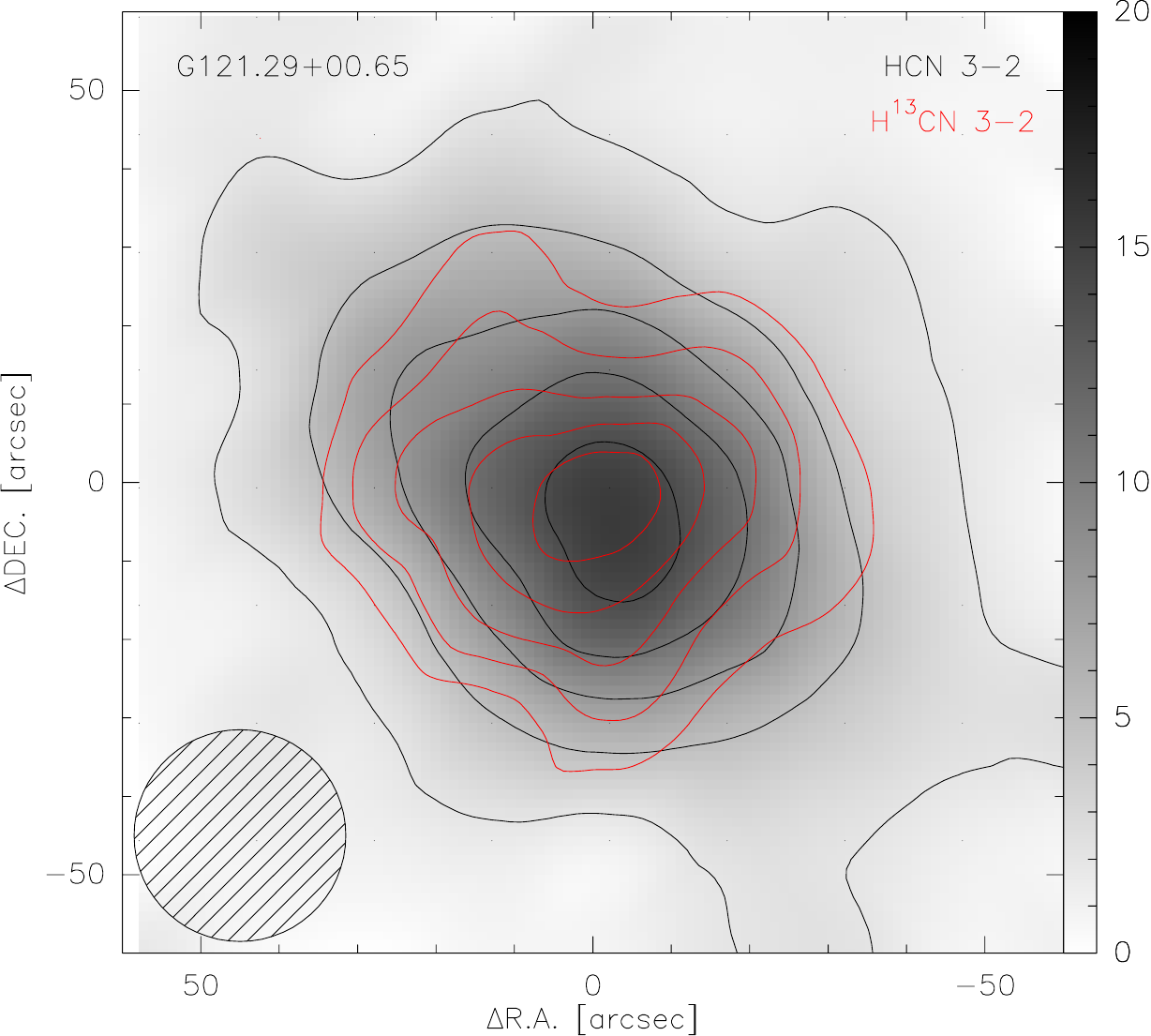}
   \includegraphics[width=3.03in]{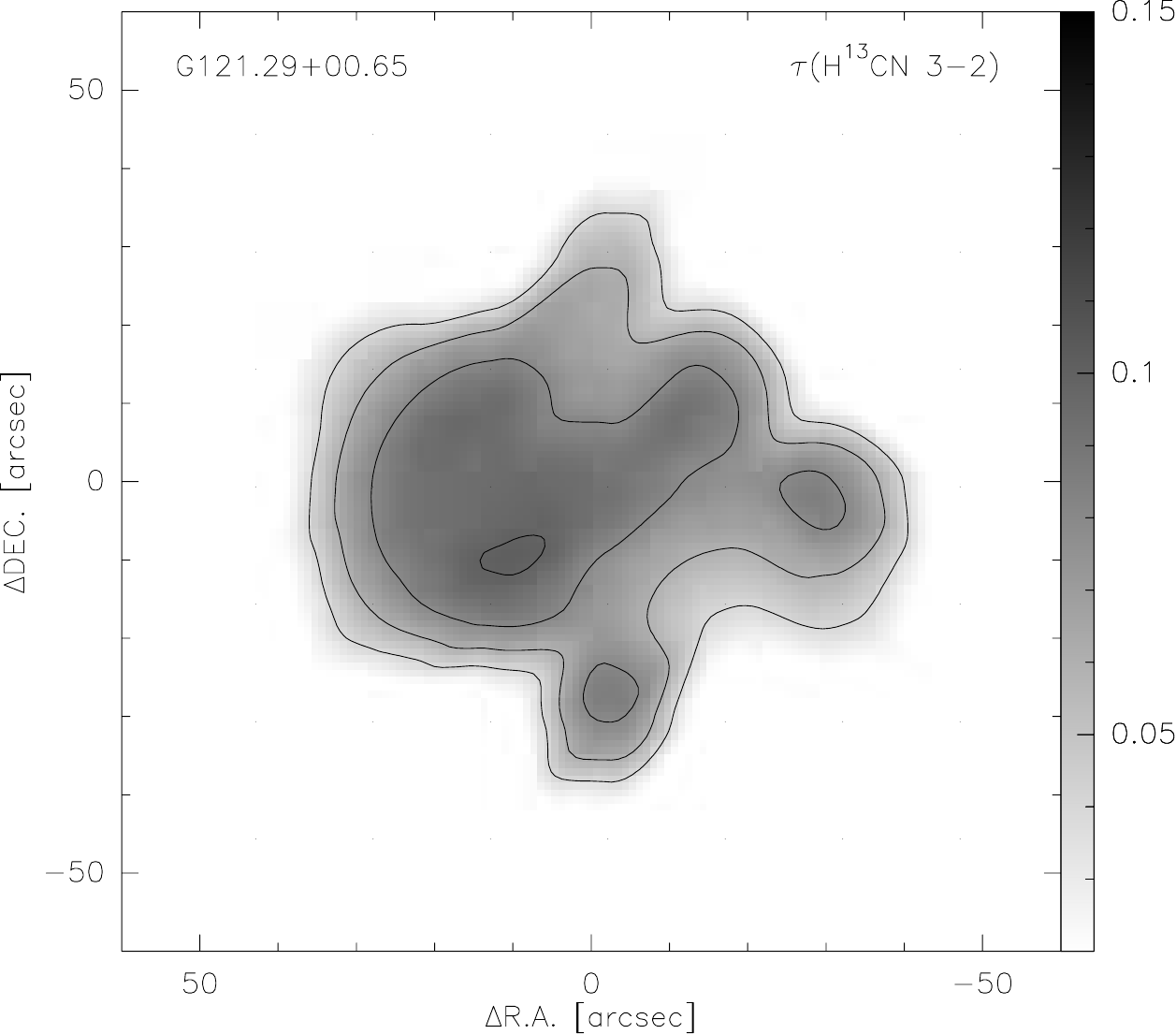}
       \includegraphics[width=3.05in]{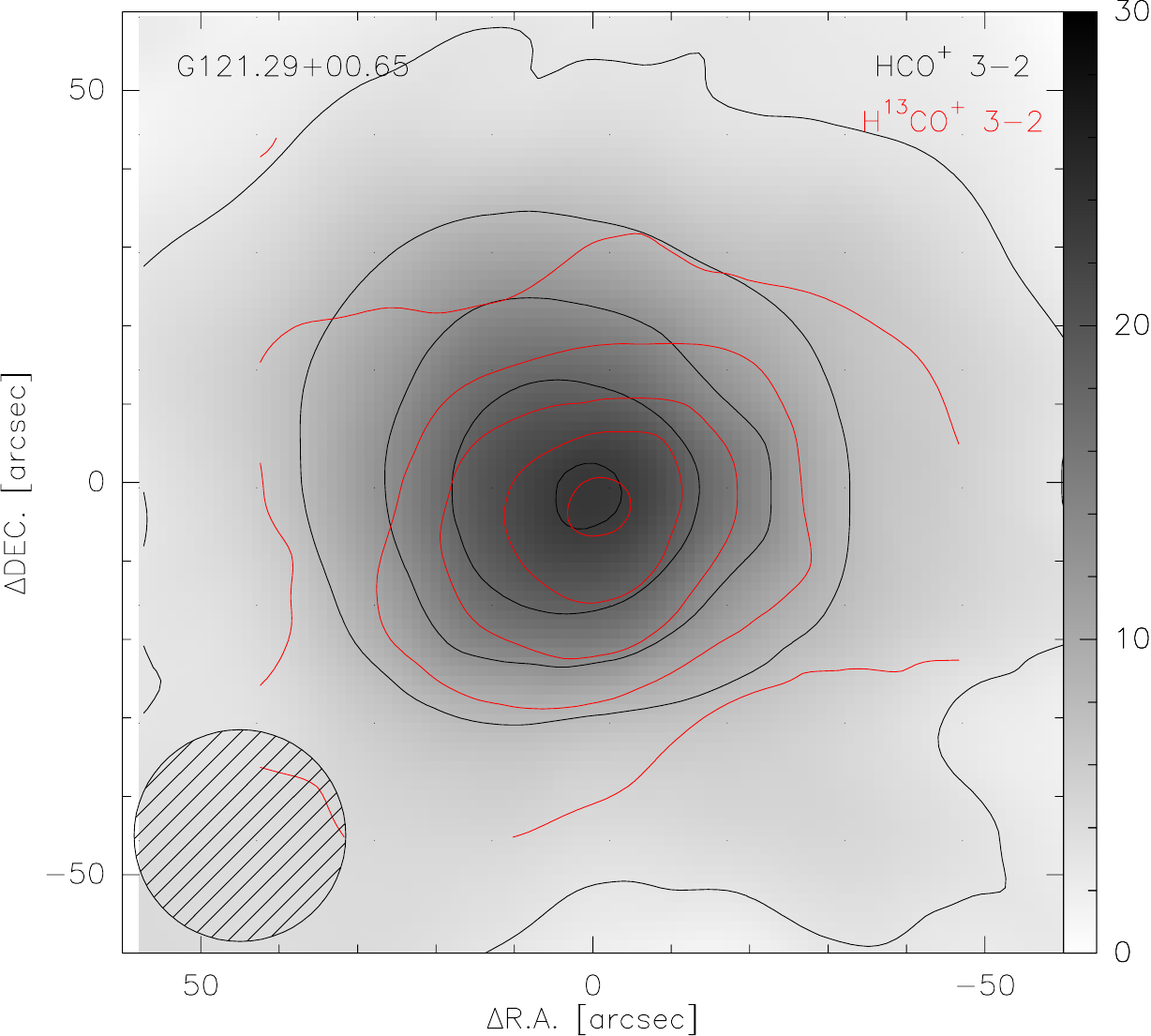}
       \includegraphics[width=3.08in]{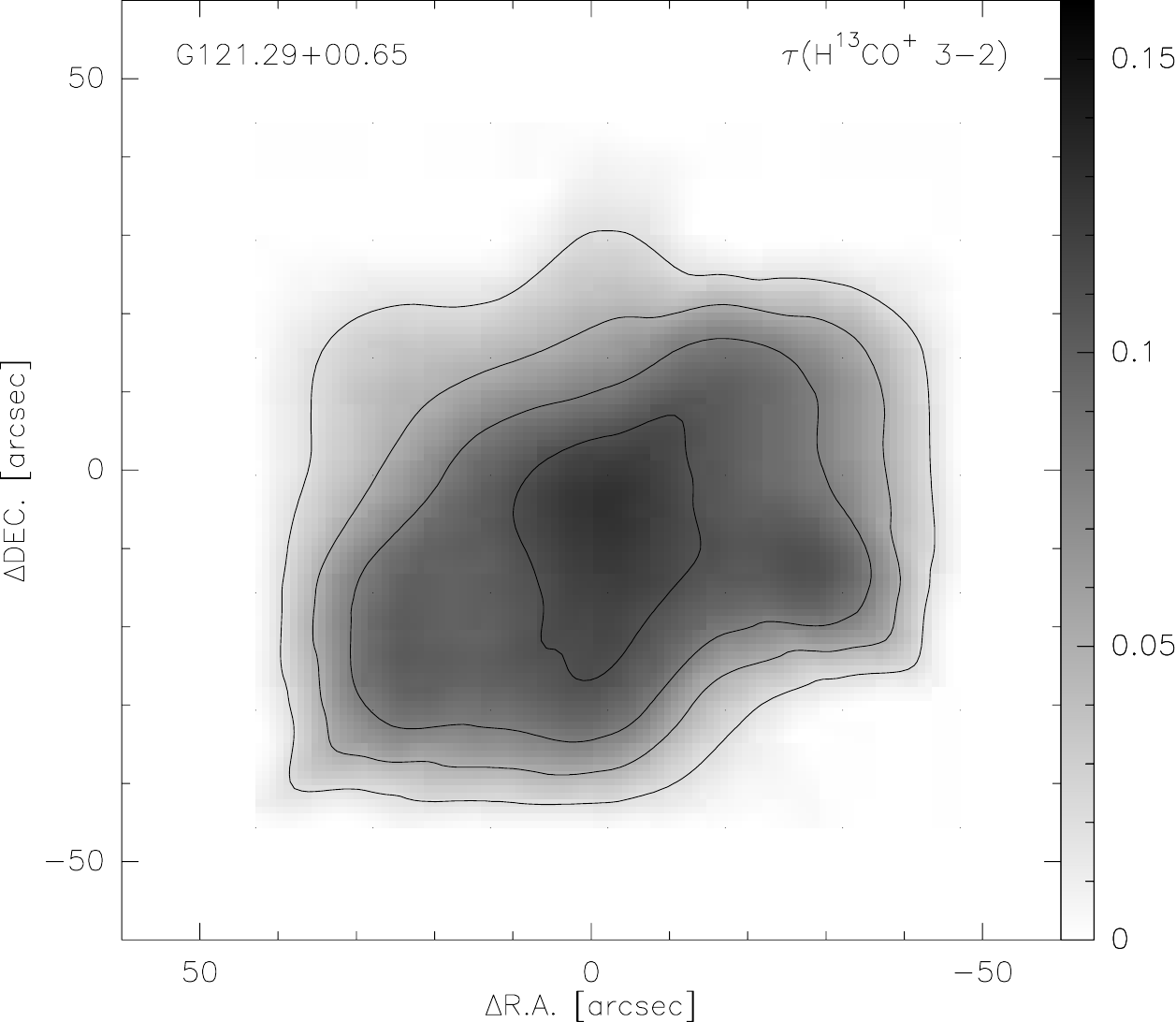}
 \caption{The data reduction results of G121.29+00.65. 
               {\it Top left:} The velocity integrated intensity maps of HCN and H$^{13}$CN 3-2. 
               The mapping size of HCN 3-2 is 2$'\times2'$, while it is 1.5$'\times1.5'$ for H$^{13}$CN 3-2, with a beam size of $\sim$ 27.8$''$.
               The grey scale and the black contour with levels starting from 2 K km s$^{-1}$ in step of 3 K km s$^{-1}$ show the observed HCN 3-2. 
               The red contour with levels starting from 0.3 K km s$^{-1}$ in step of 0.2 K km s$^{-1}$ represents H$^{13}$CN 3-2.
               {\it Top right:} The spatially resolved $\tau(\textrm{H}^{13}\textrm{CN})$ of G121.29+00.65 is demonstrated by black contour with levels 
               starting from 0.04 in step of 0.02. 
                {\it Bottom left:} The velocity integrated intensity maps of HCO$^+$ and H$^{13}$CO$^+$ 3-2. 
               The mapping size of HCO$^+$ 3-2 is 2$'\times2'$, while it is 1.5$'\times1.5'$ for H$^{13}$CO$^+$ 3-2, with a beam size of $\sim$ 27.8$''$.
               The grey scale and the black contour with levels starting from 3 K km s$^{-1}$ in step of 5 K km s$^{-1}$ show the observed HCO$^+$ 3-2. 
               The red contour with levels starting from 0.3 K km s$^{-1}$ in step of 0.6 K km s$^{-1}$ represents H$^{13}$CO$^+$ 3-2.
                {\it Bottom right:} The spatially resolved $\tau(\textrm{H}^{13}\textrm{CO$^+$})$ of G121.29+00.65 is demonstrated by black contour with levels 
               starting from 0.02 in step of 0.03. 
                }       
 \label{fig:g12129}
\end{figure*}


 \begin{figure*} 
    \centering
  \includegraphics[width=3.05in]{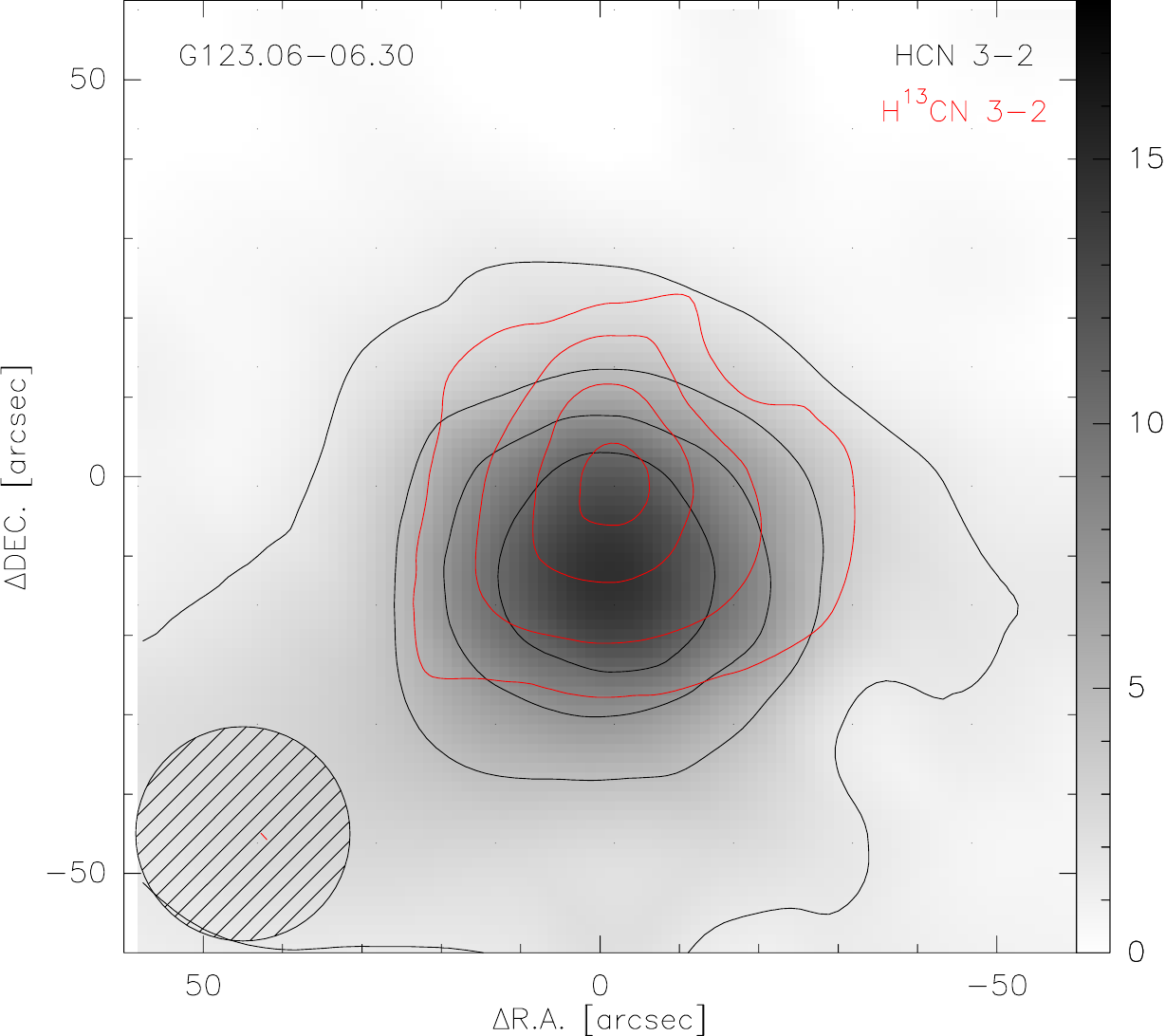}
   \includegraphics[width=3.03in]{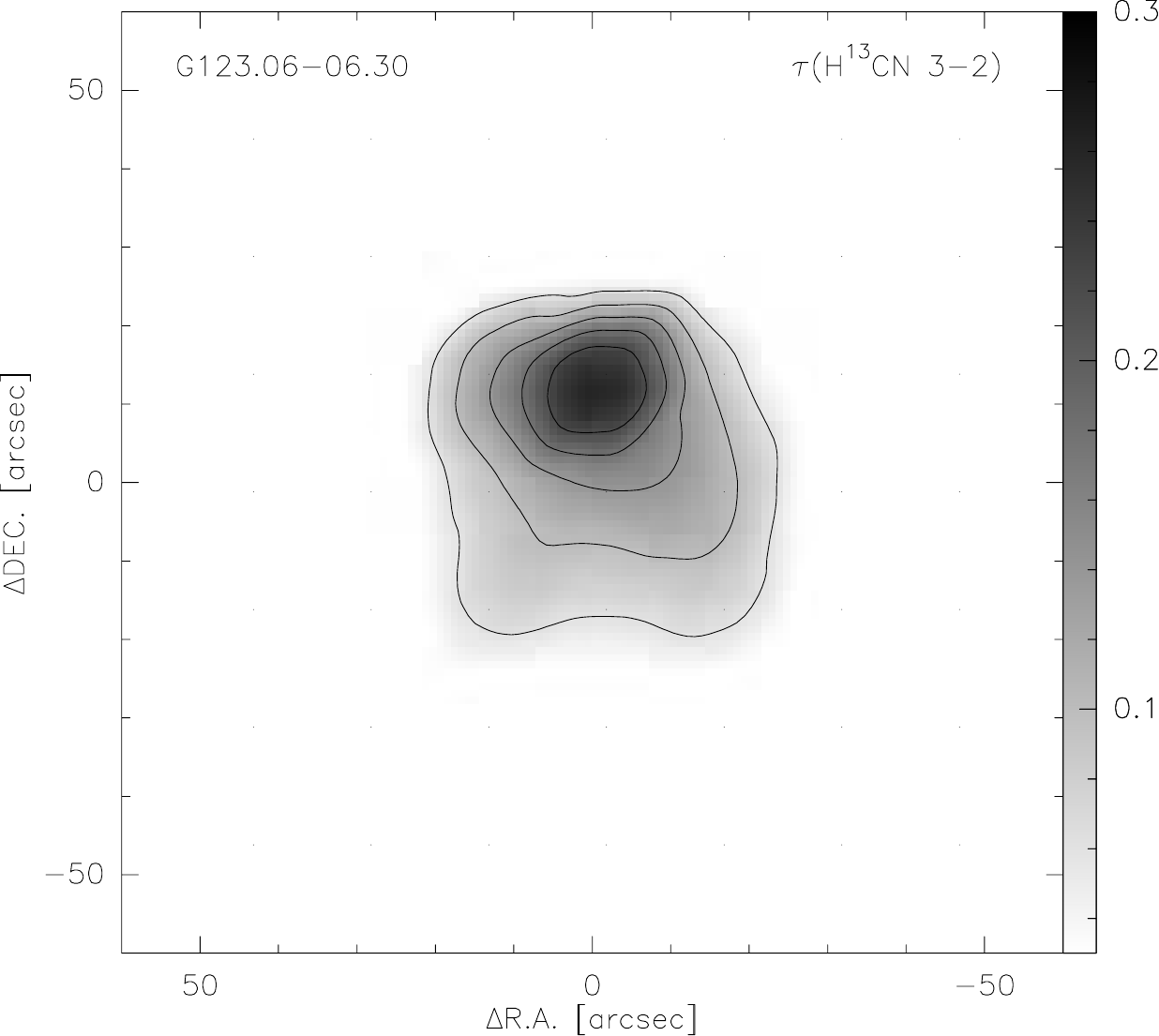}
       \includegraphics[width=3.05in]{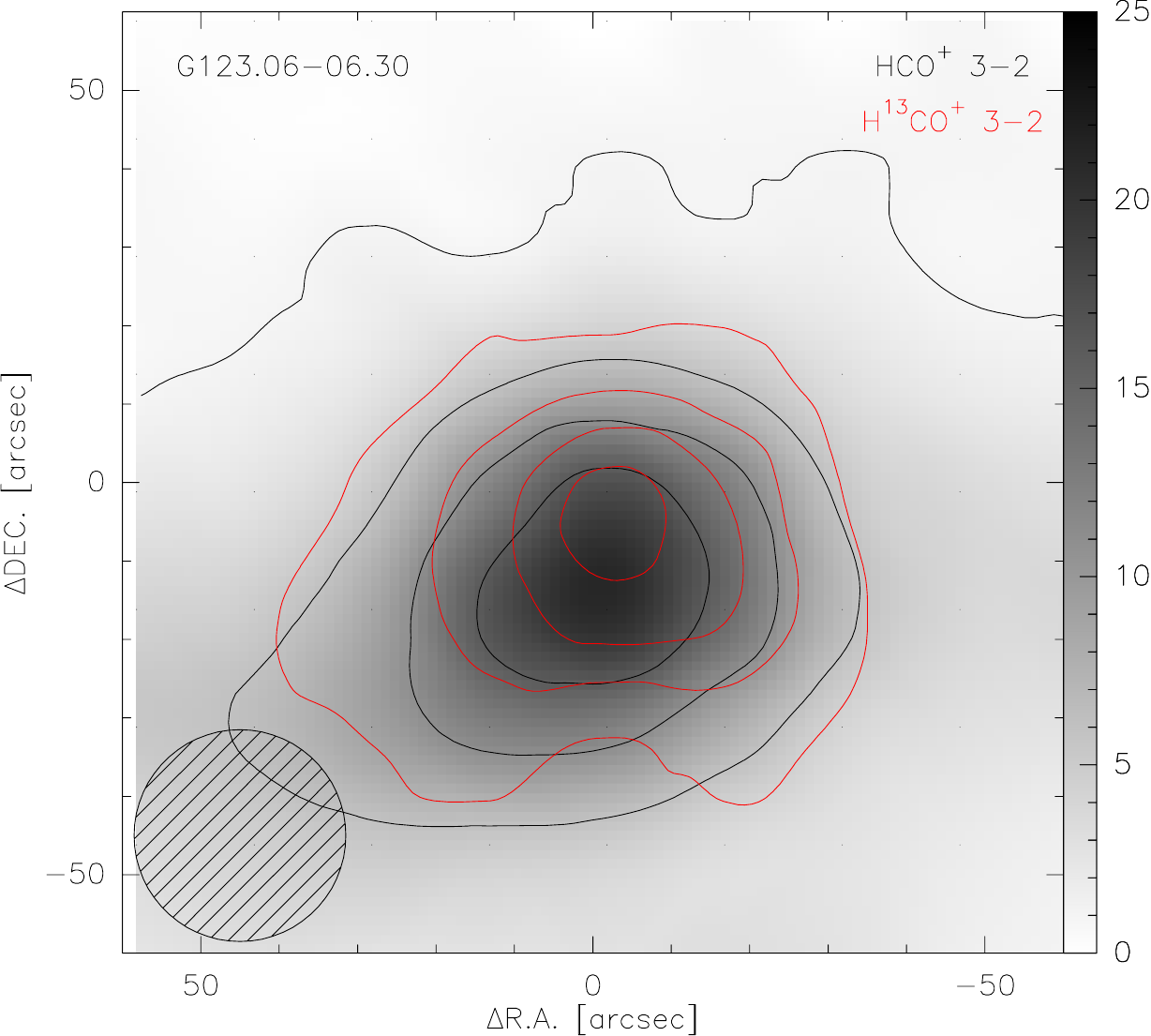}
       \includegraphics[width=3.08in]{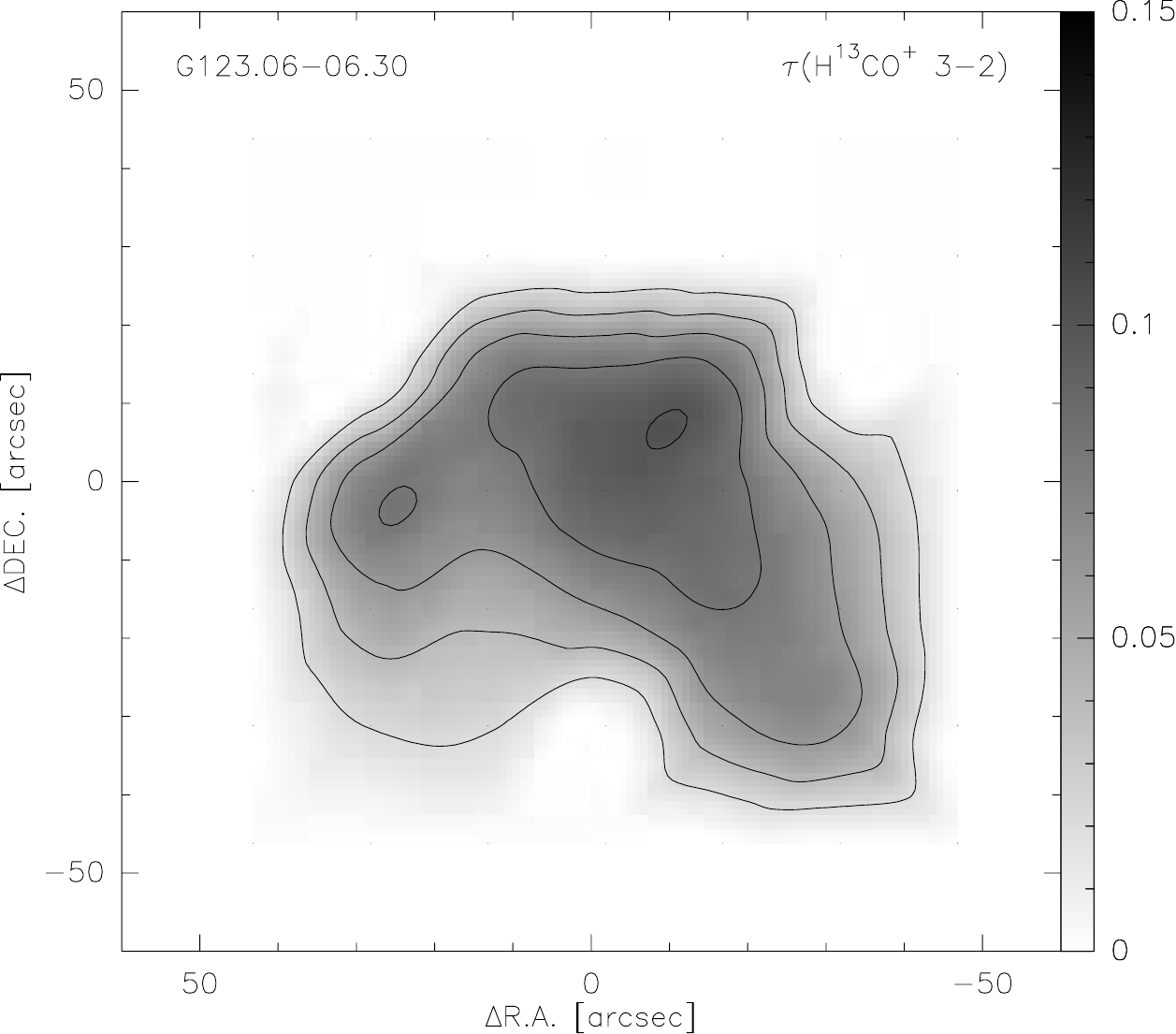}
 \caption{The data reduction results of G123.06-06.30. 
               {\it Top left:} The velocity integrated intensity maps of HCN and H$^{13}$CN 3-2. 
               The mapping size of HCN 3-2 is 2$'\times2'$, while it is 1.5$'\times1.5'$ for H$^{13}$CN 3-2, with a beam size of $\sim$ 27.8$''$.
               The grey scale and the black contour with levels starting from 1.5 K km s$^{-1}$ in step of 3 K km s$^{-1}$ show the observed HCN 3-2. 
               The red contour with levels starting from 0.3 K km s$^{-1}$ in step of 0.4 K km s$^{-1}$ represents H$^{13}$CN 3-2.
               {\it Top right:} The spatially resolved $\tau(\textrm{H}^{13}\textrm{CN})$ of G123.06-06.30 is demonstrated by black contour with levels 
               starting from 0.06 in step of 0.04. 
                {\it Bottom left:} The velocity integrated intensity maps of HCO$^+$ and H$^{13}$CO$^+$ 3-2. 
               The mapping size of HCO$^+$ 3-2 is 2$'\times2'$, while it is 1.5$'\times1.5'$ for H$^{13}$CO$^+$ 3-2, with a beam size of $\sim$ 27.8$''$.
               The grey scale and the black contour with levels starting from 1 K km s$^{-1}$ in step of 5 K km s$^{-1}$ show the observed HCO$^+$ 3-2. 
               The red contour with levels starting from 0.3 K km s$^{-1}$ in step of 0.4 K km s$^{-1}$ represents H$^{13}$CO$^+$ 3-2.
                {\it Bottom right:} The spatially resolved $\tau(\textrm{H}^{13}\textrm{CO$^+$})$ of G123.06-06.30 is demonstrated by black contour with levels 
               starting from 0.02 in step of 0.02. 
                }       
 \label{fig:g12306}
\end{figure*}


 \begin{figure*} 
    \centering
  \includegraphics[width=3.05in]{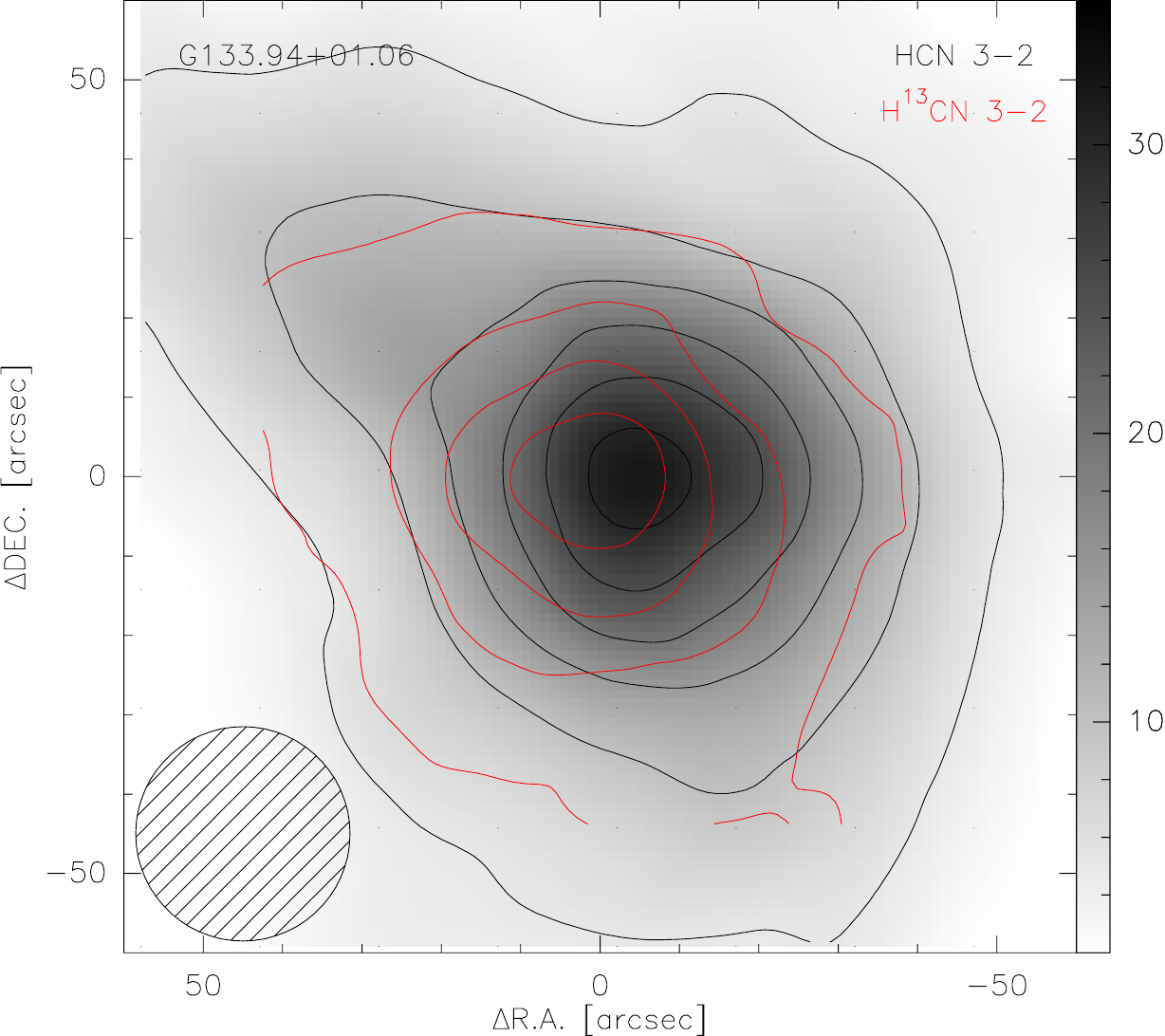}
   \includegraphics[width=3.03in]{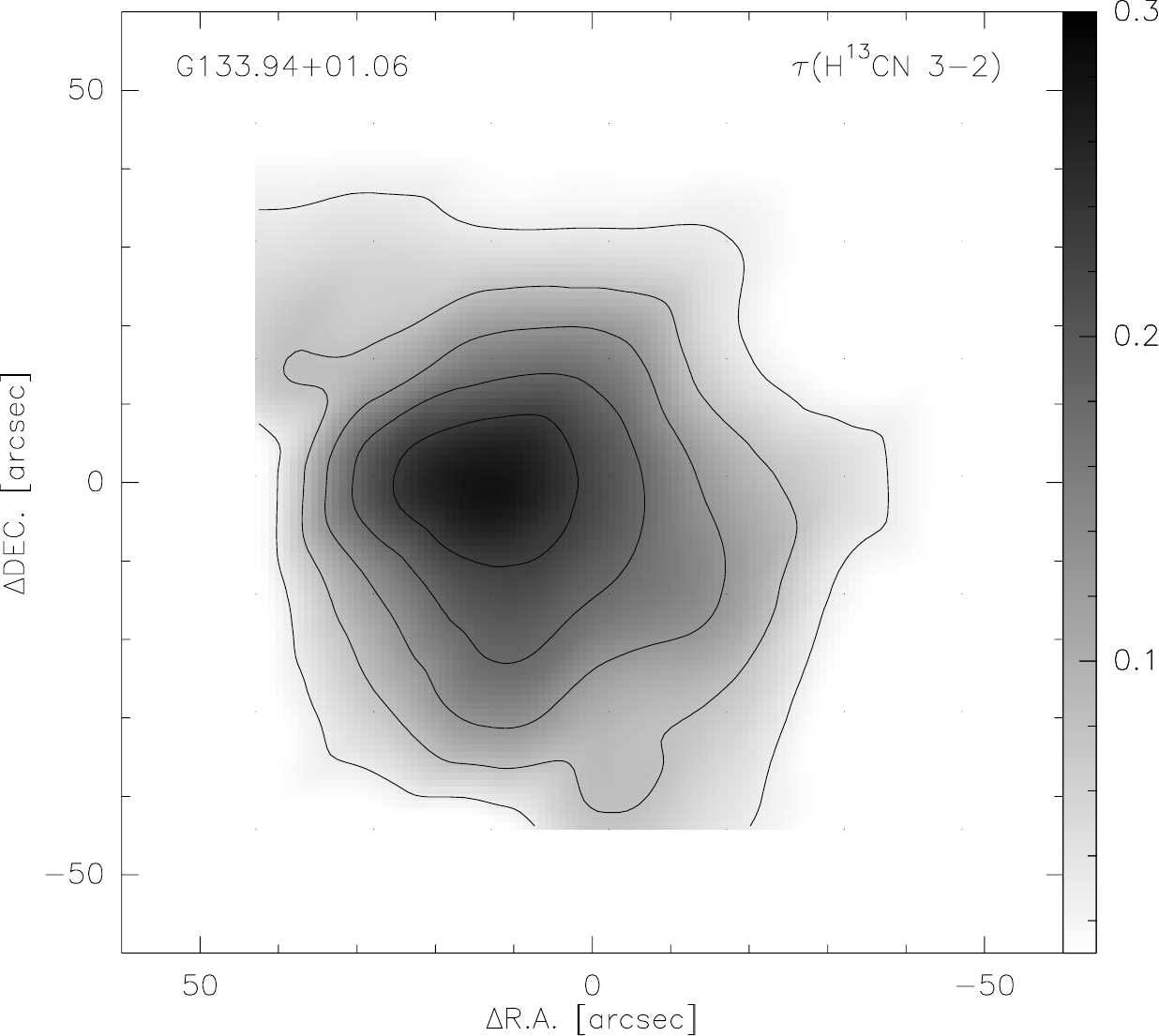}
       \includegraphics[width=3.05in]{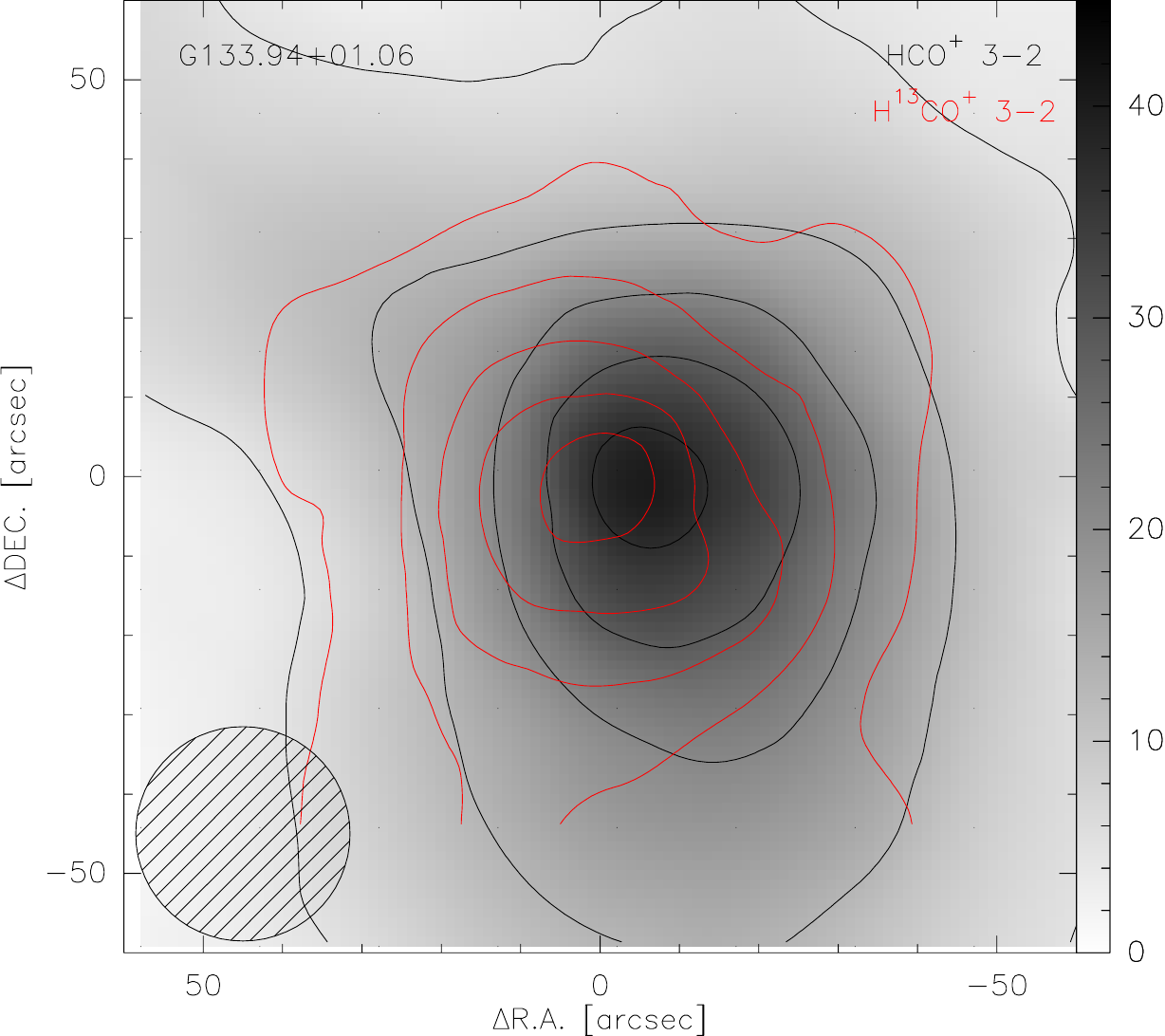}
       \includegraphics[width=3.08in]{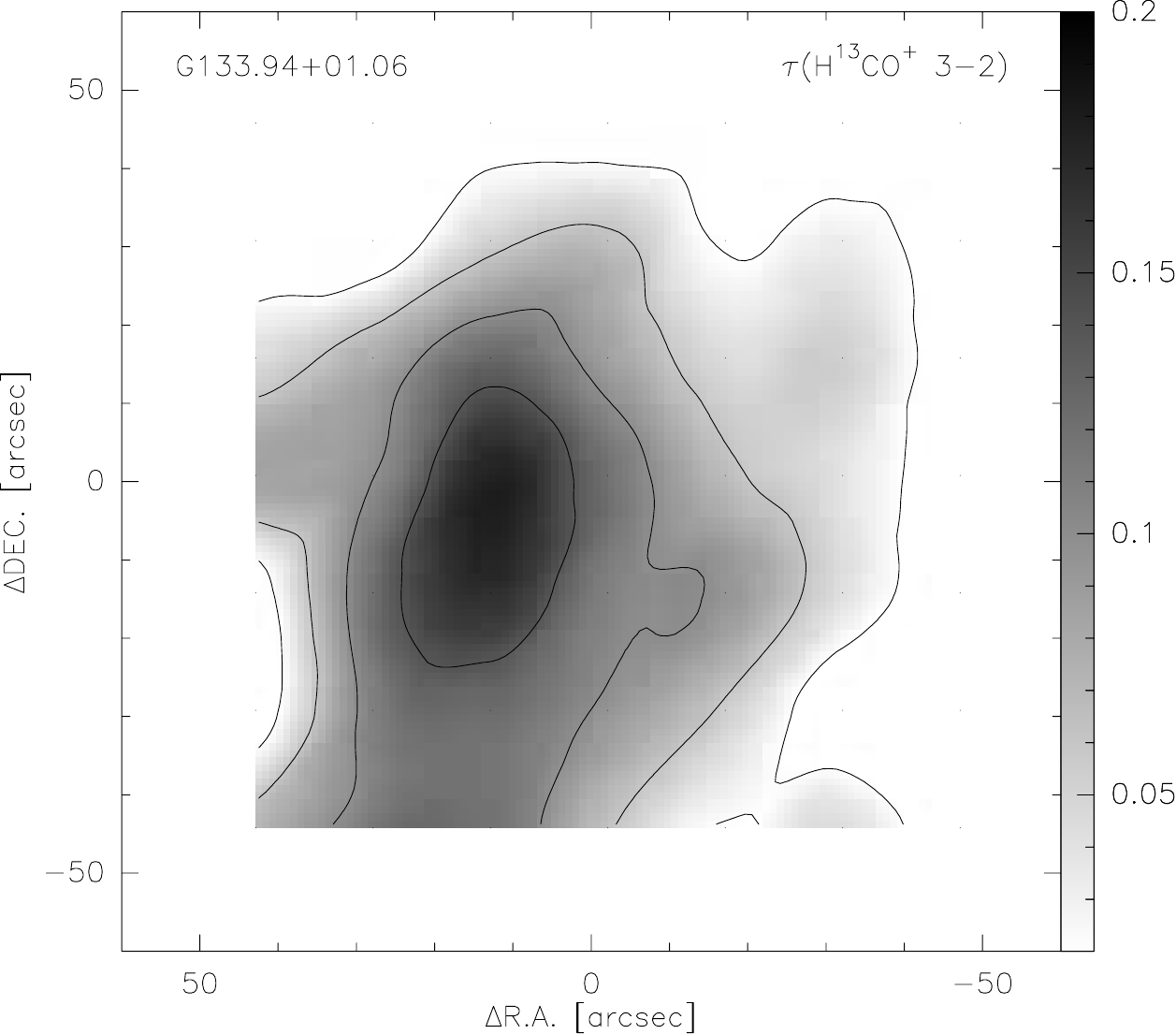}
 \caption{The data reduction results of G133.94+01.06. 
               {\it Top left:} The velocity integrated intensity maps of HCN and H$^{13}$CN 3-2. 
               The mapping size of HCN 3-2 is 2$'\times2'$, while it is 1.5$'\times1.5'$ for H$^{13}$CN 3-2, with a beam size of $\sim$ 27.8$''$.
               The grey scale and the black contour with levels starting from 5 K km s$^{-1}$ in step of 5 K km s$^{-1}$ show the observed HCN 3-2. 
               The red contour with levels starting from 0.45 K km s$^{-1}$ in step of 1.5 K km s$^{-1}$ represents H$^{13}$CN 3-2.
               {\it Top right:} The spatially resolved $\tau(\textrm{H}^{13}\textrm{CN})$ of G133.94+01.06 is demonstrated by black contour with levels 
               starting from 0.03 in step of 0.05. 
                {\it Bottom left:} The velocity integrated intensity maps of HCO$^+$ and H$^{13}$CO$^+$ 3-2. 
               The mapping size of HCO$^+$ 3-2 is 2$'\times2'$, while it is 1.5$'\times1.5'$ for H$^{13}$CO$^+$ 3-2, with a beam size of $\sim$ 27.8$''$.
               The grey scale and the black contour with levels starting from 5 K km s$^{-1}$ in step of 8 K km s$^{-1}$ show the observed HCO$^+$ 3-2. 
               The red contour with levels starting from 0.5 K km s$^{-1}$ in step of 0.9 K km s$^{-1}$ represents H$^{13}$CO$^+$ 3-2.
                {\it Bottom right:} The spatially resolved $\tau(\textrm{H}^{13}\textrm{CO$^+$})$ of G133.94+01.06 is demonstrated by black contour with levels 
               starting from 0.02 in step of 0.04. 
                }       
 \label{fig:g13394}
\end{figure*}


 \begin{figure*} 
    \centering
  \includegraphics[width=3.05in]{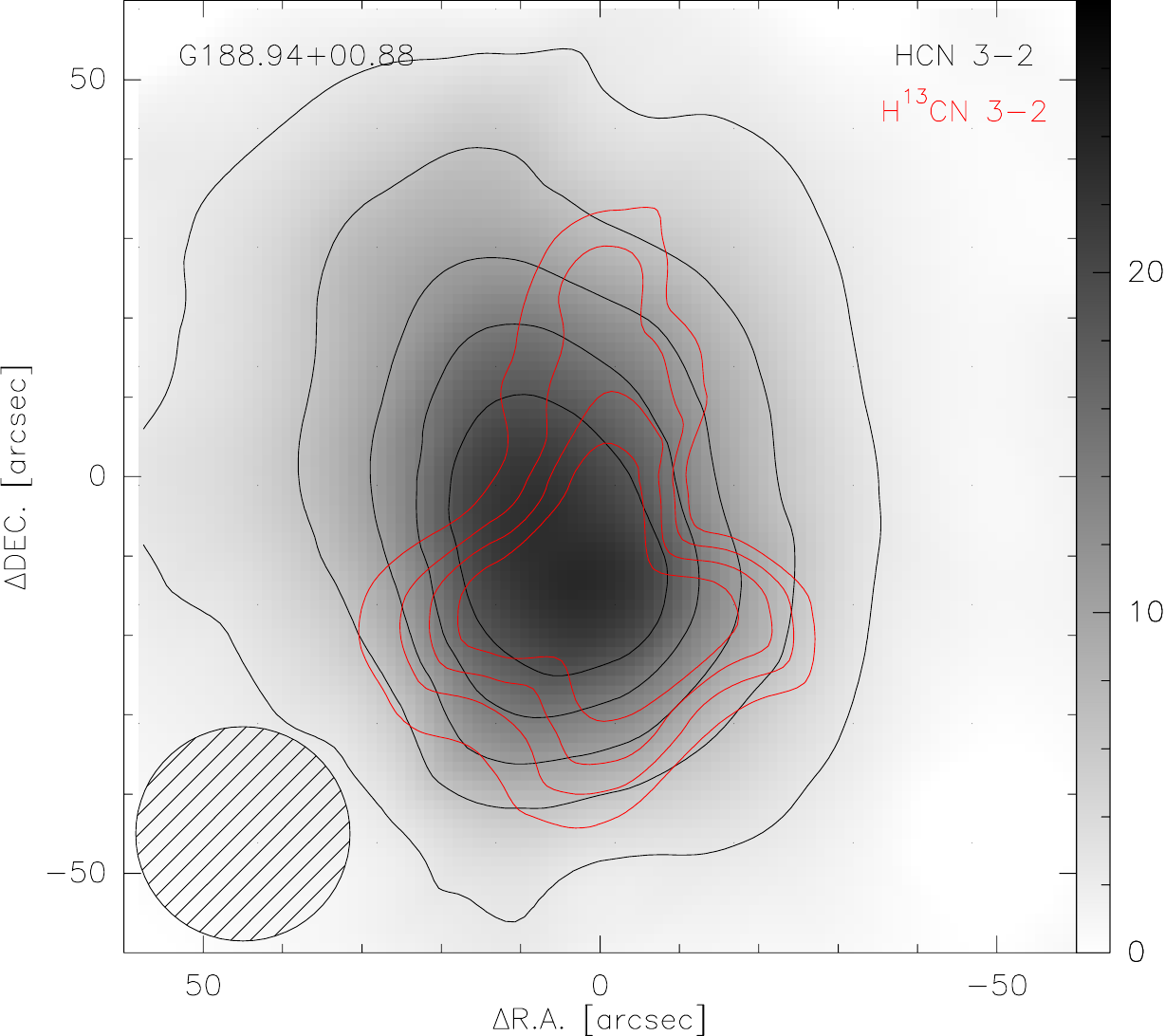}
   \includegraphics[width=3.03in]{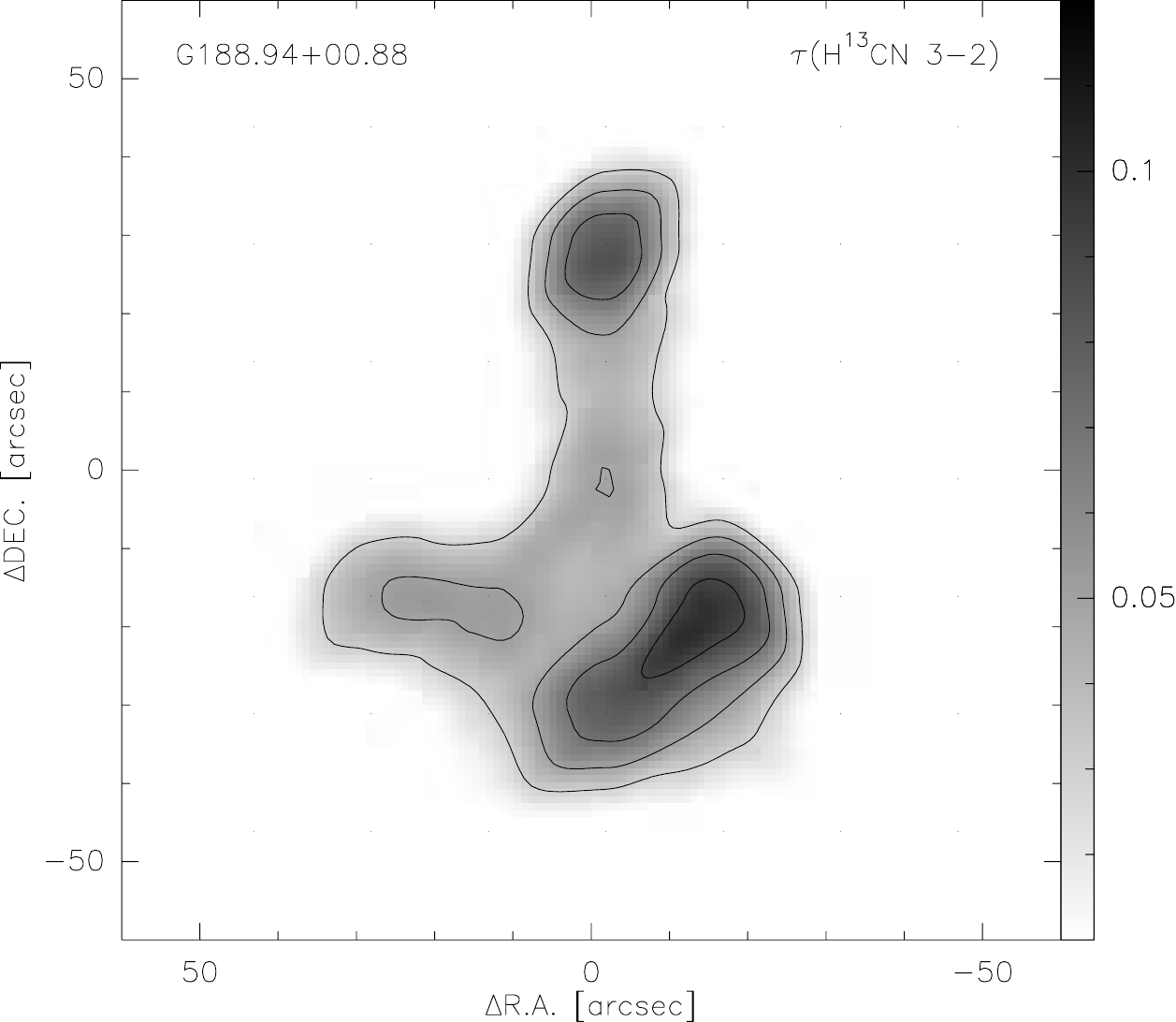}
       \includegraphics[width=3.05in]{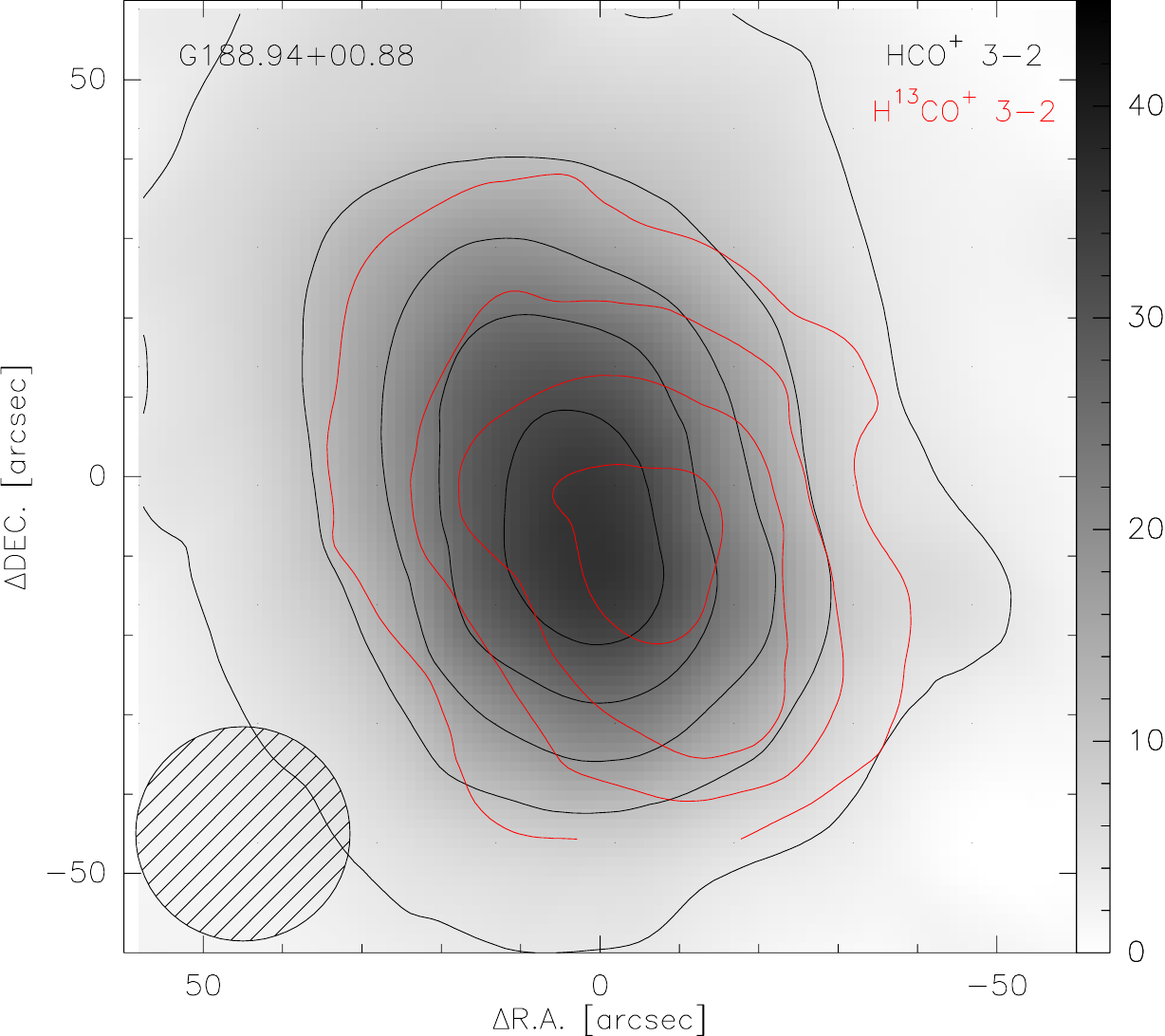}
       \includegraphics[width=3.08in]{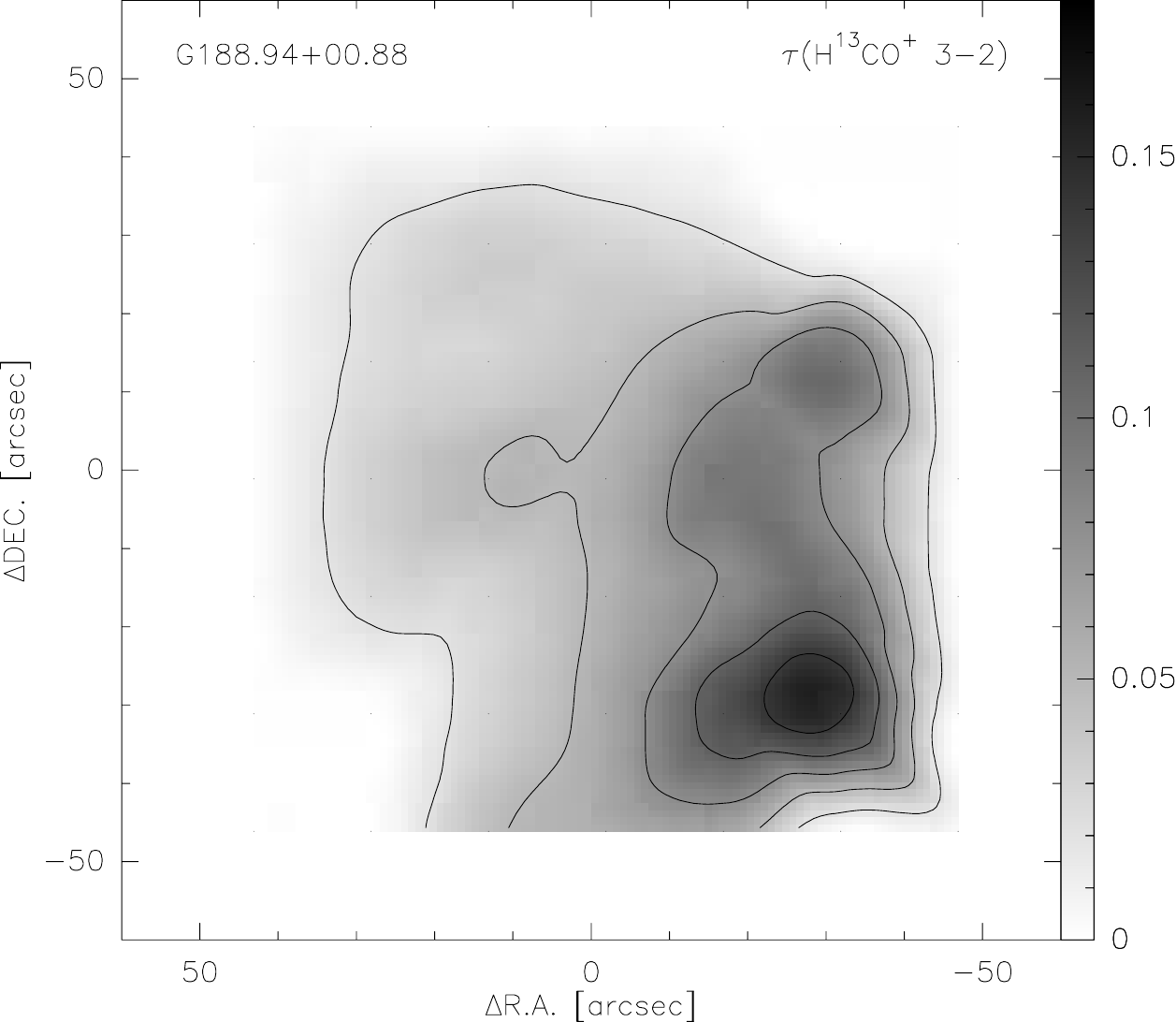}
 \caption{The data reduction results of G188.94+00.88. 
               {\it Top left:} The velocity integrated intensity maps of HCN and H$^{13}$CN 3-2. 
               The mapping size of HCN 3-2 is 2$'\times2'$, while it is 1.5$'\times1.5'$ for H$^{13}$CN 3-2, with a beam size of $\sim$ 27.8$''$.
               The grey scale and the black contour with levels starting from 2.5 K km s$^{-1}$ in step of 4 K km s$^{-1}$ show the observed HCN 3-2. 
               The red contour with levels starting from 0.4 K km s$^{-1}$ in step of 0.2 K km s$^{-1}$ represents H$^{13}$CN 3-2.
               {\it Top right:} The spatially resolved $\tau(\textrm{H}^{13}\textrm{CN})$ of G188.94+00.886 is demonstrated by black contour with levels 
               starting from 0.03 in step of 0.02. 
                {\it Bottom left:} The velocity integrated intensity maps of HCO$^+$ and H$^{13}$CO$^+$ 3-2. 
               The mapping size of HCO$^+$ 3-2 is 2$'\times2'$, while it is 1.5$'\times1.5'$ for H$^{13}$CO$^+$ 3-2, with a beam size of $\sim$ 27.8$''$.
               The grey scale and the black contour with levels starting from 4 K km s$^{-1}$ in step of 7 K km s$^{-1}$ show the observed HCO$^+$ 3-2. 
               The red contour with levels starting from 0.5 K km s$^{-1}$ in step of 0.5 K km s$^{-1}$ represents H$^{13}$CO$^+$ 3-2.
                {\it Bottom right:} The spatially resolved $\tau(\textrm{H}^{13}\textrm{CO$^+$})$ of G188.94+00.88 is demonstrated by black contour with levels 
               starting from 0.02 in step of 0.03. 
                }       
 \label{fig:g13394}
\end{figure*}


 \begin{figure*} 
    \centering
  \includegraphics[width=3.05in]{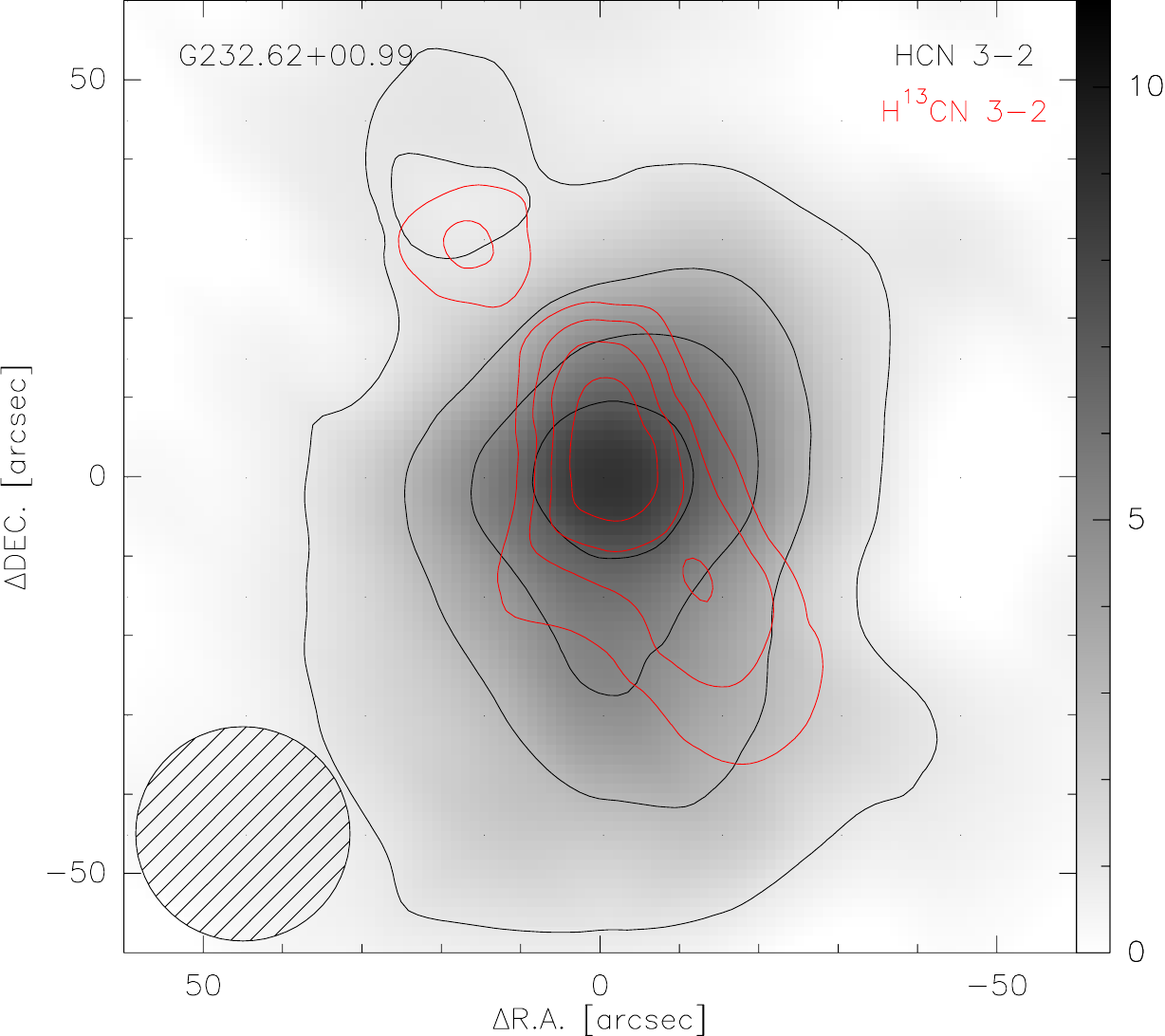}
   \includegraphics[width=3.03in]{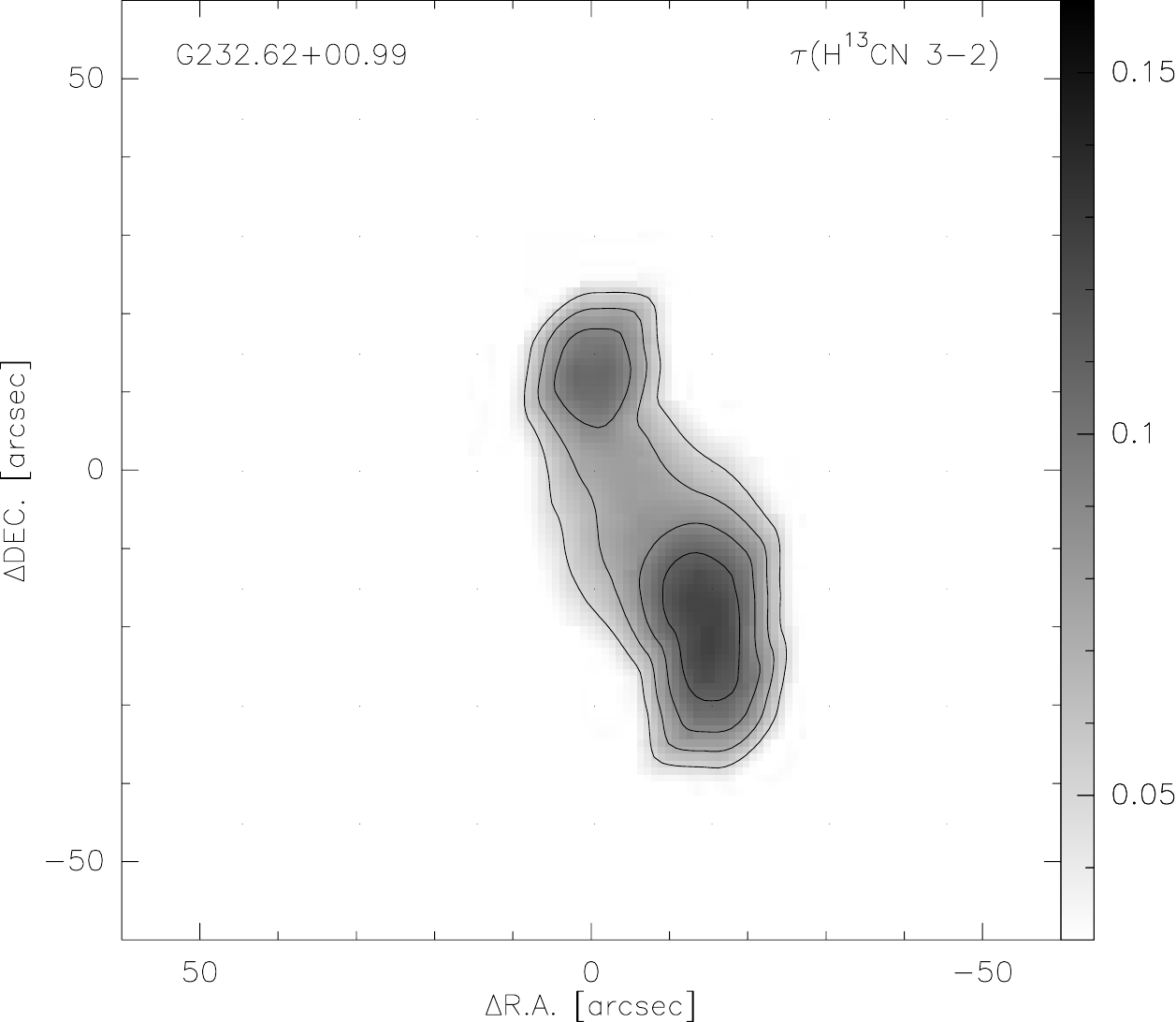}
       \includegraphics[width=3.05in]{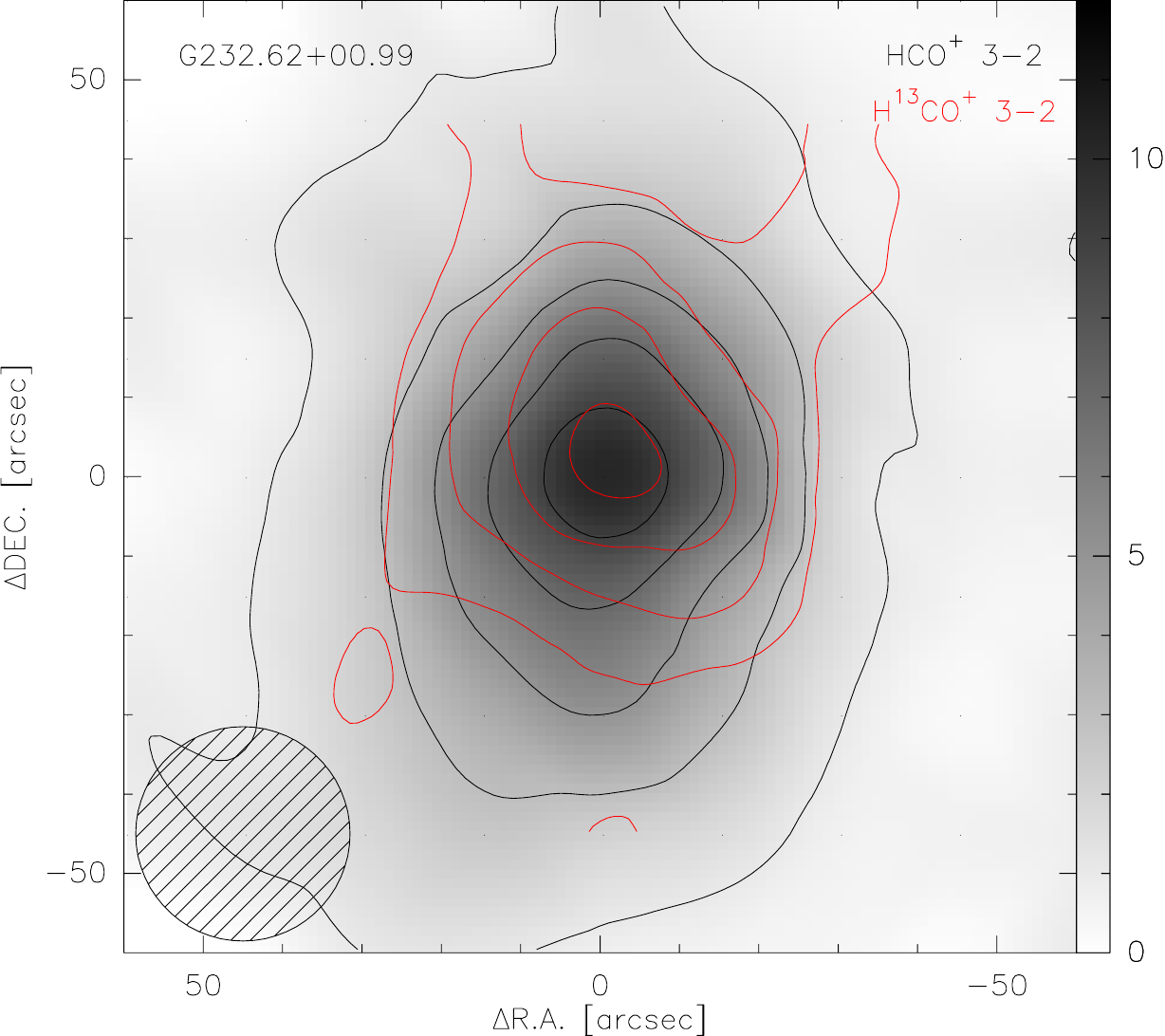}
       \includegraphics[width=3.08in]{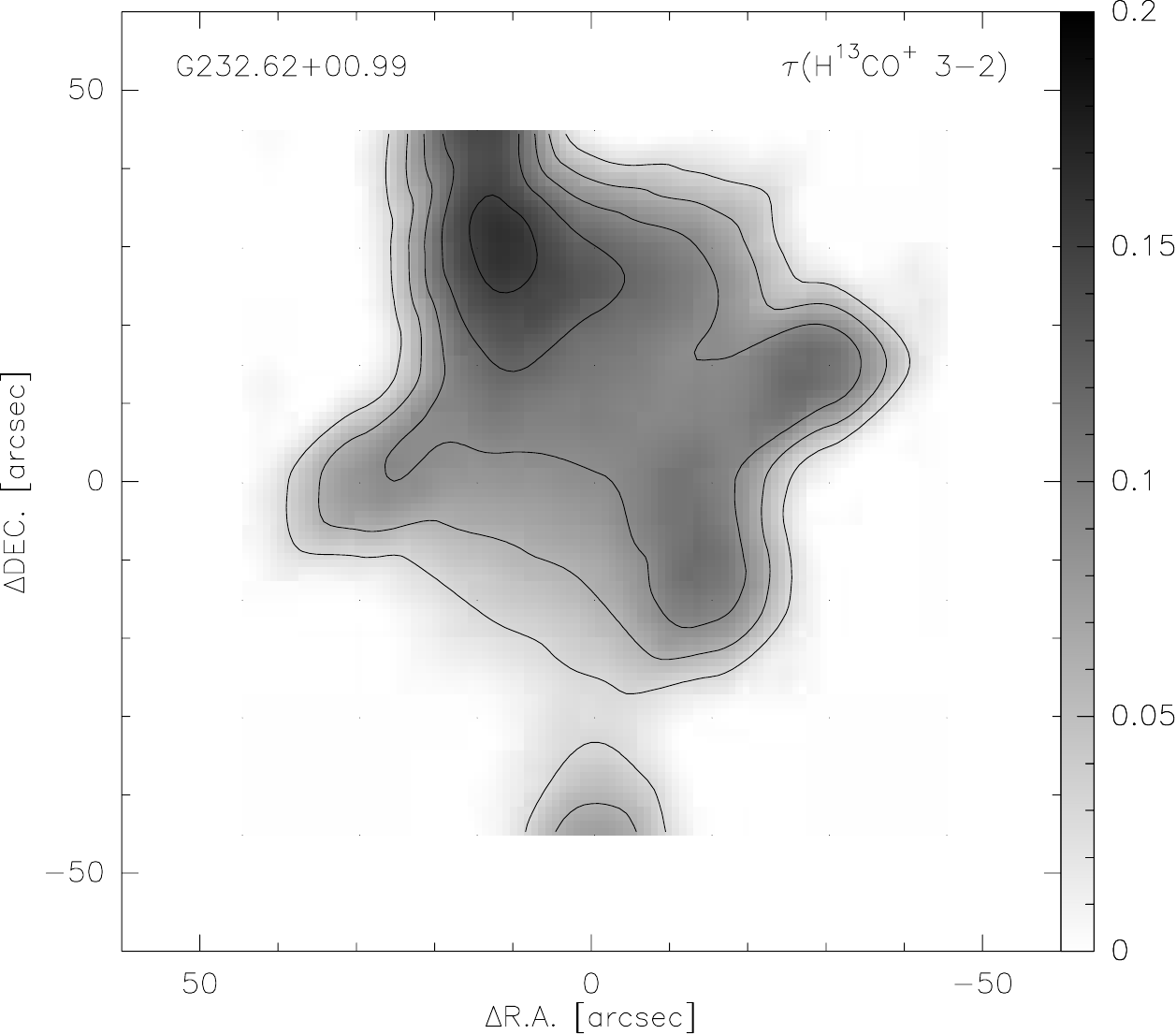}
 \caption{The data reduction results of G232.62+00.99. 
               {\it Top left:} The velocity integrated intensity maps of HCN and H$^{13}$CN 3-2. 
               The mapping size of HCN 3-2 is 2$'\times2'$, while it is 1.5$'\times1.5'$ for H$^{13}$CN 3-2, with a beam size of $\sim$ 27.8$''$.
               The grey scale and the black contour with levels starting from 1 K km s$^{-1}$ in step of 2 K km s$^{-1}$ show the observed HCN 3-2. 
               The red contour with levels starting from 0.3 K km s$^{-1}$ in step of 0.1 K km s$^{-1}$ represents H$^{13}$CN 3-2.
               {\it Top right:} The spatially resolved $\tau(\textrm{H}^{13}\textrm{CN})$ of G232.62+00.99 is demonstrated by black contour with levels 
               starting from 0.05 in step of 0.02. 
                {\it Bottom left:} The velocity integrated intensity maps of HCO$^+$ and H$^{13}$CO$^+$ 3-2. 
               The mapping size of HCO$^+$ 3-2 is 2$'\times2'$, while it is 1.5$'\times1.5'$ for H$^{13}$CO$^+$ 3-2, with a beam size of $\sim$ 27.8$''$.
               The grey scale and the black contour with levels starting from 1 K km s$^{-1}$ in step of 2 K km s$^{-1}$ show the observed HCO$^+$ 3-2. 
               The red contour with levels starting from 0.2 K km s$^{-1}$ in step of 0.2 K km s$^{-1}$ represents H$^{13}$CO$^+$ 3-2.
                {\it Bottom right:} The spatially resolved $\tau(\textrm{H}^{13}\textrm{CO$^+$})$ of G232.62+00.99 is demonstrated by black contour with levels 
               starting from 0.03 in step of 0.03. 
                }       
 \label{fig:g23262}
\end{figure*}


\bsp	
\label{lastpage}
\end{document}